\documentclass[english,12pt]{article}
\pdfoutput=1
\usepackage[a4paper,bottom=4.2cm,top=1cm,head=3cm,width=18cm,dvipdfm]{geometry}
\evensidemargin=\oddsidemargin

\usepackage[T1]{fontenc}
\usepackage[latin9]{inputenc}
\usepackage{amstext}
\usepackage{babel}
\usepackage{verbatim}
\usepackage{amsfonts}
\usepackage{mathrsfs}
\usepackage{amsmath}
\usepackage{amssymb}
\usepackage{bm}
\usepackage{extarrows}
\usepackage{slashed}
\usepackage{isodateo}
\usepackage{xcolor}
\usepackage{graphicx}
\usepackage[low-sup]{subdepth}
\usepackage{float}
\usepackage{isodateo}
\usepackage{tensor}
\usepackage{multirow}
\usepackage{enumitem}
\usepackage{ytableau}
\usepackage[linktocpage=true]{hyperref}
\makeatletter
\newsavebox\myboxA
\newsavebox\myboxB
\newlength\mylenA
\newcommand*\xoverline[2][0.7]{%
	\sbox{\myboxA}{$\m@th#2$}%
	\setbox\myboxB\null
	\ht\myboxB=\ht\myboxA%
	\dp\myboxB=\dp\myboxA%
	\wd\myboxB=#1\wd\myboxA
	\sbox\myboxB{$\m@th\overline{\copy\myboxB}$}
	\setlength\mylenA{\the\wd\myboxA}
	\addtolength\mylenA{-\the\wd\myboxB}%
	\ifdim\wd\myboxB<\wd\myboxA%
	\rlap{\hskip 0.8\mylenA\usebox\myboxB}{\usebox\myboxA}%
	\else
	\hskip -0.5\mylenA\rlap{\usebox\myboxA}{\hskip 0.5\mylenA\usebox\myboxB}%
	\fi}
\hypersetup{
colorlinks=true,
citecolor=blue,
linkcolor=blue,
urlcolor=blue,
pdfauthor={},
pdftitle={},
pdfsubject={}
}

\numberwithin{equation}{section}
\makeatletter

\newcommand{\red}{\color{red}}

\newcommand{\eqrefe}{Eq.\eqref}
\newcommand{\balign}{\begin{align}}
\newcommand{\ealign}{\end{align}}	
\newcommand{\beq}{\begin{equation}}
\newcommand{\eeq}{\end{equation}}
\newcommand{\baa}{\begin{array}}
\newcommand{\eaa}{\end{array}}
\newcommand{\beqa}{\begin{eqnarray}}
\newcommand{\eeqa}{\end{eqnarray}}
\newcommand{\beqs}{\begin{subequations}}
\newcommand{\eeqs}{\end{subequations}}
\def\dis{\displaystyle}
\newcommand{\la}{\langle}
\newcommand{\ra}{\rangle}
\newcommand{\fr}[2]{\mbox{$\frac{\,{#1}\,}{#2}$}}
\renewcommand{\rm}{\mathrm}

\def\LB{\left[}
\def\RB{\right]}

\def\leqq{\leqslant}
\def\geqq{\geqslant}

\def\({\left(}
\def\){\right)}
\def\[{\left[\,}
\def\]{\,\right]}
\def\LB{\left\{}
\def\RB{\right\}}
\def\nn{\nonumber}
\def\pd{\partial}
\def\d{\text{d}}
\def\th{\tensor{h}}
\def\thh{\tensor{\hat{h}}}
\def\thG{\tensor{\hat{\Gamma}}}
\def\pp{\prime}
\def\to{\rightarrow}
\def\ito{\!\rightarrow\!}
\def\To{\Rightarrow}

\def\over{\overline}
\def\EE{\mathcal{E}}
\def\VV{\mathcal{V}}
\def\Sb{\mathbb{S}}
\def\MD{\mathcal{M}_{\Delta}^{}}
\def\VT{\widetilde{V}}
\def\TT{\mathcal{T}}
\def\XX{\mathbb{X}}
\def\dT{\delta\mathcal{T}}
\def\M{\mathcal{M}}
\def\MT{\widetilde{\mathcal{M}}}
\def\dM{\delta\mathcal{M}}
\def\dMT{\delta\widetilde{\mathcal{M}}}

\def\zT{\tilde{z}}
\def\KK{\mathcal{K}}
\def\KKt{\widetilde{\mathcal{K}}}
\def\dKK{\delta\mathcal{K}}
\def\dKKt{\delta\widetilde{\mathcal{K}}}
\def\NN{\mathcal{N}}
\def\NNt{\widetilde{\mathcal{N}}}
\def\dNN{\delta\mathcal{N}}
\def\dNNt{\delta\widetilde{\mathcal{N}}}
\def\chit{\widetilde{\chi}}
\def\DEn{\widetilde{\Delta}_n^{}}
\def\DEt{\widetilde{\Delta}}
\def\hSS{\hat{\mathcal{S}}}
\def\mS{\mathcal{S}}

\def\th{\tilde{h}}
\def\SS{\mathcal{S}}
\def\MP{M_{\text{Pl}}^{}}

\def\An{\mathcal{A}_n^\mu}
\def\AA{\mathcal{A}}

\def\phin{\phi_n^{}}

\def\A{\mathcal{A}}

\def\CC{\mathcal{C}}
\def\D{\mathcal{D}}
\def\F{\mathcal{F}}
\def\La{\mathcal{L}}
\def\mO{\mathcal{O}}

\def\T{\mathcal{T}}
\def\tT{\widetilde{\mathcal{T}}}

\def\hh{\hat{h}}
\def\hg{\hat{g}}
\def\phih{\hat{\phi}}
\def\ZZ{\mathbb{Z}_2^{}}
\def\ii{\text{i}}

\def\OmBn{\overline{\Omega}_n^{}}
\def\bOme{\overline{\Omega}}
\def\FFn{\mathcal{F}_n^{}}
\def\FF{\mathcal{F}}
\def\cv{c_0^{}}
\def\vn{v_n^{}}
\def\xin{\xi_n^{}}
\def\al{\alpha}
\def\alp{\alpha'}
\def\be{\beta}
\def\ga{\gamma}
\def\Ga{\Gamma}
\def\ka{\kappa}
\def\ab{\alpha\beta}
\def\hab{\hat\alpha\hat\beta}

\def\mn{\mu\nu}
\def\hmn{\hat\mu\hat\nu}

\def\dnm{\delta_{nm}^{}}
\def\da{\delta}
\def\td{\text{d}}
\def\heta{\hat{\eta}}
\def\ep{\epsilon}
\def\vep{\varepsilon}
\def\lam{\lambda}

\def\hka{\hat{\kappa}}
\def\si{\sigma}
\def\hhs{\hat{s}}
\def\hht{\hat{t}}
\def\hhu{\hat{u}}

\def\hA{\hat{A}}
\def\hF{\hat{F}}
\def\hT{\hat{T}}

\def\bs{\bar{s}}
\def\bsz{\bar{s}_0^{}}

\def\Qt{\widetilde{Q}}

\def\vnt{\tilde{v}_n^{}}
\def\vt{\tilde{v}}
\def\tX{\widetilde{X}}
\def\fP{\mathsf{P}}
\def\ct{c_\theta}
\def\st{s_\theta}
\def\cct{c_\theta^2}

\def\ctt{c_{2\theta}}
\def\cttt{c_{3\theta}}
\def\ctf{c_{4\theta}}
\def\ctfif{c_{5\theta}}
\def\cts{c_{6\theta}}

\def\Tr{\text{Tr}}

\def\Rxi{R_{\xi}^{}}
\def\sz{s_0^{}}
\def\tz{t_0^{}}
\def\uz{u_0^{}}
\def\hLn{h_L^n}
\def\ALn{A_L^n}
\def\Afn{A_5^n}
\def\Mn{M_n^{}}
\def\Mnn{M_n^2}
\def\ALnnnn{A^{an}_L A^{bn}_L \ito A^{cn}_L A^{dn}_L}
\def\Afnnnn{A^{an}_{5}A^{bn}_{5} \ito A^{cn}_{5} A^{dn}_{5}}

\def\ALnkml{A^{an}_L A^{bk}_L \ito A^{cm}_L A^{d\ell}_L}
\def\Afnkml{A^{an}_{5}A^{bk}_{5} \ito A^{cm}_{5} A^{d\ell}_{5}}
\def\hLnnnn{h^{n}_L h^{n}_L \ito h^{n}_L h^{n}_L}
\def\pnnnn{\phi_n^{}\phi_n^{} \!\ito \phi_n^{}\phi_n^{} }

\def\pnnmm{\phi_n^{}\phi_n^{}  \ito \phi_m^{}\phi_m^{} }
\def\hLnkml{h^{n}_L h^{k}_L \ito h^{m}_L h^{\ell}_L}
\def\pnkml{\phi_n^{}\phi_k^{} \ito \phi_m^{}\phi_{\ell}^{}}
\def\hs{\hspace*{0.3mm}}

\def\hsm{\hspace*{-0.3mm}}
\def\hsmx{\hspace*{-0.5mm}}

\allowdisplaybreaks
\newlength{\halfpagewidth}
\divide\halfpagewidth by 2

\def\End{\end{document}}
\renewcommand{\thefootnote}{\fnsymbol{footnote}}
\makeatother

\usepackage{titlesec}
\titleformat*{\section}{\large\bfseries}
\titleformat*{\subsection}{\normalsize\bfseries}
\titleformat*{\subsubsection}{\small\bfseries}

\usepackage{xcolor}
\definecolor{Green}{RGB}{199,238,206}
\definecolor{magenta}{RGB}{255,0,10}

\begin{document}
\thispagestyle{empty}

\vspace*{-15mm}

\begin{center}
{\large\bf 
Structure of Kaluza-Klein Graviton Scattering Amplitudes
\\[1.5mm]
from Gravitational Equivalence Theorem and Double-Copy}


\vspace*{10mm}

{\sc Yan-Feng Hang}\,$^{a}$\footnote{\href{yfhang@sjtu.edu.cn}{yfhang@sjtu.edu.cn}}
~and~
{\sc Hong-Jian He}\,$^{a,b}$\footnote{\href{hjhe@sjtu.edu.cn}{hjhe@sjtu.edu.cn}}

\vspace*{3mm}
$^a$\,Tsung-Dao~Lee Institute $\&$ School of Physics and Astronomy, \\
Key Laboratory for Particle Astrophysics and Cosmology (MOE),\\
Shanghai Key Laboratory for Particle Physics and Cosmology,\\
Shanghai Jiao Tong University, Shanghai, China
\\[1mm]
$^b$\,Institute of Modern Physics $\&$ Physics Department,
Tsinghua University, Beijing, China;
\\
Center for High Energy Physics, Peking University, Beijing, China

\vspace*{8mm}

\end{center}

\begin{abstract}
\baselineskip 15pt
\noindent
We study the structure of scattering amplitudes of the Kaluza-Klein
(KK) gravitons and of the gravitational KK Goldstone bosons in the
compactified 5d General Relativity (GR).
We analyze the geometric ``Higgs'' mechanism
for mass-generation of KK gravitons under compactification with
a general $\Rxi$ gauge-fixing,
which we prove to be free from the vDVZ discontinuity.\
With these, we formulate the Gravitational Equivalence Theorem (GRET)
to connect the longitudinal KK graviton amplitudes to the
corresponding KK Goldstone amplitudes,
which is a manifestation of the geometric ``Higgs'' mechanism
at $S$-matrix level. We directly compute the gravitational KK
Goldstone amplitudes at tree level and show that they
equal the corresponding longitudinal KK graviton
amplitudes in the high energy limit.\ 
We further extend the double-copy
method with color-kinematics duality to reconstruct
the massive KK longitudinal graviton (Goldstone) amplitudes from
the KK longitudinal gauge boson (Goldstone) amplitudes
in the compactified 5d Yang-Mills (YM) gauge theory 
under high energy expansion.\ 
From these, we reconstruct the GRET of the KK longitudinal graviton
(Goldstone) amplitudes in the 5d GR theory
from the KK longitudinal gauge boson (Goldstone) amplitudes
in the 5d YM theory.\ Using either the GRET or the double-copy
reconstruction, we provide a theoretical mechanism showing that the
sum of all the energy-power terms up to $\mO(E^{10})$
in the high-energy scattering amplitudes of
four longitudinal KK gravitons
must cancel down to $\mO(E^{2})$ as enforced by matching the
energy dependence of the corresponding KK Goldstone amplitudes
or by matching that of the double-copy amplitudes from the
KK YM theory. With the double-copy approach,
we establish {\it a new correspondence between the two
energy-cancellations in the four-point
longitudinal KK scattering amplitudes:
$\,E^4\!\ito E^0\,$ in the 5d KK YM theory
and $\,E^{10}\!\!\to\! E^2\,$ in the 5d KK GR theory.}
We further analyze the structure of the residual term
in the GRET and uncover a new energy-cancellation
mechanism therein.
\\[3mm]
{Phys.\ Rev.\ D 105 (2022) 084005, 
no.8
$[$arXiv:2106.04568\,[hep-th]].}

\end{abstract}

\newpage
\baselineskip 18pt
\linespread{.8}

\tableofcontents


\setcounter{footnote}{0}
\renewcommand{\thefootnote}{\arabic{footnote}}

\vspace{5mm}
\section{\hspace{-3mm}Introduction}
\label{sec:1}

The world is apparently four-dimensional, but it could be only part of a higher dimensional space-time structure, with all the extra spatial dimensions compactified at the boundaries and with their sizes much smaller than the present observational limits. The first of such theories was proposed a century ago by Kaluza and Klein in an attempt to unify the gravitational and electromagnetic forces with a compactified fifth dimension (5d)\,\cite{KK}.\ 
This intriguing avenue was subsequently extended and explored 
in various contexts, including the
(super)\,string/M theories\,\cite{string} and extra dimensional field theories with large or small extra dimensions\,\cite{Exd0}\cite{Exd}\cite{ExdRS}.

\vspace*{1mm}

The Kaluza-Klein (KK) compactification of an extra dimension leads to an infinite tower of massive KK states in the low energy 4d effective field theory for each type of particles that propagate into the extra dimension.\ On one hand, the low-lying KK states in such extra dimensional KK theories have intrigued much phenomenological and experimental efforts over the past two decades\,\cite{exdpheno}\cite{exdtest}, as they may provide the first signatures for the new physics beyond the standard model (SM), ranging from the KK states of the SM particles to the spin-2 KK gravitons and possible dark matter candidate.
On the other hand, the mass generation of these KK states has important implications for the theory side because it is realized by a geometric ``Higgs'' mechanism through compactification itself and
{\it without} invoking any additional Higgs boson of the
conventional Higgs mechanism\,\cite{higgsM}.

\vspace*{1mm}

For the compactified 5d KK Yang-Mills (YM) gauge theories, it was realized\,\cite{5DYM2002} that each massive KK gauge boson $A_n^{a\mu}$ of KK level-$n$ acquires its mass by absorbing the fifth component $A_n^{a5}$ (Goldstone boson) of the 5d gauge field. This geometric KK ``Higgs'' mechanism is quantitatively described 
by the KK equivalence theorem for compactified gauge theories 
(KK GAET) at the $S$-matrix level\,\cite{5DYM2002}, stating 
that the scattering amplitude of the longitudinally-polarized KK gauge bosons ($A^{an}_{L}$) equals that of the corresponding KK Goldstone bosons in the high energy limit. This is a direct consequence of the spontaneous geometric breaking  of the 5d gauge symmetry down to the 4d gauge symmetry via KK compactification\,\cite{5DYM2002}\cite{KK-ET-He2004}. It was proven that the nontrivial cancellation of energy-power terms of $\mO(E^4)\ito\mO(E^0)$ in the four longitudinal KK gauge boson scattering amplitude in the high energy limit
is generally guaranteed by the KK GAET under which the corresponding KK Goldstone boson amplitude is manifestly of $\mO(E^0)$ \cite{5DYM2002}. The extension of KK GAET to quantum loop level via BRST quantization was given in Ref.\,\cite{KK-ET-He2004}.
It was realized that the KK GAET
(which ensures the energy-cancellation
of $\,E^4\!\ito E^0\hs$)\,\cite{5DYM2002}\cite{KK-ET-He2004}
originates from the 5d gauge
symmetry under compactification and the resulting BRST identity.
The 5d KK gauge boson scattering amplitudes were further studied in the context of the deconstructed 5d YM 
theories\,\cite{5DYM2002-2}\cite{KK-ET-He2004} and
the compactified 5d SM\,\cite{5dSM}.

\vspace*{1mm}

It was realized even earlier that the compactified 5d General Relativity (GR) also exhibits a geometric mechanism for the mass generation of KK gravitons.\ Refs.\,\cite{GHiggs}\cite{GHiggs2} gave formal discussions of such geometric breaking by formulating an infinite-parameter Virasoro-Kac-Moody group for the 4d effective KK theory which is spontaneously broken down to the four-dimensional translations and the U(1) gauge group by the 5d periodic boundary conditions. It is expected that the 5d gravitational diffeomorphism invariance of the Einstein-Hilbert (EH) action
is spontaneously broken by the boundary conditions to that of the 4d KK theory via a geometric breaking mechanism, where at each KK level-$n$ the helicity $\pm 1$ components (${h}^{\mu 5}_n$) and the helicity-0 component (${h}^{55}_n$) of the 5d spin-2 graviton ($\hat{h}^{AB}$) are supposed to be absorbed by the KK graviton (${h}^{\mu\nu}_n$) via geometric ``Higgs'' mechanism under the 5d compactification. However, there is no quantitative formulation of this gravitational KK
``Higgs'' mechanism at the $S$-matrix level so far.
There are recent works\,\cite{Chivukula:2020S}\cite{Chivukula:2020L} which gave direct calculations of the four-point
scattering amplitudes of  (helicity-zero) longitudinal
5d KK gravitons at tree level, and explicitly showed large energy cancellations among the individual contributions of
$\mO(E^{10})\!\to\mO(E^{2})$ for flat or warped 5d model.\ 
Following Ref.\,\cite{Chivukula:2020S},
the authors of Ref.\,\cite{Kurt-2019}
used Hodge and eigenfunction decompositions\,\cite{Kurt-2013}
to show that at tree level
such energy cancellations of four-point KK graviton
amplitudes occur for compactification
on general closed Ricci-flat manifolds.\
While showing such intricate large energy cancellations
in the tree-level amplitudes of four KK gravitons
are interesting and valuable,
it remains to be understood quantitatively
why such nontrivial cancellations must occur
at the tree level and even loop levels
for the $N$-particle KK amplitudes ($N\!\geqq\! 4$)
{\it in connection to the compactified diffeomorphism
(gauge) symmetry with geometric breaking}
in the 5d KK GR or in the 5d KK YM gauge theory.

\vspace*{1mm}

In this work, we present a general formulation of the
geometric ``Higgs'' mechanism for the compactified 5d GR in the $R_\xi^{}$ gauge, at both the Lagrangian level and scattering $S$-matrix level. For this geometric ``Higgs'' mechanism, we will formulate a KK Gravitational Equivalence Theorem (GRET) which quantitatively connects each scattering amplitude of longitudinally-polarized KK gravitons to that of the corresponding gravitational KK Goldstone bosons.\
The formulation of the KK GRET is highly nontrivial and differs from the KK GAET of the 5d KK gauge theories\,\cite{5DYM2002}, because the gravitational Goldstone bosons contain both spin-0 and spin-1 components.\
By inspecting the spin-0 gravitational KK Goldstone scattering amplitudes and the residual term of the GRET,
we show that they are manifestly of $\mO(E^{2})$ in the high energy regime without invoking any extra energy-power cancellation.\
Using the GRET (based on BRST quantization),
we provide a theoretical mechanism showing that the sum of all the energy-power terms [up to $\mO(E^{10})$]
in the four longitudinal KK graviton scattering amplitude
must cancel down to $\mO(E^{2})$ at tree level as enforced
by matching the energy-power dependence in the corresponding
KK Goldstone amplitude (and residual term).
We will also extend this conclusion to the case of
$N$-point longitudinal KK graviton scattering amplitudes and
up to any loop level, where we prove that each $N$-point longitudinal KK graviton amplitude has large energy cancellations
by a power of $\hs E^{2N}$ ($N\!\!\geqq\! 4$).\ 
This is in contrast to the case of the Fierz-Pauli (FP) gravity
and alike\,\cite{PF}\cite{Hinterbichler:2012} where the
four-point massive longitudinal graviton
scattering amplitudes generally scale as
$\,E^{10}$ \cite{PF-E10}. By including additional non-linear
polynomial interaction terms in the literature,
the high energy behavior of the massive graviton amplitudes
could be improved to no better than
$\,E^{6}$ \cite{Cheung2016}\cite{Kurt-E6},
which is still much worse than the final energy-dependence of
$\mO(E^{2})$ in the massive KK graviton scattering amplitudes
as mentioned above.

\vspace*{1mm}

In addition, using our general $R_\xi^{}$ gauge formulation of the massive KK graviton propagator, we further demonstrate that the spontaneous breaking of the 5d gravitational diffeomorphism invariance of the EH action under geometric ``Higgs'' mechanism will ensure
{\it the absence of the vDVZ (van\,Dam-Veltman and Zakharov) discontinuity}\,\cite{vDVZ} under the massless limit,
in contrast to the case of the Fierz-Pauli (FP) gravity
and alike\,\cite{PF}\cite{Hinterbichler:2012}
which are plagued by the longstanding problem of 
the vDVZ discontinuity.

\vspace*{1mm}

Furthermore, we attempt to reconstruct the 5d KK graviton scattering amplitudes from the corresponding 5d KK gauge boson scattering amplitudes\,\cite{5DYM2002} {under high energy expansion} to the leading order (LO) and the next-to-leading order (NLO) contributions,
by extending the conventional double-copy method of the color-kinematics (CK) duality of
Bern-Carrasco-Johansson (BCJ)\,\cite{BCJ:2008}\cite{BCJ:2019}
which was proposed for connecting the massless gauge theories to the  massless gravity.\ The BCJ method was inspired by
the Kawai-Lewellen-Tye (KLT)\,\cite{KLT} relation which connects
the product of the scattering amplitudes of two open strings to that of the closed string at tree level.\ Analyzing the properties of the
heterotic string and open string amplitudes can prove and refine
parts of the BCJ conjecture\,\cite{Tye-2010}.\ 
The conventional double-copy formulation reveals a deep connection between the GR theory with massless spin-2 gravitons
and the YM theory with massless spin-1 gauge bosons.
This may be schematically presented as follows\,\cite{Elvang:2013}:
\begin{equation}
\rm{GR} = (\rm{Gauge\,Theory})^2 \,.
\end{equation}
We extend the double-copy method to the massive 5d KK gravity and KK gauge theories, and compute the LO and NLO
four-particle scattering amplitudes
under the high energy expansion.\ 
This provides an extremely simple and efficient way
to construct the complicated KK graviton amplitudes from
the 5d KK gauge boson amplitudes.
Indeed, we find that our LO longitudinal KK graviton amplitudes
as reconstructed from the LO amplitudes of
5d KK gauge bosons\,\cite{5DYM2002}
are equal to the KK graviton amplitudes as obtained
by the lengthy direct calculations of \cite{Chivukula:2020S}\cite{Chivukula:2020L}.\ 
Because the 5d KK gauge boson amplitudes\,\cite{5DYM2002} are of $\mO(E^0M_n^0)$, our double-copy approach shows that the reconstructed KK graviton amplitudes must be of $\mO(E^2M_n^0)$,
where $\Mn$ denotes the relevant KK mass.\ 
Moreover, we use the KK Goldstone amplitudes of the 5d YM theory
[which are manifestly of $\mO(E^0M_n^0)$] to reconstruct the
corresponding gravitational KK Goldstone amplitudes by the
double-copy method, and find that these gravitational
KK Goldstone amplitudes must be of $\mO(E^2M_n^0)$.
We further compare the reconstructed gravitational
KK Goldstone amplitudes with
the reconstructed longitudinal KK graviton amplitudes
under the high energy expansion, and find that they are equal to
each other at the leading order of $\,\mO(E^2M_n^0)$\,
and their difference is only
$\,\mO(E^0\Mnn )$\,.
Hence, for the four-particle scattering processes,
we establish the GRET in the 5d KK GR theory
from the KK GAET in the 5d YM theory\,\cite{5DYM2002}
by using the extended double-copy reconstruction method.\ 
By doing so, we will demonstrate
{\it a nontrivial new correspondence from the
energy-cancellation of $\,E^4\!\ito E^0$
in the four-particle amplitudes for longitudinal KK
gauge bosons of the 5d KK YM theory (YM5) to the
energy-cancellation of $\,E^{10}\!\ito E^2\,$
in the four-particle amplitudes for longitudinal
KK gravitons of the 5d KK GR theory (GR5).}
Schematically, we illustrate this correspondence
between the two energy-cancellations as follows:
\begin{equation}
\label{eq:1E4_0-E10_2}
E^4\hsm\ito E^0\,\rm{(YM5)} ~~\Longrightarrow~~
E^{10}\!\hsm\to\! E^2\,\rm{(GR5)}\,,
\end{equation}
which will be established later in Eq.\eqref{eq:E4_0-E10_2} of
section\,\ref{sec:5.2}.
In addition, using the double-copy approach, we
analyze the structure of the residual terms
in the GRET and further uncover a new
energy-cancellation mechanism of $\,E^2\!\ito E^0\,$
therein.
It is clear that
the GRET and its reconstruction from the 5d KK YM gauge theory
via double-copy can provide a deep quantitative understanding on the structure of the KK graviton (Goldstone) scattering amplitudes
and thus the realization of the geometric ``Higgs'' mechanism of
KK compactification.

\vspace*{1mm}

This paper is organized as follows.\ 
In section\,\ref{sec:2}, we present the general $R_\xi^{}$ gauge
quantization for the 5d KK GR. We derive the propagators
for the KK graviton and KK Goldstone bosons.\ 
We will show that the KK graviton propagator in the $R_\xi^{}$ gauge
is free from the vDVZ discontinuity, in contrast to that of the
Fierz-Pauli gravity.
In section\,\ref{sec:3}, we present the formulation of the GRET and
use it to establish a theoretical mechanism which ensures the
nontrivial energy cancellations 
in the $N$-point longitudinal KK graviton scattering amplitudes.\  
This cancellation mechanism holds not only
for the four-particle amplitudes at tree level,
but also can be applied to the general $N$-particle
amplitudes ($N\!\!\geqq\! 4$) and up to loop levels in principle.\
In section\,\ref{sec:3.1},
we first derive the formulation of the GRET,
which nontrivially differs from the KK GAET
of the 5d KK gauge theories\,\cite{5DYM2002}.\
Then, in section\,\ref{sec:3.2}
we present a general method of energy power counting
to determine the leading energy dependence of
the high energy scattering amplitudes in the KK GR theory and
in the KK YM theory.
In section\,\ref{sec:4}, we present the explicit analyses of the scattering amplitudes of longitudinal KK gravitons and of the corresponding gravitational KK Goldstone bosons
to demonstrate how the GRET works.\ 
In section\,\ref{sec:5}, we establish the double-copy constructions of the longitudinal KK graviton scattering amplitudes and the corresponding
KK Goldstone scattering amplitudes.\  
We give in section\,\ref{sec:5.1}
the full scattering amplitudes of the longitudinal KK gauge boson
amplitudes and the KK Goldstone amplitudes, and derive their
LO and NLO contributions under high energy expansion.\ 
Then, in section\,\ref{sec:5.2}, we use the double-copy approach
to reconstruct the LO KK graviton amplitudes and KK Goldstone amplitudes.\ 
With these, we establish the GRET in the 5d KK GR theory
from the KK GAET in the 5d YM gauge theory at the LO.
In section\,\ref{sec:5.3},
we study the double-copy construction of the 
NLO gravitational KK amplitudes of $\mO(E^0\Mnn)$.\   
Based on the recent first principle approach 
of the KK string theory, 
we further propose an improved double-copy construction
to derive the exact NLO KK graviton amplitudes.\
In section\,\ref{sec:5.4}, we analyze the structure and
size of the residual term in the GRET, and establish the
correspondence from the KK GAET to the KK GRET.
We conclude in section\,\ref{sec:6}. Finally, the Appendices\,\ref{app:A}-\ref{app:G} present
a number of analyses used for the text discussions.

\vspace*{2mm}
\section{\hspace{-3mm}Gauge-Fixing and Propagators without vDVZ Discontinuity}
\label{sec:2}

In this section, we first setup the 5d compactification
under the $S^1/\ZZ$ orbifold, including the notations and
KK expansions.\ Then, we present the quadratic Lagrangian
terms from the 5d EH action, construct a general
$R_\xi^{}$ gauge-fixing, and also derive the relevant KK graviton and
KK Goldstone propagators. Finally, we prove that the
massive KK graviton propagator is naturally free from the
longstanding puzzle of the vDVZ discontinuity which plagues the
Pauli-Fierz gravity and alike\,\cite{PF}\cite{Hinterbichler:2012}.

\vspace*{1mm}
\subsection{\hspace*{-3mm}Setup and Weak Field Expansion in 5d}
\label{sec:2.1}
For the current study, we consider the five-dimensional general relativity on a compactified flat space under orbifold
$S^1\!/\ZZ$\,.\footnote{%
The extension of our present study to the case of non-flat 5d space (such as warped 5d\,\cite{RS}) does not cause any conceptual difference regarding all the major conclusions in this work, which will be addressed elsewhere.}\!\!\!
Thus, the compactified fifth dimension is a line segment with
$\,0 \!\leqslant\! x^5 \!\leqslant\! \pi r_c$\,, where $r_c^{}$ stands for the compactification radius.
Based on this, the 5d Einstein-Hilbert (EH) action is given by
\begin{equation}
\label{eq:SEH}
S_{\rm{EH}}^{} \,=\, \int\! \d^5{x} \,\frac{2}{\,\hka^2\,} \sqrt{\!-\hat{g}\,} \hat{R} \,,
\end{equation}
where $\hat{R}$ is the 5d Ricci scalar curvature, $\,\hka\,$ is the 5d gravitational coupling with mass-dimension $-\fr{3}{2}$\, and it is related to the 5d Newton constant $\,\hat{G}\,$ via
$\,\hka = \!\sqrt{32\pi\hat{G}\,}\,$. The 5d metric tensor is $\hg_{AB}^{}$ ($A,B=0,1,2,3,5$) and its determinant is given by $\,\hg=\det(\hg _{AB}^{})\,$. We also adopt the metric signature $(-,+,+,+,+)$. In addition, we denote the 4d Lorentz indices by the lowercase Greek letters (such as $\mu = 0, 1, 2, 3$), and the 5d Lorentz indices by the uppercase Latin letters (such as $A= \mu , 5$).

\vspace*{1mm}

We make the following weak field expansion of the 5d EH action \eqref{eq:SEH} around the flat Minkowski metric $\heta_{AB}^{}$\,:
\begin{equation}
\label{eq:g}
\hg_{AB}^{} =\heta_{AB}^{} + \hka \hh_{AB}^{}\,,
\end{equation}
where the graviton field $\hh_{AB}$ has the mass-dimension
$\fr{3}{2}$. Then, it is straightforward to derive
\beqs
\begin{align}
\label{eq:gInverse}
\hspace*{-3mm}
\hg^{A B} &\,=\,
\heta^{A B}\! - \hka \hh^{A B}\!
+\hka^{2} \hh^{A C} \thh{_C^B}\!
- \hka^{3} \hh^{A C} \hh_{C D} \hh^{D B}\!
+ \mO (\hh^4)   \,,
\\[1mm]
\label{eq:gDeterm}
\hspace*{-3mm}
\sqrt{-\hg\,} &\,=\,
1\!+\!\frac{\hka}{2} \hh \!+\! \frac{\hat{\ka}^2}{8}
(\hh^2\!-\!2\thh{_{\!A}_B}^{}\thh{^{\!A}^B})
\!+\!\frac{\,\hka^{3}}{\,48\,}(\hh^{3}\!
-6\hat{h}\thh{_{\!A}_B}\thh{^{\!A}^B}
\!+ 8\thh{_{\!A}_B}\thh{^{\!B}^C}
\thh{_{C}^A} )+ \mO (\hh^4),
\end{align}
\eeqs
where we have defined
$\,\hh=\heta^{AB}\hh_{AB}$\,.
Now, the 5d scalar curvature $\hat{R}$
can be decomposed in terms of the metric tensors $\hg_{AB}^{}$ and $\hg^{AB}$
as follows:
\beqs
\label{eq:RS-RT-CS}
\begin{align}
\hat{R} \
&\,=\, \hg^{AB} \hat{R}_{AB} =\hg^{AB}\tensor{\hat{R}}{_{A}_C_B^C},
\\[1mm]
\tensor{\hat{R}}{_A_C _B^C} \!&\,=\, \pd_C^{}\thG{^C_{A}_B}
\!-\pd_A\thG{^C_{C}_B}  \!+\thG{^D_{A}_B}\thG{^C_{D}_C}
\!-\thG{^D_{C}_B}\thG{^C_{D}_A}  \,,
\\[1mm]
\thG{^C_{\!A}_B} &\,=\, \fr{1}{2}\,{\hg^{CD}
(\pd_B^{} \hg_{DA}^{}\!+ \pd_{A}^{}\hg_{BD}^{}
\!- \pd_D \hg_{AB})}
\,.
\end{align}
\eeqs


With the above formulas, we can expand the 5d EH action
$\,S_{\rm{EH}}^{}\!=\! \int\!\td^5 {x}\,\hat\La_{\rm{EH}}^{}\,$ shown
in Eq.\eqref{eq:SEH} as
\begin{equation}
\label{eq:LagEHExpansion}
\hat\La_{\rm{EH}}^{} \,=\,
\hat\La_0^{}+ \hka\,\hat\La_{1}^{} +\hka^2 \hat\La_{2}^{}
+\hka^3\hat\La_{3}^{}+ \cdots   \,,
\end{equation}
where each expanded Lagrangian term
$\,\hat\La_j^{}$ $(j=0,1,\cdots)$\,
contains $\,j+2$\, graviton fields.
The effective 4d Lagrangian is obtained by integrating over the
extra dimension coordinate $x^5$ under proper compactification:
\begin{equation}
\La_{\rm{eff}}^{} \,=~
\sum_{j=0}^{\infty} \int_{0}^{L} \!\! \td x^5  \,
\hka^j_{} \hat\La_j^{}  \,.
\end{equation}
The realization of 5d compactification will be given
in the next subsection.
Finally, the corresponding effective 4d coupling
$\,\ka = \!\sqrt{32\pi G\,}\,$
is connected to the $\hka$ and the reduced Planck mass $M_{\rm{Pl}}$ via
\begin{equation}
\label{eqNewtonConst5D}
\ka = \frac{ \hka}{\sqrt{L\,}\,} = \frac{2}{M_{\text{Pl}}}   \,,
\end{equation}
where we have denoted $\,L = \pi r_c^{}\,$
as the length of the 5th dimension
under the compactification of $S^1\!/\ZZ\,$,
and the reduced Planck mass is represented as
$\MP \!= (8\pi G)^{-1/2}\,$.

\vspace*{1mm}
\subsection{\hspace*{-3mm}Geometric Higgs Mechanism and Gauge Fixing under KK Compactification}
\label{sec:2.2}

In this subsection, we will make KK compactification of the 5d EH action. This can be realized for the 5d obifold compactification
$S^1\!/\ZZ$ with proper boundary conditions, and the resulting 4d effective KK theory contains the KK tower of massive graviton states.
The 5d gravitational diffeomorphism invariance of the EH action is expected to be spontaneously broken by the boundary conditions to that of the 4d KK theory via a geometric breaking mechanism, where at each KK level-$n$ the vector components ($h^{\mu 5}_n$) and the scalar component ($h^{55}_n$) of the 5d spin-2 graviton ($\hh^{AB}$) are supposed to be absorbed by the KK graviton ($h^{\mn}_n$).
There are formal discussions of such geometric breaking in the literature\,\cite{GHiggs}\cite{GHiggs2}, by formulating an infinite-parameter Virasoro-Kac-Moody group for the 4d effective KK theory which is spontaneously broken down to the four-dimensional translations and the U(1) gauge group. These formal discussions\,\cite{GHiggs}\cite{GHiggs2} did not provide a practical formulation as needed
for our current study of perturbative KK theory and
for the scattering amplitudes at the $S$-matrix level.

\vspace*{1mm}

In the following, we present an explicit formulation of this geometric ``Higgs'' mechanism at the Lagrangian level, and then at the $S$-matrix level via the GRET (section\,\ref{sec:3}).  The 5d geometric ``Higgs'' mechanism was previously established for the compactified 5d Yang-Mills theories in Ref.\,\cite{5DYM2002}.\footnote{%
The extension to the deconstructed 5d YM theories was given in Ref.\,\cite{5DYM2002-2} and to the compactified 5d SM was given in Ref.\,\cite{5dSM}.}
In this study, we present an explicit formulation of the
5d geometric ``Higgs'' mechanism for the 5d Einstein gravity,
with which we will identify the gravitational Goldstone bosons
$(h_n^{\mu 5},\,h_n^{55})$
for each massive KK graviton $h_n^{\mn}$.\,
Then, we explicitly construct the $R_\xi^{}$ gauge-fixing term and
derive the propagators for KK gravitons and their corresponding Goldstone bosons.

\vspace*{1mm}

The 5d graviton field $\hh_{AB}$ can be parametrized as
\begin{equation}
\label{eq:hDecom}
\hh_{AB}=
\begin{pmatrix}
\hh_{\mn}^{} \!+ w \eta_{\mn} \phih &~ \hh_{\mu 5}^{}
\\[2mm]
\hh_{5\nu}^{} &~ \phih
\end{pmatrix} \!,
\end{equation}
where the (1,1) block is the 4d component of $\,\hh_{AB}$\,
and the additional term $\,w \eta_{\mn} \phih \,$
corresponds to a {Weyl transformation}\footnote{%
More precisely, under the Weyl transformation the 4d metric is rescaled as
$\,\hg_{\mn}^{} \ito \hg_{\mn}^\pp = e^{w\hka\phih}\hg_{\mn}^{}\,$.}
with a nonzero coefficient $w$\,.\footnote{%
In Ref.\cite{Hinterbichler:2012}, $w$ is expressed as
$\,w=2/(d-2)\,$,
which gives $w=1$ in 4d and $\,w=2/3\,$ in 5d.
We will determine the value of $\,w\,$ from a consistency requirement
in the following analysis.}
The (2,2) block of $\hh_{AB}^{}$ is a scalar field known as the radion field (\,$\phih\equiv \hh_{55}^{}$\,). The blocks (1,2) and (2,1) correspond to the vector component of the 5d graviton field $\hh_{AB}^{}$\,.

\vspace*{1mm}

With the 5d metric tensor \eqref{eq:g} and the 5d graviton field \eqref{eq:hDecom}, we derive the squared 5d interval
\begin{equation}
\td \hat{s}^2 = [ \eta_{\mn}\!+\hka(\hh_{\mn}\!+\!
w \eta_{\mn} \phih)]
\td x^\mu \td x^\nu
\!+ 2 \hat{\ka}\, \hh_{\mu 5} \, \td x^\mu \td x^5
\!+ (1\!+\!\hat{\ka}\phih ) \td x^5 \td x^5 \,.
\end{equation}
We compactify the 5d space under $S^1/\ZZ$
orbifold and require $\,\td \hat{s}^2\,$ to be invariant under a $\mathbb{Z}_2$ orbifold reflection $\,x_5 \ito - x_5$.\,
Hence, this requires that the graviton's tensor component
$\,\hh_{\mn}^{}\,$ and
the scalar component $\,\phih\,$ to be even under $\ZZ$ symmetry,
while the vector component $\hh_{\mu 5}^{}$
should be $\ZZ$ odd:
\\[-10mm]
\beqs
\begin{align}
\hh_{\mn}(x_\rho^{} , x_5^{}) \,&=\,
\hh_{\mn}(x_\rho^{} , - x_5^{})   \,,
\label{eq:OFh}
\\[1mm]
\hh_{\mu 5}(x_\rho^{} , x_5^{}) \,&=\,
- \hh_{\mu 5}^{}(x_\rho , - x_5^{}) \,,
\label{eq:OFA}
\\[1mm]
\phih (x_\rho^{}, x_5^{}) \,&=\, \phih (x_\rho^{}, -x_5^{})  \,.
\label{eq:OFPhi}
\end{align}
\eeqs
This is equivalent to imposing the Neumann boundary conditions on $\,\hh_{\mn}$ and $\,\phih\,$ at the ends of the 5d interval
$[0,\,L]$, and imposing the Dirichlet boundary condition on
$\hh_{\mu 5}^{}$\,,
\begin{equation}
\label{eq:BC}
\left. \pd_5^{}\hh_{\mn}^{} \right|_{x_5 \,=\, 0,L}^{}\! = 0  \,,
\hspace*{10mm}
\left. \pd_5^{}\phih^{}\right|_{x_5\,=\,0,L}^{} \!= 0  \,,
\hspace*{10mm}
\left. \hh_{\mu 5}^{} \right|_{x_5\,=\,0,L}^{} \!= 0  \,.
\end{equation}
With these, we can make the following KK expansions for the 5d
graviton fields via Fourier series in terms of their zero-modes
and KK states,
\beqs
\label{eq:Fourier}
\begin{align}
\hh^{\mn}(x^\rho, x^5) &\,=\,
\frac{1}{\sqrt{L\,}\,} \! \[  h^{\mn}_{0} (x^\rho)+
\sqrt{2} \sum_{n=1}^{\infty} h^{\mn}_{n}(x^\rho)
\cos\!\frac{n\pi x^5}{L} \!\] \!,
\label{eqHExp}
\\[1mm]
\hh^{\mu 5}(x^\rho, x^5) &\,= \,
\sqrt{\frac{2}{L}\,}
\sum_{n=1}^{\infty} h^{\mu 5}_{n}(x^\rho)
\sin\!\frac{n\pi x^5}{L}  \,,
\label{eqAExp}
\\[1mm]
\phih (x^\rho, x^5) &\,=\,
\frac{1}{\sqrt{L\,}\,} \[\phi_0 (x^\rho)+
\sqrt{2}\sum_{n=1}^{\infty} \phi_{n}(x^\rho)
\cos\!\frac{n \pi x^5}{L} \!\] \!.
\label{eqPhiExp}
\end{align}
\eeqs
Then, we examine the quadratic Lagrangian $\hat\La_0^{}$,
which takes the following form:
\begin{equation}
\label{eq:LagLO}
\hat\La_0^{} \,=\,
\fr{1}{2} (\pd_A^{} \hh)^2 - \fr{1}{2} (\pd_C \hh_{AB})^2  - \pd_A \hh^{AB} \pd_B \hh + \pd_A \hh^{AC} \pd^B \hh_{BC}  \,.
\end{equation}
Substituting \eqrefe{eq:hDecom} into the quadratic Lagrangian
\eqref{eq:LagLO}, we thus derive
\begin{align}
\label{eqLagLOExp}
\hat\La_0^{} \, =& \
\fr{1}{2}(\pd_\mu \hh)^2 \!+\!\fr{1}{2}(\pd_5^{}\hh)^2 \!-\! \fr{1}{2}(\pd_\rho \hh_{\mn})^2
\!-\! \fr{1}{2}(\pd_5^{} \hh_{\mn})^2 \!-\! (\pd_\mu \hh_{\nu 5})^2
\!+\! (\pd_\mu \hh^{\mu 5} )^2
\!+\! 3w(w\!+\!1)(\pd_\mu \phih )^2
\nn\\[1mm]
& + 6w^2 (\pd_5^{} \phih )^2\! - \pd_\mu \hh^{\mn} \pd_\nu\hh
+ \pd_\mu \hh^{\mu\rho} \pd^\nu\hh_{\nu\rho}^{}
\!- (2w\!+\!1)(\pd_\mu \hh^{\mn}\pd_\nu\phi
  -\pd_\mu\hh\,\pd^\mu\phih )
\nn\\[1mm]
& - 2 \pd_\mu \hh^{\mu 5}\pd_5^{}\hh +
2\pd^5\hh^{\mn}\pd_\mu \hh_{\nu_5}\!
- 6w\, \pd_\mu^{}\hh^{\mu 5} \pd_5^{}\phih
+ 3w\, \pd_5^{}\hh \pd^5 \phih \,.
\end{align}
In terms of the KK expansions \eqref{eq:Fourier} and integrating over $x^5$, we can further expand the Lagrangian \eqref{eqLagLOExp}
as follows:
\begin{align}
\label{eq:LagLOKK}
\La_0^{} \,=\,& \,
\sum_{n=0}^{\infty}  \left[
\fr{1}{2}(\pd_\mu h_n)^2\! +\! \fr{1}{2} M_n^2(h_n)^2
\!-\!\fr{1}{2}(\pd^\rho h^{\mn}_n)^2
\!-\! \fr{1}{2} M_n^2(h^{\mn}_n)^2\!
-\! (\pd^\mu \A^{\nu}_n)^2\!
+ (\pd_\mu \A_{n}^{\mu})^2  \right.
\nn\\[-1mm]
& \hspace{2.6em}
+3w(w \! +\!1)(\pd_{\mu}\phin)^2\! + 6w^2 M_n^2\phi_n^2\!
- \pd_\mu h^{\mn}_n\pd_\nu h_n + \pd_\mu h^{\mu\rho}_n\pd^\nu h_{\nu\rho,n}^{}
\nn\\[1mm]
& \hspace{2.6em}
-(2w\!+\!1)(\pd_\mu h^{\mn}_n\pd_\nu\phin\! - \pd_\mu h_n \pd^\mu \phin)\!
+ 2 M_n h_n \pd_\mu \A^{\mu}_n\!  - 2 M_n  h^{\mn}_n \pd_\mu \A_{\nu,n}
\nn\\
& \hspace{2.6em}\left.
+ 3w M_n^2 \, h_n^{} \phin + 6w M_n \pd_\mu \A^{\mu}_n\,\phin\,
\right] \!,
\end{align}
where for convenience we have denoted the vector field as $\,\A_n^{\mu}\!\equiv h^{\mu5}_n$, and
$\,M_n \!=\! n\pi/L$\,
stands for the mass of KK states of level-$n$.

\vspace*{1mm}

Inspecting the Lagrangian \eqref{eq:LagLOKK}, we set $\,w\!=\!-\fr{1}{2}$\, to remove
the two undesirable mixing terms in its 3rd line.
We can further eliminate the rest of the mixing terms
in the 3rd and 4th lines of Eq.\eqref{eq:LagLOKK}
by introducing the following $R_{\xi}$-type gauge-fixing terms,
\begin{equation}
\hspace*{-4mm}
\La_{\rm{GF}} = -\!\sum_{n=0}^{\infty}\!
\LB \frac{1}{\xi_n}\!\!\[\! \pd_{\nu} h_n^{\mn} \!-\!
\(\! 1 \!-\! \frac{1}{2 \xi_n}\)\!\pd^{\mu} h_n \!+
\xi_n M_n \A^{\mu}_n \!\]^{\!2}
\!+\frac{\,M_n^2\,}{\,4\xi_n^{}\,}\!\!
\(\! h_n \!\!-\! 3 \xi_n \phi_n \!+\! \frac{2 \pd_{\!\mu} \A_n^{\mu}}{M_n}\)^{\!\!2} \RB \!,
\label{eq:GF}
\end{equation}
where $\,\xi_n^{}\,$ is the gauge-fixing parameter
for the zero-mode gravitons ($n=0$) and KK gravitons
($n\geqq 1$)\,.
By imposing the gauge-fixing term \eqref{eq:GF} to remove the
quadratic mixing terms, we explicitly verify that both the
vector component $\,\A_n^\mu\,$ and scalar component $\,\phin\,$
are absorbed (``eaten'') by the KK graviton $h^{\mn}_n$,\,
and identify them as the gravitational KK Goldstone fields,
{\it which are the direct outcome of realizing the
5d geometric KK ``Higgs'' mechanism.}

\vspace*{1mm}

From the above, we can explicitly integrate over $x^5$ and derive
the effective 4d KK action at the quadratic order:
\begin{equation}
\label{eq:Action}
S_{\rm{eff}} = \int\!\! \td^4 x  \, \sum_{n=0}^{\infty}
\frac{1}{2}\!\(  h^{\mn}_n \D^{-1}_{\mn\ab,nn} h^{\ab}_n + \A^{\mu}_n \D_{\mn,nn}^{-1} \A^{\nu}_n + \phi_n \D^{-1}_{nn}  \phi_n  \)  ,
\end{equation}
where the inverse KK propagators take the following forms:
\beqs
\label{eq:PropInverse}
\begin{align}
\label{eq:GravPropInver}
\D^{-1}_{\mn\ab,nn}=&  -\!\[\! 1\!-\!\frac{2}{\xi_n}\!
\(\! 1-\frac{1}{2\xi_n} \)^{\!\!2} \] \! \eta_{\mn} \eta_{\ab} \pd^{2}
+  \(\! 1 \!-\! \frac{1}{2\xi_n}\)\eta_{\mn} \eta_{\ab} M_n^2
\nn \\[1mm]
& + \frac{1}{2}\(\eta_{\mu\al}\eta_{\nu\be}\!+\eta_{\mu \be}\eta_{\nu \al}\) \!
\(\pd^{2}\!-\!M_n^2 \)
+ \(\frac{1}{\xi_n}\!-\!1\!\)^{\!\!2} \! \(\eta_{\mn} \pd_{\al} \pd_{\be}
\!+\!\eta_{\ab} \pd_{\mu} \pd_{\nu} \)
\nn\\[1mm]
&-\frac{1}{2} \!\(\! 1\!-\!\frac{1}{\xi_n}\)\!
\(\eta_{\mu \al} \pd_{\nu} \pd_{\be}+\eta_{\mu \be} \pd_{\nu} \pd_{\al}
+ \eta_{\nu \al}^{} \pd_{\mu} \pd_{\be}+\eta_{\nu \be} \pd_{\mu} \pd_{\al}\) ,
\\[1mm]
\D^{-1}_{\mn,nn} =&\  \eta_{\mn} (  \pd^2\! - \xi_n M_n^2)
+ \frac{\,1\!-\xi_n^{}\,}{\xi_n} \, \pd_\mu \pd_\nu \,,
\hspace*{10mm}
\\[1mm]
\D^{-1}_{nn} =&\  \pd^2 \!- ( 3 \xi_n^{}\!-2)  M_n^2 \,,
\end{align}
\eeqs
and we have also rescaled the vector and scalar fields by
\begin{align}
\label{eq:rescaling-An-phin}
\A^{\mu}_n \to \frac{1}{\sqrt{2\,}\,} \A^{\mu}_n  \,, \quad~~~
\phi_n  \to \sqrt{\frac{2}{3}\,}\phi_n   \,,
\end{align}
which ensure that their kinematic terms have the correct normalization factor $\fr{1}{2}$\,.
Furthermore, the propagators of the KK graviton and KK Goldstone bosons
are the inverse of \eqrefe{eq:PropInverse} and satisfy the
following conditions:
\beqs
\label{eq:Dx-DInv-ALL}
\begin{align}
\label{eq:DxDinv}
\int\! \td^4 z\, \D_{\mn\ab,nn}^{-1}(x,z) \D^{\ab\rho\si}_{nn}(z,y)
&\,=\,
\frac{\ii}{2} ( {\delta}{_\mu^\rho}{\delta}{_\nu^\si}
+ {\delta}{_\mu^\si} {\delta}{_\nu^\rho} )
\delta^{(4)}(x-y) \,,
\\[1mm]
\int\! \td^4 z\, \D_{\mn,nn}^{-1}(x,z) \D^{\nu\rho}_{nn}(z,y)
&\,=\, \ii\, {\delta}{_\mu^\rho} \delta^{(4)}(x-y) \,,
\\[1mm]
\int\! \td^4 z\, \D_{nn}^{-1}(x,z) \D_{nn}(z,y)
&\,=\, \ii\,  \delta^{(4)}(x-y) \,.
\end{align}
\eeqs
Substituting \eqrefe{eq:PropInverse} into Eq.\eqref{eq:Dx-DInv-ALL},
we finally derive the following compact form of the propagators
for the KK gravitons $(h_n^{\mn})$
and for the KK Goldstone bosons ($\A^{\mu}_n$ and $\phin$)
in momentum space:
\beqs
\label{eq:KKpropagator-Rxi}
\begin{align}
\D_{nm}^{\mn\ab}(p) \,=\,&
-\frac{\,\ii\delta_{nm}^{}\,}{2}\left\{
\frac{\,(\eta^{\mu \al}\eta^{\nu \be}\!+\!\eta^{\mu \be}\eta^{\nu\al}
\!-\!\eta^{\mu\nu}\eta^{\al\be})\,}{\,p^{2}\!+\!M_{n}^{2}\,}
\right.
\nn\\[1mm]
& +\frac{1}{3}
\!\[\!\frac{1}{\,p^2\!+\!M_n^2\,}-
\frac{1}{\,p^{2}\!+\!(3\xi_n\!-\!2)M_n^{2}\,}
\!\] \!\!
\(\! \eta^{\mn}\!-\!\frac{\,2p^{\mu}_{}p^{\nu}_{}}{M_n^{2}} \!\)\!\!
\(\!\eta^{\ab}\!-\!\frac{\,2p^{\al}_{}p^{\be}_{}}{M_n^{2}} \!\)
\nn\\[1mm]
& +\frac{1}{\,M_n^2\,}\!
\[\!\frac{1}{\,p^2\!+\!M_n^2} -
\frac{1}{\,p^{2}\!+\!\xi_n M_n^{2}\,}
\!\]\!
(\eta^{\mu\al}_{}p^{\nu}_{}p^{\be}_{}\!
+\eta^{\mu\be}_{}p^{\nu}_{}p^{\al}_{}\!
+\eta^{\nu\al}_{}p^{\mu}_{}p^{\be}_{}\!
+\eta^{\nu\be}_{}p^{\mu}_{}p^{\al}_{})
\nn\\[1mm]
& \left.
+\frac{\,4p^\mu_{}p^\nu_{}p^\al_{}p^\be_{}\,}{\xi_n^{}M_n^4}
\!\(\!\frac{1}{\,p^2\!+\!\xi_n^2M_n^2\,}
-\frac{1}{\,p^2\!+\!\xi_nM_n^2\,} \!\) \!\right\}\!,
\label{eq:Dnn-Rxi}
\\[2mm]
\D^{\mn}_{nm}(p)  \,=\,&
\frac{-\ii\delta_{nm}^{}}{\,p^{2}\!+\!\xi_nM_{n}^{2}\,}\!
\left[\eta^{\mn} \!\!-\! \frac{\,p^{\mu} p^{\nu} (1\!-\!\xi_n)\,}
{\,p^{2}\!+\xi_n^2 M_{n}^{2}} \right]\!,
\label{eq:Dnn-A}
\\[2mm]
\label{eq:Dnn-phi}
\D_{nm}^{}(p)  \,=\,& \frac{-\ii\delta_{nm}^{}}{\,p^2 \!+\! (3 \xi_n \!\!-\!2)M_n^2\,} \,.
\end{align}
\eeqs
The Faddeev-Popov ghosts can be further included for the loop analysis although this is not needed for our present study of KK scattering amplitudes at tree level. The unphysical states of the massive KK gravitons correspond to the spin-0 and spin-1 Goldstone bosons,
and we see that the above Goldstone propagators \eqref{eq:Dnn-A} and \eqref{eq:Dnn-phi} have the same $\xi_n^{}$-dependent unphysical mass poles as those of the KK graviton propagator \eqref{eq:Dnn-Rxi}.

\vspace*{1mm}

It is instructive to consider the Feynman-'t\,Hooft gauge with $\,\xi_n^{}\!=\!1\,$. In this gauge, the above $R_\xi^{}$-gauge propagators take the following simple forms:
\beqs
\label{eq:KKpropagator-FHooft}
\begin{align}
\D_{nm}^{\mn\ab}(p)
&= -\frac{\,\ii\dnm\,}{2}
\frac{\,\eta^{\mu \al}\eta^{\nu \be}\!+\!\eta^{\mu \be}\eta^{\nu\al}\!-\!\eta^{\mu\nu}\eta^{\al\be}\,}{\,p^{2}\!+\!M_{n}^{2}\,} \,,
\label{eq:Prophh-xi=1}
\\
\label{eq:Prophmu5-xi=1}
\D^{\mn}_{nm}(p) &=
-\frac{\,\ii\eta^{\mn}\dnm\,}{\,p^{2}\!+\!M_{n}^{2}\,} \,,
\\
\label{eq:Prop55-xi=1}
\D_{nm}^{}(p) &=
-\frac{\,\ii \dnm\,}{~p^2 \!+\! M_n^2~} \,.
\end{align}
\eeqs
We can find that all the mass poles are identical to
$p^2\!=\!-M_n^2$\,.
Then, we take the limit $\,\xi_n^{}\ito \infty\,$
and derive the propagator under unitary gauge:
\begin{equation}
\label{eq:UGauge}
\D^{\mn\ab}_{nm,\rm{UG}}(p) \,=\,	
-\frac{\ii\dnm\,}
{\,2\,}\frac{~\bar{\eta}^{\mu\al}\bar{\eta}^{\nu\be} \!+\! \bar{\eta}^{\mu\be}\bar{\eta}^{\nu\al}\!-\! \frac{2}{3} \bar{\eta}^{\mn}\bar{\eta}^{\ab}~}
{\,p^2+ M_n^2\,} \,,
\end{equation}
where $\,\bar{\eta}^{\mn} \!= \eta^{\mn} \!+ p^\mu p^\nu\!/\!M_n^2\,$. As we will discuss in section\,\ref{sec:2.3}, this just coincides with the massive graviton propagator \eqref{eq:DG-PF} of the 4d Fierz-Pauli Lagrangian. Appendix\,\ref{app:B} gives more detailed discussions about the graviton propagator under the unitary gauge.

\vspace*{1mm}
\subsection{\hspace*{-3mm}Massless Limit and Absence of vDVZ Discontinuity in \boldmath{$R_\xi^{}$} Gauge}
\label{sec:2.3}

In this subsection, we examine the massless limit $\,M_n \ito 0\,$ under the $R_\xi^{}$ gauge as constructed in section\,\ref{sec:2.2}. We will demonstrate that our $R_\xi^{}$ propagator \eqref{eq:Dnn-Rxi} of KK gravitons has a smooth massless limit and is free from the conventional vDVZ (van\,Dam-Veltman and Zakharov) discontinuity\,\cite{vDVZ} of the Fierz-Pauli massive gravity\,\cite{PF}\cite{Hinterbichler:2012}.

\vspace*{1mm}

We recall the 4d Fierz-Pauli Lagrangian for massive graviton fields $h^{\mn}$ with mass $M$ \cite{PF}\cite{Hinterbichler:2012}
\begin{equation}
\La_{\rm{FP}}=
\fr{1}{2}(\pd_\mu h)^2 \!-\! \fr{1}{2}(\pd_\al h_{\mn})^2
- \pd_\mu h^{\mn}\pd_\nu h + \pd_\mu h^{\mu\al}\pd^\nu h_{\nu\al}
\!+\! \fr{1}{2}M^2(h^2 - h_{\mn}^2) \,,
\end{equation}
which has the following propagator
\begin{equation}
\label{eq:DG-PF}
\D^{\mn\ab}_{\text{FP}}(p) \,=\,	
-\frac{\ii}{\,2\,}\frac{~\bar{\eta}^{\mu\al}\bar{\eta}^{\nu\be} \!+\! \bar{\eta}^{\mu\be}\bar{\eta}^{\nu\al}
\!-\! \frac{2}{3} \bar{\eta}^{\mu\nu}\bar{\eta}^{\al\be}~}
{\,p^{2} + M^{2}\,}
\,,
\end{equation}
where
$\,\bar{\eta}^{\mn} \!= \eta^{\mn} \!+\! p^\mu p^\nu\!/\!M^2$.
In comparison, for the 4d Einstein gravity
under a harmonic gauge-fixing
\begin{equation}
\label{eq:GF-4d}
\La_{\rm{GF}}=
\frac{1}{\,\xi\,}\!\(\pd_\nu^{}h^{\mu\nu}-\fr{1}{2}\pd^\mu h\)^{\!2},
\end{equation}
the massless graviton propagator is given by
\beqs
\label{eq:DG00-4d}
\begin{align}
\label{eq:DG00-xi}
\D_{00}^{\mn\ab}(p) &=
-\frac{\ii}{\,2\,}\!\[\!
\frac{\,\eta^{\mu \al}\eta^{\nu \be}\!+\!\eta^{\mu \be}\eta^{\nu\al} \!-\!\eta^{\mu\nu}\eta^{\al\be}\,}{p^2}
+(\xi\!-\!1)\frac{\,\eta^{\mu\al}_{}p^{\nu}_{}p^{\be}_{}\!
+\!\eta^{\mu\be}_{}p^{\nu}_{}p^{\al}_{}\!
+\!\eta^{\nu\al}_{}p^{\mu}_{}p^{\be}_{}\!
+\!\eta^{\nu\be}_{}p^{\mu}_{}p^{\al}_{}\,}{p^4}\!\]
\nn\\
&&\\[-1.5mm]
\label{eq:DG00-xi=1}
&=
-\frac{\ii}{\,2\,}
\frac{\,\eta^{\mu \al}\eta^{\nu \be}\!+\!\eta^{\mu \be}\eta^{\nu\al}
\!-\!\eta^{\mu\nu}\eta^{\al\be}\,}{p^2}\,,
\qquad(\text{for}~\xi\!=\!1).
\end{align}
\eeqs
This can also describe the propagator
for the zero-mode gravitons in the KK theory under the harmonic gauge-fixing \eqref{eq:GF-4d}. We inspect the massless limit
$\,M \!\ito 0\,$ of the massive graviton propagator
\eqref{eq:DG-PF} of Fierz-Pauli.
In the massless limit, we note the following features of the
numerator in Eq.\eqref{eq:DG-PF}:
(i).\,the graviton propagator \eqref{eq:DG-PF} has {\it singularities} from all the mass-dependent terms like $\,p^\mu p^\nu\!/M^2\,$
inside those $\bar{\eta}^{\mn}$\,'s;
(ii).\,the coefficient $-\fr{2}{3}$ of the pure metric term $\,{\eta}^{\mn}{\eta}^{\ab}\,$ in the numerator
{\it does not match} the coefficient $-1$ of the corresponding term in  the massless graviton propagator \eqref{eq:DG00-xi=1},
which is the so-called vDVZ discontinuity\,\cite{vDVZ}.
This discontinuity is unique for dealing with the spin-2
massive gravitons \`{a} la Fierz-Pauli.
We note that the origin for such vDVZ discontinuity is due to the
{\it mismatch of physical degrees of freedom} between the massive
gravitons in the Fierz-Pauli gravity and the massless gravitons
in GR: the massive graviton has 5 helicity states
$(\lambda =\pm 2,\,\pm 1,\,0)$, while the massless graviton only has
two ($\lambda =\pm 2$), namely, $5\neq 2\,$.

\vspace*{1mm}

For the singularities mentioned above, we note that similar singularity exists for the spin-1 gauge fields in of the massive Yang-Mills theory (as well as the Maxwell theory with a massive photon) when considering the massless limit.\
To see this, we recall the propagator of the spin-1 massive gauge fields $A^a_{\mu}$\,:
\begin{equation}
\label{eq:D-Mspin1}
\D^{\mn} (p) \,=\, -\ii
\frac{~\eta^{\mu\nu}\!+p^\mu p^\nu\!/M^2~}{\,p^2\!+\!M^2\,}
\,,
\end{equation}
where the term $\,p^\mu p^\nu\!/M^2$ becomes singular in the massless limit.\ The appearance of the singularities in the massive graviton propagator and massive gauge boson propagator is also due to the {\it mismatch of physical degrees of freedom.} In the case of massive spin-1 gauge field $A^{a\mu}$, it has 3 helicity states $\lam \!=\! \pm1,0$, whereas the massless gauge field only has 2 helicity states $\lam \!=\! \pm 1\,$. This mismatch is the cause of the singular term $\,p^\mu p^\nu\!/M^2\,$ in the massless limit.
But in the $R_\xi^{}$ gauge of the spontaneously broken gauge
theories with the conventional 4d Higgs mechanism\,\cite{higgsM}
or with the geometric ``Higgs'' mechanism under
compactification\,\cite{5DYM2002}, the propagator of
a massive gauge boson $A^{a\mu}$ (with mass $M$) can smoothly
reduce to the massless gauge boson propagator
under the limit $\,M\!\ito 0\,$ without causing
any singularity or discontinuity. This is because the massive
gauge field $A^{a\mu}$ (with $M\!\neq\!0$) has 3 physical degrees
of freedom, and in the massless limit $\,M\ito 0\,$
the physical states of $A^{a\mu}$ reduces to two
transverse polarization states and its longitudinal component
disappears while the ``eaten'' would-be Goldstone boson becomes
a physical massless scalar.
Hence, the physical degrees of freedom remain conserved,
$\,3=2+1\,$,\, before and after taking the massless limit.

\vspace*{1mm}

Then, we examine the massless limit for the propagators of massive
KK gravitons. For this, we take the massless limit \,$M_n^{}\ito0$\,
for the $R_\xi^{}$ gauge propagator \eqref{eq:Dnn-Rxi} and expand it
up to the zeroth order of $M_n^{}$. We find that under the limit
$\,M_n^{} \ito 0\,$, the sum of all the negative powers of $M_n^{}$ vanishes, and the remaining nonzero part takes the form:
\beqs
\begin{align}
\hspace*{-3mm}
\D_{nm}^{\mn\ab}(p)
=&\, -\!\frac{\,\ii\dnm\,}{\,2\,}\!\[\!
\frac{\,(\eta^{\mu\al}\eta^{\nu\be}\!+\! \eta^{\mu\be}\eta^{\nu\al}
\!-\!\eta^{\mu\nu}\eta^{\al\be})\,}{p^2}
\!-\!\frac{\,1\!-\!\xi_n\,}{p^4}\!
\(\eta^{\mu\al}p^{\nu}p^{\be}\!+\!\eta^{\mu\be} p^{\nu} p^{\al}
\!+\!\eta^{\nu \al} p^{\mu} p^{\be}
\right.\right.
\nn\\[1mm]
\label{eq:Dnn-M=0}
& \left.\left. \hspace{6mm}
+\,\eta^{\nu\be} p^{\mu} p^{\al}\!-\! 2\eta^{\mn}p^\al p^\be
\!-\! 2\eta^{\ab} p^\mu p^\nu  \) \!-4(1\!-\!\xi_n)^3
\frac{\,p^\mu p^\nu p^\al p^\be\,}{p^6} \!\]\!,
\\[1mm]
=& -\!\frac{\,\ii\dnm\,}{\,2\,}
\frac{\,\eta^{\mu \al}\eta^{\nu \be}\!+\!\eta^{\mu \be}\eta^{\nu\al}
\!-\!\eta^{\mu\nu}\eta^{\al\be}\,}{p^2}\,,
\hspace*{5mm} (\text{for}~\xin\!=\!1).
\label{eq:Dnn-M=0-xi=1}
\end{align}
\eeqs
From the above, we see that under the massless limit there is {\it no singular term,} and the pure metric terms $(\eta^{\mu \al}\eta^{\nu \be}\!+\!\eta^{\mu \be}\eta^{\nu\al}\!-\!\eta^{\mn}\eta^{\ab})$ in the numerator agree with the massless graviton propagator \eqref{eq:DG00-4d} (the $\xin\!\!=\!1$ part) in the conventional 4d Einstein gravity. Hence, it is impressive to see that
{\it in the massless limit the $R_\xi^{}$ gauge propagator \eqref{eq:Dnn-Rxi} of massive KK gravitons is free from singularity and the vDVZ discontinuity.} Our $R_\xi^{}$ gauge formulation of the KK theory has a well-defined massless limit because {\it the physical degrees of freedom are conserved before and after taking the massless limit under the geometric ``Higgs'' mechanism.}
A massive KK graviton $h^{\mn}_n$ (having 5 helicity states $\lam \!=\! \pm2,\pm1,0$) acquires its mass via the geometric ``Higgs'' mechanism (compactification) by absorbing (``eating'') the corresponding
vector-Goldstone component $\A^\mu_n$ (having 2 helicity states
$\lam \!=\!\pm 1$) and scalar-Goldstone component $\phi_n^{}$
(having helicity $\lam \!=\!0$) of the 5d graviton field $\,\hat{h}^{AB}$.
In the massless limit, $h^{\mn}_n$ becomes massless (having only 2 helicities $\lambda \!=\! \pm2$), and the vector and scalar Goldstone bosons $(\A^\mu_n,\,\phi_n)$ become massless physical states (having $2\!+\!1$ helicities $\lam \!=\!\pm 1,\,0$). Namely, each massive KK graviton $h^{\mn}_n$ has its 3 extra helicity states
($\lam \!=\!\pm 1,0$) originate from those of the vector component
$\A^\mu_n$ ($\lam \!=\!\pm 1$) and the scalar component
$\phi_n^{}$ ($\lam =0$).
Hence, we see that the total physical degrees of freedom remain conserved before and after taking the massless limit:
$\,5=2+2+1\,$.
This shows that {\it the compactified KK GR theory provides a
consistent description of the massive spin-2 gravitons
and is free from the vDVZ discontinuity
as well as singularities under the massless limit,}
because the KK gravitons acquire their masses via the geometric
``Higgs'' mechanism without explicitly breaking the
diffeomorphism invariance in the 5d bulk
(except realizing the compactification at the 5d boundaries).

\vspace*{1mm}

Finally, we also note that the
$\,\xi_n^{}\!\neq 1\,$ part of our KK graviton propagator \eqref{eq:Dnn-M=0} differs from the conventional massless graviton propagator \eqref{eq:DG00-xi} under the harmonic gauge-fixing \eqref{eq:GF-4d}. This is because under the massless limit our $R_\xi^{}$ gauge-fixing term \eqref{eq:GF} reduces to
\begin{equation}
\La_{\rm{GF}}^{} ~\longrightarrow~
- \sum_{n=0}^{\infty} \frac{1}{\,\xi_n} \[\!
\pd_{\nu} h_n^{\mn} -
\(\! 1 - \frac{1}{2 \xi_n}\)\! \pd^{\mu} h_n^{}
\!\]^{\!2} ,
\end{equation}
where the coefficient
\,$( 1 \!-\! \fr{1}{2 \xi_n})$\,
differs from that of the conventional harmonic gauge-fixing \eqref{eq:GF-4d} except for the case $\,\xi_n^{}\!=\!1\,$.

\vspace*{2mm}
\section{\hspace*{-3mm}Formulation of
Gravitational Equivalence Theorem  \\
\hspace*{-3mm}and the Energy Cancellation Mechanism}
\label{sec:3}

In the previous section, we have presented the $R_\xi^{}$ gauge formulation of the
geometric ``Higgs'' mechanism for massive KK gravitons $h^{\mn}_n$
and the corresponding KK Goldstone bosons
$\An\,(=h^{\mu5}_n)$ and $\phin\,(=h_n^{55})$,
under which we can derive the propagators.

\vspace*{1mm}

In the subsection\,\ref{sec:3.1},
we apply our $R_\xi^{}$ gauge formulation in section\,\ref{sec:2.2} to establish a Gravitational Equivalence Theorem (GRET)
for the 5d KK GR theory, which quantitatively connects the
high-energy scattering amplitude of the (helicity-zero)
longitudinal KK gravitons
$h^n_L$ to that of the corresponding KK Goldstone bosons
$\phi_n^{}$\,.
Then, in the subsection\,\ref{sec:3.2},
we will show that the GRET identity
provides a theoretical mechanism which guarantees
the longitudinal KK graviton scattering amplitudes to have
nontrivial energy-cancellations, such as
$\,E^{10}\!\ito E^2\,$ for the four-particle amplitudes and
$\,E^{2N+2}\!\ito E^2\,$ for the $N$-particle amplitudes.
We derive a generalized naive power counting method
(\`{a} la Weinberg\,\cite{weinbergPC}) on the leading
energy-dependence of the scattering amplitudes, and apply this
to analyze the leading energy-dependence of the relevant amplitudes
on both sides of the GRET identity \eqref{eq:GET-ID}.
With these, we can demonstrate the above-mentioned
nontrivial energy-cancellations in the longitudinal KK graviton
scattering amplitudes.

\vspace*{1mm}
\subsection{\hspace*{-3mm}Formulation of
	Gravitational Equivalence Theorem}
\label{sec:3.1}

We first express the $R_\xi^{}$ gauge-fixing term \eqref{eq:GF} in the following form:
\beqs
\begin{align}
\label{eq:Lgf}
\La_{\rm{GF}}^{} &\,=\,
-\sum_{n=0}^{\infty} \frac{1}{\,\xi_n^{}}\!\(F_n^A\)^2
= -\sum_{n=0}^{\infty}\frac{1}{\,\xi_n^{}}
\!\[\!\(F_n^\mu\)^2 + \(F_n^5\)^2\!\] ,
\\[2mm]
\label{eq:Fmu}
F_n^\mu &\,=\,   \pd_\nu h^{\mn}_n \!-
\!\(\! 1 \!-\! \frac{1}{\,2\xi_n}\)\!
\pd^\mu h_n^{} +\xi_n^{}M_n^{}\An \,,
\\[1mm]
\label{eq:F5}
F_n^5 &\,=\, \frac{1}{\,2\,}\!
\( M_n^{} h_n^{}\!- 3 \xi_n^{} M_n^{}\phi_n^{}\!
+ 2 \pd_{\mu}^{}\A_n^{\mu} \) \!.
\end{align}
\eeqs
Accordingly, we can write down the Faddeev-Popov ghost term
$\La_{\rm{FP}}^{}$ and the BRST
(Becchi-Rouet-Stora-Tyutin)\,\cite{BRST}
transformations. With these and using the method of
Ref.\,\cite{ET94} (cf.\ Appendix A of the first paper therein),
we can derive a Slavnov-Taylor-type identity
\begin{equation}
\label{eq:F-identity}
\la0|
\hat{T} F_{n_1}^{\mu_1}(x_1^{})F_{n_2}^{\mu_2}(x_2^{})\cdots
F_{m_1}^{5}(y_1^{})F_{m_2}^{5}(y_2^{})\cdots \Phi
|0\ra
\,=\, 0  \,,
\end{equation}
where $\Phi$ denotes any other on-shell physical fields
after the LSZ (Lehmann-Symanzik-Zimmermann) amputation.
In the momentum space, the identity \eqref{eq:F-identity} takes the form:
\begin{equation}
\label{eq:F-identityP}
\la 0|F_{n_1}^{\mu_1}(k_1^{})F_{n_2}^{\mu_2}(k_2^{})\cdots
F_{m_1}^{5}(p_1^{})F_{m_2}^{5}(p_2^{})\cdots \Phi
|0\ra
\,=\, 0\,,
\end{equation}
where we will set each external momentum be on-shell (according to the mass of the corresponding physical KK graviton $h^{\mn}_n$):
$k_j^2=\!-M_{n_j}^2$ and
$p_j^2=\!-M_{m_j}^2$
(with $j=1,2,\cdots\,$).
For the case of just one external line of $F_n^\mu$ or $F_n^5$,
we obtain the following identities of the scattering amplitudes:
\begin{equation}
\label{eq:F-ID-n=1}
\M [F_n^\mu (k),\Phi] = 0 \,,
\qquad
\M [F_n^5 (k),\Phi] = 0 \,,
\end{equation}
where we have not yet imposed the LSZ amputation on
the external line $F_n^\mu$ or $F_n^5$.

\vspace*{1mm}

Now, combining Eqs.\eqref{eq:Fmu} with \eqref{eq:F5}, we can eliminate
the vector Goldstone field $\,\A_n^{\mu}\,$ and derive the expression:
\begin{equation}
\label{eq:Fmu-F5-0}
\pd_\mu^{}F_n^\mu \!-\xi_n^{}M_n^{}F_n^5 \,=\,
\pd_\mu^{}\pd_\nu^{} h_n^{\mn}
\!-\fr{1}{2}\!\[\!(2\!-\xi_n^{-1})\pd^2\!+\xi_n^{}\Mnn\]\!h_n^{}
\!+\!\fr{3}{2}\xi_n^2\Mnn\phi_n^{} \,.
\end{equation}
Then, choosing the Feynman-'t\,Hooft gauge
$\,\xi_n^{}\!=\!1\,$ for simplicity
and imposing the on-shell condition
$\,k^2\!=\!-M_n^2$\, in momentum space,
we derive the following formula:
\beqs
\label{eq:Fmu-F5-FFn}
\begin{align}
\label{eq:kFmu}
& \ii k_\mu^{}F_n^\mu\! +M_n^{}F_n^5
\,= \sqrt{\frac{3}{2}\,}M_n^2\FFn \,,
\\
\label{eq:FFn}
& ~\FFn \,\equiv\,
\sqrt{\frac{2}{3}\,}\frac{\,k_\mu^{}k_\nu^{}\,}{M_n^2}h_n^{\mn}\!
+\sqrt{\frac{2}{3}\,}h_n^{}\!
-\phi_n^{} ,
\end{align}
\eeqs
where we have made the rescaling \eqref{eq:rescaling-An-phin} for
$\phin\,$ and defined the external momentum $k^\mu$
to be incoming in \eqrefe{eq:kFmu}.
For the longitudinal polarization tensor
$\,\vep_L^{\mn}\,$ of the massive
KK graviton, we make the high-energy expansion under
$E=k^0\!\gg\! M_n^{}$\,,
\beqa
\label{eq:epLmunu}
\vep_L^{\mn} =\frac{1}{\sqrt{6\,}\,}\!
\(\ep_+^\mu\ep_-^\nu \!+\ep_-^\mu\ep_+^\nu \!+ 2\ep_L^\mu\ep_L^\nu\)
\equiv \sqrt{\frac{2}{3}\,}\frac{\,k^\mu_{}k^\nu_{}}{M_n^2} + \vt^{\mn}
= \sqrt{\frac{2}{3}\,}\vep_S^{\mn} + \vt^{\mn},
\eeqa
where the longitudinal polarization vector
$\ep_L^\mu = (k^0/M_n)(|\vec{k}|/k^0,\,\vec{k}/|\vec{k}|)
=\ep_S^{\mu}+v^\mu$\, with $\,\ep_S^\mu =k^\mu/M_n\,$
and $\,v^\mu \!=\! \mO(M_n/E_n)\,$.\!\!
In the above, the scalar-polarization tensor is defined to be
$\,\vep_S^{\mn} \!=\! \ep_S^\mu\ep_S^\nu = {k^\mu k^\nu}\!/{M_n^2}\,$
and the residual term has the energy scale $\,\vt^{\mn}\!\!=\!\mO (E^0)$.
Thus, we can further express Eq.\eqref{eq:FFn} as
%
\beqs
\label{eq:kFmu-all}
\begin{align}
\label{eq:kFmu-2}
\F_n^{}& = \tilde{h}_n^S - \OmBn
= h_n^L - \Omega_n^{} \,,
\\[1mm]
\OmBn &= \phin\!-\!\th_n^{} ,
~~~
\Omega_n^{} = \OmBn \!+ \vnt
=\phin \!+\widetilde{\Delta}_n^{},~~~
\widetilde{\Delta}_n^{} \!= \vnt \!-\th_n^{} ,
\hspace*{10mm}
\\[1mm]
h_n^S &=\vep^S_{\mn}h^{\mn}_n\,,~~~\dis
\tilde{h}_n^S =\sqrt{\frac{2}{3}\,}h_n^S,~~~
\th_n^{\mn} = \sqrt{\frac{2}{3}\,}h_n^{\mn},~~~
\tilde{h}_n^{} =\eta_{\mn}^{}\th_n^{\mn},
\\[1mm]
h_n^L &=
\vep^L_{\mn}h^{\mn}_n=\tilde{h}_n^S +\tilde{v}_n^{} ,~~~
\vnt = \vt_{\mn}h^{\mn}_n,
\label{eq:hL-hS-vn}
\\[1mm]
\ep_S^\mu &= \frac{\,k^\mu}{M_n},~~~~
\vep_S^{\mn} = \ep_S^\mu\ep_S^\nu
= \frac{\,k^\mu k^\nu\,}{M_n^2}\,.
\hspace*{15mm}
\end{align}
\eeqs
Then, using Eqs.\eqref{eq:F-ID-n=1} and \eqref{eq:kFmu},
we deduce
\begin{equation}
\label{eq:Fn-ID-n=1}
\M [\FFn (k),\,\Phi] = 0 \,,
\end{equation}
for one external $\FFn$ line.
In the Feynman-'t\,Hooft gauge, all the KK fields of level-$n$ have mass-pole $\,k^2\!=\!-M_n^2$\,.
Also, due to our $R_\xi^{}$ gauge-fixing \eqref{eq:Lgf}
or \eqref{eq:GF}, all the KK fields have diagonal propagators at tree level. So we can amputate the external line $\FFn$ \`{a} la LSZ by multiplying
the propagator-inverse $(k^2\!+\!M_n^2)\ito 0$\,.
Thus, the amplitude in Eq.\eqref{eq:Fn-ID-n=1} will take the same form except that the external line $\FFn$ is amputated.
After this, we can rewrite the identity \eqref{eq:Fn-ID-n=1} as follows:
\begin{equation}
\label{eq:GET-N1}
\M [\tilde{h}_n^S(k),\,\Phi] \,=\,
\M [\over\Omega_{n}^{}(k),\Phi] \,,
\end{equation}
or, equivalently,
\vspace*{-4.mm}
\beqs
\begin{align}
\label{eq:GETv-N1-hLhS}
\M [h_n^L(k),\Phi] &\,=\,
\M [\over\Omega_{n}^{}(k),\Phi]
+\M [\vnt (k),\Phi]
\\[1.5mm]
&\,=\,
\M [\phin (k),\Phi]
+\M [\DEn (k),\Phi] \,,
\label{eq:GET-N1-hLhS}
\end{align}
\eeqs
where
$\,\OmBn = \phin\!-\!\th_n^{}\,$ and
$\,\widetilde{\Delta}_n^{} \!= \vnt -\th_n^{}\,$.

\vspace*{1mm}

For the $N$ external $\FFn$ lines, we thus deduce the following identity with all $\FFn$ lines amputated and on-shell
\begin{equation}
\label{eq:Fn-ID-N-LSZ}
\M [\FF_{n_1}^{}\!(k_1^{}),\FF_{n_2}^{}\!(k_2^{}),
\cdots\!, \FF_{n_N}^{}\!(k_N^{}),
\Phi] \,=\, 0 \,,
\end{equation}
where $\,\FFn\!=\! h_n^L - \Omega_n^{}\,$ and $\,\Phi\,$ denotes any possible amputated on-shell external physical fields. Then, we derive an identity for the scattering amplitude of
$N$ longitudinally-polarized KK gravitons:
\begin{equation}
\label{eq:GET-ID00}
\M [h^L_{n_1}(k_1^{}),
\cdots\!, h^L_{n_N}(k_N^{}),\Phi]
\,=\,
\M [\Omega_{n_1}^{}(k_1^{}),
\cdots\!, \Omega_{n_N^{}}^{}(k_N^{}),\Phi]\,.
\end{equation}
Using the identity \eqref{eq:Fn-ID-N-LSZ},
we can prove the GRET identity \eqref{eq:GET-ID00} directly
by computing its righ-hand-side (RHS)
\begin{align}
\M [\Omega_{n_1}^{}(k_1^{}),\cdots\!, \Omega_{n_N}^{}(k_N^{}),\Phi]
&= \M [h_{n_1}^L\!(k_1^{})\!-\!\FF_{n_1}^{}\!(k_1^{}),
\cdots\!, h_{n_N}^L\!(k_N^{})\!-\!\FF_{n_N}^{}\!(k_N^{}), \Phi]
\nn\\[1.5mm]
&= \M [h_{n_1}^L(k_1^{}),\cdots\!, h_{n_N}^L(k_N^{}), \Phi] \,.
\end{align}
In the last step of the above derivation, we have used the fact that an amplitude including one (or more) external $\FFn$ line plus any other external on-shell physical fields must vanish according to the identity \eqref{eq:Fn-ID-N-LSZ}.

\vspace*{1mm}

Expanding the RHS of Eq.\eqref{eq:GET-ID00}, we can derive an identity
that connects the longitudinal KK graviton amplitude to
the corresponding KK Goldstone boson amplitude and will be
called the GRET identity
hereafter:
\beqs
\label{eq:GET-ID}
\begin{align}
\hspace*{-5mm}
& \M [h^L_{n_1}(k_1^{}),
\cdots\!, h^L_{n_N}(k_N^{}),\Phi] \,=\,
\M [\phi_{n_1}^{}(k_1^{}),
\cdots\!, \phi_{n_N^{}}^{}(k_N^{}),\Phi] +
\M_{\!\Delta}^{}\,,
\label{eq:GET-ID0}
\\[2mm]
& \M_{\!\Delta}^{} \,\equiv\,
\sum_{1\leqq j\leqq N}\!\!\!\!
\M [\{\widetilde{\Delta}_{n_j}^{},\phi_{n_{\!j'}}^{}\},\Phi]\,,
\label{eq:RT-0}
\end{align}
\eeqs
\\[-4mm]
where
$\,\widetilde{\Delta}_n^{} \!= \vnt \!-\th_n^{}\,$
with the notations $\,\vnt\!=\!\vt_{\mn}^{}h_n^{\mn}$\,
and $\,\th_n^{}\!=\!\eta_{\mn}^{}\th^{\mn}_n\,$.
The last term $\,\M_{\!\Delta}^{}\,$
on the RHS of Eq.\eqref{eq:GET-ID} denotes
the residual term of GRET which is the sum of individual amplitudes
where each amplitude
$\,\M [\{\widetilde{\Delta}_{n_j}^{},\phi_{n_{\!j'}}^{}\},\Phi]\,$
contains $\,j\,$ external states of $\,\DEt_{n_j^{}}^{}$ with
$\,j^{}\!\in\! \{1,2,\cdots\!,N\}\,$
and $\,{j'}\,(=\!N\!\!-\!j^{})\,$
external states of $\phi_{n_{\!j'}}^{}$\,.
We note that an on-shell KK graviton
has five physical helicity states
($\lambda \!=\!\pm 2,\,\pm 1,\,0$)
and their polarization tensors, as given by Eq.\eqref{eq:hn-Pols}
of Appendix\,\ref{app:A}, are all traceless.
Hence, the external KK graviton $\,\th_n^{}\,$
is an unphysical state.
This means that the amplitudes containing one or more external
$\,\th_n^{}$ state(s) are unphysical amplitudes.
This is why we arrange all the $\,\th_n^{}$-related amplitudes
on the RHS of the GRET identity \eqref{eq:GET-ID} as part of the
summed residual term $\,\M_{\!\Delta}^{}\,$.

\vspace*{1mm}

Besides, we can further extend the above proof of
the GRET identity \eqref{eq:GET-ID} beyond tree-level
and to be valid for all $R_\xi^{}$ gauges
by using the gravitational BRST identities.
Then, each external Goldstone boson state $\phin$ in
the amplitudes on the RHS of Eq.\eqref{eq:GET-ID}
will receive a multiplicative modification factor
$\,C_{\rm{mod}}^{}\!=\!1+\mO (\rm{loop})$,\,
which is {\it energy-independent}
and similar to the case of the KK GAET formulation
in the compactified 5d YM theories\,\cite{KK-ET-He2004}
and in the 4d SM\,\cite{ET94}\cite{ET96}\cite{ET-Rev}.\footnote{%
Our KK GRET formulation is based on the quantized BRST symmetry and
thus can be readily extended up to loop levels. This means that
our new mechanism of energy-cancellation based on the KK GRET or
KK GAET (cf.\ sections\,\ref{sec:4}-\ref{sec:5})
will generally hold up to loop orders,
which differs from the recent literatures for
the explicit verifications of
energy-cancellations in the tree-level KK graviton
amplitudes\cite{Chivukula:2020S}\cite{Chivukula:2020L}\cite{Kurt-2019}.}
So, such energy-independent factor $\,C_{\rm{mod}}^{}$
does not affect the energy-power counting of the (Goldstone-related) amplitudes of Eq.\eqref{eq:GET-ID} at loop levels.
Since we focus on the scattering amplitudes
and the application of GRET at tree level for the current study,
we will present a generalized loop-level
formulation elsewhere\,\cite{GET-2}.\footnote{%
The 4d ET in the presence of the Higgs-gravity interactions
was established in Refs.\,\cite{GET4d-1}\cite{GET4d-2}
which can be applied to studying cosmological models (such as the
Higgs inflation\,\cite{GET4d-2}\cite{HiggsInfx}\cite{HiggsInf})
or to testing self-interactions of weak gauge bosons
and Higgs bosons\,\cite{GET4d-1}\cite{GET4d-2}\cite{HRY}.}

\vspace*{1mm}

Next, inspecting both sides of the GRET identity \eqref{eq:GET-ID0},
we can readily make naive power counting on the energy-dependence
of the individual Feynman diagrams for each scattering amplitude.
For the four-particle scattering at tree level,
the longitudinal KK graviton amplitude
on the LHS of the identity \eqref{eq:GET-ID0}
contains the contributions by individual diagrams via
quartic contact interactions or via exchanging KK
(or zero-mode) gravitons. Since each exteral longitudinal
KK graviton has polarization tensor \eqref{eq:epLmunu}
scales like
$\,\vep_L^{\mn}\!\propto\! k^\mu k^\nu/\Mnn\,$
in the high energy limit, the contribution by each
individual diagram behaves as $\mO(E^{10})$,
where the energy-power $\,10=8+2$\,
contains the energy-power of $\,8=2\!\times\!4\,$ arising from
the four external longitudinal KK gravitons and
the energy-power \,2\, contributed by the internal couplings
and propagators.
On the other hand, we can make naive power counting on the
energy-dependence of the individual diagrams in each amplitude
of the RHS of Eq.\eqref{eq:GET-ID0}.
Because the external states (either the KK Goldstone boson
$\phin$, or, the KK gravitons such as
$\,\vnt\!=\!\vt_{\mn}^{}h_n^{\mn}$\,
or $\,\th_n^{}\!=\!\eta_{\mn}^{}\th^{\mn}_n\,$)
in all such amplitudes have no extra enhancement or suppression
factor, we can readily make naive power counting on their
energy-dependence and deduce that they all behave as
$\mO(E^2)$ under the high energy expansion.
Hence, the GRET identity \eqref{eq:GET-ID0} provides a
general mechanism for the energy-power cancellation
of $\,E^{10}\!\ito E^2\,$ in the longitudinal KK graviton
scattering amplitudes at tree level.

\vspace*{1mm}

We note that on the RHS of Eq.\eqref{eq:GET-ID0}
the residual term $\,\MD\,$ contains individual amplitude
$\,\M [\{\DEt_{n_j}^{},\phi_{n_{\!j'}}^{}\},\Phi]\,$
with external states of the type
$\,\DEn \!=\vnt -\th_n^{}\,$.
The external state $\,\vnt \!\!=\!\vt_{\mn}^{}h^{\mn}_n\,$
is not suppressed under high energy expansion due to
$\,\vt^{\mn}\!\!=\!\mO(E^0)\,$,\,
and the external state
$\,\th_n^{}\!=\!\eta_{\mn}^{}\th_n^{\mn}\,$
is unsuppressed either by any factor of $\Mn/E$\,.
Thus, there is no apparent ``equivalence'' between
the (helicity-zero) longitudinal KK graviton $\hLn$-amplitude
and the KK Goldstone $\phin$-amplitude
in Eq.\eqref{eq:GET-ID0} under the high energy expansion.
This differs essentially from the conventional
equivalence theorem (ET) for the spin-1 massive gauge bosons
in the SM and in the compactified KK gauge theory,
where the residual term is suppressed in the high energy limit
because of the corresponding residual factor
$\,v^\mu \!=\!\ep_L^\mu\!-\!\ep_S^\mu \!=\!\mO(M_n/E_n)\,$.
In fact, we observe that
the GRET residual term $\,\MD\,$ in Eq.\eqref{eq:RT-0}
is given by the sum of amplitudes like
$\,\M [\{\DEt_{n}^{},\phi_{n}^{}\},\Phi]\,$
with $\,\DEn \!=\vnt \!-\th_n^{}\,$
containing both the external fields
$\,\vnt\,$ and $\,\th_n^{}\,$, which do not receive additional
suppression under the high energy expansion.
As we will show in sections\,\ref{sec:4.2} and \ref{sec:5.4}
for the four longitudinal KK graviton scattering,
the residual term $\,\MD\,$ as a sum of the
$\DEn$-dependent individual amplitudes
in Eq.\eqref{eq:GET-ID} has $\mO(E^2)$ by the naive power counting
and will be further cancelled down to $\mO(E^0)$
in comparison with the leading Goldstone $\phin$-amplitude
of $\mO(E^2)$  under the high energy expansion.

\vspace*{1mm}

With the above observations, we can express the GRET as follows:
\begin{equation}
\label{eq:GET}
\hspace*{-10mm}
\M [h^L_{n_1}(k_1^{}), 
\cdots\!, h^L_{n_N}(k_N^{}),\Phi]
\,=\, \M [\phi_{n_1}^{}(k_1^{}), 
\cdots\!, \phi_{n_N}^{}(k_N^{}), \Phi] \,+\,
\mO (\DEn) \,,
\end{equation}
where the residual term $\,\MD\,$ is denoted by $\mO(\DEn)$
summing up all the remaining
amplitudes with at least one external state being $\,\DEn \,$.
We will demonstrate later
in sections\,\ref{sec:4.2} and \ref{sec:5.4} that
{\it the sum of residual terms $\mO(\DEn)$ is indeed suppressed
by $\Mn/E$\, factors relative to the leading Goldstone amplitude}
on the RHS of the GRET \eqref{eq:GET} for the high energy scattering
processes (with two or more external longitudinal KK gravitons).

\vspace*{1mm}

In principle, the GRET identity \eqref{eq:GET-ID0} and the
GRET \eqref{eq:GET} hold for any number of external longitudinal
KK graviton states, although in the above we take the case of
four longitudinal KK graviton scattering ($N\!\!=\!4$) at tree level
as an important example for discussing the naive energy-power-counting and energy cancellations.
In the following, we will extend the above naive power counting
analysis on energy-dependence of
the longitudinal KK graviton amplitudes,
the KK Goldstone amplitudes and the residual-term amplitudes
in the GRET identity \eqref{eq:GET-ID0}
to the general case of $\,N\!\!\geqq\! 4$\, and up to loop levels.


\vspace*{1mm}
\subsection{\hspace*{-3mm}Energy Cancellation Mechanism
for KK Graviton Scattering Amplitudes}
\label{sec:3.2}

We recall that Steven Weinberg\footnote{%
On July\,23, 2021, we reposted an update version of this paper to
arXiv:2106.04568 in which we systematically presented this generalized
power counting method for the compactified KK gauge theories and
KK GR theories in section\,\ref{sec:3.2}.\  On the following day
we learnt the sad news that Steven Weinberg passed away on July\,23.
One of us (HJH) wishes to express his deep gratitude to Steve
for his inspirations over the years
(including the discussion of his original work
on the power counting analysis\,\cite{weinbergPC}),
especially during his times at UT Austin where his office
was only a few doors away from that of Steve.}
originally derived a power counting rule
of energy dependence for the ungauged nonlinear $\sigma$-model
as a description of low energy QCD interactions\,\cite{weinbergPC}.
This power counting rule has two major ingredients:\
{(i).}\,The total mass-dimension $D_{\mathbb{S}}^{}$ of
a scattering $S$-matrix element $\,\mathbb{S}\,$ is determined
by the number of external states ($\mathcal{E}$)
and the spacetime dimension, namely,
$\,D_{\mathbb{S}}^{}=4-\mathcal{E}\,$,
for 4d field theories.\
{(ii).}\,Consider that the typical scattering energy $E$ is much
larger than all the relevant mass-poles
in the internal propagators of the scattering amplitude
$\,{\mathbb{S}}^{}$\,.
Then the total mass-dimension $D_C^{}$ of the $E$-independent coupling constants contained in the amplitude $\mathbb{S}$\,
can be directly counted according to the type of vertices therein.
With these, one can deduce the total energy-power dependence
$D_E^{}$ of the amplitude $\,\mathbb{S}\,$ as
$\,D_E^{}= D_{\mathbb{S}}^{} -D_C^{}\,$.
We note that the point\,(i) is fully general,
and the point\,(ii) holds for any field theory
in which the particle masses are much smaller
than the scattering energy $E$\, and
the nontrivial energy-dependence of the polarization tensors
(vectors) for the possible longitudinally polarized KK gravitons
(gauge bosons) can be properly taken into account.
Hence, we can generalize Weinberg's power counting rule
to the compactified 5d theories\footnote{%
Weinberg's power counting rule was extended
previously\,\cite{ET-Rev}\cite{ETPC-97}
to the 4d gauge theories including the SM, the SM effective theory
(SMEFT), and the electroweak chiral Lagrangian.}
including KK graviton (Goldstone) fields and/or KK gauge (Goldstone)
fields, and study the high energy scattering amplitudes of KK particles
whose masses are much smaller than the scattering energy $E$\,.

\vspace*{1mm}

Consider a scattering $S$-matrix element $\,\mathbb{S}\,$ having
$\,\EE\,$ external states and $L$ loops ($L\!\geqq\! 0$).
Thus, the amplitude $\,\Sb\,$ has a mass-dimension:
\begin{equation}
\label{eq:DS}
D_{\mathbb{S}}^{} \,=\,4 - \mathcal{E}\,,
\end{equation}
where the number of external states
$\,\EE \!=\EE_B^{}+\EE_F^{}\,$,
with $\,\EE_B^{}\,(\EE_F^{})\,$ being the number of
external bosonic (fermionic) states. For the fermions, we only
consider the SM fermions whose masses are much smaller than the
scattering energy $E$\,.\,
We denote the number of vertices of type-$j$ as $\VV_j^{}$\,.
Each vertex of type-$j$ contains $\,d_j^{}\,$ derivatives,
$\,b_j^{}\,$ bosonic lines and $\,f_j^{}\,$ fermionic lines.
Then, the energy-independent effective coupling constant in
the amplitude $\,\mathbb{S}\,$ is given by
\begin{equation}
\label{eq:DC}
D_C^{} \,=\, \sum_j \VV_j^{}\!
\(4-d_j^{}\!-b_j^{}\!-\fr{3}{2}f_j^{}\) \!.
\end{equation}
For each Feynman diagram in the scattering amplitude
\,$\mathbb{S}$\,,\,
we denote the number of the internal lines as
$\,I=I_B^{}+I_F^{}\,$ with
$\,I_B^{}$ ($\,I_F^{}\,$) being the number of the internal
bosonic (fermionic) lines. Thus, we have the following general
relations:
\begin{equation}
\label{eq:L-V-I}
L = 1+I-\VV\,,~~~~~
\sum_j \VV_j^{}b_j^{} = 2I_B^{}+\EE_B^{}\,,~~~~~
\sum_j \VV_j^{}f_j^{} = 2I_F^{}+\EE_F^{}\,,
\end{equation}
where $\,\VV=\sum_j\!\VV_j^{}\,$
is the total number of vertices in a given Feynman diagram.
The amplitude $\,\mathbb{S}\,$ may include $\,\EE_{h_L}^{}\!$
external longitudinal KK graviton states.
Then, using Eqs.\eqref{eq:DS}-\eqref{eq:L-V-I}, we deduce
the leading energy-power dependence
$\,D_E^{}= D_{\mathbb{S}}^{} -D_C^{}\,$
of the high energy scattering amplitude $\Sb$\, as follows:
\begin{equation}
\label{eq:D_E}
D_E^{} ~=~ 2\EE_{h_L}^{} \!\!+
(2L\!+2)+\sum_j  \VV_j^{}\!
\(d_j^{}\!-2+\!\fr{1}{2}f_j^{}\) .
\end{equation}

Then, we consider the pure 5d KK GR theory without
involving any matter fields.
Thus, for the pure longitudinal KK graviton scattering amplitude
with $\,N\,$ external states
$\,\Sb\!=\!\M [h^L_{n_1},\cdots\!, h^L_{n_N}]\,$,\,
we have $\,\EE_{h_L}^{}\!\!\!=\!N\,$ and $\,f_j^{}\!=\!0\,$.
Each pure KK graviton vertex always contains two partial derivatives
and thus $\,d_j^{}\!=2\,$.
For the loop level ($L\geqq 1$),
the amplitude may contain gravitational ghost loop which
involves graviton-ghost-antighost vertex, but the number of partial
derivatives $\,d_j^{}$ should be no more than two.
This means that the leading energy dependence is always given
by the diagrams containing only
the KK gravitons and/or zero-mode gravitons.
Hence, to count the leading energy dependence of the
pure longitudinal KK graviton scattering amplitudes,
we can further derive the power counting
formula \eqref{eq:D_E} as
\begin{equation}
\label{eq:DE-hL}
D_E^{}[Nh^L_n] \,=\, 2(N\!+\!1)\!+2L\,,
\end{equation}
where the notation $\,[Nh^L_n]\,$
just denotes the $N$ external longitudinal KK
graviton states ($h^L_n$) whose KK indices can differ from each other
in an inelastic scattering amplitude. Similar notations,
such as $\,[N\phin]$\, for $N$ external KK Goldstone states and so on,
will be used for other amplitudes.

\vspace*{1mm}

Next, we consider the corresponding gravitational
KK Goldstone boson scattering
amplitude $\M [\phi_{n_1}^{},\cdots\!, \phi_{n_N}^{}]\,$
with $N$ external states.
Its leading energy dependence is given by the diagrams
containing $\phi_n^{}$-$\phi_m^{}$-$h^{\mn}_\ell$ type of
cubic vertices
and the pure (KK) graviton self-interaction vertices, where
each of these vertices includes two derivatives
($d_j^{}\!=2$).
Hence, to count the leading energy dependence,
we can further derive the power counting
formula \eqref{eq:D_E} as follows:
\begin{equation}
\label{eq:DE-phin}
D_E^{}[N\phin] \,=\, 2+2L\,.
\end{equation}
Here we also note that each external Goldstone boson state $\phin$ in
the amplitudes on the RHS of Eq.\eqref{eq:GET-ID0}
will receive a multiplicative modification factor
$\,C_{\rm{mod}}^{}\!=\!1\!+\mO (\rm{loop})$\, at loop level,
which is {\it energy-independent} as mentioned earlier.
Hence such loop factor $\,C_{\rm{mod}}^{}$ will not affect the
energy power counting of the Goldstone $\phin$-amplitudes.
Comparing the energy power counting formulas \eqref{eq:DE-hL} and \eqref{eq:DE-phin},
we note that their difference arises from
the leading energy-dependence of the polarization tensors
$\,\vep_L^{\mn}\!\!\sim\!k^\mu k^\nu/\Mnn\,$ for the $N$ external
longitudinal KK gravitons in the high energy scattering:
\begin{equation}
\label{eq:DEL-DEphi}
D_E^{}[Nh^L_n] - D_E^{}[N\phin] \,=\, 2N \,.
\end{equation}
We further examine the leading $E$-power dependence of the individual
amplitudes in the residual term $\,\MD\,$ of the GRET \eqref{eq:GET-ID}.
A typical leading amplitude can be
$\M [\vt_{n_1}^{},\cdots\!, \vt_{n_N}^{}]\,$,\,
in which all the external states are KK gravitons
contracted with the tensor
$\,\vt^{\mn}\!=\vep_L^{\mn}\!\!-\vep_S^{\mn}\!=\mO(E^0)\,$,
such as $\,\vt_n^{}\!=\vt_{\mn}h_n^{\mn}\,$.\,
Hence, we can count the leading energy dependence of
this amplitude in the same way as Eq.\eqref{eq:DE-hL} for
the longitudinal KK graviton amplitude
$\M [h^L_{n_1},\cdots\!, h^L_{n_N}]\,$ except taking out
the energy-enhancement factor $E^2$ from each external longitudinal
polarization tensor $\vep_L^{\mn}$.\, Then, we deduce the
following energy power dependence of the leading residual amplitude
$\M [\vt_{n_1}^{},\cdots\!, \vt_{n_N}^{}]\,$:
\begin{equation}
\label{eq:DE-vn}
D_E^{}[N\vt_n^{}] \,=\, 2+2L\,,
\end{equation}
which gives the same energy power dependence as Eq.\eqref{eq:DE-phin}
for the leading scattering amplitude of $N$ KK Goldstone bosons.
We will establish a further energy cancellation in the residual
term $\MD$ in section\,\ref{sec:5.4} based upon the
double-copy construction.

\vspace*{1mm}

Applying the leading energy-power counting results
\eqref{eq:DE-hL}-\eqref{eq:DE-vn} to both sides
of the GRET identity \eqref{eq:GET-ID0},
we thus establish an energy cancellation
by $\,E^{2N}\,$
in a scattering amplitude of $N$ longitudinal KK gravitons
$\M [h^L_{n_1},\cdots\!, h^L_{n_N}]\,$.
For the case of four longitudinal KK graviton scattering
amplitudes ($N\!\! =\!4$) at tree level ($L\!=\!0$),\,
we can deduce the energy power cancellation
$\,E^{10}\!\ito E^2\,$, which reduces the energy powers by
$\,(10-2)\!=8\,$,\, as we mentioned earlier.
For another case of four KK graviton scattering amplitudes
containing two external longitudinal KK gravitons and
two external transverse KK gravitons
($\,\EE_{h_L}^{}\!\!=\!2\,$),\,
we have the $E$-power counting
$\,D_E^{}[2h_L^{n}\!+2h_T^{n}]=6+2L\,$.
For the corresponding KK Goldstone amplitudes, we have
energy counting
$\,D_E^{}[2\phin +\!2h_T^{n}]= 2+\!2L\,$.
The leading residual term contains the amplitudes such as
$\M [\vt_{n_1}^{},\vt_{n_2}^{},h^T_{n_3},h^T_{n_4}]\,$,
which has the same energy-power dependence as the
residual term amplitude with all external states being
$\vt_n^{}$'s [cf.\ Eq.\eqref{eq:DE-vn}].
Namely, we can deduce
$\,D_E^{}[2\vt_n^{}\!+\!2h_T^{n}]=2+2L\,$.\,
Hence, from the GRET identity \eqref{eq:GET-ID0},
we deduce that the KK graviton amplitude
$\M [h^L_{n_1},h^L_{n_2},h^T_{n_3},h^T_{n_4}]\,$
has an energy cancellation down by a factor of $\,E^4\,$.
This energy mechanism holds not only for the tree level,
but also for the loop levels ($L\!\geqq\! 1$)\,
since, as we noted earlier,
the loop-induced multiplicative modification factor
$\,C_{\rm{mod}}^{}\!=\!1\!+\mO (\rm{loop})$
associated with each external KK Goldstone state
is {\it energy-independent} and thus does not affect
the naive energy-power counting on the RHS of
Eq.\eqref{eq:GET-ID}.

\vspace*{1mm}

In the rest of this subsection,
we consider the energy power counting in
the compactified 5d KK YM theory (YM5)
under $S^1/\ZZ$ \cite{5DYM2002}.
For a scattering amplitude containing $\,\EE_{\!A_L^n}^{}\!$
external longitudinal KK gauge bosons $A_L^{an}$
and $\,\EE_{v}^{}\,$ external KK gauge bosons
$\,v_n^a\!=\!v_\mu^{}A^{a\mu}\,$
(with $v^\mu \!=\!\ep_L^\mu\!-\ep_S^\mu\,$),
we can derive the following leading energy dependence
$\,D_E^{}\!=\! D_{\mathbb{S}}^{}\! -\!D_C^{}\,$
from Eqs.\eqref{eq:DS}-\eqref{eq:L-V-I},
\\[-5mm]
\begin{equation}
\label{eq:D_EAL}
D_E^{} \,=\, \EE_{\!A_L^n}^{} \!-\EE_{v}^{}\!+
(2L\!+2)+\sum_j \VV_j^{}\!
\(d_j^{}\!-2+\!\fr{1}{2}f_j^{}\) \!.
\end{equation}
Inspecting the interaction Lagrangian of the zero-modes
and KK-modes of gauge bosons, we note that it contains
only cubic and quartic vertices. Some of the cubic vertices
contain one partial derivative and others do not (including
all quartic gauge boson vertices).
For notational convenience, we denote the gauge fields
$\,V_0^{}\!=\!A_0^{a\mu}$,\, $V_n^{}\!=\!A_n^{a\mu}$,\,
and $\,\VT_n^{}\!=\!A_n^{a5}$.
After the BRST quantization, the ghost term contains
the cubic interactions between KK ghost-antighost
$(c_n^a,\,\bar{c}_m^b)$ and
KK gauge bosons with one partial derivative
in each vertex\,\cite{KK-ET-He2004}.
Thus, the cubic vertices with one partial derivative
have the types of
$\,(V_0^{}V_n^{}V_n^{},\, V_n^{}V_m^{}V_\ell^{},\,
 c_n^{}\bar{c}_m^{}V_\ell^{})$\, and
$(V_0^{}\VT_n^{}\VT_n^{},\,V_n^{}\VT_m^{}\VT_\ell^{})$.
Hence, we have
\beqs
\label{eq:Vd}
\begin{align}
& \sum_j \VV_j^{}d_j^{} \,=\,\VV_d^{}\,,
\\[-1mm]
& \VV_d^{}\,=\,
\VV_3^{}(V_0^{}V_n^{}V_n^{})+
\VV_3^{}(V_n^{}V_m^{}V_\ell^{})+
\VV_3^{}(c_n^{}\bar{c}_m^{}V_\ell^{})+
\VV_3^{}(V_0^{}\VT_n^{}\VT_n^{})+
\VV_3^{}(V_n^{}\VT_m^{}\VT_\ell^{})\,,
\hspace*{5mm}
\end{align}
\eeqs
where $\,\VV_d^{}$\, denotes the number of all cubic vertices including
one partial derivative and $\,\VV_3^{}(XYZ)$\,
denotes the number of
cubic vertices of type $XYZ$.
For the YM5 theory, we further have the following relations:
\\[-9mm]
\beqs
\label{eq:V=V3+V4}
\begin{align}
\VV \,=\,&~ \sum_j\VV_j^{} \,=\, \VV_3^{} + \VV_4^{}\,,
\\
\VV_3^{} \,=\,&~ \VV_d^{}+\VV_F^{}+\over{\VV}_3^{}\,,
\\
\VV_F^{} \,=\,&~ \VV_3^{}(V_0^{}f_n^{}\bar{f}_n^{})+
\VV_3^{}(V_n^{}f_m^{}\bar{f}_\ell^{})+
\VV_3^{}(\VT_n^{}f_m^{}\bar{f}_\ell^{})\,,
\\[1mm]
\over{\VV}_3^{} \,=\,&~ \VV_3^{}(V_0^{}V_n^{}\VT_n^{})+
\VV_3^{}(V_n^{}V_m^{}\VT_\ell^{}) +
\VV_3^{}(c_n^{}\bar{c}_m^{}\VT_\ell^{})\,,
\\[1mm]
\VV_4^{} \,=\,&~ \VV_4^{}(V_0^{}V_0^{}V_n^{}V_n^{})+
\VV_4^{}(V_0^{}V_n^{}V_m^{}V_\ell^{})+
\VV_4^{}(V_n^{}V_m^{}V_k^{}V_\ell^{})
\nn\\
&~ +\VV_4^{}(V_0^{}V_0^{}\VT_n^{}\VT_n^{})+
\VV_4^{}(V_0^{}V_n^{}\VT_m^{}\VT_\ell^{})+
\VV_4^{}(V_n^{}V_m^{}\VT_k^{}\VT_\ell^{})\,,
\end{align}
\eeqs
where the possible fermions and their KK states are included
although they are not needed for analyzing the pure KK gauge theory
in the present work. Using Eqs.\eqref{eq:Vd}-\eqref{eq:V=V3+V4},
we further derive the leading energy-power dependence
\eqref{eq:D_EAL} as follows:
\begin{equation}
\label{eq:D_EAL2}
D_E^{} \,=~ \EE_{\!A_L^n}^{} \!-\EE_{v}^{}\!+
(2L\!+\!2)-(\VV_d^{}\!+\!\VV_F^{}\!+2\over{\VV}_3^{}\!+2\VV_4^{})
\,.
\end{equation}
Then, using the general relation $\,L\!=I\!+1\!-\VV\,$
given by Eq.\eqref{eq:L-V-I}
and the following relation of the YM5 theory
\begin{equation}
\label{eq:2i+e}
2I+\EE \,=\, 3\VV_3^{}+4\VV_4^{}\,,
\end{equation}
we can express the leading energey dependence \eqref{eq:D_EAL2}
as
\begin{equation}
\label{eq:D_E-YM5}
D_E^{}\,=\, (4-\EE)+(\EE_{\!A_L^n}^{} \!-\EE_{v}^{})
-\over{\VV}_3^{}\,,
\end{equation}
where $\,\EE$\, stands for the total number of the external states
and $\,\over{\VV}_3^{}\,$ denotes the number of cubic vertices
containing no partial derivative.
In Eq.\eqref{eq:D_E-YM5}, $\EE_{v}^{}$ denotes of number of external
KK gauge bosons contracted with the vector
$\,v^\mu\!=\ep_L^\mu\!-\ep_S^\mu = \mO(\Mn/E)\,$.
So each external state $\,v_n^a =v_\mu^{}A_n^{a\mu}\,$
contributes an energy suppression factor $\,E^{-1}$\,.
The naive power counting formula \eqref{eq:D_E-YM5}
does not depend on the loop number $L$ and
takes similar form to that of the SM case\,\cite{ETPC-97},
because the structure of each individual vertex of the KK YM5 theory
is similar to that of the SM while the non-renormalizability nature of
the KK YM5 theory is reflected by its infinite tower of KK states.

\vspace*{1mm}

Inspecting Eq.\eqref{eq:D_E-YM5}, we note that for the
pure longitudinal KK gauge boson scattering amplitude
with $\,\EE\!=\EE_{\!A_L^n}^{} \!=\!N(\geqq\! 4)\,$
and $\,\EE_{v}^{}\!=\!0\,$,\,
the leading energy dependence is given by
\begin{equation}
\label{eq:DE-NAL}
D_E^{}[N\!A_L^n]\,=\,4\,,
\end{equation}
which corresponds to $\,\over{\VV}_3^{}\!\!=\!0\,$.
This means that the leading energy-power dependence
of the pure longitudinal KK gauge boson scattering is always
given by the diagrams containing only
cubic derivative gauge vertices and/or
quartic gauge vertices.
We stress that the leading energy dependence
$\,D_E^{}=4\,$ does not depend on the number
of external longitudinal KK gauge bosons
($\,\EE_{\!A_L^n}^{}\!=\!N$\,).
The case of $\,N\!=4\,$ scattering amplitudes
was studied before\,\cite{5DYM2002}.
Then, we consider the scattering amplitudes of
pure KK Goldstone bosons ($A_n^{a5}$)  with
 $\,\EE\!=\EE_{\!A_5^n}^{} \!=\!N\,$
external $A_n^{a5}$ states. This also means
$\,\EE_{\!A_L^n}^{} \!\!=0\,$ and
$\,\EE_{v}^{}\!=0\,$.
Thus, using Eq.\eqref{eq:D_E-YM5}, we deduce the
leading energy dependence of $N$ KK Goldstone boson
scattering amplitude as
\begin{equation}
\label{eq:DE-NA5}
D_E^{}[N\!A_5^n] \,=\, 4-N-\over{\VV}_3^{\min}\,,
\end{equation}
where the number of the external KK Goldstone states
$\,N\!\!\geqq\! 4$\, and
the involved minimal number of non-derivative cubic vertices
$\,\over{\VV}_3^{\min}=0\,(1)$\,
for $\,N\!=\,$even (odd).

\vspace*{1mm}

It was established\,\cite{5DYM2002}\cite{KK-ET-He2004}
that the longitudinal KK gauge boson scattering amplitude and
the corresponding KK Goldstone boson scattering amplitude are
connected by the KK equivalence theorem for gauge theory (KK GAET) under the
high energy expansion:
\beqs
\label{eq:KK-ET-N}
\begin{align}
& \T[A_L^{a_1^{}n_1^{}},\cdots\!,A_L^{a_N^{}n_N^{}},\Phi]
~=~ C_{\rm{mod}}^{}
\T[A_5^{a_1^{}n_1^{}},\cdots\!,A_5^{a_N^{}n_N^{}},\Phi]
\,+\, \T_v^{}\,,
\label{eq:KK-ET1-N}
\\[1mm]
& \T_v^{} \,=\,\sum_{\ell=1}^N\! C_{\rm{mod}}'
\T[v^{a_1^{}n_1^{}},\cdots\!,v^{a_\ell^{}n_\ell^{}},
A_5^{a_{\ell+1}^{}n_{\ell+1}^{}},\cdots\!,A_5^{a_N^{}n_N^{}},\Phi]
\,=\, \mO(\Mn/E)\,,
\label{eq:KK-ET1-Tv}
\end{align}
\eeqs
where $\,\Phi\,$ denotes any other external physical state(s).
The modification factors
$\,C_{\rm{mod}}^{},C_{\rm{mod}}'\!=1\!+\mO(\rm{loop})\,$
are energy-independent constants and do not affect the energy power counting,
which are generated at loop level\,\cite{KK-ET-He2004}\cite{ET-Rev}
and are not needed for the tree-level analysis in the current study.

\vspace*{1mm}

Then, we consider the scattering amplitudes of
$N$ longitudinal KK
gauge bosons and of the corresponding $N$ KK Goldstone bosons.
Their leading energy powers are given by
Eqs.\eqref{eq:DE-NAL} and \eqref{eq:DE-NA5}.
Thus, we deduce the following difference
between their leading energy powers:
\begin{equation}
\label{eq:DEAL-DEA5}
D_E^{}[N\!A^L_n]-D_E^{}[N\!A_5^n] \,=\, N+\over{\VV}_3^{\min}\,,
\end{equation}
where $\,\over{\VV}_3^{\min}$\, denotes
the involved minimal number of non-derivative cubic vertices
in the KK Goldstone amplitude and
$\,\over{\VV}_3^{\min}=0\,(1)$\,
for $\,N\!=\,$even\,(odd)\,.
Next, we make naive energy counting on the residual term
$\,\T_v^{}\,$ of the KK GAET \eqref{eq:KK-ET-N}.
To extract the leading energy dependence,
we start with the pure KK Goldstone amplitude
$\,\T[A_5^{a_1^{}n_1^{}},\cdots\!,A_5^{a_N^{}n_N^{}}]\,$
and replace one external KK Goldstone state
(say, $A_5^{a_1^{}n_1^{}}$)
by the KK gauge boson contracted with the
$\,v^\mu$ factor ($v^\mu\! A_\mu^{a_1^{}n_1^{}}\!=v^{a_1^{}n_1^{}}$).
For the case of $\,N\!=\rm{even}$,\,
this means to replace a derivative vertex
by a non-derivative vertex and add the factor $v^\mu$,\,
so the leading energy dependence $D_E^{}$
will be reduced by $E^{-2}$.
For the case of $\,N\!=\rm{odd}$,\,
this means to replace a non-derivative cubic vertex
by a derivative cubic vertex and add a $v^\mu$ factor.
So the leading energy dependence $D_E^{}$ will not change.
Thus, we conclude that the leading energy dependence of the
residual term \eqref{eq:KK-ET1-Tv} is given by
\beqs
\begin{align}
D_E^{}[\T_v^{}] \,&=\, 2-N\,, \qquad (\text{for}~N\!=\rm{even}),
\\
D_E^{}[\T_v^{}] \,&=\, 3-N\,, \qquad (\text{for}~N\!=\rm{odd}).
\end{align}
\eeqs
Comparing this with the leading energy-power counting
\eqref{eq:DE-NA5} of the $N$ KK Goldstone boson amplitudes
in the high energy scattering,
we deduce that for the case of
$\,N\!=\!\rm{even}$\, the residual term
\eqref{eq:KK-ET1-Tv} is suppressed by $\Mnn/E^2\,$ factor
relative to the leading KK Goldstone amplitude on the RHS of
the KK GAET \eqref{eq:KK-ET1-N}
and thus can be ignored, while for the case of $\,N\!=\rm{odd}$\,
the residual term \eqref{eq:KK-ET1-Tv} has the same
leading energy dependence as that of the
leading KK Goldstone amplitude.
In either case,
the KK GAET \eqref{eq:KK-ET-N} guarantees that
the leading energy dependence $\,E^4\,$ of
the pure longitudinal KK gauge boson amplitudes
in Eq.\eqref{eq:DE-NAL}
has to be cancelled down to the leading energy dependence of
the corresponding KK Goldstone amplitudes in Eq.\eqref{eq:DE-NA5}.
This energy cancellation shows that
even though the $N$-particle longitudinal KK gauge boson
scattering amplitudes have superficial leading energy dependence
$E^4$ as contributed by individual Feynman diagrams,
these must be cancelled down by an energy factor
$\,E^{\delta D_{\!E}^{}}\,$ to match the leading energy dependence
of the corresponding KK Goldstone boson amplitudes, where the
energy power factor changes by
\begin{equation}
\delta D_E^{} \,=\, N+\frac{\,1\!-\!(-1)^N\,}{2} \,.
\end{equation}
This energy cancellation of $\delta D_E^{}$ coincides with
the above formula \eqref{eq:DEAL-DEA5}.
For the case of four longitudinal KK gauge boson scattering amplitudes
($N=4$), it was proven\,\cite{5DYM2002} that
the leading energy cancellation $\,E^4\ito E^0\,$
is guaranteed by the KK GAET
to match the leading energy dependence of the corresponding
KK Goldstone boson amplitudes.
This fully agrees with the above general
analysis for the $N$-particle scattering amplitudes.
In the following, we will focus on the four-particle KK amplitudes
($N\!=4$) for the explicit analysis of the GRET in section\,\ref{sec:4}
and for the double-copy construction in section\,\ref{sec:5}.
We will pursue the analysis of $N\!>\!4\,$ case in
future works\,\cite{GET-2}.

\section{\hspace*{-3mm}Structure of KK Graviton Scattering Amplitudes\\
\hspace*{-4.5mm}
from Gravitational Equivalence Theorem}
\label{sec:4}
The compactified five-dimensional Yang-Mills theory under orbifold $S^1/\ZZ$ generates a tower of massive gauge bosons via KK construction. The KK gauge boson mass-generation can
be formulated by the geometric ``Higgs'' mechanism in a generic $R_\xi$ gauge\,\cite{5DYM2002}, where each massive longitudinal KK gauge boson $A^{a\mu}_n$ acquires its mass by
absorbing the corresponding KK-state Goldstone $A^{a5}_n$ from the fifth component of the 5d gauge field.
Ref.\,\cite{5DYM2002} has established the KK GAET which states that each on-shell scattering amplitude of the longitudinal KK gauge bosons ($A^{aL}_n$) equals the amplitude of the
corresponding Goldstone bosons ($A^{a5}_n$) down to $\mO (E^0)$ under the high energy expansion,
\begin{equation}
\label{eq:KK-ET22}
\T\!\[\!A^{aL}_{n_1} A^{bL}_{n_2}
\ito A^{cL}_{n_3} A^{dL}_{n_4} \]  \,=\,
\T\!\[\!A^{a5}_{n_1} A^{b5}_{n_2} \ito A^{c5}_{n_3} A^{d5}_{n_4}\]
+ \mO (M_{n_i}^2/E^2)     \,.
\end{equation}
This formulation was extended to gauge theories in deconstructed extra
dimension\,\cite{5DYM2002-2} and to the realistic
compactified 5d standard model\,\cite{5dSM}.

\vspace*{1mm}

In this section, we will systematically compute the $2 \ito 2$ scattering amplitudes of gravitational KK Goldstone bosons
for the first time. Then, we will explicitly demonstrate the validity
of the GRET by comparing our gravitational KK Goldstone amplitudes
with the corresponding helicity-zero KK graviton amplitudes
obtained in \cite{Chivukula:2020L}.
For the case of $2 \ito 2$ scattering,
we first deduce the GRET identity from \eqrefe{eq:GET-ID0}:
\begin{equation}
\label{eq:GET-ID22}
\M \!\[\! h^L_{n_1}h^L_{n_2} \!\ito h^L_{n_3}h^L_{n_4} \!\]
\,=\,
\M \!\[\!\Omega_{n_1}^{}\Omega_{n_2}^{}
\!\ito \Omega_{n_3}^{}\Omega_{n_4}^{} \!\]
\,,
\end{equation}
where
$\,\Omega_n^{}\!=\phin \!+\DEn$\, and
$\,\DEn \!= \vnt \!-\th_n^{}\,$.
Furthermore, according to \eqrefe{eq:GET}, we reexpress our four-point GRET identity \eqref{eq:GET-ID22} as
\begin{equation}
\label{eq:GET22}
\M \!\[\!h^L_{n_1}h^L_{n_2}\!\ito h^L_{n_3}h^L_{n_4} \!\]
\,=\,
\M\!\[\!\phi_{n_1}^{}\phi_{n_2}^{} \!\ito
\phi_{n_3}^{}\phi_{n_4}^{} \!\]
\,+\, \mO (\DEn) \,.
\end{equation}

As we will show in the following section\,\ref{sec:4.2},
the leading gravitational KK
Goldstone amplitude on the RHS of the GRET \eqref{eq:GET22}
is of $\mO(E^2)$ and equals the corresponding
leading longitudinal KK graviton amplitude
on the LHS of Eq.\eqref{eq:GET22}.
However, it is highly nontrivial to demonstrate that the
full residual term $\,\mO (\DEn)\!=\!\mO(E^0)\,$ actually holds
and thus can be neglected
relative to the leading gravitational KK Goldstone amplitude
on the RHS of the GRET \eqref{eq:GET22}. This is because
the naive power counting shows each individual amplitude in
the residual term $\,\mO (\DEn)\,$ is of $\,\mO (E^2)\,$.
This can be understood by noting that the tensor
$\,\vt^{\mn}\!\!=\!\mO (E^0)$\, and thus the external state
$\,\vnt =\vt^{}_{\mn}h_n^{\mn}\,$
is unsuppressed under high energy expansion.
The same is true for the external state
$\,\th_n^{}=\eta_{\mn}^{}\th_n^{\mn}\,$
which has no extra suppression factor.
Thus, by naive power counting of energy,
each individual residual term $\,\mO (\DEn)\!=\!\mO(E^2)\,$
which has the same energy-dependence as the leading Goldstone
amplitude and is not superficially suppressed.
This is an {\it essential difference} from
the KK GAET\,\cite{5DYM2002} of the compactified 5d KK gauge
theories\,\cite{5DYM2002}, where the
residual term is suppressed by the vector
$\,v^\mu \!= \ep^\mu_L \!- \ep^\mu_S= \mO(M_n/E)\,$
and thus is of $\,\mO(M_n^2/E^2)$\, for the case of
four-particle scattering process
as shown in Eq.\eqref{eq:KK-ET22}.\footnote{%
The residual term of $\mO(v_n^{})$
is defined as the difference between
the longitudinal gauge boson amplitude and the corresponding
Goldstone amplitude.
In the 5d KK GAET for spin-1 KK
gauge bosons\,\cite{5DYM2002},  the residual term has the size
of $\mO(M_n^2/E^2)$, which is similar to that of the
conventional ET of 4d gauge theories\,\cite{ET94b}.}
We will demonstrate this additional energy-cancellation
of $\,E^2\ito E^0\,$ in the residual term $\,\mO (\DEn)\,$
in section\,\ref{sec:4.2} by the explicit calculations
and in section\,\ref{sec:5.4} by the double-copy construction
from the KK GAET of 5d YM theory.

\vspace*{1mm}
\subsection{\hspace*{-3mm}GRET for the 5d Gravitational Scalar QED}
\label{sec:4.1}
In this subsection, we first consider the 5d gravitational scalar QED
(GSQED5) compactified under $S^1/\ZZ$,
as an example to explicitly test the GRET.
This will provide important insights for our general formulation
of the GRET and double-copy reconstruction analysis
in section\,\ref{sec:5}.

\vspace*{1mm}

In this GSQED5,  both graviton and scalar fields live in the 5d bulk. Therefore, we can write down the 5d action for the matter part, including a general gauge-fixing term for the gauge field,
\begin{equation}
S_{\rm{m}} = \int \! \td^{5} x \sqrt{\!-\hat{g}\,}
\LB -\frac{1}{4} \hg^{MP} \hg^{NQ}\hF_{MN}^{} \hF_{PQ}^{}
- \frac{1}{2\zeta}(\pd_M^{} \hA^M)^2
+ |D_M \hat{\SS}|^2+m_0^{2}|\hat{\SS}|^2 \RB ,
\label{eq:5DQED}
\end{equation}
where $\,\hF_{MN} = \pd_M^{} \hA_N^{} - \pd_N^{} \hA_M^{}$
and $\,D_M^{} = \pd_M^{} + \ii \hat{e}
\hat{A}_M^{}$.\footnote{%
In Eq.\eqref{eq:5DQED}, we have imposed a minimal gauge-fixing term
for photon field with gauge-fixing function
$\,(\pd_M^{}\hA^M)\,$. One could optionally choose the usual covariant
gauge-fixing function for photon $\,(\nabla_M^{}\hA^M)\,$
\cite{CoGF},
which contains additional interaction vertices proportional to
$1/\zeta$ and will not affect physics. We have explicitly verified
that for the scattering amplitudes of relevant physical processes,
the sum of all $\zeta$-dependent contributions vanishes at
tree level, as expected.}
From this, we derive the action of the graviton-matter interactions:
\begin{align}
S_{\rm{int}} &\,=\, -\frac{\hka}{2} \int\! \td^5 x  \!
\(\hat{h}^{MN} \hT_{MN}\)
\nn\\
&\,=\,	-\frac{\hka}{4} \int\! \td^5 x \!
\[ 2\hh^{\mn}\hT_{\mn} \!+ 4 \hh^{\mu5}\hT_{\mu5} - \hh^{55}\,
(\tensor{\hT}{^\mu_\mu}\!\!-2\hT_{55})\!\]\!,
\end{align}
where the 5d energy-momentum tensor is defined as
\begin{equation}
\label{eq:SETensor}
\hT_{M N}\,=\, \left.
\frac{2}{\sqrt{-\hat{g}\,}\,}
\frac{\delta S_{\rm{m}}}{\,\delta \hat{g}^{MN}\,}
\right|_{\hg\to\heta}^{}  \,.
\end{equation}
Therefore, we can derive the energy-momentum tensors for both the photon field and scalar field as follows:
\beqs
\begin{align}
\hT_{MN}^{A} \,=\, & \
\frac{\heta_{MN}}{4} \hF_{PQ}^2 + \hF_{M}^{\ \ P} \hF_{PN} + \frac{\heta_{MN} }{2\zeta}  (\pd^{P} \!\hA_{P})^{2} \,,
\\[1.mm]
\hT_{MN}^{\mS} \,=\,& \
(D_M \hSS )^* D_N \hSS + (D_N \hSS)^* D_M \hSS -  \heta_{MN}  (|D_P \hSS |^2 + m_0^2 |\hSS |^2 )  \,.
\end{align}
\eeqs
Then, we make KK expansions for the 5d photon field and scalar
field, under the boundary conditions of the orbifold
$S^1\!/\ZZ$
\beqs
\begin{align}
\hA^{\mu} (x^\nu, x^5) &\,=\,
\frac{1}{\sqrt{L\,}\,}\!\[\! A^{\mu}_{0} (x^\nu)+\sqrt{2}
\sum_{n=1}^{\infty}\! A^{\mu}_{n}(x^\nu)
\cos\!\frac{\,n\pi x^5}{L}\] \!,
\label{eq:AExp}
\\[1mm]
\hA^{5} (x^\nu, x^5) &\,= \,\sqrt{\frac{2}{L}\,}
\sum_{n=1}^{\infty}
A^{5}_{n}(x^\nu) \sin\!\frac{\,n\pi x^5\,}{L}
\,,
\label{eq:A5Exp}
\\[1mm]
\hat\mS (x^\nu, x^5) &\,=\,
\frac{1}{\sqrt{L\,}\,} \!\[  \!
\mS_{0} (x^\nu)+\sqrt{2}\sum_{n=1}^{\infty} \mS_{n} (x^\nu) \cos\!\frac{\,n\pi x^5\,}{L} \] \!.
\end{align}
\eeqs
With these, we can derive the effective KK Lagrangian in 4d
and obtain the corresponding Feynman rules, which are
presented in Appendix\,\ref{app:C}.

\vspace*{1mm}

\begin{figure}[t]
\centering
\includegraphics[height=3.4cm,width=15.6cm]{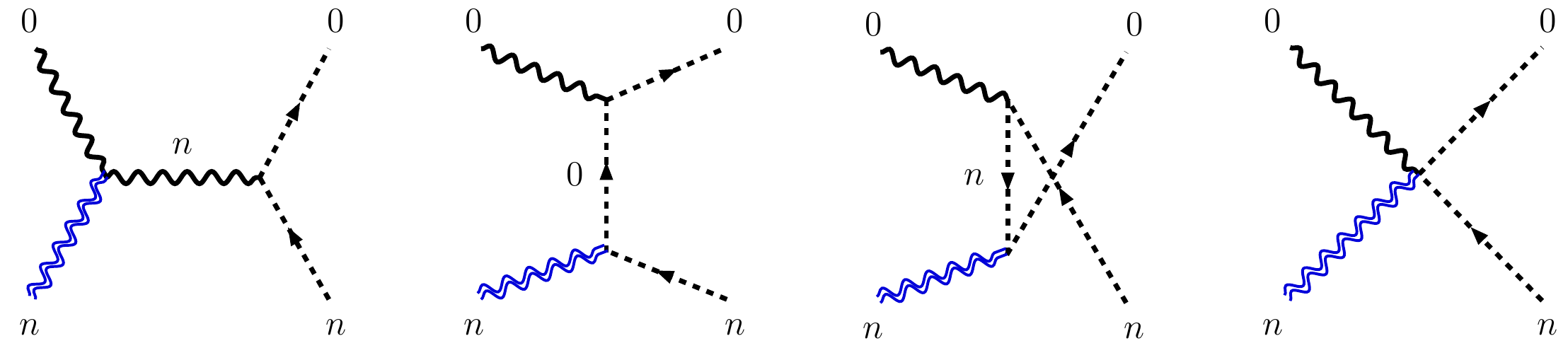}
\caption{\small{Scattering processes of zero-mode photon and
longitudinally (scalar)-polarized KK graviton,
$\,h_n^L A^T_0 \ito \SS^-_0 \SS^+_n$ and
$\,\tilde{h}_n^S A^T_0 \ito \SS^-_0 \SS^+_n$,
via the $(s,t,u)$-channels and contact interactions.
Here the blue double-waved-line denotes the KK graviton
$h_n^{\mn}$, the black waved-line denotes zero-mode
photon $A_0^\mu$\,,
and the black dashed line denotes the zero-mode scalar
$\SS_0^{}$ or KK scalar $\SS_n^{}$.} }
\label{fig:hAff}
\label{fig:1}
\vspace*{1mm}
\end{figure}

To test the GRET explicitly, we consider the scattering
of zero-mode photon and KK graviton into a pair of scalar bosons,
$h_n^L(p_1^{})A_0^T(p_2^{}) \ito
\SS^-_0(p_3^{})\SS^+_n(p_4^{})$
and
$\tilde{h}_n^S(p_1^{})A_0^T (p_2^{}) \ito
\SS^-_0(p_3^{})\SS^+_n(p_4^{})$,
where the initial state KK graviton
is either longitudinally-polarized
$\,h_n^L$
or scalar-polarized
$\,\tilde{h}_n^S$,\,
and the zero-mode photon $\,A_0^T=\ep_\mu^T A^\mu_0$\,
is massless.
The final state includes the zero-mode scalar boson
$\SS^-_0$ and the KK scalar boson $\SS^+_n$.
We present the relevant Feynman diagrams in Fig.\,\ref{fig:1}.

\begin{figure}[t]
\centering
\includegraphics[width=11.2cm]{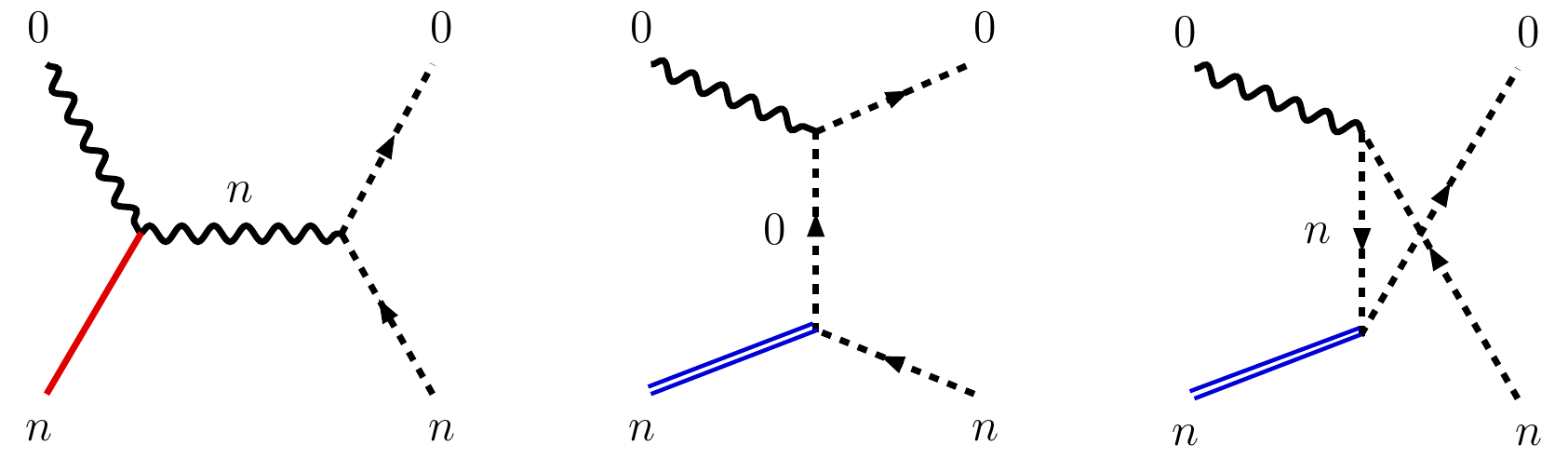}
\caption{\small{Scattering processes of the
zero-mode photon and the KK gravitational Goldstone boson
or the trace-part of KK graviton,
$\,\phi_n^{} A^T_0  \!\ito \SS^-_0 \SS^+_n$
and $\,\tilde{h}_n A^T_0  \!\ito \SS^-_0 \SS^+_n$\,,
via the $(s,t,u)$-channels, where the red solid-line denotes
the KK gravitational scalar Goldstone $\phi_n^{}$, the
blue double-line denotes the trace-part of the KK graviton $\tilde{h}_n$.}
}
\vspace*{1mm}
\label{fig:2}
\end{figure}

We first compute the diagrams in Fig.\,\ref{fig:1}
for the initial state with scalar-polarized KK graviton
$\tilde{h}_n^S$\,.
Thus, the scattering amplitude is derived as
\begin{equation}
\label{eq:T-hnS}
\M [\tilde{h}_n^S] \,=\,
- \sqrt{\frac{3}{2}\,} e\ka\,
(\hs p_3^{} \!\cdot\hsm \ep_2^\pm ) \,.
\end{equation}
Then, we consider the corresponding scattering amplitudes 
$\,\phi_n^{} A^T_0  \ito \SS^-_0 \SS^+_n$
and $\,\tilde{h}_n A^T_0  \ito \SS^-_0 \SS^+_n$,
as shown in Fig.\,\ref{fig:2}.
From Fig.\,\ref{fig:2}, we compute the scattering amplitudes
with initial state KK Goldstone boson
$\phi_n^{}$ and the unphysical trace-part of the KK graviton field $\tilde{h}_n^{}$, respectively.
We further derive their summed scattering amplitude.
Now, these scattering amplitudes are presented as follows:
\beqs
\begin{align}
\M [\phi_n^{}] &=
\frac{1}{2}\sqrt{\frac{2}{3}}\, e\ka \,
(\hs p_3^{} \!\cdot\hsm \ep_2^\pm ) \,, \qquad
\M [\tilde{h}_n] = 2\sqrt{\frac{2}{3}}\, e\ka \,
(\hs p_3^{} \!\cdot\hsm \ep_2^\pm )\,,
\\
\label{eq:T-Omega_n}
\M [\,\bOme_n^{} ] &=
\M [\phi_n^{}]- \M [\tilde{h}_n]
\,=\,  - \sqrt{\frac{3}{2}}\, e\ka \,
(\hs p_3^{} \!\cdot\hsm \ep_2^\pm ) \,,
\end{align}
\eeqs
where the notation
$\,\OmBn \!=\phin\!-\tilde{h}_n$
was introduced in Eq.\eqref{eq:kFmu-all}.
Inspecting the scalar-polarized KK graviton amplitude
\eqref{eq:T-hnS} and the summed amplitude \eqref{eq:T-Omega_n},
we deduce an equality, 
%
\begin{equation}
\label{eq:GET-N=1}
\M [\tilde{h}_n^S] \,=\, \M [\,\bOme_n^{} ]\,,
\end{equation}
which explicitly verifies the GRET identity \eqref{eq:GET-N1}.
We also note that for the current scattering process, the $\bOme_n$-amplitude contains contributions by both the gravitational KK Goldstone boson
$\,\phi_n^{}\,$ and the trace-part of
the KK graviton $\,\tilde{h}_n\,$,\,
which are of the same order of magnitude.
This shows an {\it essential difference} from the case of
the pure KK gauge theories (without gravity),
where for each longitudinal KK gauge boson $A_n^L$, its
corresponding KK Goldstone boson is just given by
the scalar component $A_n^5$ \cite{5DYM2002}.

\vspace*{1mm}

Then, in order to compute the scattering amplitudes explicitly, we choose the momenta in the center-of-mass frame and make the initial state particles move along the $z$-axis.
Then, the momenta for the initial state particles
and final state particles
are given by
\begin{alignat}{3}
p_{1}^{\mu} =&  - \!E (1, 0, 0, \be ) ,
&&  \hspace*{10mm}
p_{2}^{\mu} = - E \be (1, 0, 0, -1 ) ,
\nn \\
p_{3}^{\mu} =&\, E \be  ( 1 , \st, 0,  \ct  ) ,
&& \hspace*{10mm}
p_{4}^{\mu} =  E  ( 1 , - \be \st, 0, -\be \ct ),
\end{alignat}
where $\,\be = \!\sqrt{1\!-\!M_n^2/E^2\,}$ and \,$(\st ,\,\ct )\!=\!(\sin\theta,\,\cos\theta)$ with $\theta$ being the scattering angle. For simplicity of illustration, we consider the zero-mode mass $\,m_0^{}\!\ll\! M_n^{}$ and thus $\,m_0^{}$\, is negligible for this analysis.
The polarization vectors of the KK graviton
$h_n^L(p_1^{})$ and zero-mode photon $A_0^T(p_2^{})$
in the initial state take the following forms:
\begin{align}
\vep_{1L}^{\mn} & =
\frac{1}{\sqrt{6\,}\,} (\ep^{\mu}_{1+}\ep^{\nu}_{1-} \!+
\ep^{\mu}_{1-}\ep^{\nu}_{1+}  \!+ 2\ep^{\mu}_{1L} \ep^{\nu}_{1L}),~~~~
\ep^{\mu}_{1\pm} = \frac{1}{\sqrt{2\,}\,}(0, \pm 1, -\ii,0 ),~~~~~~
\nn\\
\ep_{1L}^{\mu} &= -\frac{E}{M_n}(\be,0,0,1 ),
~~\quad
\ep^{\mu}_{2\pm} = -\frac{1}{\sqrt{2\,}\,}(0, \pm 1, \ii,0 ) \,.
\end{align}

With the above, we compute explicitly the scattering amplitudes of
$\,h_n^LA_0^T \!\ito\SS^-_0\SS^+_n$ and
$\,\tilde{h}_n^SA_0^T \ito\SS^-_0\SS^+_n$
under the high energy expansion:
\beqs
\label{eq:Amp1-hnL-phin}
\begin{align}
\label{eq:Amp1-hnL}
\M[h_n^L] &\,=\,  -\frac{5\sqrt{3}\,e\ka}{6} (E \st ) +\frac{\,\sqrt{3}\,e\ka\,}{6}
\frac{\,M_n^2(4\!-\!\ct)\,}{\,E\tan(\theta/2)\,}
+ \mO (E^{-3}) \,,
\\[1mm]
\label{eq:Amp1-phin}
\M[\phin] &\,=\,
-\frac{\,\sqrt{3}\,e\ka\,}{6} (E \st)  +
\frac{\,\sqrt{3}\,e\ka\,}{12}\frac{\,M_n^2\,\st\,}{E}
+ \mO (E^{-3}) \,,
\end{align}
\eeqs
where we have chosen the transverse polarization
$\ep^{\mu}_{2+}$ for the initial state photon $A_0^T$.\,
For the other transverse polarization
$\ep^{\mu}_{2-}$ of $A_0^T$,\, all of the corresponding
amplitudes will flip an overall sign.

\vspace*{1mm}

According to the GRET identities
\eqref{eq:GET-N1} and \eqref{eq:GET-N1-hLhS},
we can compute the residual term:
\beqa
\label{eq:RT1}
\M [\DEn] \,=\, \M [\vnt]-\M [\th_n^{}]
\,=\,\M [h_n^L]-\M [\phin]\,,
\eeqa
where we have used the abbreviations
$\,\M [\DEn]\!\equiv\!\M [\DEn A_0^T \!\ito\SS^-_0\SS^+_n]$,
$\,\M [\vnt]\!\equiv\!\M [\vnt A_0^T \!\ito\SS^-_0\SS^+_n]$,\,
and
$\,\M [\th_n^{}]\!\equiv\!\M [\th_n^{} A_0^T \!\ito\SS^-_0\SS^+_n]\,$.
Using the longitudinal KK graviton amplitude \eqref{eq:Amp1-hnL}
and KK Goldstone amplitude \eqref{eq:Amp1-phin}, we derive
the residual term \eqref{eq:RT1} as follows:
\begin{equation}
\label{eq:RT1b}
\M [\DEn] \,=\, -\frac{\,2\sqrt{3}\,e\ka\,}{3} (E \st)
+ \mO\!\(\!\frac{\Mnn}{E}\) \!,
\end{equation}
which has the same energy order as the longitudinal KK graviton
amplitude $\M [h_n^L]$\,.
This demonstrates that for the case of one external KK graviton
line, although the GRET identity \eqref{eq:GET-N=1}
holds as expected,
\begin{equation}
\label{eq:GET-hL-GBv-N1}
\M [h_n^L]  \,=\, \M [\,\bOme_n^{} ] + \M [\tilde{v}_n^{}]
\,=\, \M [\phin] + \M [\DEn]
\,,
\end{equation}
the GRET itself no longer holds. This is because the residual term
$\,\M [\DEn]\,$ in Eq.\eqref{eq:RT1b}
has the same order of magnitude as the longitudinal KK graviton
amplitude $\,\M [h_n^L]\,$ or the KK Goldstone amplitude
$\,\M [\phin ]\,$ in Eq.\eqref{eq:Amp1-hnL-phin}
under the high energy expansion.

\vspace*{1mm}
\subsection{\hspace*{-3mm}Gravitational KK Goldstone Boson Scattering Amplitudes}
\label{sec:4.2}
In this subsection, we explicitly compute the elastic and inelastic
scattering amplitudes of four gravitational KK Goldstone bosons
in the compactified 5d GR,
which will be compared quantitatively with the corresponding
longitudinal (helicity-zero) KK graviton scattering amplitudes.
\subsubsection{\hspace*{-3mm}Elastic Gravitational KK
Goldstone Boson Scattering Amplitudes}
\label{sec:4.2.1}
To compute the scattering amplitudes of the gravitational KK Goldstone
bosons, we first derive the relevant interaction vertices.
We will show that the leading contributions arise from the Feynman diagrams
with zero-mode graviton and KK graviton exchanges.
For the cubic interaction vertices containing one graviton and
two KK scalar-Goldstone bosons,
we expand EH Lagrangian up to $\mO(\hat{\ka}^3)$,
denoted as $\,\hat{\La}_1 [\hh\phih^2 ]$\,.
We inspect the structure of  $\,\hat{\La}_1 [ \hh\phih^2 ]$
and classify it into 12  Lorentz-invariant terms,
as presented in Table\,\ref{tab:1}.

\linespread{1.5}
\begin{table}[t]
\centering
\begin{tabular}{l|c|c|c|c|c|c}
\hline\hline
\multirow{2}{*}{$\hat{\La}_1 [\hh\phih^2]$\vspace{-1mm}}	
& {\red $\hh^{\mn} \pd_\mu \phih \, \pd_\nu \phih$}
& {\red $\hh^{\mn} \phih  \, \pd_\mu  \pd_\nu \phih $}
& {\red $\hh\,\pd_\mu \phih \, \pd^\mu \phih$}
& {\red $\hh\, \phih\, \pd_\mu^2 \phih $}
& {\red $\hh \, \pd_5 \phih \, \pd^5 \phih$}
& {\red $\hh \, \phih \, \pd_5^2 \phih $}
\\ \cline{2-7}\cline{2-7}
& $ \pd_\mu \hh^{\mn}\phih \, \pd_\nu \phih$
& $ (\pd_\mu \pd_\nu \hh^{\mn}) \phih^2$
& $\pd_\mu \hh\, \phih \, \pd^\mu \phih$
& $(\pd_\mu^2 \hh)  \phih^2$
& $ \pd_5 \hh \, \phih \, \pd^5 \phih$
& $(\pd_5^2 \hh) \phih^2$
\\ \hline \hline
\end{tabular}
\caption{%
\small \baselineskip 15pt
Classification of the 12 Lorentz-invariant interaction vertices
in $\hat{\La}_1 [\hh\phih^2]$, where the 6 operators in the second row
(black color) can be converted into the combinations of the operators
in the first row (red color) via integration by parts.}
\label{tab:LagHPP}
\label{tab:1}
\end{table}
\baselineskip 18pt


We note that in Table\,\ref{tab:1}
all the 6 operators in the second row
(black color) contain partial derivatives acting on the graviton fields,
but we can always shift the partial derivatives on to the scalar fields
via integration by parts, and thus they can be converted into
combinations of the 6 operators in the first row (red color).
In this way, we can organize the cubic vertices in
the Lagrangian $\,\hat{\La}_1 [\hh\phih^2]$\,
as follows:
\begin{align}
\hat{\La}_1 [ \hh\phih^2  ] \,=\,~&
a_1^{} \, \hh^{\mn} \pd_\mu \phih \, \pd_\nu \phih
+a_2^{} \, \hh^{\mn} \phih \, \pd_\mu \pd_\nu \phih
+a_3^{} \, \hh (\pd_\mu \phih)^2
+a_4^{} \, \hh\,\phih \,\pd_\mu^2 \phih
\nn\\
& + a_5^{} \, \hh (\pd_5 \phih)^2
+ a_6^{} \, \hh \,  \phih \, \pd_5^2 \phih   \,,
\label{eqLagHPP}
\end{align}
where the coefficients are given by
\begin{equation}
(\hs a_1^{},\,a_2^{},\,a_3^{},\,a_4^{},\,a_5^{},\,a_6^{}\hs )
\,=\,
\(\! -\frac{1}{\,2\,},\, -1,\,\frac{3}{\,4\,},\,1,\,
-\frac{1}{\,2\,},\,-\frac{1}{\,2\,}\hs\) \!.
\label{eqCoeValue}
\end{equation}

\begin{figure}[t]
\begin{center}
\includegraphics[height=3.6cm]{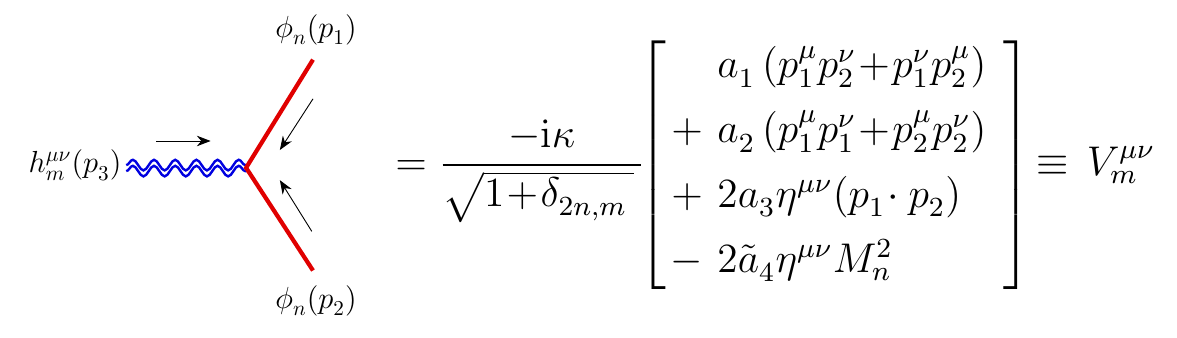}
\vspace*{-3mm}
\caption{\small\baselineskip 15pt{%
Feynman rule for the cubic interaction vertex between KK graviton
and gravitational KK Goldstone bosons,
where we define,
$\,\tilde{a}_4 = a_4^{} + (-1)^{\delta_{2n,m}} a_5^{}
-a_6^{}\hs$, with $m\!=0,\,2n$\,.}}
\label{fig:VertexHPP}
\label{fig:3new}
\end{center}
\end{figure}

Next, by substituting Eqs.\eqref{eqHExp}-\eqref{eqPhiExp} into
the Lagrangian \eqref{eqLagHPP} and integrating over $x^5$,
we derive the corresponding effective Lagrangian in 4d,
\begin{align}
&\La_1 [ h\phi^2]  = \frac{\ka}{\sqrt{2}}  \sum_{n,m,\ell = 1}^{\infty}  \!\biggl\{ \!
\nn\\
& a_1^{} \! \left[\! \sqrt{2} \(  h_0^{\mn}  \pd_\mu \phi_0  \pd_\nu \phi_0  + h_0^{\mn}\pd_\mu \phi_m \pd_\nu \phi_{\ell}  \delta_{m\ell} +  h_n^{\mn} \pd_\mu \phi_m  \pd_\nu \phi_0  \delta_{n m} +  h_n^{\mn}\pd_\mu \phi_0 \pd_\nu \phi_{\ell}  \delta_{n\ell} \) \right.  \nn\\
& 
\left. +  \,  h_n^{\mn} \pd_\mu \phi_m \pd_\nu\phi_{\ell}  \Delta_3(n,m,\ell) \right]  
+ a_2^{} \left[\!\sqrt{2} \(  h_0^{\mn}  \phi_0  \pd_\mu\pd_\nu \phi_0 + h_0^{\mn} \phi_m  \pd_\mu\pd_\nu \phi_{\ell}  \delta_{m\ell} +  h_n^{\mn}  \phi_m  \pd_\mu\pd_\nu \phi_0  \delta_{nm} \right. \right.  \nn\\[3pt]
&
\left. \left. +\,  h_n^{\mn}  \phi_0  \pd_\mu\pd_\nu \phi_{\ell}  \delta_{n\ell} \) +  h_n^{\mn} \phi_m \pd_\mu\pd_\nu \phi_{\ell} \Delta_3(n,m,\ell) \right]
+ a_3^{} \[\!\!\sqrt{2} \( h_0 \pd_\mu \phi_0 \pd^\mu \phi_0  + h_0 \pd_\mu \phi_m \pd^\mu \phi_{\ell} \delta_{m\ell}
\right.\right. \nn\\[3pt]
& \left.\left. +\,  h_n \pd_\mu \phi_m \pd^\mu \phi_0 \delta_{n m} +  h_n \pd_\mu \phi_0 \pd^\mu \phi_{\ell}  \delta_{n\ell} \)  +  h_n \pd_\mu \phi_m \pd^\mu\phi_{\ell}  \Delta_3(n,m,\ell) \right]
+a_4^{} \left[\!\sqrt{2} \( h_0  \phi_0  \pd^2_\mu \phi_0 \right.\right.
\nn  \\[3pt]
&\left. \left.  +\, h_0  \phi_m \pd^2_\mu \phi_{\ell} \delta_{m\ell} + h_{n} \phi_m  \pd^2_\mu\phi_0  \delta_{nm} + h_{n}   \phi_0  \pd^2_\mu \phi_{\ell}  \delta_{n\ell}  \)
+ h_{n} \phi_m  \pd^2_\mu \phi_{\ell} \Delta_3(n,m,\ell)\right]
\nn\\[4pt]
& + a_5^{} M_m M_\ell \[\!\!\sqrt{2} h_0 \phi_m \phi_{\ell} \delta_{m\ell}   h_n\phi_m\phi_{\ell} \widetilde{\Delta}_3(n,m,\ell) \!\]
- a_6^{} M_\ell^2 \[\!\!\sqrt{2} \( h_0 \phi_m  \phi_{\ell}
\delta_{m\ell} + h_n \phi_0 \phi_{\ell} \delta_{n\ell} \) \right.
\nn\\[-1pt]
&\left.  +\,   h_n\phi_m\phi_{\ell} \Delta_3 (n,m,\ell) \]\biggr\},
\label{eqLagHPPKK}
\end{align}
where $\Delta_3(n,m,\ell)$ and $\widetilde{ \Delta}_3(n,m,\ell)$ are given by
\beqs
\begin{align}
\Delta_3(n,m,\ell)  & \,=\, \delta(n+m-\ell)
+\delta(n-m-\ell)+\delta(n-m+\ell)\,,
\\[1mm]
\widetilde{\Delta}_3(n,m,\ell) & \,=\, \delta(n+m-\ell)-\delta(n-m-\ell)+\delta(n-m+\ell)  \,.
\end{align}
\eeqs
Hence, using \eqrefe{eqLagHPPKK}, we can derive the Feynman rule for graviton-scalar-scalar interactions
as shown in Fig.\,\ref{fig:3new}.

\begin{figure}[t]
\begin{center}
\includegraphics[height=3.4cm,width=12.3cm]{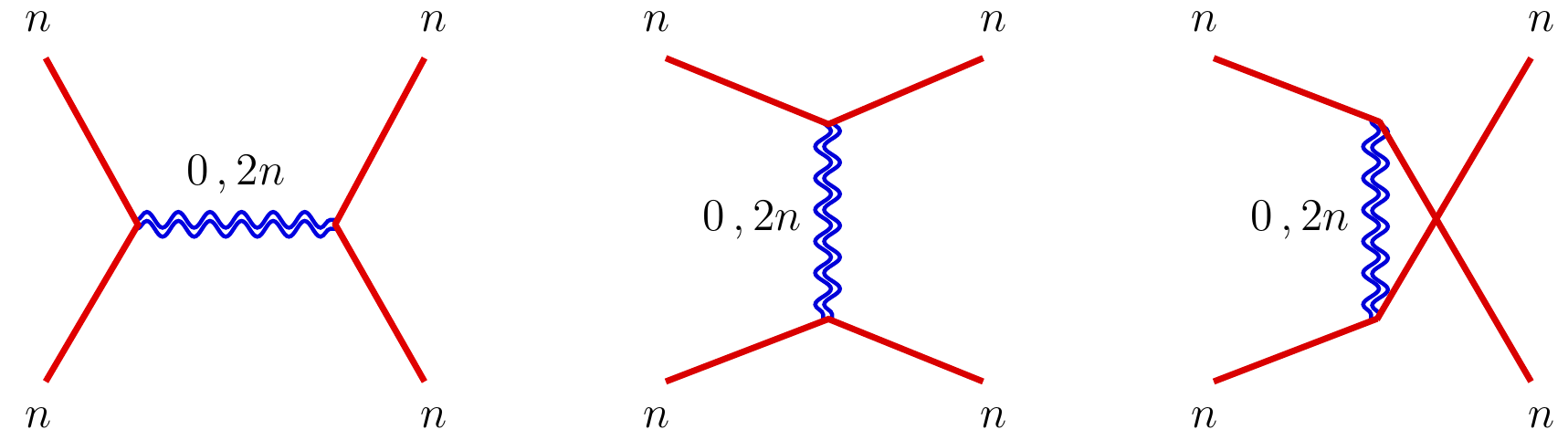}
\caption{\small\baselineskip 15pt{%
Elastic scattering of gravitational KK Goldstone bosons,
\,$\pnnnn$\,,
via $(s,\,t,\,u)$-channels mediated
by the zero-mode graviton and by the KK graviton of level-$2n$.}}
\label{fig:nn-nn-h}
\label{fig:3}
\vspace*{-3mm}
\end{center}
\end{figure}

With the above,
we are ready to analyze the elastic scattering of the
gravitational KK Goldstone bosons,
\,$\phi_n\phi_n \ito \phi_n\phi_n$\,.
Fig.\,\ref{fig:3} shows the Feynman diagrams at the tree level,
which include the scattering via
the zero-mode graviton exchange and
the KK graviton exchange at level-$2n$.
By straightforward power counting, we find that each diagram
in Fig.\,\ref{fig:3} has the leading contribution of
$\,\mO(E^2)\,$ in the high energy limit.
We stress that {\it our gravitational KK Goldstone boson scattering
amplitudes in our study do not invoke any energy cancellation
among the individual diagrams and the leading energy dependence
of $\mO(E^2)$ is manifest in each diagram.}
This feature is an {\it essential difference from the
longitudinal KK graviton amplitudes} which involve a complicated
large energy cancellations from $\mO(E^{10})$ to $\mO(E^2)$
as in \cite{Chivukula:2020S}\cite{Chivukula:2020L}.
In fact, as we will demonstrate, our formulation of the GRET
(section\,\ref{sec:3}) together with the double-copy construction
(section\,\ref{sec:5}) can provide a general mechanism for these
large energy cancellations.

\vspace*{1mm}

By using the trilinear interaction vertices
Fig.\,\ref{fig:VertexHPP} and
the KK graviton propagator \eqref{eq:Prophh-xi=1}
as well as the kinematics defined in Appendix\,\ref{app:A},
we can compute all the Feynman diagrams of Fig.\,\ref{fig:3}
in a straightforward way.
Summing up the individual diagrams,
we derive the elastic scattering amplitude of
\,$\pnnnn$\, to the leading order (LO)
of $\mO(E^2)$ under the high energy expansion:
\beqs
\label{eq:AmpLOE2-phinnnn}
\begin{align}
\label{eq:AmpE2-phinnnn}
\M[\pnnnn] &~=~
\frac{\,3\ka^2\,}{\,32}\!
\[\!\!\frac{~(3+ \cos^2\!\theta)^2\,}{\sin^2\!\theta}\!\!\]\! \sz \,,
\\[2mm]
&~=~
\frac{\,3\ka^2\,}{\,128}\!
\[\!\!\frac{~(7\!+ \cos 2\theta)^2\,}{\sin^2\!\theta}\!\!\]\!
\sz\,.
\label{eq:AmpE2-phinnnn2}
\end{align}
\label{eq:GAmpE2-phinnnn}
\eeqs
Then, the expansion to the next-to-leading order (NLO) gives
the subleading amplitude:
\begin{equation}
\label{eq:phi-nnnn-NLO}
\dM [\pnnnn] \,=\,
-\frac{\,\ka^2\Mnn\,}{128}(-1318 +2865 \ctt\! -522 \ctf\! - \cts) \csc^4\!\theta \,,
\hspace*{10mm}
\end{equation}
which is {\it mass-dependent} contribution of $\,\mO(E^0\Mnn)\,$.
We see that this NLO amplitude \eqref{eq:phi-nnnn-NLO}
is much smaller than the LO amplitude
\eqref{eq:AmpLOE2-phinnnn} of $\mO(E^2M_n^0)$
in the high energy scattering.

\vspace*{1mm}

In order to explicitly demonstrate our GRET, we will first compare
our gravitational KK Goldstone boson amplitude
\eqref{eq:AmpE2-phinnnn2}
with the corresponding longitudinal KK graviton amplitude
$\M [\hLnnnn]$ as given in
Ref.\,\cite{Chivukula:2020L} (cf.\ its Eq.(70)).
For this comparison, we note a notational difference:
our 4d gravitational coupling constant $\ka$
is defined in Eq.\eqref{eqNewtonConst5D} as
$\,\ka \!=\! {\hka}/{\sqrt{L\,}}$\, and differs from that of
Ref.\,\cite{Chivukula:2020L} by a factor
$\,\fr{1}{\sqrt{2}}\,$ since their definition leads to
$\,\ka \!=\! \hat{\ka}/\!\sqrt{2L\,}$.
Hence, our KK Goldstone amplitude \eqref{eq:AmpE2-phinnnn2}
should be rescaled by a factor $\fr{1}{2}$ for the comparison:
\begin{align}
\M  \longrightarrow
\M  \times \frac{1}{2}
=\frac{3 \ka^2}{\,256}\!
\[\!\! \frac{\,(7\!+ \cos 2\theta)^2\,}{\sin^2\!\theta} \!\!\]\!
\sz \,,
\end{align}
which equals the KK graviton amplitude in its Eq.(70) of
Ref.\,\cite{Chivukula:2020L}.
This is truly impressive because our independent computation
of the KK Goldstone amplitude \eqref{eq:AmpE2-phinnnn2} fully
differs from that of the KK graviton amplitude
which contains much more complicated energy cancellations
from $\mO(E^{10})$ to $\mO(E^2)$.
Naively and intuitively, this equivalence seems quite expected
for us because the scalar component of the KK graviton field
$\,{\phi_n^{}}\,(\equiv\! {h}^{55}_n)$
should be converted to the degree of freedom of
the helicity-zero longitudinal component of the KK graviton,
and thus we would have
\begin{equation}
\label{eq:GET-hL-phi}
\M [\hLnnnn] \,=\,
\M [\pnnnn]  +\mO(M_n^2 E^0)\,.
\end{equation}
However, in the actual situation it is far more nontrivial to
quantitatively demonstrate the equivalence between the two
amplitudes in the high energy limit. This is because
our quantative formulation of the GRET \eqref{eq:GET22}
(as systematically presented in section\,\ref{sec:3})
shows that the second term on the RHS
of the GRET contains a combination of both the KK Goldstone bosons
$\,\phi_n^{}\,$ and trace-part of graviton $\tilde{h}_n$\,
due to the structure of our $R_\xi^{}$ gauge-fixing functions
in Eqs.\eqref{eq:Fmu}-\eqref{eq:F5} and \eqref{eq:kFmu}.
To fully demonstrate such an equivalence as in Eq.\eqref{eq:GET-hL-phi},
we have to further show that all the $\tilde{h}_n$-related Goldstone amplitudes
on the RHS of the GRET \eqref{eq:GET22} together with
the $\mO(\tilde{v}_n^{})$ amplitudes could be of $\mO(M_n^2E^0)$
at most. We will present this nontrivial demonstration
in section\,\ref{sec:5} based on our double-copy construction.

\vspace*{1mm}

\begin{figure}[t]
\centering
\includegraphics[height=7cm]{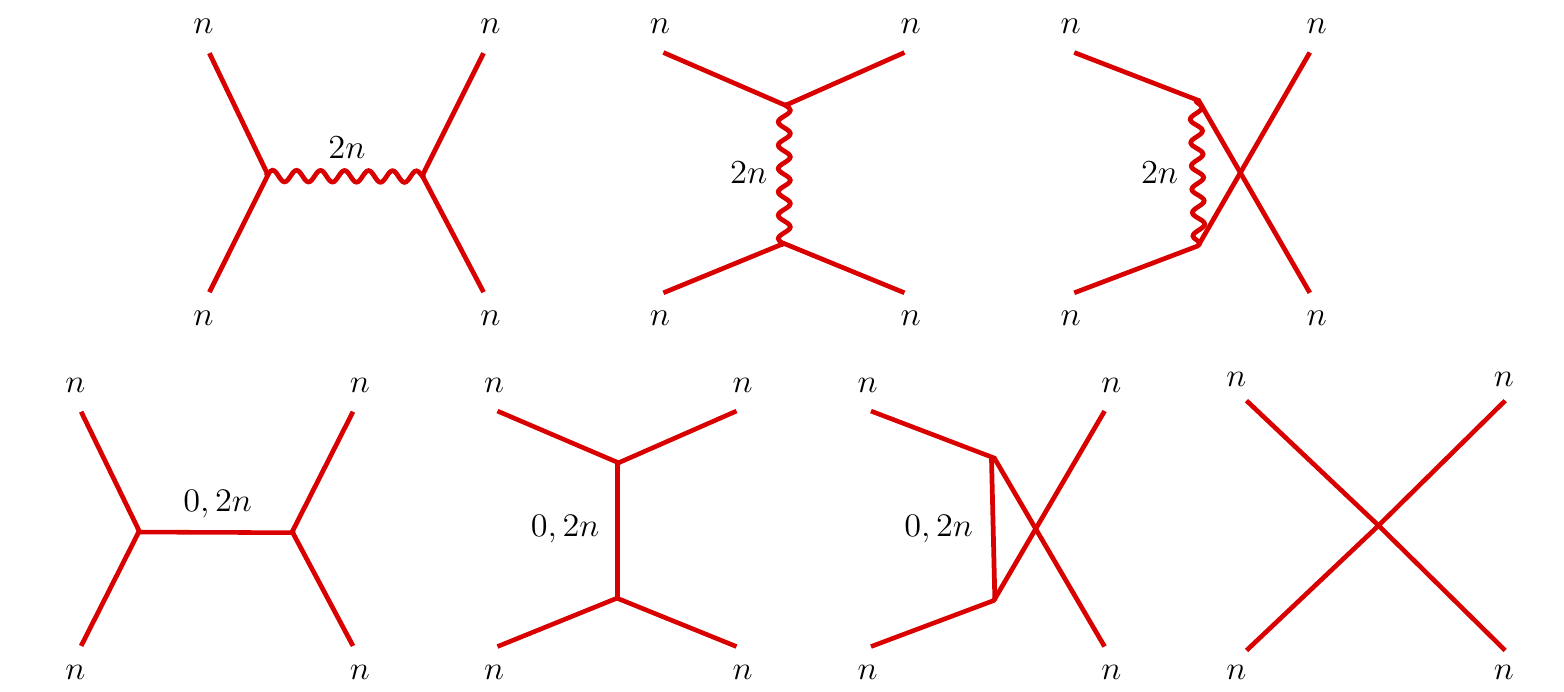}
\caption{\small\baselineskip 15pt
Gravitational KK Goldstone boson scattering
$\,\pnnnn$ from Feynman diagrams of subleading contributions.
The diagrams in the first row arise from exchanging
the KK vector $\A_{2n}^\mu$ via $(s,t,u)$-channels, while the
diagrams in the second row arise from exchanging both the zero-mode
and KK scalar ($\phi_0^{}\,, \phi_{2n}^{}$) via $(s,t,u)$-channels
and a KK scalar-Goldstone contact interaction.}
\label{fig:4}
\vspace{3mm}
\end{figure}

Next, we compute the subleading contributions to the elastic
KK Goldstone amplitude $\,\phin\phin\ito \phin\phin\,$
as shown in Fig.\,\ref{fig:4}, where the relevant Feynman rules
are presented in Appendix\,\ref{app:D}.
These include the subleading contributions via $(s,t,u)$-channels
mediated by a vector $\A_{2n}^\mu$ (the first row), a scalar
$\phi_0^{}$ or $\phi_{2n}$ (the second row), and a contact interaction
(the second row). Thus, we derive the following three kinds of subleading contributions accordingly under the high energy expansion:
\\[-7mm]
\beqs
\label{eq:AmpExGoldstone}
\begin{align}
\M_{\A}^{}[\phin\phin \ito \phin\phin]
&~=~ \frac{\,3\,}{4} \ka^2 M_n^2\,,
\\[1.5mm]
\M_{\phi}^{}[\phin\phin \ito \phin\phin]
&~=~ -18\ka^2 M_n^2\,,
\\[1mm]
\M_{c}^{}[\phin\phin  \ito \phin\phin]
&~=~ \frac{\,9\,}{2}  \ka^2 M_n^2 \,.
\end{align}
\eeqs
Their sum is given by
\begin{equation}
\label{eq:fig4abc-sum}
\M_{\A}^{} + \M_{\phi}^{}+\M_c^{} \,=\,
-\frac{\,51\,}{4}\ka^2\Mnn \,.
\end{equation}
We see that the above subleading contributions are all
of $\mO(E^0M_n^2)$. The same feature also holds for
the subleading contributions to the inelastic channels.

\vspace*{1mm}

Finally, we sum up the contributions of both
Fig.\,\ref{fig:3} and Fig.\,\ref{fig:4},
and derive the complete elastic scattering amplitude
of KK scalar-Goldstone bosons without energy expansion:
\begin{equation}
\label{eq:Amp-4phinnnn-full}
\M[\phin\phin\!\ito\phin\phin] \,=\,
\frac{\,\ka^2 M_n^2(\tX^0_0 \!+\! \tX^0_2\ctt \!+\! \tX^0_4\ctf
\!+\! \tX^0_6\cts )}
{~512\bs_0^{}(\bs_0^{}\!+\!4)
\[2\st^2\bs_0^2 \!+\! 32\bs_0^{} \!+\! 128\]\!\st^2~} \,,
\hspace*{10mm}
\end{equation}
where $\,\bs_0^{}=\sz /M_n^2$\,,\,
$\bs =s/M_n^2=\bs_0^{}\!+\!4\,$,\, and
\beqs
\begin{align}
\widetilde{X}^0_0 &= 2 (255 \bs_0^5 +\! 13348 \bs_0^4 + 168624 \bs_0^3 \!+\! 984384 \bs_0^2 + 3514368 \bs_0 + 6012928) \hs ,
\\[1mm]
\widetilde{X}^0_2 &= -429 \bs_0^5 - 12732 \bs_0^4 - 156288 \bs_0^3-777728 \bs_0^2-2572288 \bs_0 - 5210112 \hs ,
\\[1mm]
\widetilde{X}^0_4 &= -2 (39 \bs_0^5 - 1212 \bs_0^4 - 7824 \bs_0^3 - 10688 \bs_0^2) \hs ,
\\[1mm]
\tX^0_6 &\,=\, -\(3\bs_0^5+\!4\bs_0^4\) \!.
\end{align}
\eeqs
From the above,
we expand the full amplitude
\eqref{eq:Amp-4phinnnn-full}
$(\M[\pnnnn]\equiv\M[4\phin])$
down to the subleading order under the high energy expansion
$\,\sz\!\gg\! M_n^2\,$ (or $\bs_0^{}\!\gg\! 1$),\footnote{%
\baselineskip 15pt
As a clarification of the notations,
in sections\,\ref{sec:3}-\ref{sec:4} we do not
put an extra ``tilde" symbol above the $\phin$-amplitude
$\M\,$ and $\dM\,$
such as those in Eqs.\eqref{eq:Amp-4phinnnn-full} and
\eqref{eq:AmpGR5-4phi-LONLO},
but we will add a ``tilde" on top of the same $\phin$-amplitude
symbols such as $\,\MT\,$ and $\,\dMT\,$
in section\,\ref{sec:5} as well as in Appendix\,\ref{app:F}
for the convenience of notations.}
\beqs
\label{eq:AmpGR5-4phi-LONLO}
\begin{align}
\label{eq:AmpGR5-4pi-nnnn}
\M[4\phin] &\,=\, \M_0[4\phin] + \delta\M[4\phin]\,,
\\[1mm]
\label{eq:AmpGR5-LO-4pi-nnnn}
\M_0[4\phin] &\,=\,\frac{\,3\ka^2\,}{\,128\,}\!
\[\!\! \frac{\,(7\!+ \cos 2\theta)^2\,}{\sin^2\!\theta} \!\!\]\!
\sz \,,
\\[1mm]
\label{eq:AmpGR5-NLO-4pi-nnnn}
\delta\M[4\phin] & \,=\, -
\frac{\,\ka^2M_n^2\,}{128}
(-706 +2049 \ctt\! -318 \ctf\! -\cts)\hsm\csc^4\!\theta \,.
\end{align}
\eeqs
We see that the above leading amplitude
$\,\M_0^{}[4\phin]\hsm =\hsm\mO(E^2M_n^0)$\,
is mass-independent and agrees with
Eq.\eqref{eq:AmpLOE2-phinnnn}, while the subleading
amplitude $\,\delta\M[4\phin]\hsm =\mO(M_n^2E^0)\,$
is mass-dependent. As a consistency check, we also note that
the above subleading amplitude $\,\delta\M[4\phin]\,$
just equals the sum of the two NLO amplitudes
\eqref{eq:phi-nnnn-NLO} and \eqref{eq:fig4abc-sum}
which are computed earlier.

\subsubsection{\hspace*{-3mm}Inelastic Gravitational KK Goldstone Boson Scattering Amplitudes}
\label{sec:4.2.2}

In this subsection, we further analyze the inelastic scattering
processes for the gravitational KK Goldstone bosons.
Based on the analysis of the previous section, we have demonstrated
that the longitudinal-Goldstone equivalence
\eqref{eq:GET-hL-phi} holds down to $\mO (E^2)$
under the high energy expansion, which is equivalent to taking
the high energy limit $\,M_n/E \ito 0$\,.

\vspace*{1mm}

From the trilinear interaction vertex Fig.\,\ref{fig:VertexHPP},
we can deduce a relation between the
$\,h_0^{\mn}$-$\phin$-$\phin\,$ coupling ($V^{\mn}_{0}$) and
$\,h_{2n}^{\mn}$-$\phin$-$\phin\,$ coupling ($V^{\mn}_{2n}$):
\begin{equation}
V^{\mn}_{0} = \sqrt{2} \,V^{\mn}_{2n}   \,.
\end{equation}
Thus, for each channel of the elastic scattering process,
the corresponding amplitudes with the exchanges of zero-mode graviton
$h_0^{\mn}$ and KK graviton $h_{2n}^{\mn}$ are connected by
the relation:
$\,\M_j^{2n}\!=\!\fr{1}{2}\M_j^{0}\,$.\,
Hence, for a given channel-$j$\,, we have
$\,\M_j^{}\!=\M_j^{0}\!+\!\M_j^{2n}\!=\fr{3}{2}\M_j^{0}\,$
in the high energy limit $\,M_n/E \ito 0$\,.
With these, we can reproduce the elastic KK Goldstone scattering
amplitude \eqref{eq:AmpLOE2-phinnnn} by
\begin{equation}
\M [\pnnnn]
\,= \, \frac{3}{2}\,  \sum_{j} \!\M_{j}^{0} \,,
\end{equation}
where $\,j\in (s,\,t,\,u)$.\,
The above amplitude $\,\M_j^{0}\,$ arises from the exchange of
zero-mode graviton and is given by
\begin{equation}
\label{eq:Mj0}
\M_j^{0} ~=\, -\ii\,
V_{\mn}^{\,0} \, \D^{\mn\ab}_{00}\, V_{\ab}^{\,0} \,.
\end{equation}

\begin{figure}[t]
\centering
\includegraphics[height=3.4cm,width=12.3cm]{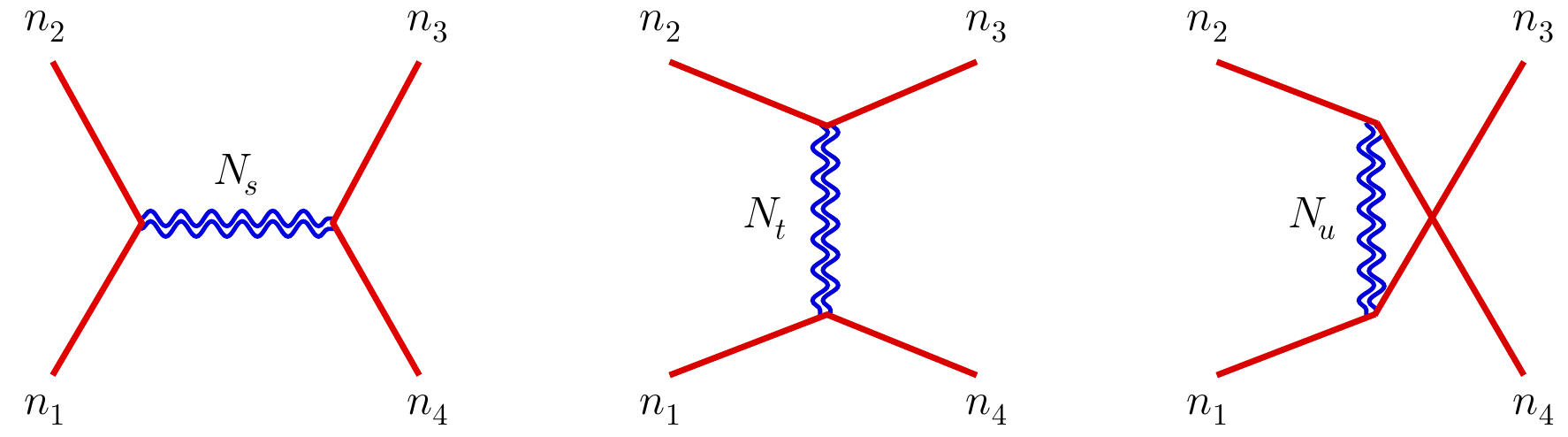}
\caption{\small\baselineskip 15pt
{General four-point scattering process of the gravitational
KK Goldstone bosons,
\,$\phi_{n_1}^{}\phi_{n_2}^{} \!\ito \phi_{n_3}^{}\phi_{n_4}^{}$\,,
via $(s,\,t,\,u)$-channels mediated
by a KK graviton of level $N_{s_j^{}}$, where
$N_{s_j^{}}\!\!\geqq 0$ and $s_j^{} \in (s,t,u)$.}}
\label{fig:inel-h}
\label{fig:5}
\vspace*{3mm}
\end{figure}

With the above, we can extend our analysis of the
elastic scattering amplitude to a general case as shown
in Fig.\,\ref{fig:inel-h},
including all the inelastic scattering channels.
In Fig.\,\ref{fig:inel-h}, the external KK Goldstone bosons have
KK-levels of $(n_1,\,n_2^{},\,n_3^{},\,n_4^{})$,
and we denote the intermediate graviton
with levels $(N_s,\, N_t,\,N_u)\!\geqq 0$, respectively.

\vspace*{2mm}

In the following, we consider two types of the
inelasic scattering processes:
\\[-9mm]
\begin{enumerate}
\item[{\bf (i)}]
For the inelasic scattering
$\pnnmm$
(with $n \!\neq\! m$), we have
\begin{align}
& n_1^{}=n_2^{}=n \,, \quad  n_3^{}=n_4^{}=m \,,
\nn\\
& N_s =0\,, \quad N_t=N_u=|n \pm m| \,,
\end{align}
where only the $s$-channel diagram includes the exchange of zero-mode
graviton because of KK number conservation. With these, we
compute the inelastic KK graviton scattering amplitude
in the high energy limit as follows:
\begin{align}
\M [\pnnmm]
&~=~ \M_{s}^0 +2\times\fr{1}{2}
\(\M_{t}^0 +\M_{u}^0 \,\)
\nn\\[1mm]
&~=~ \fr{2}{3} \, \M [\pnnnn]  \,,
\label{eq:AmpNNMM}
\end{align}
where $\,\M_j^0\,$ is defined in Eq.\eqref{eq:Mj0}
and equals the elastic amplitude of $\,h_0^{\mn}$ exchange
in the channel-$j$\,.

\item[{\bf (ii)}]
For the inelasic scattering
$\pnkml$
(with $n \neq k \neq m\neq \ell$), we have
\begin{align}
& n_1^{}=n \,, \quad n_2^{}=k\,, \quad
n_3^{}=m\,, \quad n_4^{}=\ell \,,
\nn\\
& N_s = |n \pm k| = |m \pm \ell|\,, \quad
N_t = |n \pm \ell| = |k \pm m| \,, \quad
N_u = |n \pm m| = |k \pm \ell| \,. \quad
\end{align}
In this case, the process of exchanging zero-mode graviton
is prohibited because of the KK number conservation,
while the process by exchanging the relevant KK gravitons
is allowed via $(s,\,t,\,u)$-channels. Thus, we have
\begin{align}
\M [\pnkml]
&~=~
\fr{1}{2}  \( \M_s^0 + \M_t^0 +\M_u^0 \)
\nn\\[1mm]
&~=~ \fr{1}{3} \, \M [\pnnnn]  \,. \label{eq:AmpNMLK}
\end{align}
\end{enumerate}
As we checked, our above inelastic KK Goldstone boson amplitudes
\eqref{eq:AmpNNMM} and \eqref{eq:AmpNMLK} also equal
the inelastic longitudinal KK graviton
amplitudes\,\cite{Chivukula:2020L} (cf.\ its Eq.(76))
after taking into account the notation difference.

\section{\hspace{-3mm}Construction of Gravitational KK Amplitudes from\\
\hspace*{-4.5mm}
Gauge KK Amplitudes with Double-Copy}
\label{sec:5}

In this section, we study the double-copy construction of
the massive gravitational KK scattering amplitudes from
the corresponding massive gauge KK scattering amplitudes
{\it under the high energy expansion.}
The conventional double-copy approaches (such as \cite{BCJ:2008}\cite{BCJ:2019}) are realized for massless
gauge theories and massless GR. The extension to
the massive YM theory and massive Fierz-Pauli gravity
is difficult without modification\,\cite{dRGT}.
We stress that the KK YM gauge theory and KK GR are truly distinctive
because they can consistently generate masses for KK gauge bosons and
KK gravitons via geometric ``Higgs'' mechanism
(under compactification) as shown in
our sections\,\ref{sec:2}-\ref{sec:3} and in
Refs.\,\cite{5DYM2002}\cite{KK-ET-He2004}\cite{GHiggs}\cite{GHiggs2}.
Hence, we expect that extending the conventional
double-copy method to the KK theories should be truly promising
even though highly challenging due to the KK mass-poles in
the scattering amplitudes. Unlike the conventional double-copy
approaches in the literature, we propose to realize the
double-copy construction {\it by using the high energy expansion
order by order,} and we will demonstrate explicitly
how such a double-copy construction can work up to
the leading order (LO) and the next-to-leading order (NLO).
We are well motivated to use this high energy expansion approach
for realizing the double-copy construction also because it
perfectly matches our KK GAET and GRET formulations.
So it should appropriately reconstruct the GRET based upon the
KK GAET.
Under the high energy expansion, we find that
the LO KK gauge boson (Goldstone) amplitudes and KK graviton
(Goldstone) amplitudes are {\it mass-independent}, so we can
directly realize the double-copy construction of the LO KK
amplitudes. Then, we show that the gauge and gravitational
KK scattering amplitudes at the NLO are {\it mass-dependent}.
We find that the double-copy construction for the mass-dependent
NLO KK-amplitudes is highly nontrivial, where the conventional
double-copy methods (such as BCJ\,\cite{BCJ:2008}\cite{BCJ:2019})
could not fully work. We will present an improved BCJ-type
double-copy construction for the KK gauge and gravitational
amplitudes at the NLO.

\vspace*{1mm}

In section\,\ref{sec:5.1}, we will first analyze the structure of
KK scattering amplitudes for the compactified
5d KK YM gauge theories without gravity.
We present the exact tree-level four-particle scattering amplitudes
of the KK longitudinal gauge bosons ($A_L^{an}$)
and of the corresponding KK Goldstone bosons  ($A_5^{an}$).
With these, we analyze the structure of the KK $A_L^{an}$-amplitudes
and KK $A_5^{an}$-amplitudes
at both the LO and NLO under the high energy expansion.
We show explicitly that the BCJ-type numerators hold the
kinematic Jacobi identity for the LO KK-amplitudes, but the
numerators of the NLO KK-amplitudes do not.
Then, we show that the NLO numerators can be properly improved
to obey the kinematic Jacobi identities.
We also show explicitly how
the KK equivalence theorem for gauge theory (KK GAET)\,\cite{5DYM2002}
is realized in such KK YM gauge theories.
Then, in section\,\ref{sec:5.2},
we demonstrate that the scattering amplitudes
of massive longitudinal KK gravitons ($h_L^n$) and
the amplitudes of their KK Goldstone bosons
($\phin$) in the 5d KK GR can be reconstructed from the
corresponding scattering amplitudes of the
massive longitudinal KK gauge bosons and KK Goldstone bosons
in the 5d KK YM gauge theory
by using the double-copy method at the LO of
the high energy expansion, where the reconstructed LO
KK-amplitudes of $h_L^{n}$ and of $\phin$
have $\mO(E^2M_n^0)$ and are {\it mass-independent.}
The reconstructed NLO gravitational KK-amplitudes have
$\mO(E^0M_n^2)$ and are {\it mass-dependent.}
We find that their double-copy construction is highly nontrivial.
In section\,\ref{sec:5.3}, we show that by direct extension
of the double-copy method to the NLO KK amplitudes, we
can reconstruct the correct kinematic structure of the
KK $h_L^{n}$-amplitude and $\phin$-amplitude,
but not their exact cofficients. For the difference
between the $h_L^{n}$-amplitude and $\phin$-amplitude,
such a naive extension fails to reproduce even the
correct structure in the original gravitational
amplitude-difference at the NLO. We will present an improved
method to realize the correct structure of the NLO gravitational
amplitude-difference, and then further demonstrate
how to fully reconstruct the exact KK $h_L^{n}$-amplitude and $\phin$-amplitude separately.
In section\,\ref{sec:5.4}, we apply the double-copy approach
of sections\,\ref{sec:5.2}-\ref{sec:5.3} to reconstruct the
residual term of the KK GRET and show it has $\mO(E^0M_n^2)$
and is indeed suppressed relatively to the leading KK
Goldstone $\phin$-amplitude. In this way,
we can build the KK GRET in the 5d KK GR theory
from the KK GAET in the 5d KK YM gauge theory.

\subsection{\hspace*{-2mm}Structure of Amplitudes
for KK Gauge Bosons and Goldstone Bosons}
\label{sec:5.1}

Consider a non-Abelian gauge group $\mathcal{G}$,
such as $\mathcal{G}=\rm{SU(N)}$,
with group structure constant $C^{abc}$.
For convenience, we denote the products of two structure constants
as

\begin{equation}
\(\, \CC_s,\, \CC_t,\, \CC_u \,\)  \equiv \(  C^{abe}C^{cde},\, C^{ade}C^{bce},\,  C^{ace}C^{dbe} \) .
\end{equation}
Thus, the Jacobi identity for the group structure constants takes
the following form:
\begin{equation}
\CC_s + \CC_t + \CC_u \,=\, 0\,.
\label{eq:Jacobi-C}
\end{equation}
This maybe called the ``color'' Jacobi identity
since it contains the gauge group's structure constants only.

\vspace*{1mm}

We compactify a 5d YM gauge theory on $S^1\!/\mathbb{Z}_2^{}$.
This 5d compactification leads to
a geometric ``Higgs'' mechanism\,\cite{5DYM2002}
for the KK gauge boson mass-generation, where the longitudinal
KK gauge boson $A^{an}_L$ arises from absorbing the fifth component
of the KK state $A^{an}_5$.
We start with the elastic scattering of longitudinal KK gauge bosons
$\ALnnnn$
and the elastic scattering of the corresponding KK Goldstone bosons
$\Afnnnn$.
For the KK Goldstone amplitude, we choose the Feynman-'t\,Hooft gauge
under which each KK Goldstone boson $A^{an}_5$ has the same mass
$M_n^{}$ as the KK gauge boson $A^{an}_\mu$.
In the center-of-mass frame of the four-particle elastic scattering,
we recall the kinematic variables defined in \eqrefe{eq:s0-t0-u0}:
\\[-7mm]
\beqs
\label{eq:s0t0u0}
\begin{align}
\sz &= 4k^2  \,,
\\[1.mm]
\tz &= -\frac{\,\sz\,}{2}(1+\ct)  \,,
\\
\uz &= -\frac{\,\sz\,}{2}(1-\ct)  \,,
\end{align}
\eeqs
where the on-shell condition
$\,k^2\!=\!E^2\!-\!M_n^2\,$
and $\,k\!=\!|\vec{p}\,|$\,.
The notations $(\sz,\,\tz,\,\uz)$
correspond to the massless limit whose sum obeys
$\,\sz +\tz +\uz =0$\,.
They are connected to the Mandelstam variables
of the massive case via
$(\sz,\,\tz,\,\uz)=(s\!-\!4\Mnn,\,t,\,u)$,
where $\,s\!+t\!+u=4M_n^2\,$.
We choose the convention that the momenta of all external particles
are outgoing and the external particle numbers (1,2,3,4) are arranged clockwise in the scattering plane.

\vspace*{1mm}

For the longitudinal KK gauge boson scattering
and the corresponding KK Goldstone boson scattering
in 5d YM under $S^1\!/\mathbb{Z}_2^{}$\,,
the leading tree-level scattering amplitudes
were given before\,\cite{5DYM2002} under high energy expansion.
For the current study,
we have further computed the exact tree-level
KK longitudinal gauge boson amplitude
$\,\TT[\ALnnnn] \!\equiv\!\TT[4A_L^n]$\,
and
KK Goldstone boson amplitude
$\,\tT[\Afnnnn]\equiv\tT[4A_5^n]$\,
as follows:
\beqs
\label{eq:TL-T5-exact}
\begin{align}
\label{eq:TL-exact}
\TT[4A_L^n]
\,=~\,& g^2 ({\CC_s \KK_s} + {\CC_t \KK_t} + {\CC_u \KK_u})\,,
\\[1.mm]
\label{eq:T5-exact}
\tT[4A_5^n]
\,=~\,& g^2 ({\CC_s \KKt_s} + {\CC_t \KKt_t} + {\CC_u \KKt_u})\,,
\end{align}
\eeqs
where
\beqs
\label{eq:K-KT-exact}
\begin{align}
\label{eq:Ks-exact}
\KK_s &\dis =
-\frac{\,(4 \bs^2_0\!+27\bs_0^{}\!+36)\ct\,}
{2(\bs_0^{}\!+\!4)} \,,
\hspace*{-5mm}
&\KKt_s &\dis = -\frac{\,(3\bs_0^{}\!+4)\ct\,}{2(\bs_0^{}\!+4)} \,,
\\[1.5mm]
\label{eq:Kt-exact}
\KK_t &\dis =
-\frac{\,Q_0^{}\!+ Q_1^{}\ct\! + Q_2^{}\ctt\!+Q_3^{}c_{3\theta}^{}\,}
{\,4\bs_0^{}[8+\bs_0^{}(1\!+\!\ct)](1\!+\!\ct)\,} \,,
\hspace*{-5mm}
&\KKt_t &\dis =
\frac{\,\Qt_0^{}\!+ \Qt_1^{}\ct\! + \Qt_2^{}\ctt\,}
{\,4\bs_0^{}[8+\bs_0^{}(1\!+\!\ct)](1\!+\!\ct)\,} \,,
\\[1.5mm]
\label{eq:Ku-exact}
\KK_u &\dis =
\frac{\,Q_0^{}\!- Q_1^{}\ct\! + Q_2^{}\ctt\!-Q_3^{}c_{3\theta}^{}\,}
{\,4\bs_0^{}[8+\bs_0^{}(1\!-\!\ct)](1\!-\!\ct)\,} \,,
\hspace*{-5mm}
&\KKt_u &\dis =
-\frac{\,\Qt_0^{}\!- \Qt_1^{}\ct\! + \Qt_2^{}\ctt\,}
{\,4\bs_0^{}[8+\bs_0^{}(1\!-\!\ct)](1\!-\!\ct)\,} \,,
\end{align}
\eeqs
with
\beqs
\label{eq:Q-QT-exact}
\begin{align}
& \bs_0^{}=\sz /M_n^2\,,~~~
\bs = s/M_n^2 = \bs_0^{}\!+\!4\,,~~~
\ctt\! =\cos 2\theta \,,~~~
\cttt\! =\cos 3\theta \,,
\hspace*{5mm}
\\
& Q_0^{} = 8 \bs_0^3\!+ 33\bs_0^2\!- 48\bs_0^{}-\!128\,,~~~~
Q_1^{} = 2 (7\bs_0^3\! +40\bs_0^{2}\! +64\bs_0^{})\,,
\\
& Q_2^{} = 8 \bs_0^3\!+ 51\bs_0^2\!+ 32\bs_0^{}\!-128 \,,
\hspace*{5.7mm}
Q_3^{} = 2 (\bs^{3}_0\! + \!2\bs_0^{2} \!- 8\bs_0^{})  \,,
\\
& \Qt_0^{}= 15\bs_0^2\! +\! 144\bs_0^{}\! +\!256\,, ~~~~
\Qt_1^{} = 4(3\bs^{2}_0\!+\!4\bs_0^{}) \,,~~~~
\Qt_2^{} = -3\bs_0^2  \,.
\end{align}
\eeqs

\vspace*{1mm}

We note that the scattering amplitudes
\eqref{eq:TL-exact}-\eqref{eq:T5-exact}
have the leading high-energy behavior of $\mO(E^0)$.
We make the high energy expansion for the amplitudes
\eqref{eq:TL-T5-exact}-\eqref{eq:Q-QT-exact}
down to the subleading order:
\beqs
\label{eq:T0-dT-4n}
\begin{alignat}{3}
&\T[4A_L^n] = \T_{0L}\!+\da\T_L^{} \,,
& \hspace*{5mm}
&\tT[4A_5^n] \,= \tT_{05}\!+\da\tT_5^{}
\label{eq:T0+dT-4n}  \,,
\\[1mm]
&\T_{0L} =\, g^2({\CC_s\KK_s^0} + {\CC_t\KK_t^0} + {\CC_u\KK_u^0}) \,,
\hspace*{1cm}
&&\tT_{05}=\, g^2 ({\CC_s\KKt_s^0} + {\CC_t\KKt_t^0} + {\CC_u\KKt_u^0}) \label{eq:TL0-T50-4n} \,,
\\[1mm]
&\da\T_L^{} =\, g^2({\CC_s\da\KK_s} + {\CC_t\da\KK_t} + {\CC_u\da\KK_u}) \,, \quad
&&\da\tT_5^{} =\, g^2({\CC_s\da\KKt_s} + {\CC_t\da\KKt_t} + {\CC_u\da\KKt_u}) \label{eq:dTL-dT5-4n} \,,
\end{alignat}
\eeqs
where $\,(\KK_j^{},\,\KKt_j^{})=(\KK_j^0\!+\!\da\KK_j^{},\,
\KKt_j^0\!+\!\da\KKt_j^{})$\, are given by
\beqs
\label{eq:K0-L5}
\begin{align}
\label{eq:K0s-L5}
\KK_s^0 &\dis \,=\,  -\frac{\,11\,}{2}\ct  \,,
\hspace*{-20mm}
&\KKt_s^0 &\dis \,=\,  -\frac{3}{\,2\,}\ct  \,,
\\[1mm]
\label{eq:K0t-L5}
\KK_t^0 &\dis \,=\, \frac{\,5-11\ct\!-4\ctt\,}{\,2(1\!+\!\ct)\,} \,,
\hspace*{-20mm}
&\KKt_t^0 &\dis \,=\,  \frac{\,3(3\!-\!\ct)\,}{\,2(1\!+\!\ct)\,}\,,
\\[1mm]
\label{eq:K0u-L5}
\KK_u^0 &\dis \,=\, -\frac{\,5+11\ct\!-4\ctt\,}{\,2(1\!-\!\ct)\,}\,,
\hspace*{-20mm}
& \KKt_u^0 &\dis \,=\, -\frac{\,3(3\!+\!\ct)\,}{\,2(1\!-\!\ct)\,}\,;
\end{align}
\eeqs
and
\\[-9mm]
\beqs
\label{eq:dK-L5}
\begin{align}
\label{eq:dK-L5s}
\da\KK_s^{}	&\dis \,=\,\frac{\,4\ct\,}{\bs_0^{}}\,,
\hspace*{-20mm}
& \da\KKt_s^{} &\dis \,=\,\frac{\,4\ct\,}{\bs_0^{}}\,,
\\[1mm]
\label{eq:dK-L5t}
\da\KK_t^{}	&\dis \,=\,
\frac{\,2(2\!-\!3\ct\!-\!2\ctt\!-\!\cttt)\,}
{(1\!+\!\ct)\,\bar{t}_0^{}}\,,
\hspace*{-20mm}
& \da\KKt_t^{} &\dis \,=\,
-\frac{\,8\ct\,}{(1\!+\!\ct)\,\bar{t}_0^{}}\,,
\\[1mm]
\label{eq:dK-L5u}
\da\KK_u^{}	&\dis \,=\,
-\frac{\,2(2\!+\!3\ct\!-\!2\ctt\!+\!\cttt)\,}
{(1\!-\!\ct)\,\bar{u}_0^{}}\,,
\hspace*{-20mm}
& \da\KKt_u^{} &\dis \,=\,
-\frac{\,8\ct\,}{(1\!-\!\ct)\,\bar{u}_0^{}}\,.
\end{align}
\eeqs
We note that the leading order amplitudes
$\,\KK_j^0,\KKt_j^0\!=\!\mO(E^0M_n^0)\,$ which are both
energy-independent and mass-independent, while the
subleading amplitudes
$\,\da\KK_j^{},\da\KKt_j^{}=\mO(M_n^2/E^2)\,$
which will vanish in the high energy limit
$M_n^2/E^2\ito 0$\,.

\vspace*{1mm}

Inspecting the leading amplitudes of $\mO(E^0M_n^0)$ as in
\eqrefe{eq:TL0-T50-4n} and
Eqs.\eqref{eq:dK-L5s}-\eqref{eq:dK-L5u},
we find that longitudinal KK gauge boson amplitude
and KK Goldstone boson amplitude differ
by the same amount in each channel:
\begin{equation}
\KK_s^0-\KKt_s^0 \,=\, \KK_t^0-\KKt_t^0 \,=\, \KK_u^0-\KKt_u^0
\,=\, -4\ct \,,
\end{equation}
which has zero contribution to the scattering amplitude due to the
color Jacobi identity \eqref{eq:Jacobi-C}.
Hence, we have explicitly demonstrated the
{\it longitudinal-Goldstone equivalence}
between the longitudinal KK gauge boson scattering amplitude
and KK Goldstone boson scattering amplitude at the leading order:
\begin{equation}
\label{eq:T0-KK-ET-nnnn}
\TT_{0L}^{} ~=~ \tT_{05}^{} \,.
\end{equation}
We note that since our above leading amplitudes are obtained
by the high energy expansion of $\,M_n^2/\sz$\,, instead of
$\,M_n^2/s$\,, our present longitudinal KK amplitude
$\,\TT_{0L}^{}\,$ differs from that of Ref.\,\cite{5DYM2002}
[in its Eq.(17)] by a common term of
$\,-8g^2\ct\,$ in each of the $(s,t,u)$ channels,
whose contribution to the amplitude vanishes due to the
Jacobi identity \eqref{eq:Jacobi-C}.
On the other hand, the leading KK Goldstone boson amplitude
$\,\tT_{05}^{}\,$ coincides with that of Ref.\,\cite{5DYM2002}
[in its Eq.(21)].
This is because there is no extra energy-cancellation in the
KK Goldstone boson amplitude and the leading Goldstone amplitude
does not depend on the choice of the expansion parameter
as $\,M_n^2/\sz$\, or $M_n^2/s$\,.\,

\vspace*{1mm}

We note that the subleading amplitudes
\eqref{eq:dTL-dT5-4n} and
\eqref{eq:dK-L5s}-\eqref{eq:dK-L5u} are of $\mO(M_n^2/E^2)$.
Thus, we deduce the KK longitudinal-Goldstone equivalence
at the LO under the high energy expansion:
\begin{equation}
\label{eq:KK-ET-nnnn}
\T[\ALnnnn]
\,=\, \tT[\Afnnnn]
+\, \mO (M_n^2/E^2)\,,
\end{equation}
which coincides with
the KK Equivalence Theorem (KK-ET)\,\cite{5DYM2002}.

\vspace*{1mm}

For the convenience of double-copy construction, we define
the notations:
\beqs
\label{eq:N-K}
\begin{align}
\label{eq:N=K*s0j}
& (\NN_s^{},\,\NN_t^{},\,\NN_u^{}) \,=\,
(\sz\hspace*{0.4pt}\KK_s^{},\,
\tz\hspace*{0.4pt}\KK_t^{},\,
\uz\hspace*{0.4pt}\KK_u^{}) \,,
\\[1mm]
\label{eq:NT=KT*s0j}
& (\NNt_s^{},\,\NNt_t^{},\,\NNt_u^{}) \,=\,
(\sz\hspace*{0.4pt}\KKt_s^{},\,
\tz\hspace*{0.4pt}\KKt_t^{},\,
\uz\hspace*{0.4pt}\KKt_u^{}) \,,
\\[1mm]
\label{eq:N=N0+dN}
& \NN_j^{} =\, \NN_j^0\!+\da\NN_j^{}
=\, s_{0j}^{}(\KK_j^0\!+\da\KK_j^{})\,,
\\[1mm]
\label{eq:NT=NT0+dNT}
& \NNt_j^{} =\, \NNt_j^0\!+\da\NNt_j^{}
=\, s_{0j}^{}(\KKt_j^0\!+\da\KKt_j^{})\,,
\end{align}
\eeqs
where $\,s_{0j}^{}\hsm\!\in\hsm\! (\sz,\,\tz,\,\uz)\,$
and $\,j\hsm\!\in\hsm\! (s,\,t,\,u)\,$.\
With these, we can reexpress the elastic KK longitudinal and
Goldstone scattering amplitudes as follows:
\beqs
\label{Amp-ALA5-nnnn}
\begin{align}
\label{Amp-AL-nnnn}
\T[4A^{n}_{L}]
&\,=\, g^2\! \(\! \frac{\,\CC_s \NN_s\,}{\sz}
+ \frac{\,\CC_t \NN_t\,}{\tz}
+ \frac{\,\CC_u \NN_u\,}{\uz}  \!\)  \!,
\\[1mm]
\label{Amp-A5-nnnn}
\tT[4A^{n}_5]
&\,=\, g^2\! \(\! \frac{\,\CC_s \NNt_s\,}{\sz}
+ \frac{\,\CC_t \NNt_t\,}{\tz}
+ \frac{\,\CC_u \NNt_u\,}{\uz}  \!\)  \!.
\end{align}
\eeqs
Inspecting the leading-order kinematic quantities
$(\NN^0_s,\, \NN^0_t,\, \NN^0_u)$ and
$(\NNt^0_s,\, \NNt^0_t,\, \NNt^0_u)$ as given
Eqs.\eqref{eq:N=K*s0j}-\eqref{eq:NT=KT*s0j} and
Eq.\eqref{eq:K0-L5},
we find that they are mass-independent and satisfy
the following kinematic Jacobi identities:
\beqs
\label{eq:KJacobi}
\begin{align}
\label{eq:KJacobi-N0}
\NN_s^0 + \NN_t^0 + \NN_u^0 &\,=\, 0 \,,
\\[1.mm]
\label{eq:KJacobi-N50}
\NNt_s^0 + \NNt_t^0 + \NNt_u^0 &\,=\, 0 \,.
\end{align}
\eeqs
We can compare the two types of
Jacobi identities \eqref{eq:Jacobi-C} and
\eqref{eq:KJacobi}: the former depends on the
{\it color factor} (group structure constants)
and the latter depends on {\it kinematics}.
Since our above kinematic Jacobi identities
\eqref{eq:KJacobi-N0}-\eqref{eq:KJacobi-N50}
are mass-independent, they bear a similarity with
the conventional
{color-kinematics duality}\,\cite{BCJ:2008}\cite{BCJ:2019}
which was constructed for the 4d massless YM gauge theory and
massless GR.

\vspace*{1mm}

Furthermore, we note that, because of the Jacobi identity
\eqref{eq:Jacobi-C}, the above amplitudes
\eqref{Amp-AL-nnnn}-\eqref{Amp-A5-nnnn}
are invariant under the shift:
\begin{equation}
\label{eq:GGT}
\NN_j^{} \,\longrightarrow\,
\NN_j^{\,\pp} = \NN_j^{} +\Delta\!\times\! s_{0j}^{}\,,
\end{equation}
where
$\,\Delta\,$ is an arbitrary local function of kinematics.
This shift may be called {\it a generalized gauge transformation}
since its form bears some similarity to the gauge transformation.
From \eqrefe{eq:GGT}, we deduce the sum relation for numerators,
$\,\sum_j\NN_j^{\,\pp}\!=\hsm \sum_j \NN_j^{}\hs$,
due to the fact of
$\,\sz\hsm +\tz\hsm +\uz\! =\hsm 0\,$.
This will no longer hold for our double-copy construction
under the $\hs 1\hsm /s\hs$ expansion in section\,\ref{sec:5.3.2}
because of $\hs s\hsm +t\hsm +u\hsm =4\Mnn\,$.

\vspace*{1mm}

Then, we compare the formulas of the leading KK longitudinal
and Goldstone boson amplitudes in Eq.\eqref{eq:K0-L5} and
Eqs.\eqref{eq:N=N0+dN}-\eqref{eq:NT=NT0+dNT}.\
With these we derive the following relations between
the two sets of kinematic quantities
$(\NN_s^0,\, \NN_t^0,\, \NN_u^0)$ and
$(\NNt_s^0,\, \NNt_t^0,\, \NNt_u^0)\hs$:
\begin{equation}
\NN_s^0 \,=\, \NNt_s^0 \!- 4\hs\ct\hs\sz \,,~~~~
\NN_t^0 \,=\, \NNt_t^0 \!- 4\hs\ct\hs\tz  \,,~~~~
\NN_s^0 \,=\, \NNt_u^0 \!- 4\hs\ct\hs\uz \,.
\label{eq:KL-K5}
\end{equation}
The above relations \eqref{eq:KL-K5} show that the leading
longitudinal KK gauge boson scattering amplitude
in Eq.\eqref{Amp-AL-nnnn}
and the leading KK Goldstone boson scattering amplitude
in Eq.\eqref{Amp-A5-nnnn}
differ by an amount
$\,-4g^2\ct (\CC_s+\CC_t+\CC_u)$,\,
which vanishes identically due to the Jacobi identity
\eqref{eq:Jacobi-C}.
As we noted earlier, this realizes the KK GAET as in
Eq.\eqref{eq:T0-KK-ET-nnnn} or \eqref{eq:KK-ET-nnnn}.

\vspace*{1mm}

We can further extend the above analysis to general processes
including the inelastic KK scattering channels
$\ALnkml$\, and $\Afnkml$\,,
where the KK numbers of the initial and final states obey the condition
$|n\!\pm\!k| = |m\!\pm\!\ell|$\,.
For this, we derive the following relations under
the high energy expansion:
\beqs
\label{eq:Amp-nkml}
\begin{align}
\T\!\[\! \ALnkml \]
&\,=\, \zeta_{nkm\ell}^{}
\T\!\[\!\ALnnnn\!\] +\mO(M_j^2/E^2)\,,
\\[1.mm]
\tT\!\[\! \Afnkml \]
&\,=\, \zeta_{nkm\ell}^{}
\tT\!\[\!\Afnnnn\!\] +\mO(M_j^2/E^2)\,,
\end{align}
\eeqs
where $\,\zeta_{nnnn}^{}\!=\!1\,$,
$\,\zeta_{nnmm}^{}\!=\!\fr{2}{3}\,$
for $n\!\neq\! m$\,,\, and
$\,\zeta_{nkm\ell}^{}\!=\!\fr{1}{3}\,$
for the cases where the KK numbers $(n,k,m,\ell)$ have
no more than one equality.
From the above, we derive the KK GAET
for general scattering processes including inelastic channels:
\begin{equation}
\label{eq:KK-ET-nkml}
\T\!\[\! \ALnkml \] =
\tT\!\[\! \Afnkml \]+\mO(M_j^2/E^2)\,,
\end{equation}
where the KK GAET for the elastic channel
($\,n\!=\!k\!=\!m\!=\!\ell\,$) and the inelastic channel
($\,n\!=\!k\neq m\!=\!\ell$\,)
were demonstrated in Ref.\,\cite{5DYM2002}.

\vspace*{1mm}

Next, we examine the subleading amplitudes in
Eqs.\eqref{eq:dK-L5s}-\eqref{eq:dK-L5u} and
Eqs.\eqref{eq:N=N0+dN}-\eqref{eq:NT=NT0+dNT}.
From these, we derive
\beqs
\label{eq:dN-dNT-sum}
\begin{align}
\label{eq:dN-dNT-sum1}
& \sum_j\da\NN_j^{} \,=\, \sum_j\da\NNt_j^{} \,=\,\chi\,,
\\
\label{eq:dN-dNT-sumX}
& \chi \,= -2(7\!+\ctt)\ct\csc^2\!\theta \,M_n^2 \,,
\end{align}
\eeqs
where $\,j\!\in\!(s,\,t,\,u)\,$.
We note that the next-to-leading-order (NLO) sums of
$\dNN_j^{}$ and $\dNNt_j^{}$ are equal and do not vanish.
Then, we compute the differences of the NLO numerators
$\,(\dNN_j^{}-\dNNt_j^{})\,$ as follows:
%
\begin{equation}
\label{eq:dNj-dNTj-stu}
\dNN_s^{}\!-\dNNt_s^{} =\,0\,, ~~~~~
\dNN_t^{}\!-\dNNt_t^{} =\,8s_{\theta}^2\Mnn\,, ~~~~~
\dNN_u^{}\!-\dNNt_u^{} =\,-8s_{\theta}^2\Mnn\,.
\end{equation}
From the above results, we find that the sum of the differences of
these NLO numerators obeys a Jacobi identity:
\begin{equation}
\label{eq:dNj-dNTj-Jacobi}
\sum_j (\da\NN_j^{} \!-\da\NNt_j^{}) =\, 0\,.
\end{equation}
This property is important for us to understand
the structure of the residual term in the GRET \eqref{eq:GET-ID}
or \eqref{eq:GET}, as will be shown in section\,\ref{sec:5.2}.
Using \eqref{eq:dNj-dNTj-stu}
and from Eqs.\eqref{eq:T0+dT-4n}\eqref{Amp-ALA5-nnnn},
we also derive the NLO amplitude difference:
\begin{equation}
\label{eq:dTL-dT5}
\dT_L^{}-\da\tT_5^{} ~=~
8g^2s_{\theta}^2\Mnn
\(\frac{\,\CC_t^{}\,}{\tz} - \frac{\,\CC_u^{}\,}{\uz}\)\!.
\end{equation}
%


As an extension,
we may make two possible re-decompositions of the sum $\chi$
into the $(s,t,u)$ channels:
\begin{equation}
\label{eq:chi=sum(chi_j)}
\chi \,\equiv\, \sum_j \chi_j^{}
\,\equiv\, \sum_j \widetilde\chi_j^{} \,,
\end{equation}
where the kinematics hold the relations
$\chi_u^{}(\theta)=-\chi_t^{}(\pi\!-\!\theta)$
and $\chit_u^{}(\theta)=-\chit_t^{}(\pi\!-\!\theta)$\,.
Then, we define the following
modified subleading numerator factors:
\begin{equation}
\label{eq:dN'-dNT'}
\da\NN_j' \,=\,  \da\NN_j^{}\!-\!\chi_j^{}\,,~~~~
\da\NNt_j' \,=\, \da\NNt_j^{}\!-\!\chit_j^{} \,,
\end{equation}
which keep Eq.\eqref{eq:dNj-dNTj-Jacobi} invariant
and satisfy the kinematic Jacobi identities separately:
\beqs
\label{eq:KJacobi-dNj'-dNTj'}
\begin{align}
\label{eq:KJacobiDiff-dNj'-dNTj'}
& \sum_j (\da\NN_j' \!-\da\NNt_j' )
=\, 0\,,
\\
\label{eq:KJacobi-dNj'dNTj'}
& \sum_j\!\da\NN_j' = 0\,, ~~~~
\sum_j\!\da\NNt_j' = 0\,.
\end{align}
\eeqs
Thus, from Eq.\eqref{Amp-ALA5-nnnn}, we define the improved
scattering amplitudes for the
KK longitudinal gauge bosons and KK Goldstone bosons:
\beqs
\label{AmpMod-ALA5-nnnn}
\begin{align}
\label{AmpMod-AL-nnnn}
\T'[4A^{n}_{L}]
&\,=\, g^2\! \(\! \frac{\,\CC_s \NN_s'\,}{\sz}
+ \frac{\,\CC_t \NN_t'\,}{\tz}
+ \frac{\,\CC_u \NN_u'\,}{\uz}  \!\)  \!,
\\[1mm]
\label{AmpMod-A5-nnnn}
\tT'[4A^{n}_5]
&\,=\, g^2\! \(\! \frac{\,\CC_s \NNt_s'\,}{\sz}
+ \frac{\,\CC_t \NNt_t'\,}{\tz}
+ \frac{\,\CC_u \NNt_u'\,}{\uz}  \!\)  \!.
\end{align}
\eeqs
We note that according to the Jacobi identities
\eqref{eq:KJacobi} and
\eqref{eq:KJacobi-dNj'dNTj'},
the improved numerators
$\,\NN_j'=\NN_j^0+\da\NN_j'\,$ and
$\,\NNt_j'=\NNt_j^0+\da\NNt_j'\,$
obey the kinematic Jacobi identities separately:
\beqs
\label{eq:KJacobiMod}
\begin{align}
\label{eq:KJacobi-N'}
\NN_s' + \NN_t' + \NN_u' &\,=\, 0 \,,
\\[0.6mm]
\label{eq:KJacobi-NT'}
\NNt_s' + \NNt_t' + \NNt_u' &\,=\, 0 \,.
\end{align}
\eeqs
Thus, the improved KK scattering amplitudes
\eqref{eq:KJacobi-N'}-\eqref{eq:KJacobi-NT'}
exhibit all the nice features required by
the conventional double-copy construction of
BCJ-type\,\cite{BCJ:2008}\cite{BCJ:2019}.
We will present such a double-copy construction
for the KK graviton scattering amplitudes and the GRET
in the next subsection. For the subleading KK YM amplitudes
and KK graviton amplitudes, our focus will be on the
residual term $\,\T_v^{}\,$ in the KK GAET identity
and the residual term $\,\M_\Delta^{}\,$ in the GRET identity,
which can be expressed respectively as the difference between the
NLO longitudinal KK amplitude and the corresponding NLO KK
Goldstone amplitude:
\beqs
\label{eq:Tv-MD}
\begin{align}
\label{eq:T_v}
\T_v^{}	&~=~ \dT_L^{}-\delta\tT_5^{}  \,,
\\
\label{eq:M_D}
\M_\Delta^{} &~=~
\delta\M - \delta\MT  \,,
\end{align}
\eeqs
where we have used the notations
$\,\delta\M \equiv\delta\M [4h_L^n]\,$ and
$\,\delta\MT\equiv\delta\MT [4\phin]\,$.
For deriving the above NLO KK GAET identity \eqref{eq:T_v}
and the NLO GRET identity \eqref{eq:M_D},
we have input the LO KK GAET identity
\eqref{eq:T0-KK-ET-nnnn} and the LO GRET identity
\eqref{eq:M0=MT0}.
The modified NLO numerators in Eq.\eqref{eq:dN'-dNT'}
give the modified NLO amplitudes as follows:
\beqs
\label{eq:KKET-NLO-mod}
\begin{align}
\label{eq:TL'}
\dT_L' &~=~ \dT_L^{} -\,\sum_j
\frac{\,\CC_j^{}\chi_j^{}\,}{s_{0j}^{}} \,,
\\
\label{eq:T5'}
\delta\tT_5' &~=~
\delta\tT_5^{} -\,\sum_j
\frac{\,\CC_j^{}\chit_j^{}\,}{s_{0j}^{}}
\,.
\end{align}
\eeqs
With the above, we can reexpress the NLO KK GAET identity
\eqref{eq:T_v} in the following form:
\begin{align}
\label{eq:Tv'}
\T_v'	~=~ \dT_L'-\delta\tT_5'  \,,
\end{align}
where $\,\T_v'\,$ denotes the modified residual term defined by
$\,\T_v'= \T_v^{}-\sum_j
\CC_j^{}(\chi_j^{}-\chit_j^{})/s_{0j}^{}\,$.
We note that even though in Eq.\eqref{eq:Tv'}
the NLO KK longitudinal and Goldstone amplitudes
$(\dT_L',\,\delta\tT_5')$
are both modified as in Eq.\eqref{eq:KKET-NLO-mod},
the residual term is also modified as $\,\T_v'\,$ accordingly.
Hence, the NLO KK GAET identity \eqref{eq:Tv'} is equivalent to
its original form \eqref{eq:T_v}, which means that the
gauge symmetry of the KK YM theory is still retained
by the identity \eqref{eq:Tv'}.

\vspace*{1mm}

With the double-copy construction, we can justify the size
of the GRET residual term
$\,\M_\Delta^{}\!=\mO(M_n^2E^0)\,$
from the KK GAET residual term
$\,\T_v^{}\!\!=\!\mO(M_n^2/E^2)\,$,
where $\hsm\T_v^{}\hsm$ is well understood.
We will demonstrate that
the connection between {\it sizes} of the two residual terms
$\,\T_v^{}\!=\!\mO(M_n^2/E^2)\,$ and
$\,\M_\Delta^{}\!\!=\!\mO(M_n^2E^0)\,$ is a general prediction
of the double-copy construction and does not depend on details
of the construction.
\vspace*{2mm}
\subsection{\hspace*{-2mm}Constructing KK Scattering Amplitudes
	and GRET by Double-Copy}
\label{sec:5.2}
\vspace*{1.5mm}

For the compactified 5d YM gauge theory and compactified 5d GR theory,
we expect the double-copy correspondence:
\\[-10mm]
\beqs
\label{eq:DCcorrespond}
\begin{align}
\label{eq:DCcorrespond-AmuAnu-hmuhnu}
A_n^{a\mu}\otimes\!A_n^{a\nu} \ &\longrightarrow\, h^{\mn}_n \,,
\\
\label{eq:DCcorrespond-AmuA5-hmuh5}
A_n^{a5}\otimes\!A_n^{a5} \ &\longrightarrow\, h^{55}_n \,,
\\
\label{eq:DCcorrespond-A5A5-h5h5}
A_n^{a\mu}\otimes\!A_n^{a5} \ &\longrightarrow\, h^{\mu 5}_n \,.
\end{align}
\eeqs
It is instructive to note that the physical spin-2 KK graviton field
$h^{\mn}_n$ arises from the double-copy of spin-1 KK gauge fields
$A_n^{a\mu}\otimes\!A_n^{a\nu}$.
On the other hand,
the $A_n^{a5}$ is the would-be KK Goldstone boson
in the compactified 5d YM gauge theory,
and the double-copy counterparts $h^{55}_n(=\phin)$
and $h^{\mu 5}_n$ just correspond to the scalar KK Goldstone boson
and vector KK Goldstone boson in the compactified 5d GR.
From Eq.\eqref{eq:DCcorrespond-AmuAnu-hmuhnu} , we further expect
the double-copy correspondence between the (helicity-zero)
longitudinal KK graviton and KK gauge boson:
$\,A_L^{an}\otimes\!A_L^{an}\!\longrightarrow h_L^n\,$.
We observe that in the high energy limit the longitudinal KK gauge
boson $A_L^{an}\!=\ep_L^\mu A^{an}_\mu\,$ has its
polarization vector $\,\ep_L^\mu\!\!\sim\! k^\mu/M_n\,$,
and the longitudinal KK graviton
$\,h_L^n\!=\!\vep_L^{\mn}h^n_{\mn}\,$ has its polarization tensor
$\,\vep_L^{\mn}\!\!\sim\! k^\mu k^\nu/M_n^2\,$.
Thus, we have
$\,\vep_L^{\mn}\!\!\sim\!\ep_L^\mu\ep_L^\nu\,$
in the high energy limit,
which also makes the longitudinal correspondence
($\,A_L^{an}\otimes\!A_L^{an}\!\longrightarrow h_L^n\,$)
well expected. The demonstration of the double-copy
correspondence between the longitudinal KK gauge boson
amplitudes and the longitudinal KK graviton amplitudes
is much more nontrivial than the above relation between
the on-shell longitudinal polarization vector/tensor,
as we will analyze further in this subsection.

\vspace*{1mm}

In this subsection, we will first demonstrate that
a double-copy construction from the KK gauge theory amplitudes
to the KK graviton amplitudes at the leading order (LO) of the
high energy expansion, which corresponds to the limit
$\,\Mn/E \ito 0\,$. We find that such leading order amplitudes
are {\it mass-independent} and their kinematic Jacobi identities
\eqref{eq:KJacobi} hold, in addition to the massless Mandelstam
relation $\,\sz\!+\tz\!+\uz\!=0\,$.
Thus, we will first extend the conventional double-copy
method\,\cite{BCJ:2008}\cite{BCJ:2019} to the LO amplitudes
in our 5d KK theory and demonstrate how it works quantitatively.

\vspace*{1mm}

We note that
the (helicity-zero) longitudinal KK gauge bosons $A_L^{an}$
and longitudinal KK gravitons $h_L^n$ are truly distinctive
in the KK theory because they do not exist in the commonly studied
massless YM gauge theory or massless GR.
Also, in the limit $M_n\ito 0$\,,
the KK Goldstone bosons $A_5^{an}$
and $\phi^n (=\!h^n_{55})$ both become massless and correspond to
the physical degrees of freedom.
But, it is important to observe that according to the KK GAET
(cf.\ section\,\ref{sec:5.1})\,\cite{5DYM2002}\cite{KK-ET-He2004}
and GRET (sections\,\ref{sec:3}-\ref{sec:4}),
the leading scattering amplitudes of the longitudinal KK gauge
bosons (KK gravitons) equal the corresponding amplitudes
of the KK Goldstone bosons and are mass-independent
(which correspond to the limit $M_n^2/E^2\ito0$\,
under high energy expansion).
Hence, we can construct a double-copy from the
leading longitudinal KK gauge boson amplitudes of $\mO(E^0)$
to the corresponding longitudinal KK graviton amplitudes
of $\mO(E^2)$, in parallel to the double-copy construction
between the KK Goldstone amplitudes in the KK YM theory and KK GR.
The KK Goldstone amplitudes are much simpler
due to the absence of any nontrivial
energy-cancellations in the KK Goldstone amplitudes.
Furthermore, since the compactified KK theories
have very different Feynman rules from
the 4d massless gauge theory or massless GR as commonly studied,
the double-copy realization in the KK theory is far from
obvious even for the leading order amplitudes before explicit
demonstration. For instance, there are highly nontrivial and
intricate energy-cancellations in the longitudinal KK gauge boson
scattering amplitudes
[from $\mO(E^4)$ down to $\mO(E^0$)]\,\cite{5DYM2002} and
in the (helicity-zero) longitudinal KK graviton scattering amplitudes
[from $\mO(E^{10})$ down to $\mO(E^2$)]\,\cite{Chivukula:2020L},
all these do not exist in the 4d massless gauge theory and massless GR.

\vspace*{1mm}

We inspect the structures of the KK longitudinal gauge boson
scattering amplitude \eqref{Amp-AL-nnnn}
and the KK corresponding Goldstone boson scattering amplitude
\eqref{Amp-A5-nnnn} in the compactified 5d YM gauge theory
under the high energy expansion.
We see from Eqs.\eqref{eq:T0-dT-4n} and
\eqref{eq:K0-L5}-\eqref{eq:dK-L5}
that under high energy expansions, the leading amplitudes
$(\T_{0L}\,, \,\tT_{05})$
are of $\,\mO(E^0)\,$ and {\it mass-independent,}
while the subleading amplitudes
$(\da\T_L^{}\,, \,\da\tT_5^{})$ are of
$\,\mO(\Mnn/E^2)\,$ and vanish in the massless limit $\Mn\ito 0$\,.
We have formally expressed these leading amplitudes in the form
the massless gauge theories with pole factors $(\sz,\,\tz,\,\uz)$
in the denominator of each channel, even though these poles
are no longer real poles under the current high energy expansion.
For the current study of the 5d KK YM gauge theories and 5d KK GR,
we present an extended formulation
of the conventional BCJ double-copy method
of the massless gauge theories\,\cite{BCJ:2008}\cite{BCJ:2019},
{\it by making the high energy expansion with
$\,\Mnn/E^2\!\ll\! 1$}\,
under which all the nonzero KK mass-poles are removed,
and the mass-dependent contributions can be treated
order by order.

\vspace*{1mm}

From the numerators of the amplitudes
\eqref{Amp-AL-nnnn}-\eqref{Amp-A5-nnnn},
we see that the kinematic factors $(\NN_s,\, \NN_t,\, \NN_u)$
and $(\NNt_s,\, \NNt_t,\, \NNt_u)$ may be viewed as
{\it dual to} the color factors $(\CC_s,\, \CC_t,\, \CC_u)$
according to the conventional double-copy method in the massless
gauge theories\,\cite{BCJ:2008}\cite{BCJ:2019}.
Thus, we attempt to construct the elastic scattering amplitude
$\,\M[\hLnnnn]\,$
of the longitudinal KK gravitons
and the gravitational KK Goldstone boson amplitude
$\,\MT[\pnnnn]\,$
from the corresponding longitudinal KK gauge boson amplitude
$\,\T[\ALnnnn]$\,
and the KK Goldstone boson amplitude
$\,\tT[\Afnnnn]$,\,
respectively.
We realize an extended double-copy construction
for the 5d KK YM gauge theory
and 5d KK GR by the following replacement:
\\[-8mm]
\beqs
\begin{align}
(\CC_s,\, \CC_t,\, \CC_u) \ & \longrightarrow \
(\NN_s,\, \NN_t,\, \NN_u)\,,
\label{eq:CKDual-ALhL}
\\[1mm]
(\CC_s,\, \CC_t,\, \CC_u) \ & \longrightarrow \
(\NNt_s,\, \NNt_t,\, \NNt_u)\,.
\label{eq:CKDual-A5h55}
\end{align}
\eeqs
Applying this duality replacement to the scattering amplitudes
of the longitudinal KK gauge bosons and KK Goldstone bosons
in Eqs.\eqref{Amp-AL-nnnn}-\eqref{Amp-A5-nnnn} and
Eqs.\eqref{eq:K0-L5}\eqref{eq:N=N0+dN}-\eqref{eq:NT=NT0+dNT},
we first construct the corresponding scattering amplitudes of
the longitudinal KK gravitons and gravitational
KK Goldstone bosons, to the nonzero leading contributions
of $\mO(E^2)$ in the high energy expansion:
\beqs
\label{AmpK0-hLh5-nnnn}
\begin{align}
\M_0^{}[\hLnnnn]
\, &= c_0^{}\,g^2\!
\[\! \frac{(\NN_s^0)^2}{\sz} +\frac{(\NN_t^0)^2}{\tz}
+\frac{(\NN_u^0)^2}{\uz} \!\]\!,
\label{AmpK-hL-nnnn}
\\[1mm]
\MT_0^{} [\pnnnn]
\, &= c_0^{}\,g^2\!
\[\! \frac{(\NNt_s^0)^2}{\sz} +\frac{(\NNt_t^0)^2}{\tz}
+\frac{(\NNt_u^0)^2}{\uz} \!\]\!,
\label{AmpK-h55-nnnn}
\end{align}
\eeqs
where the overall coefficient $c_0^{}$ is a conversion constant
due to replacing the gauge coupling $\hs g\hs$
by gravitational coupling
$\ka$\,. The constant $c_0^{}$ is not known a priority
before a unified UV theory of gauge and gravitational forces
becomes available.

\vspace*{1mm}

Then, substituting Eqs.\eqref{eq:K0s-L5}-\eqref{eq:K0u-L5}
into Eqs.\eqref{AmpK-hL-nnnn}-\eqref{AmpK-h55-nnnn},
we explicitly reconstruct the longitudinal KK graviton
scattering amplitude and the gravitational KK Goldstone
scattering amplitude as follows:
\beqs
\label{AmpDC-hLphi-nnnn}
\begin{align}
\M_0^{} [\hLnnnn]
&=
\MT_0^{} [\pnnnn]
\nn\\[1.5mm]
& 
= \(\!-\frac{\,9c_0^{}g^2\,}{\,4\,}\!\)
\!\[\!\frac{\,(3+\cos^2\!\theta)^2\,}{\sin^2\!\theta}\!\]\!\sz
\label{AmpDC-phinnnn-1}
\\[1mm]
& 
= \(\!-\frac{\,9c_0^{}g^2\,}{\,16\,}\!\)\!
\left[\(7\!+\cos2\theta\)^2\!\csc^2\!\theta\,\right]\!\sz
\label{AmpDC-phinnnn-2}
\\[1mm]
& 
= \(\!-\frac{\,9c_0^{}g^2\,}{\,4\,}\!\)\!\!
\left[\frac{\,(s_0^2 + t_0^2+ u_0^2)^2\,}
{\sz\hspace*{1.5pt}\tz\hspace*{1.5pt}\uz}\right] \!,
\label{AmpDC-phinnnn-3}
\end{align}
\eeqs
where we have dropped the mass-dependent subleading term
of $\mO(M_n^2)$ which is much smaller than the above
leading $\mO(E^2)$ amplitude in the high energy scattering.

\vspace*{1mm}

Strikingly, we find that our above
leading amplitudes of the longitudinal KK graviton
and the gravitational KK Goldstone boson
in Eq.\eqref{AmpDC-hLphi-nnnn},
as constructed by the double-copy method, perfectly agree to
the gravitational KK Goldstone amplitude \eqref{eq:GAmpE2-phinnnn}
at $\mO(E^2)$
which we computed directly
from the KK theory of compactified 5d GR.

\vspace*{1mm}

Eq.\eqref{AmpDC-hLphi-nnnn} also explicitly
establishes the {\it equivalence} between the longitudinal
KK graviton amplitude and the corresponding gravitational KK
Goldstone boson amplitude.
In fact, we can demonstrate this equivalence in
a more elegant and transparent way,
by making use of the relation \eqref{eq:KL-K5}.
With this, we can express the KK graviton amplitude \eqref{AmpK-hL-nnnn}
in terms of the gravitational KK Goldstone boson amplitude:
%
\begin{align}
& \M_0^{} [\hLnnnn]
\nn\\[1mm]
&= c_0^{}g^2\!\LB\!
\[\! \frac{(\NNt_s^0)^2}{\sz}\!+\!\frac{(\NNt_t^0)^2}{\tz}\! +\!\frac{(\NNt_u^0)^2}{\uz}\!\]\!
\!-8\ct (\NNt_s^0 \!+ \NNt_t^0 \!+ \NNt_u^0)
\!+16\cct (\sz\! +\tz\! +\uz) \RB
\hspace*{10mm} 	
\nn\\[1mm]
&= \MT_0^{}[\pnnnn] ,
\label{AmpE2-hL-h55-nnnn}
\end{align}
%
where in the last step we have made use of the
kinematic Jacobi identity \eqref{eq:KJacobi-N50}
and the Mandelstam relation $\,\sz +\tz+\uz\! =0\,$.
We see that the longitudinal KK graviton scattering amplitude
equals the gravitational KK Goldstone scattering amplitude
at the leading $\mO (E_n^2)$ and they differ only by
subleading terms of $\mO(E^{0} \Mnn)$.
The above Eq.\eqref{AmpE2-hL-h55-nnnn} just demonstrates
that the GRET holds for the longitudinal KK graviton scattering
amplitude and the corresponding KK Goldstone scattering amplitude
down to $\mO (E^2M_n^0)$ under the high energy expansion,
\begin{equation}
\label{eq:GET-nnnn}
\M[\hLnnnn]
\,=\, \MT[\pnnnn]
+\mO(E^{0} \Mnn)\,.
\end{equation}

It is truly impressive to see that
building upon the longitudinal-Goldstone equivalence
of the KK GAET \eqref{eq:KK-ET-nnnn} [or \eqref{eq:KK-ET-nkml}],
we have established the corresponding
longitudinal-Goldstone equivalence
of the GRET for the amplitudes of the
longitudinal KK graviton scattering and of the gravitational
KK Goldstone scattering as in the above Eq.\eqref{eq:GET-nnnn}
by using the double-copy construction.
Hence, {\it this demonstrates a double-copy correspondence between
the KK GAET in the compactified 5d YM gauge theory and the KK GRET
in the compactified 5d GR.}

\vspace*{1mm}

We have the following comments in order:
\vspace*{-1.5mm}
\begin{enumerate} 
	\item[{\bf (i)}]
	Impressively, we find that our reconstructed gravitational KK Goldstone
	$\phin\,(=h^{55}_n)$ amplitude
	$\MT_0[\pnnnn]$
	in Eqs.\eqref{AmpK-h55-nnnn}\eqref{AmpDC-hLphi-nnnn}
	from the KK Goldstone $A_5^{an}$-amplitude
	$\tT_0[\Afnnnn]$
	in Eqs.\eqref{Amp-A5-nnnn}\eqref{eq:K0s-L5}-\eqref{eq:K0u-L5}
	in the compactified 5d YM gauge theory via the double-copy approach
	has exactly the same energy and angular dependence as what
	we obtained by directly computing the $\phin$-amplitude \eqref{eq:GAmpE2-phinnnn} in the compactified 5d KK GR theory.
	This double-copy reconstruction is naturally expected
	via the correspondence
	$\,A_5^{an}\otimes\!A_5^{an}\!\longrightarrow h_{55}^n\,$
	where both the KK Goldstone bosons $A_5^{an}$
	and $h^n_{55}(=\!\phin)$ become effectively massless
	in the high energy limit $M_n^2/E^2\ito 0$\,.
	We note that both the leading gravitational KK Goldstone
	amplitude \eqref{AmpK-h55-nnnn}\eqref{AmpDC-hLphi-nnnn}
	of $\mO(E^2)$
	and the leading gauge-theory KK Goldstone amplitude
	\eqref{Amp-A5-nnnn}\eqref{eq:K0s-L5}-\eqref{eq:K0u-L5} of $\mO(E^0)$
	are {\it mass-independent.} Hence, their structures reflect
	the 5d gauge symmetry of the KK YM theory
	and the 5d diffeomorphism invariance of the KK
	GR theory.

	\item[{\bf (ii)}]
	Note that the (helicity-zero) longitudinal KK gauge bosons $A_L^{an}$
	and longitudinal KK gravitons $h_L^n$ do not exist in the
	massless YM gauge theory or massless GR. Hence they
	are truly distinctive in the KK theories.
	As we observe, {\it the key point} is that
	according to the KK GAET (section\,\ref{sec:5.1})
	\cite{5DYM2002}\cite{KK-ET-He2004}
	and GRET (sections\,\ref{sec:3}-\ref{sec:4}),
	the leading longitudinal scattering amplitudes of
	$\,A_L^{an}\,$ and $\,h_L^n\,$ equal the corresponding amplitudes
	of the KK Goldstone bosons ($A_5^{an}$ and $h_{55}^n$)
	and are mass-independent
	(corresponding to the limit $M_n^2/E^2\ito0$\,
	under high energy expansion), despite that the longitudinal
	polarization vector $\ep_L^\mu$
	(tensor $\vep_L^{\mn}$) of $A_L^{an}$ ($h_L^n$) has
	explicit mass-dependence.
	This is why we can construct a similar double-copy from the
	leading longitudinal KK gauge boson amplitudes of
	$\mO(E^0M_n^0)$
	to the corresponding longitudinal KK graviton amplitudes
	of $\mO(E^2M_n^0)$.
	The above also explains that even though the original double-copy formulation\,\cite{BCJ:2008}\cite{BCJ:2019} was shown to hold
	in the massless theory, we can still extend it to our current
	double-copy construction for the compactified massive KK theories
	to the leading order amplitude of $\mO(E^2M_n^0)$, which is
	{\it mass-independent}.
	All the mass-dependent terms belong to the subleading order of
	$\mO(E^0M_n^2)$ and are of the same order as the residual term
	in the GRET, as we will analyze further
	in sections\,\ref{sec:5.3}-\ref{sec:5.4}.

	We stress that our double-copy construction guarantees that
	{\it the leading longitudinal KK graviton (Goldstone) amplitude
	\eqref{AmpK0-hLh5-nnnn}-\eqref{AmpDC-hLphi-nnnn}
	must scale as $\,\mO(E^2M_n^0)\,$
	under the high energy expansion.}
	According to our double-copy construction,
	this $\,\mO(E^2M_n^0)\,$ high energy behavior just
	corresponds to the \,$\mO(E^0M_n^0)\,$
	leading energy behavior of the KK gauge (Goldstone) boson amplitude \eqref{Amp-ALA5-nnnn}, {\it which are both mass-independent.}
    In fact, our double-copy construction
	(based on the scattering amplitudes of 5d YM gauge theory
	and the KK GAET\,\cite{5DYM2002}\cite{KK-ET-He2004})
	gives an independent proof
	that the longitudinal KK graviton scattering amplitudes
	must have large energy-cancellations of
	$\mO(E^{10})\ito \mO(E^2)$.
	We achieve this by establishing a new correspondence
	between the two energy-cancellations of
	the four-particle longitudinal KK
	scattering amplitudes:
	$\,E^4\!\ito E^0\,$ in the 5d KK YM theory (YM5)
	and $\,E^{10}\!\!\to\! E^2\,$ in the 5d KK GR (GR5).
	Here, with the double-copy construction,
	we use the first energy-cancellation of $\,E^4\!\ito E^0\,$
	(YM5) to deduce the second energy-cancellation of
	$\,E^{10}\!\!\to\! E^2\,$ (GR5).
	Thus, we may present schematically this new correspondence
	between the two energy-cancellations as follows:
\beqa
\label{eq:E4_0-E10_2}
E^4\!\ito E^0\,\rm{(YM5)} ~~\Longrightarrow~~
E^{10}\!\!\to\! E^2\,\rm{(GR5)}\,.
\eeqa
In passing, some recent literature on the double-copy
construction for certain specific KK models
appeared\,\cite{xDC-KK1}\cite{xDC-KK2},
in which \cite{xDC-KK1} briefly discussed
a scalar model compactified on
$\mathbb{R}^4\!\times\!S^1$
with an extra spectral condition imposed on
the KK mass-spectrum, and \cite{xDC-KK2}
discussed a KK inspired action
with extra global U(1) symmetry to have certain
special mass-condition for double-copy.
But these special KK models differ from the standard KK theory
with obifold $S^1\!/\ZZ$ in our study and their methods do not
apply to our case, so they do not overlap with our current study.

\item[{\bf (iii)}]
Our reconstructed gravitational KK Goldstone boson scattering
amplitude \eqref{AmpK-h55-nnnn}\eqref{AmpDC-hLphi-nnnn}
by double-copy method
is confirmed by our direct computation of the gravitational KK
Goldstone amplitude in Eq.\eqref{eq:GAmpE2-phinnnn},
which also equals our reconstructed (helicity-zero)
longitudinal KK graviton amplitude \eqref{AmpDC-hLphi-nnnn}. 	
In addition, we find that our longitudinal KK graviton
amplitude in Eq.\eqref{AmpDC-hLphi-nnnn} as reconstructed from
our longitudinal KK gauge boson amplitude \eqref{Amp-AL-nnnn}
has exactly the same energy and angular dependence as those obtained
by direct Feynman-diagram calculations of the longitudinal KK
graviton amplitudes in
Refs.\,\cite{Chivukula:2020S}\cite{ Chivukula:2020L}.%
\footnote{%
\baselineskip 15pt
After submitting this paper to arXiv:2106.04568,
we learnt from colleagues Sekhar Chivukula and Elizabeth Simmons
via private communication that their postdoc Xing Wang also checked
that the double-copy gave the correct expression for
massive KK graviton scattering in the case of orbifolded torus.}

\item[{\bf (iv)}]
The amplitudes \eqref{AmpK-hL-nnnn}-\eqref{AmpK-h55-nnnn}
	have no double poles, so its denominator
	should be proportional to the product
	$\,\sz\hspace*{1pt}\tz\hspace*{1pt}\uz\,$,\,
	which is permutation invariant among $(\sz,\,\tz,\,\uz)$.
	We note that for the elastic scattering
	$\,(n,n) \!\ito\! (n,n)\,$,
	the above amplitude should be invariant under all possible permutations,
	so the structure of this amplitude should take the form of
	$\,(\sz\hspace*{1pt}\tz\hspace*{1pt}\uz)^{a}
	(s_0^2\!+\!t_0^2\!+\!u_0^2)^{b}\,$
	with $(a,\,b)$ being certain integers.
	Since the denominator of the scattering amplitude 
    should scale like
	$\,\sz\hspace*{1pt}\tz\hspace*{1pt}\uz\!\propto\! s_0^3\,$,\,
	so we have $\,a=-1\,$.
	Note that the whole amplitude is expected to scale
	like  $\mO (s^1)$,\, so the numerator has to scale as $\mO (s^4)$\,.
	This means that the only possibility for the numerator is to scale
	as  $\,(s_0^2+t_0^2+u_0^2)^2\,$ with $\,b\!=\!2\,$.
	With these, we can generally deduce
	that the kinematic structure of the amplitude
	\eqref{AmpK0-hLh5-nnnn}
	behaves as
	$\,(s_0^2\!+\!t_0^2\!+\!u_0^2)^2/
	(\sz\hspace*{1pt}\tz\hspace*{1pt}\uz)\,$,\,
	which explains why our explicit construction
	should lead to the formula of \eqref{AmpDC-phinnnn-3} indeed.

\item[{\bf (v)}]
The overall conversion constant $c_0^{}$ in
Eqs.\eqref{AmpK0-hLh5-nnnn}-\eqref{AmpDC-hLphi-nnnn}
is undetermined by the double-copy construction itself,
but is expected to be universal at least for each given spacetime dimension.
To match our double-copy result
\eqref{AmpDC-hLphi-nnnn}
with the gravitational KK Goldstone amplitude
\eqref{eq:GAmpE2-phinnnn},
we choose the following conversion constant:
\begin{equation}
\label{eq:c0}
\cv \,=\, -\frac{\ka^2}{\,24g^2\,}  \,.
\end{equation}
	We also notice that in the traditional massless 4d theory,
	the graviton amplitude reconstructed from the BCJ double-copy
	can fully match the graviton amplitude in 4d massless
	GR with the conversion constant
\begin{equation}
\tilde{c}_0  \,=\, \frac{\ka^2}{\,16g^2\,}  \,.
\label{eq:c0-4d}
\end{equation}
Our definition of the group structure constant $C^{abc}$ differs from
the group structure constant $f^{abc}$ of Refs.\cite{BCJ:2008}\cite{BCJ:2019} by a simple normalization factor:
\begin{equation}
C^{abc} =  \frac{1}{\sqrt{2\,}\,}f^{abc}
= - \frac{\ii}{\sqrt{2\,}\,} \Tr\( [T^a,T^b]\, T^c \) \,,
\end{equation}
where $T^a$ is the generator of SU(N) group.
	
\end{enumerate}

\vspace*{1mm}


Next, we further extend the above analysis to general processes
$\ALnkml$\,
and
$\Afnkml$\,
including the inelastic KK scattering channels.
According to the Eqs.\eqref{Amp-AL-nnnn}-\eqref{Amp-A5-nnnn} and
Eq.\eqref{eq:Amp-nkml},
we write the LO inelastic scattering amplitudes as follows:
\beqs
\begin{align}
\T_0^{}\!\[\!\ALnkml\]
&=
g^2\zeta_{nkm\ell}^{} \!\left(\!
\frac{\,\CC_s \NN_s^0\,}{\sz} +
\frac{\,\CC_t \NN_t^0\,}{\tz} +
\frac{\,\CC_u \NN_u^0\,}{\uz}\!\right) \!,
\label{Amp-AL-nkml}
\\[1.mm]
\tT_0^{}\!\[\!\Afnkml\]
&=
g^2\zeta_{nkm\ell}^{} \!\left(\!
\frac{\,\CC_s \NNt_s^0\,}{\sz} +
\frac{\,\CC_t \NNt_t^0\,}{\tz} +
\frac{\,\CC_u \NNt_u^0\,}{\uz}\!\right) \!,
\end{align}
\eeqs
where $\,\zeta_{nnnn}^{}\!\!=\!\!1\,$,
$\,\zeta_{nnmm}^{}\!=\!\fr{2}{3}\,$
for $n\!\neq\! m$\,,\, and
$\,\zeta_{nkm\ell}^{}\!=\!\fr{1}{3}\,$
for $(n,k,m,\ell)$ having
no more than one equality.
Thus, using the color-kinematics duality relations
\eqref{eq:CKDual-ALhL}-\eqref{eq:CKDual-A5h55}
and up to an overall conversion constant $c_0^{}$\,,
we can further reconstruct the
general scattering amplitude of longitudinal KK gravitons
and the scattering amplitude of the corresponding gravitational
KK Goldstone bosons by using the following relations:
\beqs
\label{eq:DC-ineAMP}
\begin{align}
\M\!\[\!\hLnkml\!\]
&\,=~ \zeta_{nkm\ell}^{}\,
\M\!\[\!\hLnnnn\!\]
+\mO (E^0 M_j^2)\,,
\\[1mm]
\MT\!\[\!\pnkml\!\]
&\,=~ \zeta_{nkm\ell}^{}\,
\MT[\pnnnn]
+\mO (E^0 M_j^2)\,.
\label{eq:hL-GB-nlmk}
\end{align}
\eeqs
Then, using Eq.\eqref{eq:GET-nnnn},
we can deduce the GRET by double-copy reconstruction
for the general scattering process:
\begin{equation}
\label{eq:GET-nkml}
\M\!\[\!\hLnkml\!\] \,=\, \MT\!\[\!\pnkml\!\]
+\mO (E^0 M_j^2)\,,
\end{equation}
where the KK numbers of the initial and final states obey
$|n\!\pm\!k| = |m\!\pm\!\ell|$\,.

\vspace*{1mm}

We observe that our double-copy constructions in
Eqs.\eqref{eq:GET-nnnn} and \eqref{eq:GET-nkml}
have explicitly established the KK GRET from the KK GAET
\eqref{eq:KK-ET-nnnn}  and \eqref{eq:KK-ET-nkml}:
{\it the leading amplitude of the longitudinal KK graviton
	scattering equals that of the gravitational KK Goldstone scattering
	at $\mO(E^2)$ (which is mass-independent)
	under the high energy expansion,
	and their difference is only of $\hs\mO(E^0 \Mnn)\hs$.}
This means that in our general formulation
of the KK GRET \eqref{eq:GET}
the sum of all the $\mO(\DEn)$ residual terms must be of
$\mO(E^0 \Mnn)$,
even though the naive power counting on their individual amplitudes
containing one or more external state of
$\,\vnt (=\vt_{\mn}^{}h^{\mn}_n)$\, or
\,$\th_n^{} (=\eta_{\mn}^{}\th_n^{\mn})$\,
gives $\mO(E^2)$.
Hence, we deduce that {\it the double-copy construction
of the KK GRET identity \eqref{eq:GET-ID} from the
KK GAET identity\,\cite{KK-ET-He2004} in
the KK YM gauge theory provides a new mechanism of
energy cancellation from $\mO(E^2)$ down to $\mO(E^0)$
in the sum of all the $\mO(\DEn)$ residual terms on the RHS of the
KK GRET \eqref{eq:GET}.}
We will further demonstrate the realization
of this new energy-cancellation mechanism of
$\,E^2\ito E^0\,$
for the residual terms of GRET in the next subsections.

\subsection{\hspace*{-2mm}Constructing Mass-Dependent KK Amplitudes
	from Double-Copy}
\label{sec:5.3}
\vspace*{1.5mm}

In this subsection, we study the extended double-copy construction
of the mass-dependent KK scattering amplitudes at the NLO.
We will make two types of high energy expansions by using the
expansion parameter $\hs 1\hsm /\sz\hs$ or $\hs 1\hsm /s\hs$,
where the Mandelstam variable
$\,s\hsm =\hsm \sz\hsm +4\Mnn\,$ for the four-point
elastic KK amplitudes.
As we will show,
the advantage of the $\hs 1\hsm /\sz\hs$ expansion is that
it can automatically ensure that the LO numerators of
the KK amplitudes are mass-independent,
but then the mass-dependent NLO numerators cannot obey the
kinematic Jacobi identity even after the
generalized gauge-transformation due to
$\,\sz\hsm +\tz\hsm +\uz=0\hs$ [cf.\ Eq.\eqref{eq:s0-t0-u0}].
So additional modifications are needed.
In contrast, under the $\hs 1\hsm /s\hs$ expansion the LO
numerators of KK amplitudes depend only on the
$\hs s\hs$ and $\hs\theta\hs$ (where the mass-dependence is contained
only in $\hs s\hs$ via $\hs s\hsm =\hsm \sz\hsm +4\Mnn\,$),
and we can make the gauge-transformed numerators obey the
kinematic Jacobi identity oredr by order in the
$\hs 1\hsm /s\hs$ expansion due to
$\,s\hsm +t\hsm +u=4\Mnn\hs$ [cf.\ Eq.\eqref{eq:s-t-u}].
Hence, the $\hs 1\hsm /s\hs$ expansion is expected to realize
the double-copy construction more successfully.
In the following,
we will present the extended double-copy constructions for
the NLO massive KK amplitudes by using the high energy expansion
of $\hs 1\hsm /\sz\hs$ in section\,\ref{sec:5.3.1}
and the high energy expansion of $\hs 1\hsm /s\hs$
in section\,\ref{sec:5.3.2}.

\subsubsection{\hspace*{-2mm}Double-Copy of NLO KK Amplitudes under \boldmath{$1\hsm /\hsm\sz$} Expansion}
\label{sec:5.3.1}
\vspace*{0.5mm}

In section\,\ref{sec:5.2},
we focused on the double-copy construction
of the KK gravitational amplitudes  \eqref{AmpK-hL-nnnn}-\eqref{AmpK-h55-nnnn}
at the leading order (LO) of the high energy expansion.
For this subsection, we study the double-copy
KK amplitudes \eqref{Amp-AL-nnnn}-\eqref{Amp-A5-nnnn}
of the 5d KK YM gauge theory up to the next-to-leading order
(NLO) of the $\hs 1\hsm /\sz\hs$ expansion.
For this, we extend the reconstructed KK gravitational
amplitudes \eqref{AmpK-hL-nnnn}-\eqref{AmpK-h55-nnnn} as follows:
\beqs
\label{AmpK-hLh5-nnnn}
\begin{align}
\M [4h_L^n]
&~=~ c_0^{}\,g^2\!
\[\frac{(\NN_s)^{2}}{\sz} +\frac{(\NN_t)^{2}}{\tz}
+\frac{(\NN_u)^{2}}{\uz}\]=\M_0 +\dM\,,
\label{AmpKF-hL-nnnn}
\\[1mm]
\label{AmpKF-h55-nnnn}
\MT [4\phin]
&~=~ c_0^{}\,g^2\!
\[ \frac{(\NNt_s)^{2}}{\sz} +\frac{(\NNt_t)^{2}}{\tz}
+\frac{(\NNt_u)^{2}}{\uz}\]= \MT_0 +\dMT\,,
\end{align}
\eeqs
where the conversion constant
$\,c_0^{}\!\!=\!\!-{\ka^2}/(24g^2)\,$
is given by Eq.\eqref{eq:c0}, as determined by matching the
corresponding leading order gravitational KK amplitude
\eqref{eq:AmpLOE2-phinnnn}.
According to Eqs.\eqref{eq:N=N0+dN}-\eqref{eq:NT=NT0+dNT},
we expand the numerator factors $(\NN_j^{},\,\NNt_j^{})$
to the NLO and naively derive the following reconstructed
subleading order gravitational KK amplitudes:
\beqs
\label{NLO-dN-hLh5-nnnn}
\begin{align}
\da\M 	&\,=~ 2c_0^{}\,g^2\!
\(\frac{\NN_s^0\dNN_s^{}}{\sz} +\frac{\NN_t^0\dNN_t^{}}{\tz}
+\frac{\NN_u^0\dNN_u^{}}{\uz}\)\!,
\label{NLO-dN-hL-nnnn}
\\[1mm]
\label{NLO-dN-h5-nnnn}
\da\MT	&\,=~ 2c_0^{}\,g^2\!
\(\frac{\NNt_s^0\dNNt_s^{}}{\sz}
+\frac{\NNt_t^0\dNNt_t^{}}{\tz}
+\frac{\NNt_u^0\dNNt_u^{}}{\uz}\)\!.
\end{align}
\eeqs
We first note that the above double-copy construction should give
the correct powers of the (energy,\,mass)-dependence of
the corresponding NLO gravitational KK amplitudes under the
high energy expansion.
The structure of the KK amplitudes in the 5d KK YM gauge theory
has been well understood as we showed in
Eqs.\eqref{eq:TL-T5-exact}-\eqref{eq:dK-L5} and
Eqs.\eqref{eq:N-K}-\eqref{Amp-ALA5-nnnn}
of section\,\ref{sec:5.1}.
We see that in the 5d KK YM gauge theory,
the LO and NLO amplitudes in each channel are
$(\KK_j^0,\,\KKt_j^0)\!=\mO(E^0M_n^0)$ and
$(\dKK_j^{},\,\dKKt_j^{})\!=\mO(\Mnn/E^2)$.
Thus, the LO and NLO numerators are
$(\NN_j^0,\,\NNt_j^0)=\mO(E^2M_n^0)$ and
$(\dNN_j^{},\,\dNNt_j^{})=\mO(E^0\Mnn)$.
Hence, we generally deduce that the reconstructed double-copy
of the LO and NLO KK amplitudes for gravitational KK scattering
should have the following power-dependence on
the (energy,\,mass):
\beqs
\label{eq:Amp-LONLO-counting}
\begin{align}
\label{eq:Amp-LO-counting}
(\M_0^{},\,\MT_0^{}) &\,=\,
\mO\!\(\!\ka^2\frac{\,(\NN_j^0)^2\!,\,(\NNt_j^0)^2\,}
{s_{0j}^{}}\!\) =\,\mO\!\(\ka^2E^2M_n^0\)\!,
\\[1.mm]
\label{eq:Amp-NLO-counting}
(\da\M,\,\da\MT) &\,=\,
\mO\!\(\!\ka^2\frac{\,\NN_j^0\dNN_j^{},\,\NNt_j^0\dNNt_j^{}\,}
{s_{0j}^{}}\!\) =\,\mO\!\(\ka^2E^0\Mnn \)\!,
\end{align}
\eeqs
where
$\,s_{0j}^{}\!\in\!(s_0^{},t_0^{},u_0^{})\!=\!\mO(E^2)\,$
and we have used $\,c_0^{}g^2\sim\!\mO(\ka^2)$
according to Eq.\eqref{eq:c0}.
The above power counting fully agrees with the explicit
calculations of the KK graviton (Goldstone) amplitudes
of the compactified 5d GR (GR5)
in Eqs.\eqref{eq:AmpGR5-4phi-LONLO} and
\eqref{eq:Amp-E-2hL}-\eqref{eq:Amp-E-2phi}.
The above general power counting results
\eqref{eq:Amp-LO-counting}-\eqref{eq:Amp-NLO-counting}
are predicted by the double-copy method based upon the
amplitude structure of the well-understood
5d KK YM gauge theory (section\,\ref{sec:5.1}).
These are important for our GRET formulation
as we will discuss further in section\,\ref{sec:5.4}.

\vspace*{1mm}

As we noted in Eq.\eqref{eq:dN-dNT-sum},
the NLO numerators  $(\dNN_j^{},\,\dNNt_j^{})$
do not satisfy the kinematic Jacobi identity.
Thus, we expect that the reconstructed NLO amplitudes
by double-copy may not exactly reproduce the corresponding
gravitational amplitudes.
Using Eqs.\eqref{eq:K0-L5}-\eqref{eq:dK-L5} and
Eqs.\eqref{eq:N=N0+dN}-\eqref{eq:NT=NT0+dNT},
we can directly compute the reconstructed NLO amplitudes
\eqref{NLO-dN-hL-nnnn}-\eqref{NLO-dN-h5-nnnn} as follows:
\beqs
\label{NLO-dN-hLh5-nnnn2}
\begin{align}
\da\M
&\,=\,
-\frac{\,\ka^2 M_n^2}{192\,}
(2050 + 959\hs\ctt\! +62\hs\ctf\!+\cts)\hsm\csc^4\!\theta \,,
\label{NLO-dN-hL-nnnn2}
\\[1mm]
\label{NLO-dN-h5-nnnn2}
\da\MT
&\,=\, -\frac{\,\ka^2 M_n^2\,}{64}
(494 + 513\hs\ctt\! + 18\hs\ctf\! -\cts)\hsm\csc^4\!\theta \,,
\end{align}
\eeqs
where the conversion constant $c_0^{}$ is given by
Eq.\eqref{eq:c0} as determined by matching the
double-copy amplitudes with the gravitational amplitudes
at the leading order.
It is instructive to compare the above NLO double-copy amplitudes
\eqref{NLO-dN-hL-nnnn2}-\eqref{NLO-dN-h5-nnnn2}
with the corresponding gravitational amplitudes
\eqref{eq:Amp-E-2hL}-\eqref{eq:Amp-E-2phi}
as directly computed from the 5d KK GR theory.
It is good to see that the reconstructed NLO
double-copy amplitudes \eqref{NLO-dN-hL-nnnn2}-\eqref{NLO-dN-h5-nnnn2}
indeed have the same kinematic structures
as that of the corresponding gravitational amplitudes
\eqref{eq:Amp-E-2hL}-\eqref{eq:Amp-E-2phi}
because they all contain the angular terms
of the type
$\,(1,\,\ctt,\,\ctf,\,\cts)\!\times\!\csc^4\!\theta\,$
though their coefficients differ.
Their differences in the coefficients are qiute expected
because our Eq.\eqref{eq:dN-dNT-sum} shows that
the NLO numerators  $(\dNN_j^{},\,\dNNt_j^{})$
do not satisfy the kinematic Jacobi identity
even though in each channel of the NLO amplitudes
\eqref{NLO-dN-hL-nnnn}-\eqref{NLO-dN-h5-nnnn} the numerator
$\,\NN_j^0\dNN_j^{}\,$ or $\,\NNt_j^0\dNNt_j^{}\,$
contains product of both LO and NLO factors where
the LO factors $(\NN_j^{0},\,\NNt_j^{0})$
still obey the kinematic Jacobi identities.
Thus, we do not expect the current BCJ-type double-copy method
would exactly hold.

\vspace*{1mm}

Next, we further compute the differences between the
NLO amplitudes of the longitudinal KK graviton scattering and
of the KK gravitational Goldstone bosons
for the original gravitational amplitudes
\eqref{eq:Amp-E-2hL}-\eqref{eq:Amp-E-2phi}
of the GR5 and
for the above reconstructed amplitudes \eqref{NLO-dN-hL-nnnn2}-\eqref{NLO-dN-h5-nnnn2}
by double-copy (DC)
at the NLO:
\beqs
\label{eq:Diff2-AmpGR5DC-nnnn}
\begin{align}
\label{eq:Diff2-AmpGR5-nnnn}
\Delta\M (\rm{GR5}) &\,=\,
\da\M -  \da\MT \,=\,
-\frac{~3\ka^2\!M_n^2~}{2}
\hs (19.5 + \ctt) \hs,
\\[1mm]
\label{eq:Diff2-AmpDC-nnnn}
\Delta\M (\rm{DC}) &\,=\,
\da\M -  \da\MT \,=\,
\frac{~\ka^2\!M_n^2~}{12}\hs
(-69 +4\hs\ctt +\ctf)\hsm\csc^2\!\theta \hs,
\end{align}
\eeqs
which exhibit different angular structures.

\vspace*{1mm}

We see that the GR5 result \eqref{eq:Diff2-AmpGR5-nnnn} contains
only the terms of $(1,\,\ctt)$ types due to rather
{\it precise cancellations of the
$(\ctf,\,\cts)\times\csc^4\!\theta$ terms} between
the amplitudes \eqref{eq:Amp-E-2hL} and \eqref{eq:Amp-E-2phi},
while the double-copy result \eqref{eq:Diff2-AmpDC-nnnn}
contains an extra non-cancelled angular term $\,\ctf\,$ and
an extra overall angular factor
$\,\csc^2\hsm\theta\,$.
This shows the failure of the double-copy result
\eqref{eq:Diff2-AmpDC-nnnn}
to correctly reconstruct even the structure of $(1,\,\ctt)$
in the original GR5 result \eqref{eq:Diff2-AmpGR5-nnnn}.
In fact, this precise cancellation is highly nontrivial
because after careful examination we observe that
this precise cancellation depends on {\it all the coefficients}
in the angular structure
$\,(1,\,\ctt,\,\ctf,\,\cts)\!\times\!\csc^4\!\theta\,$
of both the original gravitational amplitudes
\eqref{eq:Amp-E-2hL} and \eqref{eq:Amp-E-2phi}.
We find that if one changes by hand
any one of these coefficients [even for the constant term
inside the parentheses of
$\,(\cdots)\!\times\csc^4\!\theta$\,]
by any small number (such as $+1$ or $-1$)
in either the KK graviton amplitude \eqref{eq:Amp-E-2hL} or
the KK Goldstone amplitude \eqref{eq:Amp-E-2phi}, then
it has to destroy this precise cancellation
in the amplitude-difference
$\,\Delta\M (\text{GR5})$\, of Eq.\eqref{eq:Diff2-AmpGR5-nnnn}
and thus all the terms of
$\,(1,\,\ctt,\,\ctf,\,\cts)\times\csc^4\!\theta\,$
in the original amplitudes have to reappear in the difference
$\,\Delta\M (\text{GR5})$\,.

\vspace*{1mm}

We can understand the failure of the correct cancellation
in the reconstructed result $\,\Delta\M (\text{DC})$\,
of Eq.\eqref{eq:Diff2-AmpDC-nnnn}
by noting the violation of the kinematic Jacobi identity
for the NLO numerators  $(\dNN_j^{},\,\dNNt_j^{})$
as shown in Eq.\eqref{eq:dN-dNT-sum}.
In fact, by inspecting
Eqs.\eqref{NLO-dN-hL-nnnn}-\eqref{NLO-dN-h5-nnnn},
we note that for each given channel the amplitude-difference
$\,\Delta\M (\rm{DC})\!=\da\M\!-\da\MT\,$ has the numerator
$\,\NN_j^0\dNN_j^{}-\NNt_j^0\dNNt_j^{}\,$
which could not even be factorized into any BCJ-type product
$\,X_j^{}Y_j^{}\,$ with each factor ($X_j^{}$ or $Y_j^{}$)
obeying the kinematic Jacobi identity separately.
Hence, it is no surprise that the reconstructed
amplitude-difference $\,\Delta\M (\rm{DC})$\,
could not even reproduce the correct structure of
the original GR5 result \eqref{eq:Diff2-AmpGR5-nnnn}.

\vspace*{1mm}

In the following, we will try to construct an improved
amplitude-difference $\,\Delta\xoverline{\M} (\rm{DC})$\,
in which the numerator of each channel can take the
BCJ-type product form $\,X_j^{}Y_j^{}\,$
with each factor ($X_j^{}$ or $Y_j^{}$)
obeying the kinematic Jacobi identity separately.
For the above purpose,  we first rewrite the reconstructed
NLO KK scattering amplitudes
\eqref{NLO-dN-hL-nnnn}-\eqref{NLO-dN-h5-nnnn}
by using the relation \eqref{eq:KL-K5}:
\\[-9mm]
\beqs
\label{eq:dM-dMT'}
\begin{align}
\label{eq:dMdM'+X}
\da\M	&\,=\, \da\M' - 8c_0^{}g^2\ct\sum_j\!\dNN_j^{}
\,=\,  \da\M' - 8c_0^{}g^2\ct\,\chi\,,
\\[1mm]
\label{eq:dMTdMT'+X}
\da\MT	&\,=\, \da\MT' + 8c_0^{}g^2\ct\sum_j\!\dNNt_j^{}
\,=\,  \da\MT' + 8c_0^{}g^2\ct\,\chi\,,
\\[0mm]
\label{eq:dMT'}
\da\M'	&\hs\equiv\,
2c_0^{}\,g^2\!\sum_j\!\frac{\,\NNt_j^0\dNN_j^{}\,}{s_{0j}^{}} \,,
\quad~~
\da\MT'	\hs\equiv\,
2c_0^{}\,g^2\!\sum_j\!\frac{\,\NN_j^0\dNNt_j^{}\,}{s_{0j}^{}} \,,
\end{align}
\eeqs
where $\,s_{0j}^{}\!\in\!(\sz,\,\tz,\,\uz)\,$,
and we have used Eq.\eqref{eq:dN-dNT-sum} in the last step of
Eqs.\eqref{eq:dMdM'+X}-\eqref{eq:dMTdMT'+X}.
It is clear that the last terms on the RHS of
Eqs.\eqref{eq:dMdM'+X}-\eqref{eq:dMTdMT'+X} are proportional to
$\,\sum_j\!\dNN_j^{}=\sum_j\!\dNNt_j^{}\!=\chi \neq 0\,$,
which violate the kinematic Jacobi identity.

\vspace*{1mm}

Then, we can compute the difference between the NLO
KK longitudinal and Goldstone amplitudes:
\beqs
\label{eq:DM1-DM2}
\begin{align}
\label{eq:DM1}
\Delta\M_1^{} &\,\equiv\,
\da\M'\!-\da\MT \,=\, 2c_0^{}g^2
\sum_j\!\frac{\,\NNt_j^0(\dNN_j^{}\!-\dNNt_j^{})\,}{s_{0j}^{}}
\,=\,-\ka^2 M_n^2( 7 \!+\! \ctt )\,,
\\[1mm]
\label{eq:DM2}
\Delta\M_2^{} &\,\equiv\,
\da\M\!-\da\MT' \,=\, 2c_0^{}g^2
\sum_j\!\frac{\,\NN_j^0(\dNN_j^{}\!-\dNNt_j^{})\,}{s_{0j}^{}}
\,=\,-\ka^2 M_n^2( 7 \!+\! \ctt )\,,
\end{align}
\eeqs
where in the last steps of Eqs.\eqref{eq:DM1}-\eqref{eq:DM2},
we have computed each sum directly by using the LO and NLO
numerators of the KK gauge (Goldstone) amplitudes
(section\,\ref{sec:5.1}) as well as Eq.\eqref{eq:c0}
for the conversion constant $c_0^{}\,$.
This explicit calculation shows an equality
$\,\Delta\M_1^{}\!=\Delta\M_2^{}\,$.
We can prove this equality in a more general way.
Using Eqs.\eqref{eq:DM1}-\eqref{eq:DM2}
and Eqs.\eqref{eq:dMdM'+X}-\eqref{eq:dMTdMT'+X},
we reexpress the difference \eqref{eq:Diff2-AmpDC-nnnn}
of the NLO double-copy amplitudes as follows:
%
\begin{equation}
\label{eq:DM(DC)-mod1}
\Delta\M (\rm{DC}) \,=\,
\da\M\! -  \da\MT \,=\,
\Delta\M_1^{}\! - 8c_0^{}g^2\ct\!\sum_j\!\dNN_j^{}
\,=\, \Delta\M_2^{}\! -8c_0^{}g^2\ct\!\sum_j\!\dNNt_j^{}\,,
\end{equation}
%
where $\,\sum_j\!\dNN_j^{}\!=\sum_j\!\dNNt_j^{}\!=\!\chi\,$
because of the equality \eqref{eq:dN-dNT-sum1}.
This leads to $\,\Delta\M_1^{} \!=\! \Delta\M_2^{}$\,,
which agrees with the explicit calculations of
Eq.\eqref{eq:DM1-DM2}.
Hence, we deduce
\beqs
\label{eq:DM(DC)-BDM(DC)-X}
\begin{align}
\label{eq:DM(DC)=BDM(DC)-X}
\Delta\M (\rm{DC}) &\,=\,
\Delta\xoverline{\M}(\rm{DC}) -\mathbb{X}\,,
\\[1mm]
\label{eq:X}
\mathbb{X} &\,\equiv\, 8c_0^{}g^2\ct\,\chi
= \frac{2}{3}\ka^2\Mnn (7+\ctt)\cot^2\!\theta \,,
\end{align}
\eeqs
and
%
\begin{align}
\label{eq:BDM(DC)}
\Delta\xoverline{\M}(\rm{DC}) &\,\equiv\,
\Delta\M_1^{} = \Delta\M_2^{}
\,=\, -\ka^2 M_n^2\,( 7 + \ctt )\,.
\end{align}
%
It is important to note that in Eq.\eqref{eq:DM(DC)=BDM(DC)-X}
we have identified and separated a special term $\XX$ from the
amplitude-difference $\Delta\M (\rm{DC})$,
where $\,\XX \!\propto\! \chi\,$ violates the kinematic Jacobi identity
at the NLO as shown in Eq.\eqref{eq:dN-dNT-sum1}.
By doing so, we observe that the improved
amplitude-difference $\,\Delta\xoverline{\M} (\rm{DC})\,$,\,
as defined in Eq.\eqref{eq:DM1-DM2},
does have a good feature, namely, each numerator
of Eq.\eqref{eq:DM1} [Eq.\eqref{eq:DM2}]
just equals the product of the LO factor
$\,\NNt_j^{0}$ ($\,\NN_j^{0}\,$)
and the NLO factor $\,\dNN_j^{}\!-\dNNt_j^{}\,$,
which satisfy separately the kinematic Jacobi identities
\eqref{eq:KJacobi} and \eqref{eq:dNj-dNTj-Jacobi}.

This is just the desired feature as required
by the conventional BCJ-type double-copy
construction\,\cite{BCJ:2008}\cite{BCJ:2019}.\ 
On the other hand, the situation of $\Delta\M (\rm{DC})$
[\eqrefe{eq:Diff2-AmpDC-nnnn}] is different
because in each channel the numerator of
$\Delta\M (\rm{DC})$
cannot be factorized into a simple product of two factors
which could hold the kinematic Jacobi identity separately.

\vspace*{1mm}

Then, it is instructive to compare our improved
amplitude-difference $\,\Delta\xoverline{\M} (\rm{DC})\,$
[Eq.\eqref{eq:BDM(DC)}] by double-copy construction
with the original gravitational amplitude-difference
$\Delta\M (\rm{GR5})$ [Eq.\eqref{eq:Diff2-AmpGR5-nnnn}]
as computed in the compactified 5d GR.
It is impressive that our improved
amplitude-difference $\,\Delta\xoverline{\M} (\rm{DC})\,$
in Eq.\eqref{eq:BDM(DC)} does have a much simpler structure
than the naive double-copy construction
$\,\Delta\M (\rm{DC})\,$ in Eq.\eqref{eq:Diff2-AmpDC-nnnn},
because the undesired extra $\,\ctf\,$ term and extra overall
factor $\,\csc^2\!\theta\,$ of $\,\Delta\M (\rm{DC})\,$
fully disappear in our improved
amplitude-difference $\,\Delta\xoverline{\M} (\rm{DC})\,$.
This comparison shows that our improved
amplitude-difference $\,\Delta\xoverline{\M} (\rm{DC})\,$
does share the same kinematic structure of $(1,\,\ctt)$
as that of $\,\Delta\M (\rm{GR5})$\, in the GR5,
although their coefficients are still different.
Given the fact that the conventional BCJ approach
was formulated only for the massless gauge and gravity theories,
it is expected that for constructing the {\it mass-dependent
scattering amplitudes} such as the NLO amplitudes
of our 5d KK theories, the conventional BCJ approach
would not exactly work.
Nevertheless, we have shown that our reconstructed
KK longitudinal graviton and Goldstone scattering amplitudes
\eqref{NLO-dN-hL-nnnn2}-\eqref{NLO-dN-h5-nnnn2}
indeed exhibit the {\it same kinematic structure}
$\,(1,\,\ctt,\,\ctf,\,\cts)\!\times\!\csc^4\!\theta\,$
as that of the corresponding gravitational KK amplitudes
\eqref{eq:Amp-E-2hL}-\eqref{eq:Amp-E-2phi}.

\vspace*{1mm}

Furthermore, the double-copy reconstruction of the
KK amplitude-difference at the NLO is much more nontrivial
because the original gravitational amplitude-difference
$\Delta\M (\rm{GR5})$ [Eq.\eqref{eq:Diff2-AmpGR5-nnnn}]
contains very {\it precise cancellations of the terms
$\,(\ctf,\,\cts)\!\times\csc^4\!\theta$\,}
between the amplitudes
\eqref{eq:Amp-E-2hL}-\eqref{eq:Amp-E-2phi}.
The naive double-copy construction of the NLO amplitude-difference
$\,\Delta\M (\rm{DC})\,$ [Eq.\eqref{eq:Diff2-AmpDC-nnnn}]
fails to reproduce the correct kinematic structure
of the $\Delta\M (\rm{GR5})$.
But, it is impressive that
after we properly define the improved amplitude-difference
$\,\Delta\xoverline{\M} (\rm{DC})\,$
as in Eq.\eqref{eq:BDM(DC)} and Eqs.\eqref{eq:DM1}-\eqref{eq:DM2}
by removing the Jacobi-violating term and
ensuring its numerator in each channel
factorized into product factors (obeying
the kinematic Jacobi identities respectively),
we find that the improved double-copy result
$\,\Delta\xoverline{\M} (\rm{DC})\,$
[Eq.\eqref{eq:BDM(DC)}]
does exhibit the {\it same kinematic structure}
as that of the original  gravitational amplitude-difference
$\Delta\M (\rm{GR5})$ [Eq.\eqref{eq:Diff2-AmpGR5-nnnn}].
This is an encouraging evidence showing that
as long as the BCJ-type numerators can be properly improved
to satisfy the kinematic Jacobi identities,
such a double-copy approach is still quite meaningful
to certain extent, {\it predicting the correct structure of
the corresponding gravitational amplitudes and
the (energy,\,mass)-dependence up to NLO,}
even for the mass-dependent amplitudes.

\vspace*{1mm}

In the rest of this subsection, we will attempt to make an
improved double-copy construction of the NLO KK amplitudes
and reproduce the original NLO KK graviton (Goldstone) amplitudes
\eqref{NLO-dN-hL-nnnn2}-\eqref{NLO-dN-h5-nnnn2}
by following our proposal of the improved NLO numerators
$(\dNN_j',\,\dNNt_j')$ in Eq.\eqref{eq:dN'-dNT'}
which have the desired property of
satisfying the kinematic Jacobi identities
\eqref{eq:KJacobi-N'}-\eqref{eq:KJacobi-NT'}.
Moreover, the corresponding improved NLO KK
longitudinal and Goldstone amplitudes
$\,(\dT_L',\,\delta\tT_5')$\, still obey
the KK GAET identity \eqref{eq:Tv'} which reflects
the KK YM gauge symmetry.

\vspace*{1mm}

Using the improved KK gauge (Goldstone) boson amplitudes
\eqref{AmpMod-AL-nnnn}-\eqref{AmpMod-A5-nnnn}, we construct
the following new NLO gravitational KK amplitudes by double-copy:
\beqs
\label{eq:DC-dM"-dMT"}
\begin{align}
\da\M'' 	&\,=~ 2c_0^{}\,g^2\!
\(\frac{\NN_s^0\dNN_s'}{\sz} +\frac{\NN_t^0\dNN_t'}{\tz}
+\frac{\NN_u^0\dNN_u'}{\uz}\)\!,
\label{eq:DC-dM"-hL}
\\[1mm]
\label{eq:DC-dMT"-h5}
\da\MT''	&\,=~ 2c_0^{}\,g^2\!
\(\frac{\NNt_s^0\dNNt_s'}{\sz}
+\frac{\NNt_t^0\dNNt_t'}{\tz}
+\frac{\NNt_u^0\dNNt_u'}{\uz}\)\!,
\end{align}
\eeqs
where the improved NLO numerators
$\,\dNN_j'\!=\dNN_j^{}\!-\chi_j^{}\,$
and $\,\dNNt_j'\!=\dNNt_j^{}\!-\chit_j^{}\,$
as defined in Eq.\eqref{eq:dN'-dNT'}.
Since the NLO gravitational KK amplitudes
\eqref{NLO-dN-hL-nnnn2}-\eqref{NLO-dN-h5-nnnn2}
only contain angular factors
$\cos m\theta$ (with $m=2,4,6$) and
$\csc^4\!\theta\!=\!1/\sin^4\!\theta\,$
which are invariant under $\,\theta\ito\pi-\theta$\,,
we may choose the decomposition terms
in Eq.\eqref{eq:chi=sum(chi_j)} as
\,$(\chi_s^{},\,\chi_t^{},\,\chi_u^{})=(\chi,\,z,\,-z)$\,
and
$\,(\chit_s^{},\,\chit_t^{},\,\chit_u^{})=(\chi,\,\zT,\,-\zT )$\,,\,
where $\,\chi\,$ is given by Eq.\eqref{eq:dN-dNT-sumX}.
Thus, using Eq.\eqref{eq:dN'-dNT'} we express
the improved NLO numerators as follows:
\beqs
\label{eq:dN'dNT'-XzzT}
\begin{align}
(\dNN_s',\,\dNN_t',\,\dNN_u') &\,=\,
(\dNN_s^{}\!-\!\chi,\,\dNN_t^{}\!-\!z,\,\dNN_u\!+\!z\,)\hs ,
\\[1mm]
(\dNNt_s',\,\dNNt_t',\,\dNNt_u') &\,=\,
(\dNNt_s^{}\!-\!\chi,\,\dNNt_t^{}\!-\!\zT,\,\dNNt_u\!+\!\zT\,)\hs .
\end{align}
\eeqs
The new parameters $(z,\,\zT)$ are functions of $\,\theta\,$
and will be determined by matching the reconstructed
NLO KK amplitudes $(\da\M'',\,\da\MT'')$
in Eq.\eqref{eq:DC-dM"-hL}-\eqref{eq:DC-dMT"-h5}
with the original NLO KK graviton (Goldstone) amplitudes
$(\da\M,\,\da\MT)$
in Eqs.\eqref{eq:Amp-E-2hL}-\eqref{eq:Amp-E-2phi}
of the 5d KK GR:
\begin{equation}
\label{eq:dM"=dM(GR5)}
\da\M'' =\, \da\M\,, ~~~~
\da\MT'' =\, \da\MT\,.
\end{equation}
Thus, we can solve the parameters $\,(z,\,\zT)$\,
from Eq.\eqref{eq:dM"=dM(GR5)} as follows:
\beqs
\label{eq:sol-z-zT}
\begin{align}
\label{eq:sol-z}
z &\,=\,
\frac{\,M_n^2\(614+371\ctt\!+42\ctf\!-3\cts\)\,}
{16\(7\!+\ctt\)\sin^2\!\theta}\hs ,
\\[2mm]
\label{eq:sol-zT}
\zT &\,=\,
\frac{\,M_n^2(1666-\!1025\ctt\!+382\ctf\!+\cts)\,}
{16\(7\!+\ctt\)\sin^2\!\theta} \hs .
\end{align}
\eeqs
Finally, by substituting the improved NLO numerators
\eqref{eq:dN'dNT'-XzzT} with Eqs.\eqref{eq:sol-z}-\eqref{eq:sol-zT}
into Eqs.\eqref{eq:DC-dM"-hL}-\eqref{eq:DC-dMT"-h5},
we obtain the reconstructed NLO KK amplitudes:
\beqs
\label{eq:AmpE-2-hL-phi}
\begin{align}
\dM''\text{(DC)} &\,=\, -\frac{\,\ka^2 M_n^2\,}{128}
\(650+261\hs\ctt+102 \ctf+11\hs\cts\)\hsm \csc^4\!\theta\hs,
\label{eq:AmpE-2hL}
\\[1mm]
\dMT''\text{(DC)} &\,=\,
- \frac{\,\ka^2 M_n^2\,}{128}
(-706 +2049\hs\ctt -318\hs\ctf -\cts )\hsm\csc^4\!\theta
\hs ,
\label{eq:AmpE-2phi}
\end{align}
\eeqs
which reproduce precisely
the original NLO gravitational KK amplitudes
$\,(\da\M ,\,\da\MT)$\, in
Eqs.\ \eqref{eq:Amp-E-2hL}-\eqref{eq:Amp-E-2phi}
of the 5d KK GR theory, as expected.\
This gives a consistency check of the above analysis.

\vspace*{1mm}

In passing, it would be useful to extend our
present LO and NLO analyses to the scattering processes
with five or more external particles in our future work.
We also note that the original BCJ conjecture was inspired
by the KLT relation that connects the amplitudes of
the massless gravity theory
to that of the massless YM gauge theory.
The KLT kernal may be further reinterpreted as
the inverse amplitude of a bi-adjoint scalar field theory\,\cite{CHY}.
In Appedix\,\ref{app:G}, we will extend the
KLT double-copy approach for constructing the four-particle
KK graviton amplitudes
and demonstrate the consistency with the above
improved BCJ construction.

\subsubsection{\hspace*{-2mm}Improved Double-Copy of NLO KK Amplitudes under \boldmath{$1\hsm /\hsm s$} Expansion}
\label{sec:5.3.2}
\vspace*{0.5mm}

In this subsection, we present an improved double-copy construction
of the KK graviton (Goldstone) amplitudes from the KK gauge boson
(Goldstone) amplitudes under the high energy expansion of
$\hs 1\hsm /s\hs$. With this we can construct improved numerators for
the KK gauge boson (Goldstone) amplitudes which can fully satisfy
the kinematic Jacobi identity.
We will show that this improved massive double-copy approach
is better than the double-copy method by using the
$\hs 1\hsm /\sz\hs$  expansion
as we gave in the previous subsection\,\ref{sec:5.3.1}.

\vspace*{1mm}

In the following, we make the high energy expansion of
$\hs 1\hsm /s\hs$, where $\hs s\hs$ is the conventional
Mandelstam variable and $\hs s\hsm =\hsm \sz + 4\Mnn\hs$
holds for the four-particle elastic scattering.
We use the notations
$\hs (\bs,\,\bsz)= (s,\,\sz )/\Mnn\,$ and thus
$\,\bs \!=\!\bsz + 4\,$.\
For the exact tree-level
KK longitudinal gauge boson amplitude $\,\TT[4A_L^n]$\,
and the KK Goldstone boson amplitude
$\,\tT[4A_5^n]$\, in Eq.\eqref{eq:TL-T5-exact},
we can reexpress their kinematic factors \eqref{eq:K-KT-exact}
in terms of the conventional Mandelstam variable
$\hs\bs\hs$ as follows:
\beqs
\label{eq:K-Kt-s}
\begin{align}
\KK_s &\dis =
-\frac{\,(4\bs^2 \!-\! 5\bs \!-\! 8)\ct\,}{2\bs} \,,
\hspace*{1mm}
&\KKt_s &\dis =
-\frac{\,(3\bs \!-\! 8)\ct\,}{2\bs} \,,
\\[1.5mm]
\KK_t &\dis =
-\frac{\,Q_0^{}\!+ Q_1^{}\ct\! + Q_2^{}\ctt\!+Q_3^{}c_{3\theta}^{}\,}
{\,2(\bs \!-\! 4)[(3\bs\!+\!4)\!+\!4\bs\hs\ct\!+\!(\bs\!-\!4)\ctt] \,} \,,
\hspace*{1mm}
&\KKt_t &\dis =
\frac{\,\Qt_0^{}\!+ \Qt_1^{}\ct\! + \Qt_2^{}\ctt\,}
{\,2(\bs \!-\! 4)[(3\bs\!+\!4)\!+\!4\bs\hs\ct\!+\!(\bs\!-\!4)\ctt] \,} \,,
\\
\KK_u &\dis =
\frac{\,Q_0^{}\!- Q_1^{}\ct\! + Q_2^{}\ctt\!-Q_3^{}c_{3\theta}^{}\,}
{\,2(\bs \!-\! 4)[(3\bs\!+\!4)\!-\!4\bs\hs\ct\!+\!(\bs\!-\!4)\ctt] \,} \,,
\hspace*{1mm}
&\KKt_u &\dis =
-\frac{\,\Qt_0^{}\!- \Qt_1^{}\ct\! + \Qt_2^{}\ctt\,}
{\,2(\bs \!-\! 4)[(3\bs\!+\!4)\!-\!4\bs\hs\ct\!+\!(\bs\!-\!4)\ctt]\,} \,,
\end{align}
\eeqs
where the functions $\{Q_j^{},\Qt_j^{}\}$ are given by
\begin{equation}
\begin{alignedat}{3}
& Q_0^{} =  8\bs^3 \!- \! 63\bs^2  \!+ \! 72\bs  \!+ \! 80 \hs, \hspace*{8mm}
&&\Qt_0^{}= 15\bs^2 \!+\! 24\bs \!-\! 80 \hs,
\\
&Q_1^{} = 2 (7\bs^3  \!- \! 44 \bs^2  \!+ \! 80\bs  \!- \! 64) \hs,
\hspace*{8mm}
&&\Qt_1^{} = 4(3\bs^2 \!-\! 20\bs \!+\! 32) \hs,
\\
& Q_2^{} = 8 \bs^3  \!- \! 45\bs^2  \!+ \! 8\bs  \!+ \! 48 \hs,
\hspace*{8mm}
&&\Qt_2^{} = -3(\bs\!-\!4)^2  \hs,
\\
&Q_3^{} = 2 \bs(\bs^2  \!- \! 10\bs  \!+ \! 24)  \hs.
\end{alignedat}
\end{equation}
Then, we make the $\hs 1\hsm /s\hs$ expansion
for \eqrefe{eq:K-Kt-s} and derive the LO expressions,
\beqs
\begin{alignat}{3}
\KK_s^0 &= \frac{\,5\ct\,}{2} \,,
\hspace*{10mm}
& \KKt_s^0  &= -\frac{\,3\ct\,}{2} \,,
\\[1mm]
\KK_t^0  &=\frac{\,13 \!+\! 5 \ct \!+\! 4 \ctt\,}{2(1 \!+\! \ct)\,}  \,, \hspace*{10mm}
&\KKt_t^0 &=\frac{\,3(3 \!-\! \ct)\,}{2(1 \!+\! \ct)\,}  \,,
\\[1mm]
\KK_u^0  &=-\frac{\,13 \!-\! 5 \ct \!+\! 4 \ctt\,}{2(1 \!-\! \ct)\,} \,, \hspace*{10mm}
&\KKt_u^0  &=-\frac{\,3(3\!+\!\ct)\,}{2(1 \!-\!\ct)\,} \,,
\end{alignat}
\eeqs
and the NLO expressions,
\beqs
\begin{alignat}{3}
\dKK_s &= \frac{4\ct}{\bs} \,, \hspace*{10mm}
&\dKKt_s & =\frac{4\ct}{\bs} \,,
\\[1mm]
\dKK_t &= -\frac{\,4 (2\!-\!3\ct\!-\!2\ctt\!-\!\cttt)\,}
{(1\!+\!\ct)^2\, \bs\,} \,, \hspace*{10mm}
&\dKKt_t&= \frac{16\hs\ct}{(1\!+\!\ct)^2\,\bs} \,,
\\[1mm]
\dKK_u &= \frac{\,4 (2\!+\!3\ct\!-\!2\ctt\!+\!\cttt)\,}
{(1\!-\!\ct)^2\,\bs\,}
\,, \hspace*{10mm}
&\dKKt_u&= \frac{16\hs\ct}{\,(1\!-\!\ct)^2\,\bs\,} \,.
\end{alignat}
\eeqs
We further define the BCJ-type numerators:
\beqs
\begin{alignat}{3}
& \NN_j =
s_j \hs \KK_j  \hs,
\hspace*{10mm}
&& \NN_j = \NN_j^0\!+\da\NN_j
= s_j (\KK_j^0\!+\da\KK_j)\hs,
\\[1mm]
& \NNt_j =
s_j \hs \KKt_j  \hs,
\hspace*{10mm}
&& \NNt_j = \NNt_j^0\!+\da\NNt_j
= s_j (\KKt_j^0\!+\da\KKt_j) \hs,
\end{alignat}
\eeqs
where the subscripts $\,j\!\in\!(s,\hs t,\hs u)\,$ and
$\,s_j^{}\!\in\!(s,\hs t,\hs u)\,$.
With these, we reformulate the amplitudes \eqref{eq:TL-T5-exact}
in the following forms:
\beqs
\label{eq:AmpNj-4AL-4A5}
\begin{align}
\TT[A_L^{a\hs n}A_L^{b\hs n}\hsm\ito A_L^{c\hs n}A_L^{d\hs n}]
\,=~\,& g^2 \!\(\!\frac{\,\CC_s\hs\NN_s\,}{s} +
\frac{\,\CC_t\hs \NN_t\,}{t}
+ \frac{\,\CC_u\,\NN_u\,}{u} \!\)\!,
\\[1.mm]
\tT[A_5^{a\hs n}A_5^{b\hs n}\hsm\ito A_5^{c\hs n}A_5^{d\hs n}]
\,=~\,& g^2 \!\(\!\hsmx
\frac{\,\CC_s\hs\NNt_s\,}{s} +
\frac{\,\CC_t\hs\NNt_t\,}{t} +
\frac{\,\CC_u\hs\NNt_u\,}{u}
\hsm\!\) \!.
\end{align}
\eeqs
We note that the newly formed LO numerators
$\hs\{\NN_j^{0},\, \NNt_j^0\hs\}$
and NLO numerators
$\hs\{\dNN_j, \, \dNNt_j\}$
are mass-dependent (through $\hs s\hs$), and
their sums violate the kinematic Jacobi identity:
\beqs
\label{eq:Nj-sum-1/s}
\begin{align}
&\sum_j \NN_j^{0} = 10\hs\ct \Mnn  \,,~~~~
\sum_j \NNt_j^{0} = -6\hs\ct \Mnn  \,,
\\[1mm]
& \sum_j \da\NN_j =
-\frac{\,\Mnn\,}{2\bs}
[13 \bs \!-\! 496 \!-\! 12 \bs \hs \ctt \!-\! (\bs \!+\! 16)\ctf]\ct \csc^4 \hsmx \theta \,,
\\
& \sum_j \da\NNt_j  =
-\frac{\,\Mnn\,}{2\bs}
[13 \bs \!-\! 448 \!-\! 4 (3\bs \!+\! 16) \ctt \!-\! \bs \hs \ctf]\ct \csc^4 \hsmx \theta \,.
\end{align}
\eeqs
The violation of the kinematic Jacobi identity
at both the LO and NLO shows that
the conventional BCJ double-copy method of the massless gauge theories
cannot be naively applied to the case of the elastic scattering
amplitudes of KK gauge (Goldstone) bosons.
But, we observe that the amplitudes \eqref{eq:AmpNj-4AL-4A5}
are invariant under the generalized gauge transformations
of the kinematic numerators:
\begin{equation}
\label{eq:GGtransf}
\NN_{\hsm j}^{\pp} =\hs \NN_{\hsm j} + s_{\hsm j}^{}\hs\Delta \,,
\hspace*{8mm}
\NNt_{\hsm j}^{\pp} =\hs \NNt_{\hsm j}  + s_{\hsm j}^{}\hs
\widetilde{\Delta} \,.
\end{equation}
Thus, we can realize the kinematic Jacobi identities
for the gauge-transformed numerators
\beqa
\label{eq:Jacobi-Nj'-Ntj'}
\sum_j\hsm\NN_{\hsm j}^{\pp}\hsm = 0\,,
\hspace*{8mm}
\sum_j\hsm\NNt_{\hsm j}^{\pp}\hsm = 0\,,
\eeqa
by generally solving the gauge parameters
$(\Delta ,\,\widetilde{\Delta})$
for the elastic amplitudes as follows:
\begin{equation}
\label{eq:sol-Delta-tDelta}
\Delta =
-\frac{1}{\,4\Mnn\,}\!\sum_j \NN_j \,,
\hspace*{8mm}
\widetilde{\Delta} =
-\frac{1}{\,4\Mnn\,}\!\sum_j \NNt_j \,.~~~
\end{equation}
Expanding both sides of \eqrefe{eq:sol-Delta-tDelta}, we derive the gauge parameters
$(\Delta,\,\widetilde{\Delta})=
(\Delta_0\hsm +\hsm\Delta_1,\,
\widetilde{\Delta}_0\hsm +\hsm\widetilde{\Delta}_1)\,
$
at the LO and NLO$\hs$:
\beqs
\label{eq:sol-Delta01}
\begin{alignat}{3}
\Delta_0 &=
\frac{1}{\,4\,}
(9\hsm +\hsm 7\hsm\ctt)\ct\csc^2\!\theta \,,
\qquad
&\widetilde{\Delta}_0 &=
\frac{1}{\,4\,}
(17\!-\hsm\ctt)\ct\csc^2\!\theta \,,
\label{eq:sol-Delta0}
\\[1mm]
\Delta_1& =
-\frac{\,(62\ct^{}\! +\hsm 5\cttt\! +\hsm c_{5\theta}^{})\hsm
\csc^4\!\theta\,}{\bs}
\,, \qquad
&\widetilde{\Delta}_1 &=
-\frac{\,(15\ct^{}\! +\hsm \cttt)\hsm\csc^4\!\theta\,}{\bs} \,.
\label{eq:sol-Delta1}
\end{alignat}
\eeqs

Using Eqs.\eqref{eq:GGtransf} and \eqref{eq:sol-Delta0},
we derive the gauge-transformed LO numerators
$(\NN_{\hsm j}^{0\hs\pp},\,\NNt_{\hsm j}^{0\hs\pp})$,
which are {\it equal} to each other:
\beqs
\label{eq:Nj'-Ntj'-LO}
\begin{align}
\NN_s^{0\hs\pp} &=\, \NNt_s^{0\hs\pp} =\,
\fr{1}{\,2\,}\hs s\hs
(7\hsm +\hsm\ctt)\hs\ct\csc^2\!\theta \,,
\\[1mm]
\NN_t^{0\hs\pp} &=\, \NN_t^{0\hs\pp} =\,
-\frac{~s\hs (42\hsm -\hsmx 15\hs\ct\hsm
+\hsm 6\hs\ctt\hsm -\hsm\cttt)~}
{16\hs (1 \!-\hsm \ct)}
\,,\hspace*{5mm}
\\
\NN_u^{0\hs\pp} &=\, \NNt_u^{0\hs\pp} =\,
\frac{~s\hs (42\hsm +\hsmx 15\hs\ct\hsm
+\hsm 6\hs\ctt\hsm +\hsm\cttt)~}
{16\hs (1 \!+\hsm \ct)} \,.
\label{eq:N0j'}
\end{align}
\eeqs
We see that the LO numerators
$(\NN_{\hsm j}^{0\hs\pp},\,\NNt_{\hsm j}^{0\hs\pp})$
are of $\mO(E^2M_n^0)$ and the LO equality holds:
$\NN_j^{0\hs\pp} \!= \NNt_j^{0\hs\pp}\hs$.
Hence, we deduce the {\it equivalence} between the two LO amplitudes
\beqa
\label{eq:LO-GAET'}
\T_{\hs 0}^{\,\prime}[4A^n_L]\,=\,\tT_{\hs 0}^{\,\prime}[4A_5^n]\,,
\eeqa
at the $\mO(E^0M_n^0)$,
which is an explicit realization of the KK GAET\,\cite{5DYM2002}.

\vspace*{1mm}

Then, substituting Eq.\eqref{eq:sol-Delta1} into
Eq.\eqref{eq:GGtransf}, we further derive the gauge-transformed NLO
numerators $\,\dNN_{\hsm j}^{\pp}\,$
for the elastic KK gauge boson amplitude:
\beqs
\label{eq:Nj'-NLO}
\begin{align}
\da\NN^{\pp}_s &\,=\,
-\fr{1}{\,4\,}\Mnn\hs
(246\hs\ct \!+\! 7\hs\cttt \!+\! 3\hs\ctfif)
\hsm\csc^4\!\theta \,,
\\[1.5mm]
\da\NN^{\pp}_t &\,=\,
\frac{~\Mnn\hs (131 \!-\! 8 \ct \!-\! 4 \ctt \!+\! 8\cttt \!+\! \ctf)~}
{8\hs (1\!-\!\ct)^2} \,,
\\
\da\NN^{\pp}_u &\,=\,
-\frac{~\Mnn\hs (131 \!+\! 8 \ct \!-\! 4 \ctt \!-\! 8\cttt \!+\! \ctf)~}
{8\hs (1\!+\!\ct)^2} \,,
\end{align}
\eeqs
and the gauge-transformed NLO numerators
$\,\dNNt_{\hsm j}^{\pp}\,$
for the corresponding KK Goldstone boson amplitude:
\\[-9mm]
\beqs
\label{eq:Ntj'-NLO}
\begin{align}
\da\NNt^{\pp}_s &\,=\,
-\fr{1}{\,4\,}\Mnn\hs
(238\hs\ct \!+\! 19\hs\cttt \!-\!  \ctfif)
\hsm\csc^4\!\theta \,,
\\[2mm]
\da\NNt^{\pp}_t &\,=\,
\frac{~\Mnn\hs (99 \hsm +\hsm 8\hs\ct \!+\! 28 \ctt
\!-\hsm 8\hs\cttt \!+\hsm \ctf)~}
{8\hs (1\!-\!\ct)^2}  \,,
\\
\da\NNt^{\pp}_u &\,=\,
-\frac{~\Mnn\hs (99 \!-\!8\hs\ct \!+\! 28\hs\ctt
\!+\! 8\hs\cttt \!+\hsm \ctf)~}
{8\hs (1\!+\!\ct)^2}\,.
\end{align}
\eeqs
We see that these NLO numerators
are of $\mO(E^0\Mnn)$, which are mass-dependent.
It is straightforward to verify explicitly that the gauge-transformed
numerators \eqref{eq:Nj'-Ntj'-LO} and
\eqref{eq:Nj'-NLO}-\eqref{eq:Ntj'-NLO} obey the
kinematic Jacobi identities \eqref{eq:Jacobi-Nj'-Ntj'}
at the LO and NLO, respectively.

\vspace*{1mm}

Since the above gauge-transformed numerators hold the
kinematic Jacobi identities \eqref{eq:Jacobi-Nj'-Ntj'},
they are expected to realize the
color-kinematics duality order by order.
Thus, we construct the following extended BCJ-type
massive double-copy formulas for the elastic scattering amplitudes
of KK gravitons and KK Goldstone bosons:
\beqs
\label{AmpK-hLh5-nnnn-LONLOx}
\begin{align}
\M^{\pp} [4h_L^n]
&\,=\, c_0^{}\,g^2\!
\[\frac{(\NN_s^{\pp})^{2}}{s} +\frac{(\NN_t^{\pp})^{2}}{t}
+\frac{(\NN_u^{\pp})^2}{u}\]
=\,\M_0^{\pp} +\dM^{\pp}\,,
\label{AmpKF-hL-nnnn-x}
\\[1mm]
\label{AmpKF-h55-nnnn-x}
\MT^{\pp} [4\phin]
&\,=\, c_0^{}\,g^2\!
\[ \frac{(\NNt_s^{\pp})^{2}}{s} +\frac{(\NNt_t^{\pp})^{2}}{t}
+\frac{(\NNt_u^{\pp})^2}{u}\]
=\, \MT_0^{\pp} +\dMT^{\pp}\,,
\end{align}
\eeqs
where the LO and NLO KK gravitational amplitudes are given by
\beqs
\label{AmpK-hLh5-nnnn-LOx}
\begin{align}
\M_0^{\pp}
&\,=\, c_0^{}\,g^2\!
\[\frac{(\NN_s^{0\hs\pp})^{2}}{s} +\frac{(\NN_t^{0\hs\pp})^{2}}{t}
+\frac{(\NN_u^{0\hs\pp})^{2}}{u}\]\!,
\label{AmpKF-hL-nnnn-LOx}
\\[1mm]
\label{AmpKF-h55-nnnn-LOx}
\MT_0^{\pp}
&\,=\, c_0^{}\,g^2\!
\[ \frac{(\NNt_s^{0\hs\pp})^2}{s} +\frac{(\NNt_t^{0\hs\pp})^2}{t}
+\frac{(\NNt_u^{0\hs\pp})^2}{u}\]\!,
\end{align}
\eeqs
and
\beqs
\label{AmpK-hLh5-nnnn-NLOx}
\begin{align}
\dM^{\pp} &\,=~ 2c_0^{}\,g^2\!
\(\frac{\NN_s^{0\hs\pp}\dNN_s^{\pp}}{s}
 +\frac{\NN_t^{0\hs\pp}\dNN_t^{\pp}}{t}
+\frac{\NN_u^{0\hs\pp}\dNN_u^{\pp}}{u}\)\!,
\label{AmpKF-hL-nnnn-NLOx}
\\[1mm]
\label{AmpKF-h55-nnnn-NLOx}
\dMT^{\pp}	&\,=~ 2c_0^{}\,g^2\!
\(\frac{\NNt_s^{0\hs\pp}\dNNt_s^{\pp}}{s}
+\frac{\NNt_t^{0\hs\pp}\dNNt_t^{\pp}}{t}
+\frac{\NNt_u^{0\hs\pp}\dNNt_u^{\pp}}{u}\)\!.
\end{align}
\eeqs
In the above the conversion constant
$\,c_0^{}\!=\!-{\ka^2}/(24g^2)\,$
is given by Eq.\eqref{eq:c0}.

\vspace*{1mm}

Using the double-copy formulas in
Eqs.\eqref{AmpK-hLh5-nnnn-LOx}-\eqref{AmpK-hLh5-nnnn-NLOx}
and
the gauge-transformed numerators
$(\NN_j^{\pp},\,\NNt_j^{\pp})$
in Eqs.\eqref{eq:Nj'-Ntj'-LO}-\eqref{eq:Ntj'-NLO},
we construct the following
LO elastic KK graviton and KK Goldstone amplitudes:
\begin{align}
\label{eq:DC-Amp-LOf}
\M_0^{\pp}(\rm{DC})
&=	\MT_0^{\pp}(\rm{DC})=
\frac{\,3\ka^2\,}{\,128\,}
\frac{~(7\!+ \cos 2\theta)^2\,}{\sin^2\!\theta}\hs s \,,
\end{align}
and NLO elastic KK graviton and KK Goldstone amplitudes:
\beqs
\label{eq:DC-Amp-NLOf}
\begin{align}
\label{eq:DC-Amp-hL-NLOf}
\da\M^{\pp}(\rm{DC})
&= -\frac{~5\ka^2 M_n^2~}{768}
(1642 \hsm +\hsm  297\hs\ctt
\hsm +\hsm 102\hs\ctf\hsm +\hsm 7\hs\cts)
\hsm \csc^4 \hsmx\theta  \,,
\\[2mm]
\label{eq:DC-Amp-phi-NLOf}
\da\MT^{\pp} (\rm{DC})
&=- \frac{~\ka^2 M_n^2~}{768}
(6386 \hsm +\hsm 3837 \hs\ctt
\hsm +\hsm 30\hs\ctf
\hsm -\hsm 13\hs\cts)\hsm\csc^4 \hsmx\theta\,,
\end{align}
\eeqs

The above Eq.\eqref{eq:DC-Amp-LOf}
shows that the reconstructed LO KK amplitude
$\,\M_0^{\pp}\,(\MT_0^{\pp})\,$
equals the LO KK Goldstone amplitude \eqref{eq:AmpE2-phinnnn2}
and the corresponding LO longitudinal KK graviton
amplitude\,\cite{Chivukula:2020S}\cite{Chivukula:2020L}.\
It is impressive that with Eq.\eqref{eq:DC-Amp-LOf}
we have proven explicitly the KK GRET relation
$\M_0^{\pp}(\text{DC}) \!=\!	\MT_0^{\pp}(\text{DC})\hs$
from the KK GAET realtion
$\,\T_{\hs 0}^{}[4A^n_L]\!=\!\tT_{\hs 0}^{}[4A_5^n]\,$
{\it by using the double-copy construction.}\
Note that we deduced the GRET relation $\,\M_0^{\pp}\!=\!\MT_0^{\pp}$\,
earlier in Eq.\eqref{eq:GET-hL-phi}
based on the direct Feynman-diagram calculations.
Because our KK GAET relation $\,\T_0^{}\hsm =\hsm\tT_0^{}\,$
generally holds for $N$-point longitudinal KK gauge (Goldstone)
amplitudes\,\cite{5DYM2002}\cite{KK-ET-He2004},
we can make double-copy on both sides of
$\,\T_0^{}\hsm =\hsm\tT_0^{}\,$
and establish the GRET,
$\hs\M_0^{\pp}(\rm{DC}) \!=\! \MT_0^{\pp}(\rm{DC})\hs$,
for $N$-point longitudinal KK graviton (Goldstone) amplitudes.

We can further compute the gravitational residual term of the GRET
from the difference between the two NLO amplitudes
\eqref{eq:DC-Amp-hL-NLOf} and \eqref{eq:DC-Amp-phi-NLOf}:
\begin{equation}
\label{eq:Rterm-DC}
\Delta\M^{\pp} (\rm{DC})
= \da\M^{\pp}(\text{DC}) - \da\MT^{\pp}(\text{DC})
= -\ka^2 M_n^2\,( 7 \!+ \ctt )    \hs .
\end{equation}
We see that the above reconstructed residual term \eqref{eq:Rterm-DC}
by the extended double-copy approach
does give the same size of $\mO(E^0\Mnn)$
and takes the same angular structure of $(1,\,\ctt)$
as the original residual term
\eqref{eq:RTerm-NLO-GR} although their numerical coefficients
still differ.
It is impressive to note that
Eq.\eqref{eq:Rterm-DC} also demonstrates a very precise cancellation
between the angular structures
$(1,\,\ctt,\,\ctf,\,\cts)\!\times\!\csc^4\!\theta\,$
of the NLO double-copied KK amplitudes
\eqref{eq:DC-Amp-hL-NLOf}-\eqref{eq:DC-Amp-phi-NLOf}
down to the much simpler angular structure
$(1,\,\ctt)\hs$. This is the same kind of angular cancellations
as what we found for the original NLO KK
graviton-Goldstone amplitudes
\eqref{eq:Amp-E-2hLxx}-\eqref{eq:Amp-E-2phixx} and
their difference \eqref{eq:RTerm-NLO-GR}.
This shows that the above double-copied NLO KK amplitudes
have captured the essential features of the original KK
graviton-Goldstone amplitudes at both the LO and NLO.

\vspace*{1mm}

Finally, as a comparison, we also note that the above
NLO amplitude-difference \eqref{eq:Rterm-DC} does agree
to our earlier independent derivation of
Eq.\eqref{eq:BDM(DC)} which was deduced under the
high energy expansion of $\hs 1\hsm /\sz\hs$
and by further removing the Jacobi-violating terms
in the NLO amplitude-difference \eqref{eq:DM1-DM2}
with the improved NLO amplitudes \eqref{eq:dM-dMT'}
by double-copy. We stress that the key advantage of the
current double-copy approach under  $\hs 1\hsm /s\hs$
expansion is that our gauge-transformed numerators
$(\NN_j^{\pp},\,\NNt_j^{\pp})$
in Eqs.\eqref{eq:GGtransf}\eqref{eq:sol-Delta-tDelta}
are generally guaranteed to obey the kinematic Jacobi identities
\eqref{eq:Jacobi-Nj'-Ntj'} under the
high energy expansion of $\hs 1\hsm /s\hs$.
Hence they are expected to naturally realize the
color-kinematics duality order by order.

\vspace*{1mm}

The above extended NLO double-copy results of
the NLO KK amplitudes
\eqref{eq:DC-Amp-hL-NLOf}-\eqref{eq:DC-Amp-phi-NLOf}
and their difference \eqref{eq:Rterm-DC} are very encouraging,
because this construction
{\it uses only the pure longitudinal KK gauge boson amplitude}
and can already give {\it the correct structure}
of the NLO KK amplitudes including {\it the precise
cancellations} of the angular dependence from
Eqs.\eqref{eq:DC-Amp-hL-NLOf}-\eqref{eq:DC-Amp-phi-NLOf}
to Eq.\eqref{eq:Rterm-DC}.\
We have made the gauge-transformed numerators
$(\NN_j^{\pp},\,\NNt_j^{\pp})$
obey the kinematic Jacobi identities, which is a necessary
condition for realizing color-kinematics duality,
although not yet sufficient.\
To further construct the precise KK graviton (Goldstone)
amplitudes at the NLO and beyond, we propose another improved double-copy method, which also uses  {\it the amplitudes of
pure longitudinal KK gauge bosons (KK Goldstone bosons) alone,}
hence it is practically simple and valuable.\ 
For this, we propose the following improved BCJ-respecting
numerators at the NLO:
\beqs
\label{eq:dN'dNT'-XzzT-s}
\begin{align}
\hspace*{-3mm}
(\dNN_s'',\,\dNN_t'',\,\dNN_u'') &\,=\,
(\dNN_s^{\pp},\,\dNN_t^{\pp}\!-\!z,\,\dNN_u^\pp\!+\!z\,)\hs ,
\\[1mm]
\hspace*{-3mm}
(\dNNt_s'',\,\dNNt_t'',\,\dNNt_u'') &\,=\,
(\dNNt_s^{\pp},\,\dNNt_t^{\pp}\!-\!\zT,\,\dNNt_u^\pp \!+\!\zT\,)\hs,
\end{align}
\eeqs
where the new parameters
$(z,\,\zT)$ are functions of $\,\theta\,$
and can be determined by matching our improved NLO KK amplitudes
of double-copy with the original NLO KK graviton (Goldstone)
amplitudes of the GR5.\
We stress that the modified numerators \eqref{eq:dN'dNT'-XzzT}
still respect the Jacobi identity because
$\,\sum\hsm \dNN_j''\hsm =\hsm \sum\hsm\dNN_j'\hsm =\hsm 0\,$
and
$\sum\hsm\dNNt_j''\!=\! \sum\hsm\dNNt_j'\hsm =0\,$.
Then, we can solve the parameters $(z,\,\zT)$ as follows:
\\[-4.5mm]
\beqs
\label{eq:sol-z-zT-s}
\begin{align}
z &\,=\,
\frac{\,M_n^2(1390 \!+\! 603\hs\ctt\!+\! 66\hs\ctf\!-11\cts)\,}
{12(13\!-\!12\hs\ctt \!-\! \ctf )}\hs,
\\[1mm]
\zT &\,=\,
\frac{\,M_n^2(4546 \!-\! 3585\hs\ctt\!+\! 1086\hs\ctf\!+\cts)\,}
{12\(13\!-\!12\hs\ctt \!-\! \ctf \)}
\hs .
\end{align}
\eeqs
%
Since the corresponding NLO KK gauge (Goldstone) boson
amplitudes $(\dT'',\,\da\tT'')$ are corrected only by terms of NLO,
we can still retain the general GAET identity
$\T'' \!\!=\!\tT''\!+\!\T_v''$
by redefining the residual term as
\begin{equation}
\TT_v''\,=\, \TT_v^{}\!-g^2
\!\(\!\frac{\,\CC_t\,}{t}-\frac{\CC_u}{u}\!\)\!
(z\!-\!\zT ) \,.
\end{equation}
Using Eqs.\eqref{eq:dN'dNT'-XzzT-s}-\eqref{eq:sol-z-zT-s},
we can reproduce the exact NLO
gravitational KK amplitudes
\eqref{eq:Amp-E-2hLxx} and \eqref{eq:Amp-E-2phixx} at tree level.
We can further apply this double-copy procedure to
higher orders (beyond NLO) when needed.

\subsubsection{\hspace*{-2mm}Exact NLO Double-Copy based on KK String Approach}
\label{sec:5.3.3}
\vspace*{0.5mm}

In this subsection, we will construct an exact double-copy of
the KK graviton scattering amplitudes at the NLO based on 
our recent first principle approach
of the KK string theory\,\cite{Li:2021yfk}.\ 
We note that the above improved NLO double-copy constructions in
Eqs.\eqref{eq:DC-Amp-NLOf}-\eqref{eq:Rterm-DC}
can already give {\it the correct structure}
of the NLO KK amplitudes including {\it the precise cancellations}
of the angular dependence as in Eq.\eqref{eq:Diff2-AmpGR5-nnnn}.\
These are indeed truly encouraging, and provide strong evidence
that our massive KK double-copy approach is on the right track.\
The importance of the current approach is twofold:\
(i).\,In practice, this KK double-copy construction is realized
{\it under high energy expansion}. Thus, the LO double-copy
is the most important part, with which we newly establish
the GRET relation $\M_0^{} \!=\! \MT_0^{}$\,
[\eqrefe{eq:DC-Amp-LOf}]
from the GAET relation
$\,\T_0^{}\hsm =\hsm\tT_0^{}\,$ [\eqrefe{eq:LO-GAET'}],
as will be stressed in \eqrefe{eq:KKET-GET}.\
We use the NLO KK graviton amplitudes only for
{\it estimating the size} of the residual term
$\M_{\!\Delta}^{}$ of the GRET \eqref{eq:GET}.\
So the precise form of $\M_{\!\Delta}^{}\!$ is not needed 
for the GRET {\it except to justify its size
$\M_{\!\Delta}^{}\!\!=\mO(E^0\Mnn)$}
by our double-copy construction [cf.\ Eq.\eqref{eq:RTerm}].\
Hence, for our GRET formulation under the high energy limit,
{\it the residual term $\M_{\!\Delta}^{}\!$
does belong to the NLO amplitudes and is thus neligible.}\
(ii).\,The current KK double-copy approach makes the
first serious attempt to construct the massive KK graviton amplitudes.
In general, this provides {strong motivation and important guideline}
for realizing the exact massive double-copy.
Our further investigation has fully clarified the mismatch
between the numerical coefficients of the
double-copied NLO amplitudes \eqref{eq:DC-Amp-NLOf}
and that of the exact NLO amplitudes
\eqref{eq:Amp-E-2hLxx}-\eqref{eq:Amp-E-2phixx}.\
One reason is because the KK amplitudes
from the orbifold compactification ($S^1/\ZZ$) contain
the double-pole structure
(including exchanges of both the zero-mode and KK-modes)
beyond the conventional massless theories.\ 
Hence our mass-dependent NLO amplitudes has extra contributions from
the KK mass-poles under the high energy expansion,
which cause a mismatch.\
Another reason is due to the fact that
the (helicity-zero) longitudinal KK graviton has
its full polarization tensor given by
$\,\vep_{L}^{\mn} \!=\!\(
\ep_{+}^{\mu} \ep_{-}^{\nu}\!+\ep_{-}^{\mu} \ep_{+}^{\nu}\!
+2 \ep_{L}^{\mu} \ep_{L}^{\nu}\)\hsm\!/\!\sqrt{6\,}$,\,
where $\,\vep_{L}^{\mn}\,$ contains both
the longitudinal product $\ep_{L}^{\mu} \ep_{L}^{\nu}$
and the transverse products
$\,\ep_{+}^{\mu} \ep_{-}^{\nu}\!+\ep_{-}^{\mu} \ep_{+}^{\nu}$\,.\,\
This means that for a full double-copy construction,
all the four-point massive KK scattering amplitudes
containing possible external transverse KK gauge boson states
have to be included in addition to
the four-longitudinal KK gauge boson amplitude
\eqref{Amp-ALA5-nnnn}.

\vspace*{1mm}

Taking these into account,
we have further proposed a first principle approach
of the KK string theory in our recent work\,\cite{Li:2021yfk}.\ 
With this approach, we derived the extended massive KLT-type relations
between the product of the KK open string amplitudes
and the KK closed string amplitude at tree level.\
Thus, under the field theory limit,
we can further derive the exact double-copy relations
between the product of the KK gauge boson amplitudes and the
corresponding KK graviton amplitude\,\cite{Li:2021yfk}.\
We note that in these exact double-copy relations
we have summed over all the relevant helicity indices
of the external KK gauge boson states to match
the corresponding polarization tensors of the external KK graviton
states.\ We can further avoid the double-pole structure
by first studying the 5d compactification under
$S^1$ (without orbifold), which conserves the KK numbers
($\pm n\!=\!\pm 1,\pm 2,\pm 3,\cdots$) and
ensures the KK amplitudes to have single-pole structure.\
With these, we then define the following KK states
$|n_\pm^{}\rangle$
with $\ZZ$ even\,(odd) parity:
\beqa
|n_\pm^{}\rangle \,=\, \fr{1}{\sqrt{2\,}\,}
(|\hsmx +\hsm n\rangle \pm |\hsmx -\hsm n\rangle)\,,
\eeqa
and derive the KK amplitudes under the compactification
of $S^1\! /\ZZ$
from the combinations of those KK amplitudes under
the $S^1$ compactification\,\cite{Li:2021yfk}.\
Hence, we can exactly reconstruct all the massive KK graviton
amplitudes at tree level by this improved massive KK double-copy
method. For instance, consider the four-point longitudinal KK
graviton amplitudes under the $S^1\! /\ZZ$ compactification.\
We can construct the (BCJ-type) exact massive double-copy formula
as follows:
\begin{equation}
\label{eq:MBCJ}
\M \,=\, -\frac{\ka^2}{\,64\,}
\sum_{\fP}\sum_{j}\sum_{\lam^{}_k,\lam'_k}
\prod_k\!
C_{\lam_k^{}\lam'_k}^{} \frac{\,N_{j}^\fP(\lam_k^{})N_{j}^\fP(\lam_k')\,}{D_j^{}}\,,
\end{equation}
where each external KK gauge boson has three helicity states
(with $\lam_k^{}\hsm\!=\!\pm 1,L$ and $k\!=\!1,2,3,4$).\
The kinematic numerators $\{N_j^{}\}$ are given by each given
scattering amplitude of KK gauge bosons
under the $S^1$ compactification.
Here we denote each possible combination of the KK numbers
for the external gauge bosons by
$\fP\!=\!\{n_1^{},n_2^{},n_3^{},n_4^{}\}$
and they obey the KK-number conservation condition,
$\sum_{k=1}^{4}\! n_k^{}\hsm =\hsm 0\hs$.
For the elastic KK scattering process, we have the following
combinations:
\begin{equation}
\fP \hs =\hs \{\pm n,\pm n,\mp n,\mp n\}, ~~
\{\pm n,\mp n,\pm n,\mp n\}, ~~
\{\pm n,\mp n,\mp n,\pm n\} \hs.
\end{equation}
In the above formula \eqref{eq:MBCJ}, we use
$\,C_{\lam_k^{}\lam'_k}^{}$ to denote the coefficients
in the longitudinal polarization tensor
of the $k$-th external KK graviton,
$\ep_{L,k}^{\mn}\!=\! \sum C_{\lam_k^{}\lam'_k}^{}
 \ep_{\lam_k}^\mu\ep_{\lam'_k}^\nu$,
where $\lam_k^{},\lam'_k\!=\!\pm 1,L$
are the helicity indices for the $k$-th external gauge boson and
$(C_{+-}^{},\,C_{-+}^{},\,C_{LL}^{})
\!=\!(1,\hs 1,\hs 2)/\sqrt{6\,}\,$,
as shown in Eq.\eqref{eq:hn-Pols}.
The denominator of \eqrefe{eq:MBCJ} is given by
$\,D_j^{}\!=\hsm s_j^{}- M_j^2$,\, where
$\,s_j^{}\!\in\!\{s,t,u\}\,$ and
$M_j^2\!\in\!\{M_{n_1+n_2}^2,M_{n_1+n_4}^2,M_{n_1+n_3}^2\}$.

\vspace*{1mm}

Next, we consider the corresponding elastic scattering amplitude of
the KK gauge bosons under $S^1$ compactification, and make the
high energy expansion of it at the LO and NLO:
\beqs
\begin{align}
\label{AmpS1-Annnn}
& \TT =\,
g^2 \sum_{j}\frac{\,\CC_j N_{\hsm j}^{\fP}}{D_{\hsm j}^{}}
= g^2 \sum_{j}\frac{\,\CC_j (N^{0,\fP}_{j} \!+ \da N^{\fP}_{j})}{s_{j}^{}}
= \TT_0^{} + \dT \,,
\\[1mm]
& \frac{s_{\hsm j}^{}}{\,D_{j}^{}\,} N_{j}^{\fP} =
{N_{j}^{0,\fP}} \!+ \da N^{\fP}_{j}\hs .
\end{align}
\eeqs
With the above, we can expand the exact double-copy formula of the
longitudinal KK graviton amplitude \eqref{eq:MBCJ}
under the high energy expansion of $1/s$\,:
\begin{align}
\label{eq:MBCJ-LO+NLO}
\M &= -\frac{\ka^2}{\,64\,}
\sum_{\fP}\sum_{j} \!\sum_{\lam^{}_k,\lam'_k}
\prod_k C_{\lam_k^{}\lam'_k}^{}
\frac{\,D_j^{}\,}{s_{j}^2}\!
\[\!N_{j}^{0,\fP}\!(\lam^{}_k)\!+\da N_{j}^{\fP}\!(\lam^{}_k)\!\]\!
\hsm\[\!N_{j}^{0,\fP}\!(\lam'_k)\!+\da N_{j}^{\fP}\!(\lam'_k)\!\]
\nn\\[1mm]
&= \,\M_0^{}+\dM\,,
\end{align}
where $\M_0^{}$ and $\dM$ are expanded up to the LO and NLO respectively,
\beqs
\label{eq:MBCJ-LO-NLO}
\begin{align}
\label{eq:MBCJ-LO}
\M_0^{} &=  -\frac{\ka^2}{\,64\,}
\sum_{\fP}\sum_{j} \!\sum_{\lam^{}_k,\lam'_k}
\prod_k C_{\lam_k^{}\lam'_k}^{}
\frac{\,N_{j}^{0,\fP}\! (\lam^{}_k)\hs N_{j}^{0,\fP}\!(\lam'_k)~}{s_j^{}} \,,
\\
\label{eq:MBCJ-NLO}
\dM &=  -\frac{\ka^2}{\,64\,}\!
\sum_{\fP}\!\sum_{j} \!\sum_{\lam^{}_k,\lam'_k}\!
\prod_k \!\hsm C_{\lam_k^{}\lam'_k}^{}\!
\frac{\,N_{j}^{0,\fP}\! (\lam^{}_k)\da N_{j}^{\fP}\!(\lam'_k)
\!+\! N_{j}^{0,\fP}\! (\lam'_k)\da N_{j}^{\fP}\!(\lam^{}_k)
\!-\! N_{j}^{0,\fP}\! (\lam^{}_k)N_{j}^{0,\fP}\!(\lam'_k)M_j^2/s_j^{}\,}{s_j^{}}\hs .
\end{align}
\eeqs
%
The above precise LO and NLO double-copy formulas 
are fully general.\ 
For the four-point elastic longitudinal KK graviton scattering, 
we can readily prove that the above double-copied LO amplitude
\eqref{eq:MBCJ-LO} is equivalent to the LO amplitude given in
Eq.\eqref{AmpKF-hL-nnnn-LOx}\,\cite{Li:2021yfk}.\
Then, we explicitly compute the LO amplitude \eqref{eq:MBCJ-LO}
and find that $\M_0^{}$ is equal to that of
Eq.\eqref{eq:DC-Amp-LOf}, as expected.\
We further compute the above double-copied NLO amplitude
\eqref{eq:MBCJ-NLO} and derive the following expression:
\begin{equation}
\label{eq:DC-Amp-hL-NLOx}
\da\M = -\frac{\,\ka^2 M_n^2\,}{256}
(1810 + 93\hs\ctt +\hsm 126\hs\ctf +\hsm 19\hs\cts) \hsm \csc^4 \hsmx\theta  \,.
\end{equation}
It is impressive that this NLO formula does fully agree to
the exact NLO elastic KK graviton amplitude \eqref{eq:Amp-E-2hLxx}
given by the direct Feynman diagram calculation.\
The above demonstration explicitly shows that
we can realize the (BCJ-type) exact massive double-copy
construction of the four-point KK graviton amplitudes
in Eq.\eqref{eq:MBCJ}, and accordingly we can successfully construct
the precise double-copy for the KK graviton amplitudes
\eqref{eq:MBCJ-LO+NLO}-\eqref{eq:DC-Amp-hL-NLOx}
at the LO and NLO under the high energy expansion.
The above success is truly encouraging and strongly motivates us to
pursue this new direction systematically in the future works.

\subsection{\hspace*{-2mm}GRET Residual Terms: Energy Cancellation
	from Double-Copy}
\label{sec:5.4}
\vspace*{1.5mm}

The main purpose of this subsection is to understand
the {\it structure of the GRET} \eqref{eq:GET-ID} or \eqref{eq:GET}
including its {\it mass-dependent residual term}
in the 5d KK GR theory from the
{\it structure of the KK GAET}
in the 5d KK YM gauge theory
by using the double-copy construction of
sections\,\ref{sec:5.2}-\ref{sec:5.3}.
This will bring us important insights
on the gravitational KK scattering amplitudes and
how the GRET actually works.

\vspace*{1mm}

We start by considering the compactified 5d YM gauge theory
and the KK GAET identity as derived in Ref.\,\cite{KK-ET-He2004}
(cf.\ its section\,3).
For the application to the current study,
we consider the four-particle scattering of longitudinal
KK gauge bosons
$\,\ALnkml\,$
and the corresponding KK Goldstone boson scattering $\Afnkml$.
According to Refs.\,\cite{5DYM2002}\cite{KK-ET-He2004},
we can write the KK GAET identity for the above
four-particle scattering process:
\begin{equation}
\label{eq:KKET-general}
\T[4\ALn] \,=\, \tT[4\Afn] + \sum\!\T[\Afn,\vn] \,,
\end{equation}
where the residual term $\T[\Afn, v_n] $ contains at least
one external field $\vn \!=\! v^{}_\mu A^{a\mu}_{n}$\,
with $\,v^\mu =\ep_L^\mu \!-\ep_S^\mu =\mO(M_n/E_n)\,$
for the high energy scattering.
In this subsection, the amplitudes such as
$\T[4\ALn] $ or $\tT[4\Afn]$ will denote either elastic or
inelastic scattering process.
In section\,\ref{sec:5.1}, we showed that
the leading longitudinal KK gauge boson amplitude
$\,\T[4\ALn]$\, and the leading KK Goldstone amplitude
$\,\tT[4\Afn]$\, are of $\,\mO(E^0M_n^0)\,$
under the high energy expansion.
In the following, we expand them symbolically to the next-to-leading
order (NLO) of $E^{-2}$\,:
\beqs
\label{eq:Amp-4AL-4A5-Expand}
\begin{align}
\T[4\ALn] &~=~
\T_{0L}^{}+ \dT_{L}^{}   \,,
\\[1mm]
\tT[4\Afn] &~=~
\tT_{05}^{} + \da \tT_{5}^{}    \,,
\end{align}
\eeqs
where the leading order (LO) amplitudes
$\,\T_{0L}^{},\tT_{05}^{}\!=\!\mO(E^0M_n^0)\,$ and
the NLO amplitudes
$\,\dT_L^{},\da\tT_{5}^{}=\mO(\Mnn/E^2)\,$.\,
In Eqs.\eqref{eq:T0-KK-ET-nnnn} and \eqref{eq:KK-ET-nkml}
of section\,\ref{sec:5.1}, we showed explicitly that
under high energy expansion, the LO KK amplitudes
obey the longitudinal-Goldstone equivalence:
\begin{equation}
\label{eq:TL0=T50}
\T_{0L} \,=\, \tT_{05}
\,=\,\mO(E^0 M_n^0) \,,
\end{equation}
which is the prediction of KK GAET\,\cite{5DYM2002}.
Thus, from the KK GAET identity \eqref{eq:KKET-general}, we can
derive the residual term as follows:
\begin{equation}
\label{eq:KKET-general2}
\T_v^{}\,\equiv\,
\sum \!\T[\Afn , v_n]  \,=\,	
\dT_L - \da \tT_5
\,=\, \mO(\Mnn/E^2) \,.
\end{equation}
In the above, each residual term $\T[\Afn, v_n]$
is no larger than $\mO(E^{-1})$ by the naive power counting.
In fact, we can explicitly compute
the above four-particle amplitudes of the
longitudinal and Goldstone boson scattering,
and our \eqrefe{eq:dTL-dT5} proves their difference is of
$\,\mO(\Mnn/E^2)\,$.
This also agrees with the
general estimate of Ref.\,\cite{ET94b}, with which
we have the following power counting formula for the residual term:
\begin{equation}
\label{eq:R-term}
\T_v^{}
~=~ \mO\!\(\!\frac{M_n^2}{E_n^2}\)\!\tT[4\Afn]
+\mO\!\(\!\frac{M_n^{}}{E_n^{}}\!\)\!\T[A_T^{n},3A_5^n]\,,
\end{equation}
where $E_n$ denotes the energy of the relevant
external KK gauge boson and $A_T^n$ denotes
a transverse KK gauge boson.\
The naive power counting shows
$\,\tT[4A_5^n]\!=\!\mO(E_n^0)\,$
and $\,\T[A_T^{n},3A_5^n]\!=\mO(M_n/E_n)\,$.
Thus, using Eq.\,\eqref{eq:R-term}, we also deduce
$\,\T_v^{}\!=\mO(\Mnn/E_n^2)\,$,
which agrees with the (mass,\,energy)-dependence
given in Eq.\eqref{eq:KKET-general2}.

\vspace*{1mm}

Next, we consider the four-particle scattering of the
longitudinal KK gravitons
$\,\hLnkml\,$
and the corresponding KK Goldstone boson scattering
$\,\pnkml\,$.
Thus, we can express the GRET identity \eqref{eq:GET-ID}
as follows:
\begin{equation}
\label{eq:GET-4P}
\M[4\hLn] ~=~ \MT[4\phin] +
\sum\!\M[\DEn,\phin] \,,
\end{equation}
where $\,\widetilde{\Delta}_n^{} \!= \vnt \!-\th_n^{}\,$
with $\,\vnt\!=\!\vt_{\mn}^{}h_n^{\mn}$\,
and $\,\th_n^{}\!=\hsm\eta_{\mn}^{}\th^{\mn}_n\hs$.
We denote the residual term on the RHS of \eqrefe{eq:GET-4P}
as $\,\M_\Delta^{}\!\equiv\sum\!\M[\DEn,\phin]$\,.
We note that each amplitude inside the residual term
contains at least one external state of $\,\DEn\,$,
which will further split into two amplitudes
with external fields $\,\vnt\,$ and $\,\th_n^{}\,$, respectively.
Since the naive power counting shows the residual term
$\,\M_\Delta^{}\!=\!\mO(E^2M_n^0)$\,
under the high energy expansion, we expect that
$\,\M_\Delta^{}$\,
should contain further nontrivial energy-cancellations
of $\,\mO(E^2M_n^0)\ito \mO(E^0M_n^2)\,$,
which we will justify shortly.

\vspace*{1mm}

For high energy scattering, we can expand the amplitudes
of the longitudinal KK gravitons and of their KK Goldstone bosons
into the LO and NLO contributions:
\beqs
\label{eq:Amp-4hL-4phi-Expand}
\begin{align}
\M[4\hLn] &\,=\,
\M_0^{} + \dM   \,,
\\[1mm]
\MT [4\phin] &\,=\,
\MT_0^{}  + \dMT  \,.
\end{align}
\eeqs
As shown explicitly in section\,\ref{sec:4.2} and
Appendix\,\ref{app:F}
for the gravitational KK scattering, the LO KK amplitudes
$\,\M_0^{}\!=\!\mO(E^2M_n^0)\,$ and
$\,\MT_0^{}\!=\!\mO(E^2M_n^0)$,\,
while the NLO KK amplitudes
$\dM=\mO(E^0M_n^2)$ and
$\dMT=\mO(E^0M_n^2)$\,.

\vspace*{1mm}

Furthermore, using the double-copy construction
from the 5d KK YM gauge theory
in sections\,\ref{sec:5.2}-\ref{sec:5.3},
we have deduced independently the magnitudes of
the LO and NLO gravitational KK amplitudes in
Eqs.\eqref{eq:Amp-LO-counting}-\eqref{eq:Amp-NLO-counting}
which agree with the direct calculations in the
5d KK GR theory.
According to Eqs.\eqref{AmpDC-hLphi-nnnn}\eqref{eq:GET-nkml}
of section\,\ref{sec:5.2},
our double-copy constructions of the LO longitudinal
KK graviton (Goldstone) amplitudes give:
\begin{equation}
\label{eq:GET-LO-E2}
\M_0^{}(\rm{DC})\,=\, \MT_0^{}(\rm{DC})
\,=\, \mO(E^2M_n^0) \,.
\end{equation}
In fact, our double-copy construction has explicitly
demonstrated in sections\,\ref{sec:5.1}-\ref{sec:5.2} that
the {\it gravitational equivalence}
\eqref{eq:GET-LO-E2}
between the two LO gravitational amplitudes
is generally built upon the KK GAET \eqref{eq:TL0=T50}.
The KK GAET identity \eqref{eq:KKET-general} can be expressed
as $\,\T_v^{}=\T[4A_L^n]-\tT[4\Afn]\,$,
and the double-copy of its left-hand-side
$\,\T_v^{}\ito \M_\Delta^{}(\rm{DC})\,$
corresponds to the double-copy of its right-hand-side:
\begin{equation}
\label{eq:KKET-DC-RHS}
\T[4A_L^n]\!-\!\tT[4\Afn] ~\longrightarrow~
\M (\rm{DC}) \!-\! \MT (\rm{DC})
= \dM (\rm{DC}) \!-\!\dMT (\rm{DC}) \,,
\,
\end{equation}
where in the last step we haved used the LO double-copy result
\eqref{eq:GET-LO-E2}.
Using this double-copy construction from the 5d KK
gauge theory amplitudes, we further demonstrated
in Eq.\eqref{eq:Amp-NLO-counting} of section\,\ref{sec:5.3}
that under the high energy expansion \eqref{eq:Amp-4AL-4A5-Expand},
the NLO gravitational KK scattering amplitudes
depend on the KK mass, but not on the energy:
\begin{equation}
\label{eq:GET-NLO-E0}
\dM(\rm{DC})  \,=\,\mO(E^0\Mnn) \,,~~~~~
\dMT(\rm{DC})   \,=\,\mO(E^0\Mnn) \,.
\end{equation}
Given this result \eqref{eq:GET-NLO-E0} and using
Eq.\eqref{eq:KKET-DC-RHS},
we deduce that the double-copy construction of the residual term
$\,\T_v^{}\to \M_\Delta^{}(\rm{DC})\,$
is given by
\begin{equation}
\label{eq:RTerm}
\T_v^{}~\longrightarrow~\M_\Delta^{}(\rm{DC})
\hsm =\dM(\rm{DC}) \hsm -\hsm \dMT(\rm{DC})
\hsm =\mO(E^0\Mnn) \,.
\end{equation}
We can extend the above estimate \eqref{eq:RTerm} of the
residual term to the general case of GRET \eqref{eq:GET}
for any KK graviton amplitude containing two or more longitudinal KK gravitons.\footnote{%
\baselineskip 15pt
We note that the special case including a single external
longitudinal KK graviton state is an exception,
where the residual term can be of the same order
as the leading KK longitudinal (Goldstone) amplitudes.
We gave an explicit example of this kind by our
GRET analysis of the SQED5 model in section\,\ref{sec:4.1}.}

\vspace*{1mm}

Because in the residual term
$\,\M_\Delta^{}\equiv\sum\!\M[\DEn,\phin]$\,
each individual amplitude
$\,\M [\DEn,\phin]=\mO(E^{2})$\, 
by naive power counting,
the conclusion of Eq.\eqref{eq:RTerm} proves that
there is in fact a nontrivial energy-cancellation
of $\,\mO(E^{2})\!\ito\mO(E^0)\,$
in the residual term of the GRET.
Hence, Eq.\eqref{eq:RTerm} ensures
our GRET to realize the {\it equivalence} between the
longitudinal KK graviton amplitude and its corresponding
KK gravitational Goldstone boson amplitude
at $\mO(E^2)$\,.

\vspace*{1mm}

In summary, based on the KK GAET identity \eqref{eq:KKET-general}
for the 5d KK YM gauge theory (YM5)
and the double-copy construction
in sections\,\ref{sec:5.2}-\ref{sec:5.3},
we have established {\it a new correspondence}
from the KK GAET of the YM5 theory to the KK GRET of the
5d KK GR theory (GR5):
\begin{equation}
\label{eq:KKET-GET}
\rm{KK \ GAET\,(YM5)}
\ \Longrightarrow \
\rm{KK \ GRET\,(GR5)}\,.
\end{equation}
We have demonstrated that {\it the residual term in the
GRET \eqref{eq:GET-4P} or \eqref{eq:GET} is indeed suppressed relative to the leading KK Goldstone $\phin$-amplitude;
and in the case of four-particle longitudinal
KK graviton	scattering, the leading (helicity-zero)
longitudinal KK graviton amplitude
and KK Goldstone amplitude scale as $\mO(E^2M_n^0)$
and are equal to each other;
while the residual term of the GRET is only of $\mO(E^0\Mnn)$,
as in Eq.\eqref{eq:RTerm},
due to a nontrivial energy-cancellation of $\,\mO(E^{2})\ito\mO(E^0)$.\,}
This conclusion can be readily extended to
other longitudinal KK graviton scattering processes
with two or more external longitudinal KK
graviton states.
As a final remark, we build the above correspondence
\eqref{eq:KKET-GET} based on our current analyses of
the tree-level scattering amplitudes, and it will be
worthwhile to further extend it to loop orders by invoking
the BRST transformations in both the 5d KK YM gauge theory
and the 5d KK GR theory\,\cite{GET-2}.
We also note that our power-counting method presented in
section\,\ref{sec:3.2} holds for the general $N$-point amplitudes
with $L$ loops ($L\!\geqq 0\hs$), so our present power-counting analysis
can be extended up to loop orders in a straightforward way.

\section{\hspace{-3mm}Conclusions}
\label{sec:6}

Studying the structure of scattering amplitudes
of Kaluza-Klein (KK) gravitons and that of the KK gauge bosons
is important for understanding the dynamics of KK theories
and the deep gauge-gravity connection.\
The KK gravitons and KK gauge bosons serve as the key ingredients
in all extra dimensional 
models\,\cite{Exd0}-\cite{exdtest}
and string theories\,\cite{string}
which attempt to resolve the naturalness problem,
the quantum gravity, and the gauge-gravity unification.

\vspace*{1mm}

In this work, we studied the structure of the scattering amplitudes
of the KK gravitons and their KK Goldstone bosons (radions)
with compactified fifth dimension.\
In section\,\ref{sec:2}, using a general $R_\xi^{}$
gauge-fixing \eqref{eq:GF} for the quantization of 5d KK General Relativity (GR),
we derived the massive KK graviton propagator
and the corresponding
Goldstone boson propagators in Eq.\eqref{eq:KKpropagator-Rxi}.\
These propagators take particularly simple forms of
\eqrefe{eq:KKpropagator-FHooft}
under the Feynman-'t\,Hooft gauge ($\,\xi_n^{}\!=\!1\,$).\
We proved that the KK graviton propagator
is naturally {\it free from the longstanding puzzle} of
the vDVZ discontinuity\,\cite{vDVZ},
in contrast to that of the Fierz-Pauli gravity\,\cite{PF}
and alike\,\cite{Hinterbichler:2012}.\ 
Hence, the KK gravity theories provide {\it a consistent
realization of mass-generations for the spin-2 KK gravitons.}\

\vspace*{1mm}

With these, we presented in section\,\ref{sec:3.1} the formulation
of the Gravitational Equivalence Theorem (GRET) to connect the
scattering amplitudes of longitudinally-polarized (helicity-zero)
KK gravitons $\,h_L^n\,$ to that of the corresponding gravitational
KK scalar Goldstone bosons $\phin \,(\equiv\! h^{55}_n)$.
The GRET is a manifestation of the geometric ``Higgs'' mechanism
at the $S$-matrix level.
Starting from the general Slavnov-Taylor-type identity
\eqref{eq:F-identityP} for the gravitational gauge-fixing functions,
we derived its LSZ amputated form \eqref{eq:Fn-ID-N-LSZ}
under the Feynman-'t\,Hooft gauge at tree level, which suffices
for the present study.\ From this we derived the key GRET identity
\eqref{eq:GET-ID0} and gave the GRET formulation in
Eqs.\eqref{eq:GET-ID} and \eqref{eq:GET}.\
Then, extending Weinberg's power counting rule 
for the low energy QCD\,\cite{weinbergPC} to the KK theories,
we derived general energy-power counting rule
\eqref{eq:D_E} for the 5d KK GR theory.\
With this we derived the leading energy-dependence of the
$N$-particle longitudinal KK graviton scattering amplitudes
and of $N$-particle KK Goldstone scattering amplitudes in
Eqs.\eqref{eq:DE-hL}-\eqref{eq:DE-phin}, namely,
$\,D_E^{}(Nh^L_n)\! = 2(N\!+\!1)\!+\hsm 2L\,$
and
$\,D_E^{}(N\phin)\!\hsm  =\! 2\hsm +\hsm 2L\hs$.\
We further counted the superficial leading energy-dependence
of the residual term $\MD$ as in Eq.\eqref{eq:DE-vn},
which gives
$D_E^{}(N\vt_n^{}) \!=\! 2+2L\,$.\
Using the GRET identity \eqref{eq:GET-ID},
we established the nontrivial energy cancellations in the
$N$-particle longitudinal KK graviton scattering amplitudes
by $\hs E^{2N}$ as in Eq.\eqref{eq:DEL-DEphi},
where the number of external KK states $N\!\!\geqq\! 4\,$.\ 
For the scattering amplitudes of $N$ longitudinal KK gravitons
at tree level ($L\!=\!0$), 
this proves the large energy cancellations
of $\,E^{2N+2}\!\ito E^2\hs$.\ 
In the case of the four longitudinal KK graviton scattering amplitudes
($N\!\!=\!4$), this establishes the energy cancellations of
$\,E^{10}\hsm\ito E^2\,$,\, which we further demonstrated
by explicit analyses in sections\,\ref{sec:4.2} and \ref{sec:5.2}.
Hence, {\it the GRET identity \eqref{eq:GET-ID}
provides a general mechanism for guaranteeing the nontrivial
large energy-cancellations in the $N$-particle longitudinal KK
graviton amplitudes by $\,E^{2N}\hsm$, where $N\!\geqq 4$\,.}
This conclusion holds up to loop levels
because the radiative multiplicative modification factor
$\,C_{\text{mod}}^{}\,$
associated with each external Goldstone state is
{\it energy-independent}.
Our present GRET formulation is highly nontrivial
because its residual term does not appear superficially suppressed
relative to the leading KK Goldstone amplitude in high energy
limit by the naive power counting. The suppression of the
residual term was further established in the following
sections\,\ref{sec:4}-\ref{sec:5}.

\vspace*{1mm}

In section\,\ref{sec:4}, we performed systematically a direct
computation of the gravitational KK Goldstone boson scattering
amplitudes at tree level. In section\,\ref{sec:4.1}, we took
a simple model of 5d gravitational scalar QED (GSQED5) as an
example and explicitly verified the GRET identity
\eqref{eq:GET-N=1} or \eqref{eq:GET-hL-GBv-N1}
for the case of including a single external KK graviton field.\ 
Our analysis showed that the GRET identity in this case
holds exactly. Then, in section\,\ref{sec:4.2},
we derived the exact four-particle
KK Goldstone boson scattering amplitude, and
expanded it to the leading order (LO) and
the next-to-leading order (NLO)
under the high energy expansion, which are given in
Eq.\eqref{eq:AmpLOE2-phinnnn} and
Eqs.\eqref{eq:Amp-4phinnnn-full}-\eqref{eq:AmpGR5-4phi-LONLO}.\
The leading energy-dependence in
these KK Goldstone amplitudes is manifestly of $\mO(E^2)$
without any extra energy cancellations 
among the individual diagrams.\ 
Hence, they are substantially simpler than those of the 
longitudinal KK graviton amplitudes
in the literatures\,\cite{Chivukula:2020S}\cite{Chivukula:2020L}
since the latter involve various intricate energy cancellations
among individual diagrams from $\mO(E^{10})$ down to $\mO(E^2)$.\
With these we proved explicitly the {\it equivalence} between
the leading $h_L^n$-amplitudes and $\phin$-amplitudes
at $\mO(E^2)$, which supports the GRET \eqref{eq:GET}.
Hence, the longitudinal-Goldstone equivalence of the GRET
guarantees the nontrivial large energy-power cancellations
in the longitudinal KK graviton amplitudes.
We further computed the difference between the exact
$h_L^n$-amplitude and $\phin$-amplitude as
in Eq.\eqref{eq:RTerm-NLO-GR}, which has $\mO(\Mnn E^0)$
and determines the size of the residual term of the GRET
to be of the NLO.

\vspace*{1mm}

In section\,\ref{sec:5},
we studied systematically the double-copy construction of
the gravitational KK scattering amplitudes by using
the corresponding KK gauge (Goldstone) boson scattering amplitudes
in the 5d KK YM gauge theory,
{under the high energy expansion.}
The conventional BCJ-type double-copy
approach\,\cite{BCJ:2008}\cite{BCJ:2019})
is given for massless gauge theories and massless GR.
Because the KK gauge theories and KK GR
can consistently generate masses for KK gauge bosons and
KK gravitons by geometric ``Higgs'' mechanism
under compactification, we expect that extending the
conventional double-copy method to the KK theories
should be truly promising even though highly challenging
due to the nontrivial KK-mass-poles in the scattering amplitudes
and the nontrivial polarization tensors for the external longitudinal
KK gravition states.\
Unlike the conventional double-copy approaches,
we proposed to realize the
double-copy construction {by using the high energy expansion
order by order.}  With this, we demonstrated explicitly
how such a double-copy construction can work at
the LO and the NLO, as in sections\,\ref{sec:5.2}-\ref{sec:5.3}.
This high energy expansion approach
for realizing our double-copy construction also
perfectly matches our KK GAET and GRET formulations.

\vspace*{1mm}

In section\,\ref{sec:5.2},
under the high energy expansion, we found that
the LO KK gauge boson (Goldstone) amplitudes have \,$\mO(E^0M_n^0)$\,
and the LO KK graviton (Goldstone) amplitudes have \,$\mO(E^2M_n^0)$,\,
which are both {\it mass-independent}.
Thus, we made an extended BCJ double-copy construction from
our LO KK gauge boson (Goldstone) amplitudes and fully reconstructed
the correct KK graviton (Goldstone) amplitudes at the LO,
as shown in Eqs.\eqref{AmpDC-hLphi-nnnn} and \eqref{eq:DC-ineAMP}.
Then, in section\,\ref{sec:5.3}, we studied the extended
double-copy constructions of the NLO KK gauge/gravity
amplitudes by making two types of high energy expansions,
under $\hs 1\hsm /\sz\hs$ expansion (section\,\ref{sec:5.3.1})
and under $\hs 1\hsm /s\hs$ expansion (section\,\ref{sec:5.3.2}),
respectively.
We showed that the NLO KK gauge (Goldstone) boson amplitudes have
\,$\mO(\Mnn /E^2)$\, and the NLO KK graviton (Goldstone) amplitudes
have \,$\mO(E^0M_n^2)$,\, which are both {\it mass-dependent}.
We demonstrated that the double-copy construction for the mass-dependent NLO KK-amplitudes is highly nontrivial,
where the conventional double-copy method could not fully work.
We found that the reason for this problem is
due to violations of the kinematic Jacobi identities \eqref{eq:dN-dNT-sum} at the NLO of
the $\hs 1\hsm /\sz\hs$ expansion, where the generalized
gauge-transformations \eqref{eq:GGT} cannot recover the
kinematic Jacobi identities as we explained below \eqrefe{eq:dN-dNT-sum}.
But, for our extended double-copy construction under
the $\hs 1\hsm /s\hs$ expansion in section\,\ref{sec:5.3.2},
we successfully recovered the kinematic Jacobi identities
as in Eq.\eqref{eq:Jacobi-Nj'-Ntj'} by using the generalized
gauge-transformations \eqref{eq:GGtransf}
and the solutions \eqref{eq:sol-Delta-tDelta}-\eqref{eq:sol-Delta01}.
With these, we derived the NLO double-copied KK graviton/Goldstone
amplitudes \eqref{eq:DC-Amp-NLOf} and their difference
\eqref{eq:Rterm-DC}, which predict the correct
angular structures as in the original exact KK graviton (Goldstone)
amplitudes \eqref{eq:Amp-E-2hLxx}-\eqref{eq:Amp-E-2phixx}
and  \eqref{eq:RTerm-NLO-GR}.
Finally, in section\,\ref{sec:5.3.3},
we constructed an exact double-copy of
the KK graviton scattering amplitudes at the NLO based on 
our recent first principle approach
of the KK string theory\,\cite{Li:2021yfk}.\ 
We presented the exact double-copy formulas for the LO and NLO
KK graviton amplitudes in 
Eqs.\eqref{eq:MBCJ-LO+NLO}-\eqref{eq:MBCJ-LO-NLO}, 
respectively.\ Applying these formulas to the four-point elastic
longitudinal KK graviton scattering, we explicitly computed 
the double-copied LO amplitude [which fully agrees to 
Eq.\eqref{eq:DC-Amp-LOf}] and the double-copied NLO amplitude 
\eqref{eq:DC-Amp-hL-NLOx} [which exactly agrees to 
Eq.\eqref{eq:Amp-E-2hLxx}].\   
The above analyses and findings are truly encouraging
and we will pursue this direction in future works.

\vspace*{1mm}

In passing,
we recently studied\,\cite{Hang:2021oso} the extended  
double-copy construction of the scattering amplitudes
of massive gauge bosons/grvitons in
the 3d topologically massive Yang-Mills theory (TMYM) and
the 3d topologically massive gravity (TMG) theory\,\cite{TMG}.\
There we proposed\,\cite{Hang:2021oso} a brand-new 
Topological Equivalence Theorem (TET) to formulate
the topological mass-generation and to uncover the nontrivial
energy cancellations in the massive Chern-Simons
gauge boson scattering amplitudes of the TMYM.\ 
Using the TET and the double-copy construction,
we further discovered\,\cite{Hang:2021oso}
the strikingly large energy cancellations
in the massive Chern-Simons graviton scattering amplitudes
of the TMG.

\vspace*{1mm}

Finally, in section\,\ref{sec:5.4},
based upon the KK GAET identity \eqref{eq:KKET-general}
in the 5d KK YM theory, we used double-copy approach to
reconstruct the KK GRET identity \eqref{eq:GET-4P},
and demonstrated a new correspondence of
\,{KK GAET~$\To$~KK GRET}\,
in Eq.\eqref{eq:KKET-GET}.
Especially, we analyzed the (energy,\,mass)-dependence of
the residual term $\M_{\!\Delta}^{}$ in the GRET
and deduced $\,\M_{\!\Delta}^{}\!=\mO(E^0M_n^2)\,$
in Eq.\eqref{eq:RTerm}.
This justifies that even though the amplitudes
in the GRET residual term $\,\M_\Delta^{}\,$ contain
individual contributions having
superficial energy-dependence of $\,\mO(E^2)\,$
by naive power counting, they are ensured to cancel down
to $\mO(E^0M_n^2)$, in agreement with our explicit computation
of $\,\M_{\!\Delta}^{}\!=\dM \!-\dMT\,$
in Eq.\eqref{eq:RTerm-NLO-GR}.

\vspace*{1mm}

In summary, it is impressive that using the double-copy approach,
we established {\it a new correspondence between the two
energy-cancellations in the four-particle longitudinal KK
scattering amplitudes:
$\,E^4\!\ito E^0\,$ in the 5d KK YM gauge theory
and $\,E^{10}\!\!\to\! E^2\,$ in the 5d KK GR theory.}\ 
This was presented schematically in Eq.\eqref{eq:E4_0-E10_2}.\
Furthermore, using the double-copy approach,
we analyzed the structure of the residual term
$\,\M_{\!\Delta}^{}$ in the GRET and further uncovered a
new energy-cancellation mechanism of $\,E^2\ito E^0\,$ 
therein. 
Some of our key analyses and findings of this substantial work
are summarized in the companion letter paper\,\cite{KK-Letter}.

\vspace*{7mm}
\noindent
{\bf\large Acknowledgements}
\\[0.8mm]
We would like to thank Song He, Henry Tye, and Yang Zhang
for discussing the double-copy approaches and KLT relations,
and thank Huan-Hang Chi for discussing the
conventional BCJ method and joining this study at an early stage.\
We thank Sekhar Chivukula and Elizabeth Simmons for correspondence
on this subject.\ We also thank Tony Zee for discussing the
geometric mass-generation of Kaluza-Klein compactification.\
This research was supported in part by National NSF of China
(under grants Nos.\,11835005, 12175136 and 11675086),
by National Key R\,\&\,D Program of China
(under grant No.\,2017YFA0402204).

\vspace*{8mm}

\appendix

\noindent
{\Large\bf Appendix:}

\vspace*{-5mm}
\section{\hspace{-2mm}Kinematics of KK Particle Scattering}
\label{app:A}

\renewcommand{\theequation}{\thesection.\arabic{equation}}

We consider $\,2\ito 2\,$ KK scattering process,
with the four-momentum of each
external state obeying the on-shell condition
$\,p^2_j \!=\! -M_j^2$, ($j=1,2,3,4$).
We number the external states clockwise, with their momenta being
outgoing. Thus, the energy-momentum conservation gives
$\,\sum p_j^{}\!=0\,$, and
the physical momenta of the two incident particles
equal $-p_1^{}$ and $-p_2^{}$, respectively.
For illustration, we take the elastic scattering
$\,X_n X_n\!\ito X_n X_n$ ($n\!\!\geqq\! 0$)
as an example,
where $X_n$ denotes any given KK state of level-$n$
and has $M_j^{}\!=\!M_n$.
For the KK theory, the external particle has
mass $M_n$ for a given KK-state of level-$n$\,.
\begin{figure}[H]
\centering
\includegraphics[width=5cm]{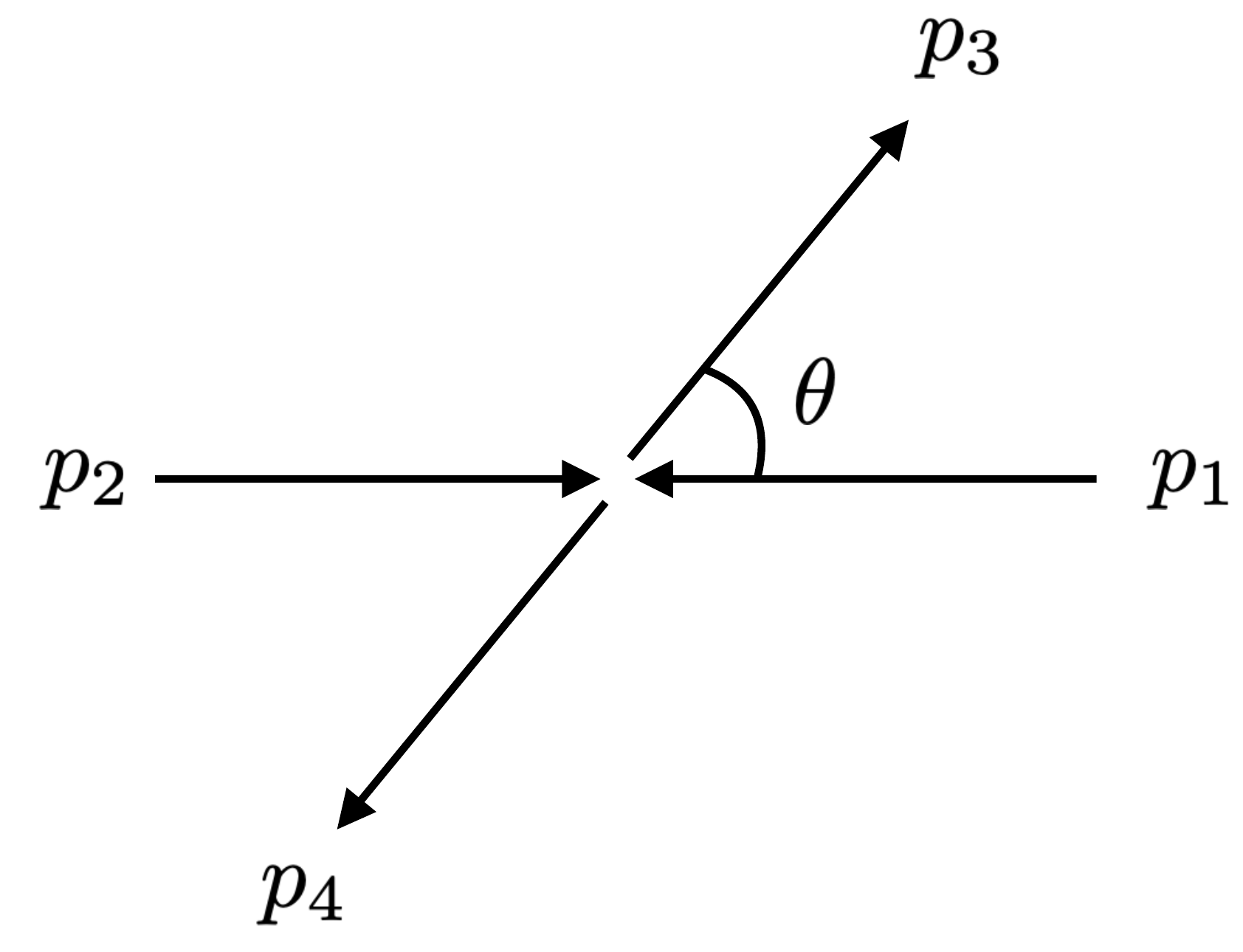}
\vspace*{-4mm}
\caption{\small{Kinematics of the $\,2 \ito 2\,$ scattering process
in the center-of-mass frame.}}
\label{fig:6}
\end{figure}

In the center-of-mass frame (Fig.\ref{fig:6}),
we define the momenta as follows:
\begin{alignat}{3}
p_{1}^{\mu} =&  - \(E, 0, 0, k \) ,
&& \quad\quad
p_{2}^{\mu} = - \(E, 0, 0, -k \) ,
\nn \\
p_{3}^{\mu} =&  \( E , k \st, 0, k \ct  \) ,
&& \quad\quad
p_{4}^{\mu} =   \( E , -k \st, 0, -k\ct \) ,
\end{alignat}
where $k=|\vec{p}\,|$.
Then, the Mandelstam variables $(s,t,u)$ take the following form:
\beqs
\label{eq:s-t-u}
\begin{align}
s\, &=-\(p_1 + p_2 \)^2 = 4E^2\,,
\\[1.mm]
t\, & =-\(p_1 + p_4 \)^2 = -\frac{s-4M_n^2}{2}(1+\ct)\,,
\\
u\, &=-\(p_1 + p_3 \)^2= -\frac{s-4M_n^2}{2}(1-\ct)\,.
\end{align}
\eeqs
Optionally, with \eqrefe{eq:s-t-u}, we can also use the relation
$\,E^2 \!= k^2 \!+\! M_n^2\,$
to define another set of Mandelstam variables $(\sz,\tz,\uz)$:
\\[-9mm]
\beqs
\label{eq:s0-t0-u0}
\begin{align}
\sz &\,=\, 4k^2  \,,
\\[1mm]
\tz &\,=\, -\frac{\,\sz\,}{2}(1+\ct)  \,,
\\[1mm]
\uz &\,=\, -\frac{\,\sz\,}{2}(1-\ct)  \,.
\end{align}
\eeqs
The summations of the Mandelstam variables \eqref{eq:s-t-u}
and \eqref{eq:s0-t0-u0} obey the identities
$\,s+t+u \!=\! 4M_n^{2}$\, and $\,\sz\!+\tz\!+\uz\! =0$\,,
respectively.

\vspace*{1mm}

Moreover,  the above formulas can be extended to the general
scattering process
$\,X_n X_k\!\ito X_m X_\ell\,$,\,
where $\,n,k,m,\ell\!\geqq 0$\,.
Thus, the sum of these Mandelstam variables $(s,t,u)$ satisfies
$\,s+t+u = M_n^2 \!+\! M_k^2 \!+\! M_m^2 \!+\! M_\ell^2$\,.
The incident and outgoing states have the following momenta
in the center of mass frame,
%
\begin{alignat}{3}
p_1^{\mu} =& -\!(E_1^{}, 0, 0, k ),
&& \hspace*{10mm}
p_2^{\mu} =  -(E_2^{}, 0, 0, -k ),
\nn \\
p_3^{\mu} =& \,(E_3^{}, k'\st, 0, k'\ct ),
&& \hspace*{10mm}
p_4^{\mu} =  (E_4^{}, -k'\st, 0, -k'\ct ),
\end{alignat}
where the energy conservation condition,
$\sqrt{s\,}\!=E_1^{}\!+\!E_2^{}=E_3^{}\!+\!E_4^{}$\,,\,
determines the momenta $k$ and $k'$ as follows:
\vspace*{-3mm}
\begin{align}
k &= \frac{1}{\,2\sqrt{s\,}\,}
\([s-(M_1\!+\!M_2)^2][s-(M_1\!-\!M_2)^2]\)^{\!{1}/{2}},
	\nn\\
k' &=\frac{1}{\,2\sqrt{s\,}\,}
\([s-(M_3\!+\!M_4)^2][s-(M_3\!-\!M_4)^2]\)^{\!{1}/{2}} .
\end{align}

Finally, as mentioned in section\,\ref{sec:2},
a massive KK graviton has 5 helicity states
($\lambda \!=\pm 2,\pm 1,0$\,). Their polarization tensors
take the following forms:
\begin{align}
\label{eq:hn-Pols}
\vep_{\pm 2}^{\mn} &=\ep_{\pm}^{\mu} \ep_{\pm}^{\nu}\,,\qquad
\vep_{\pm 1}^{\mn} =
\frac{1}{\sqrt{2\,}\,}\!\(\ep_{\pm}^{\mu}\ep_{L}^{\nu}
\!+\ep_{L}^{\mu} \ep_{\pm}^{\nu} \) \!,
\qquad
\vep_{L}^{\mn} \!=\frac{1}{\sqrt{6\,}\,}\!\( \ep_{+}^{\mu} \ep_{-}^{\nu}\!+\ep_{-}^{\mu} \ep_{+}^{\nu}\!
+2 \ep_{L}^{\mu} \ep_{L}^{\nu}\) \!,
\end{align}
where $(\ep_{\pm}^{\mu},\,\ep_{L}^{\mu})$
denote the (transverse,\,longitudinal) polarization vectors
of a vector boson with the same 4-momentum $p^\mu$.
These polarization tensors satisfy the traceless and orthonormal
conditions. They are also orthogonal to the 4-momentum $p^\mu$
of the KK graviton. Thus, the following conditions hold:
\begin{equation}
\eta_{\mn}^{}\vep^{\mn}=0 \,, \qquad
\vep_\lambda^{\mn} \vep_{\lambda'\!,\,\mn}^*
= \delta_{\lambda\lambda'}^{} \,,\qquad
p_\mu^{} \vep^{\mn}=0 \,,
\end{equation}
where
$\,\lambda,\lambda'\,(=\pm 2,\,\pm 1,\,0)\,$
are the helicity indices of the KK graviton.

\section{\hspace{-2mm}From \boldmath{$R_\xi^{}$} Gauge to Unitary Gauge}
\label{app:B}
\label{sec:unitarygauge}

We note that the KK graviton propagator \eqref{eq:Dnn-Rxi}
in the general $R_\xi^{}$ gauge can be decomposed into
the unitary gauge propagator \eqref{eq:UGauge} plus the
$\xi_n^{}$-dependent part.
In momentum space, we present this decomposition
in the following form:
\beqs
\label{eq:Dhh}
\begin{align}
\label{eq:Dhh-sum}
\hspace*{-20mm}
\D^{\mn\ab}_{nm}(p) \ =&\
\D_{nm,{\rm{UG}}}^{\mn\ab}(p)
+\D_{nm,\,\xi}^{\mn\ab}(p) \,,
\\[2mm]
\label{eq:D-UG1}
\hspace*{-20mm}
\D_{nm,{\rm{UG}}}^{\mn\ab}(p)  \ =&\
-\frac{\ii \dnm}
{2}\frac{~\bar{\eta}^{\mu\al}\bar{\eta}^{\nu\be}\!+\! \bar{\eta}^{\mu\be}\bar{\eta}^{\nu\al}\!-\! \frac{2}{3} \bar{\eta}^{\mu\nu}\bar{\eta}^{\al\be}~}
{\,p^2\!+\! M_n^2\,}
\,,  \hspace*{-20mm}
\\[2mm]
\label{eq:D-xi}
\D_{nm,\,\xi}^{\mn\ab}(p) \ =&\
\frac{\ii \dnm/2}{\,p^{2} + \xi_n M_n^{2}\,}
\biggl[ \!\(\!\eta^{\mu\al} \!+\! \frac{p^{\mu}p^{\al}}{\xi_n M_n^{2}\,}\!\) \!
\frac{p^{\nu}p^{\be}}{\,\xi_n M_n^{2}\,}
\!+\! \(\!\eta^{\mu\be} \!+\! \frac{p^{\mu}p^{\be}}{\xi_n M_n^{2}\,}\!\) \! \frac{p^{\nu}p^{\al}}{\,\xi_n M_n^{2}\,}
\nn\\[1.5mm]
& \
+\! \(\!\eta^{\nu\al} \!+\! \frac{p^{\nu}p^{\al}}{\xi_n M_n^{2}\,}\!\) \!\!
\frac{p^{\mu}p^{\be}}{\,\xi_n M_n^{2}\,}
\!+\! \(\!\eta^{\nu\be} \!+\! \frac{p^{\nu}p^{\be}}{\,\xi_n M_n^{2}\,}\!\) \!
\frac{p^{\mu}p^{\al}}{\,\xi_n M_n^{2}\,}\!\biggr]
-\frac{\ii \dnm 2p^{\mu}p^{\nu}p^{\al}p^{\be^{}}}
{\,(p^2\!+\!\xi_n^{2}M_n^{2} )\,\xi_n M_n^{4}\,}
\nn\\[1.5mm]
& \
+\frac{\ii \dnm /6}{\,p^{2}\! +\! (3\xi_n \!-\! 2)M_n^{2}\,}\!
\(\!\eta^{\mn}_{}\!\!-\!\frac{\,2 p^{\mu}p^{\nu}\,}{M_n^{2}}\!\) \!\!
\(\!\eta^{\ab} \!-\!\frac{\,2 p^{\al}p^{\be}\,}{M_n^{2}}\!\) \!,
\end{align}
\eeqs
where
$\,\bar{\eta}^{\mn}\!=\!\eta^{\mn} \!+ p^\mu p^\nu\!/\!M^2$.
We see that the $\xi_n^{}$-dependent part
$\D_{nm,\,\xi}^{\mn\ab}(p)$ vanishes under
$\,\xi_n^{}\!\ito\infty$. So, the propagator
$\,\D^{\mn\ab}_{nm}(p)\,$ will reduce to the unitary gauge
form $\,\D_{nm,{\rm{UG}}}^{\mn\ab}(p)\,$
in this limit.
Also, the gravitational KK
Goldstone propagators $\D_{nm}^{\mn}(p)$ and $\D_{nm}^{}(p)$
in Eqs.\eqref{eq:Dnn-A}-\eqref{eq:Dnn-phi}
vanish in this limit $\,\xi_n^{}\!\ito\infty\,$,
which removes the unphysical KK Goldstone bosons
in the unitary gauge as expected.
For the Feynman-'t\,Hooft gauge ($\hs\xi_n^{}\!=1$), we find that
the $\xi_n^{}$-dependent part of the KK graviton propagator
takes a much simpler form:
\begin{align}
\D_{nm,\,\xi}^{\mn\ab}(p)\Big|_{\xi=1}^{}
=& \
\frac{\ii\dnm/6}{~p^2 \!+\! M_n^2~}\!
\(\!\eta^{\mn}_{}\!\!-\!\frac{\,2 p^{\mu}p^{\nu}\,}{M_n^{2}}\!\) \!\!
\(\!\eta^{\ab} \!-\!\frac{\,2 p^{\al}p^{\be}\,}{M_n^{2}}\!\)
\nn\\[1.5mm]
&\ + \frac{\ii\dnm/2}{\,(p^{2} \!+\! M_n^{2})M_n^2\,} \! \(
p^\mu p^\al\eta^{\nu\be}\!+p^\nu p^\al\eta^{\mu\be}\!
+p^\mu p^\be \eta^{\nu\al}\!+p^\nu p^\be\eta^{\mu\al}\) \!.
\hspace*{10mm}	
\end{align}
In passing, a $R_\xi^{}$ gauge-fixing was considered\,\cite{japan}
in the Randall-Sundrum model with warped 5d which contains additional
terms related to warp parameter; we note that their KK graviton
propagator was written in a rather different form, but can be
converted into the form consistent with our \eqref{eq:Dhh}.

\vspace*{1mm}

We note that the Lagrangian \eqref{eq:LagLO} is invariant under the general coordinate transformation (gauge transformation):
\begin{equation}
\hh_{AB}^{} \,\to~
\hh_{AB}^{\pp} = \hh_{AB}^{}
-2\pd_{(A}^{}\hat{\chi}_{B)}^{} \,,
\end{equation}
where $\hat{\chi}_{A}^{}(x)$\, is an infinitesimal translation which refers to a vector field generating a one-parameter diffeomorphism group in the background spacetime.

\vspace*{1mm}

Taking \eqrefe{eq:hDecom} under the KK expansion
\eqref{eq:Fourier}, we derive the following gauge transformations
for the KK fields:
\beqs
\begin{align}
h^{\mn}_n & \,~\to~~ h^{\pp\,\mn}_{n}
=\, h^{\mn}_n-2\pd^{(\mu} \chi^{\nu)}_n
- M_{n} \eta^{\mn} \chi^{5}_{n} \,,
\\[1mm]
\A^{\mu}_n & \,~\to~~ \A^{\pp\,\mu}_{n}
=\,\A^{\mu}_n-\pd^{\mu} \chi^{5}_n+M_{n}\chi^{\mu}_{n}\,,
\\[1mm]
\phi_n^{} & \,~\to~~ \phi^{\pp}_n=\,\phi_n^{} - 2M_n \chi^5_n  \,.
\end{align}
\eeqs
In the above the group parameters
$(\chi_n^{\mu},\,\chi_n^5)$ arise from the following
KK expansions of the corresponding 5d parameters
$(\hat\chi^{\mu},\,\hat\chi^5)$:
\beqs
\begin{align}
\hat\chi^\mu(x^\nu, x^5)
&\,=\,  \frac{1}{\sqrt{L}} \[ \chi^\mu_0 (x^\nu)+ \sqrt{2} \sum_{n=1}^{\infty} \chi^\mu_n (x^\nu)
\cos\!\frac{\,n \pi x^5\,}{L} \] \!,
\\[1.5mm]
\hat\chi^5(x^\nu, x^5)
&\,=\,  \sqrt{\frac{2}{L}} \, \sum_{n=1}^{\infty} \chi^5_{n}(x^\nu)\sin\!\frac{\,n \pi x^5\,}{L}   \,,
\end{align}
\eeqs
where we set $\,(\hat\chi^{\mu},\,\hat\chi^5)$\,
as \,(even,\,odd)\, under the $\ZZ$ reflection of 5d orbifold.

\vspace*{1mm}

To transform into unitary gauge, we choose the
gauge parameters as follows:
\begin{align}
\chi^\mu_n = -\frac{1}{M_n} \!
\(\!\A^{\mu}_n - \frac{\,\pd^\mu \phi_n^{}\,}{2M_n} \)\!,
\quad~~~
\chi^5_n = \frac{\phi_n^{}}{2M_n} \,.
\end{align}
Then, we derive the field transformations to
the unitary gauge:
\beqs
\begin{align}
h^{\mn}_n & \,~\to~~ h^{\pp\,\mn}_n\!
=\, h^{\mn}_n +\frac{2}{M_n} \, \pd^{(\mu}\!\A^{\nu)}_n
-\frac{1}{2}\!\(\!\eta^{\mn} \!+\! \frac{2\pd^{\mu} \pd^{\nu}}{M_{n}^{2}}\)\!\phi_n^{} \,,
\\
\A^{\mu}_n & \,~\to~~  \A^{\pp\,\mu}_{n}=\,0 \,,
\\[1.mm]
\phi_n^{} & \,~\to~~ \phi_n^{\pp}=\,0 \,.
\end{align}
\eeqs
Thus, under the unitary gauge, both the KK Goldstone states
$\,\A^{\mu}_n\,$ and $\,\phi_n^{}$ ($n\!>\!0$)
are gauged away, so the 4d action
of the Lagrangian \eqref{eq:LagLOKK} becomes:
\begin{align}
S_{\rm{eff}}  &\,=
\int\!\td^4 x  \! \sum_{n=0}^{\infty}
\bigg\{\!\! -\fr{1}{2}(\pd^\mu h_n^{})^2
+\fr{1}{2}(\pd^\rho h_n^{\mn})^2
+\pd_{\mu}h^{\mn}_n \pd_{\nu}h_n^{}
-\pd^{}_\mu h^{\mu\rho}_n \pd^\nu h_{\nu\rho,n}^{}  \nn\\
& \hspace{6em}
-\fr{1}{2}M^2_n \[ h_n^2 \!-\! (h^{\mn}_n)^2 \, \]\!\bigg\}
+\fr{3}{4} (\pd_\mu \phi^{}_0)^2   \,.
\end{align}
%

\section{\hspace{-2mm}Feynman Rules for KK Graviton Interaction
	with Matter}
\label{app:FRmatter}
\label{app:C}

In this Appendix,
we present the relevant Feynman rules of the
5d gravitational scalar QED (GSQED5) as studied in section\,\ref{sec:4.1},
including the propagators of the matter fields
and the vertices for the KK graviton (Goldstone) interaction
with the matter fields.
All the Feynman Rules are derived in the Feynman-'t\,Hooft gauge ($\zeta_n=1$).

\vspace*{1mm}

We first present the photon propagator
and scalar propagator as follows:
\begin{equation}
\D_{nm}^{\mn}(p) = \frac{-\ii \eta^{\mn}\dnm }{p^2 + M_n^2} \,, \qquad
\D_{nm}(p) =  \frac{\ii \dnm }{p^2 + m_n^2}   \,,
\end{equation}
where the KK number \,$n \!\geqq\!0$\, and
the KK mass for the scalar field is
$\,m_n^{} \!=\! \sqrt{m_0^2\!+\!M_n^2\,}\,$.

\vspace*{1mm}

\begin{figure}[t]
\centering
\includegraphics[height=14cm]{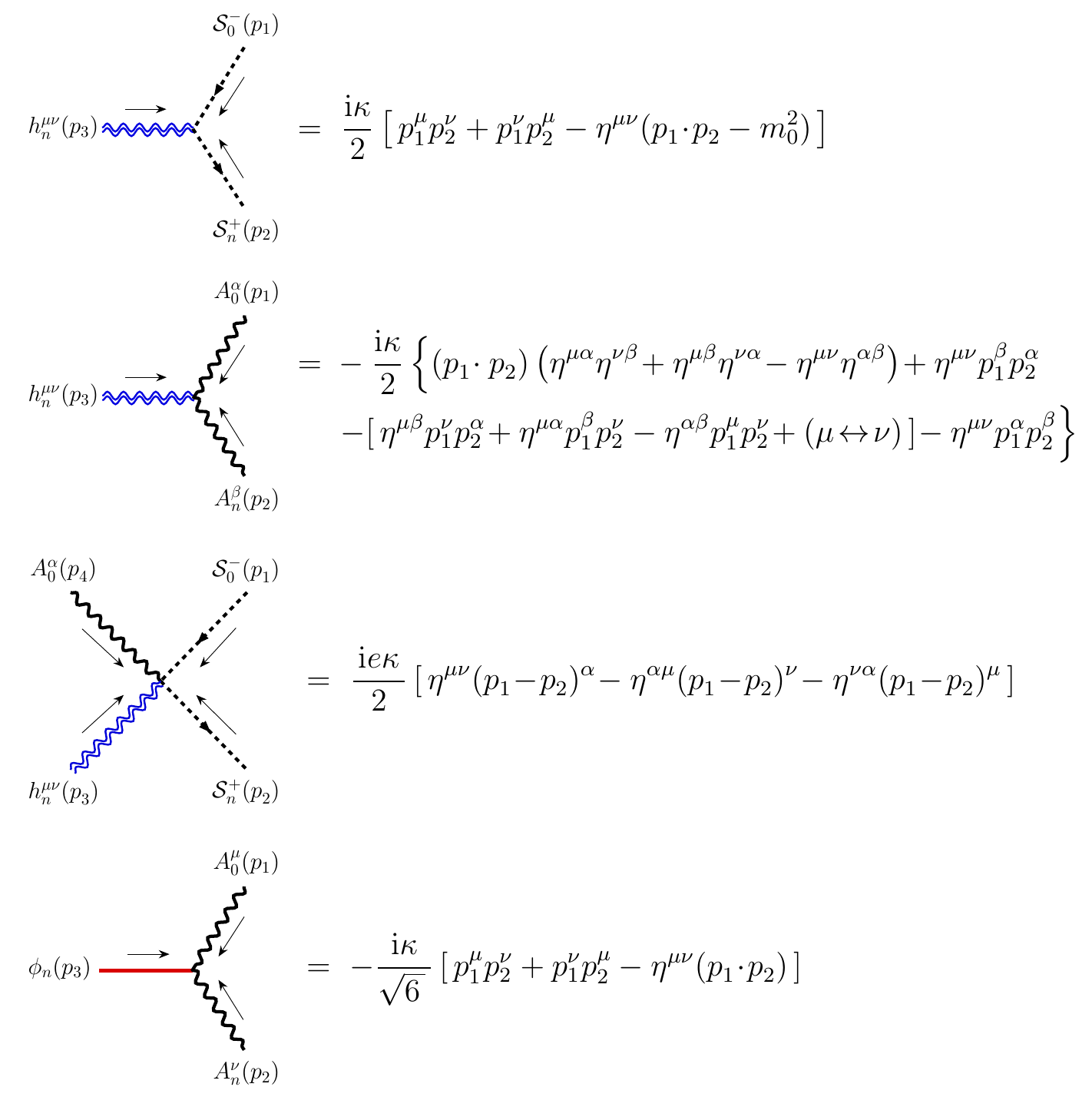}
\vspace*{-3mm}
\caption{\small\baselineskip 15pt
{Feynman rules for the cubic and quartic interaction vertices
between the KK graviton (KK Goldstone boson) and matter fields.
}}
\label{fig:8new}
\end{figure}

Then, with the relations between the 4d coupling constants
and 5d couplings $\,e = \hat e /\!\sqrt{L\,}$\, and
$\,\ka = \hka / \!\sqrt{L\,}$\,,\,
we derive the cubic and quartic interaction vertices
for the KK graviton (Goldstone) interactions with matter,
which are presented in Fig.\,\ref{fig:8new}.

\section{\hspace{-2mm}Gravitational KK Goldstone Amplitudes
	from Goldstone Exchanges\\
	and Contact Interactions}
\label{app:D}

In this Appendix, we derive the Feynman rules
of the KK Goldstone self-interactions
which we will use to compute the subleading diagrams
in Fig.\,\ref{fig:4}, for the analysis of section\,\ref{sec:4.2}.

\linespread{1.5}
\begin{table}[t]
\centering
\begin{tabular}{c|c|c|c|c|c}
\hline\hline
$\hat{\La}_1 [\hat{\A}\phih^2]$
& {\red $\hat{\A}^\mu \pd_\mu\phih\, \pd_5 \phih $}
& {\red $\hat{\A}^\mu (\pd_\mu \pd_5 \phih)\phih $}
&  $\pd_{\mu} \hat{\A}^\mu \pd_5 \phih \,\phih$
&  $\pd_5 \hat{\A}^\mu \pd_\mu \phih \,\phih$
&  \multicolumn{1}{c}{$\pd_{\mu} \pd_5 \hat{\A}^\mu \phih^2$} \\ \hline
$\hat{\La}_1 [\phih^3]$
& {\red $\phih  (\pd_\mu \phih)^2$}
& {\red $\phih  (\pd_5 \phih)^2$}
&  $\phih^2 \pd_\mu^2 \phih$
&  $\phih^2 \pd_5^2 \phih$
& \multirow{2}{*}{}    \\ \cline{1-5}
$\hat{\La}_2 [\phih^4]$
& {\red $\phih^2  (\pd_\mu \phih)^2$}
& {\red $\phih^2  (\pd_5 \phih)^2$}
&  $\phih^3 \pd_\mu^2 \phih$
&  $\phih^3 \pd_5^2 \phih$
&
\\ \hline \hline
\end{tabular}
\caption{\small Lorentz-invariant vertices in the 5d
Lagrangian terms $\hat{\La}_1 [\hat{\A}\phih^2]$,\,
$\hat{\La}_1[\phih^3]$, and $\,\hat{\La}_2 [\phih^4]$.}
\label{tab:Goldstone}
\end{table}

Under the basis defined in Table\,\ref{tab:Goldstone},
we can expand the Lorentz-invariant structure of 5d Lagrangian terms  $\hat{\La}_1[\hat{\A}\phih^2]$, \,$\hat{\La}_1[\phih^3]$,\,
and $\hat{\La}_2[\phih^4]$ as follows:
%
\beqs
\begin{align}
\hat{\La}_1[\hat{\A}\phih^2] &\,=\, b_1\, \hat{\A}^\mu \pd_\mu \phih \, \pd_5 \phih + b_2\, \hat{\A}^\mu \phih\, \pd_\mu  \pd_5 \phih \,,
\\[1mm]
\hat{\La}_1[ \phih^3 ] &\,=\, c_1 \, \phih \, (\pd_\mu \phih)^2 + c_2 \, \phih \, (\pd_5 \phih)^2 \,,
\\[1mm]
\hat{\La}_2[ \phih^4 ] &\,=\, d_1 \, \phih^2  (\pd_\mu \phih)^2 + d_2 \, \phih^2  (\pd_5 \phih)^2  \,,
\end{align}
\eeqs
where we have computed systematically
the coefficients
$(b_1^{},\, b_2^{},\, c_1^{},\, c_2^{}, \, d_1^{},\, d_2^{})$
as follows:
\beqa
(b_1^{},\, b_2^{},\, c_1^{},\, c_2^{},\, d_1^{},\,d_2^{}) \,=
\(\hsm 0 , \,  \frac{1}{\sqrt{2\,}\,},\, \sqrt{\frac{2}{3}\,},\,  -\frac{1}{\sqrt{6\,}\,} \,, -\frac{1}{\,8\,},\, \frac{1}{\,4\,}\) \!.
\eeqa

\vspace*{1mm}

\begin{figure}[t]
\vspace*{-5mm}
\centering
\includegraphics[height=10cm]{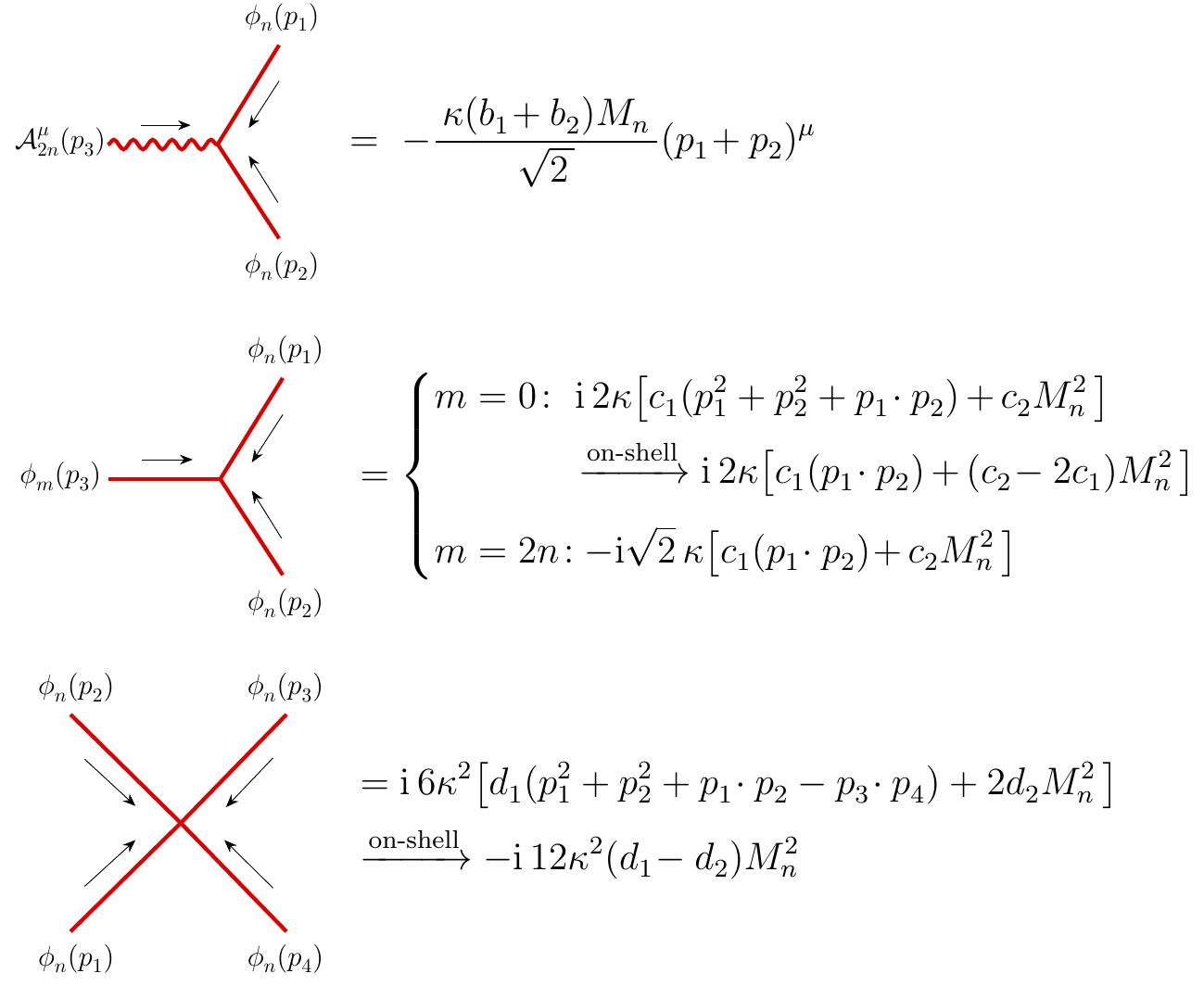}
\vspace*{-3mm}
\caption{\small\baselineskip 15pt
{Feynman rules for the cubic and quartic interaction
vertices among the KK gravitational Goldstone bosons $\phin$
and $\AA^\mu_n$\,.
}}
\label{fig:9new}
\end{figure}

Then, by integrating over $x^5$ on the 5d interval $[0,\, L]$,
we derive the 4d effective KK Lagrangian terms as follows:
\beqs
\begin{align}
\La_1 [\A\phi^2]
&= - \frac{\ka}{\sqrt{2\,}}   \sum_{n,m,\ell=1}^{\infty} \! \biggl\{  b_1 M_\ell [  \sqrt{2}  \A^{\mu}_n  \pd_\mu \phi_0  \phi_\ell \delta_{n\ell} +  \A^{\mu}_n  \pd_\mu \phi_m  \phi_\ell \widetilde{\Delta}_3(m,n,\ell) ]
\nn\\[-1mm]
&
+  b_2 M_\ell [ \sqrt{2} \A^{\mu}_n \phi_0  \pd_\mu \phi_\ell  \delta_{n\ell} + \A^{\mu}_n  \phi_m  \pd_\mu \phi_\ell \widetilde{\Delta}_3(m,n,\ell)] \biggr\}  ,
\\[2mm]
\La_1[ \phi^3 ]  &=
\frac{\ka}{\sqrt{2\,}} \sum_{n,m,\ell=1}^{\infty} \bigg\{
c_1 [\sqrt{2} ( \phi_0 (\pd_\mu\phi_0)^2  + \phi_0 \pd_\mu \phi_m \pd^\mu \phi_\ell \, \delta_{m\ell} + \phi_n \pd_\mu \phi_0 \pd^\mu \phi_m \delta_{n m}
\nn\\
&  +  \phi_n  \pd_\mu \phi_0 \pd^\mu \phi_\ell \, \delta_{n\ell}) + \phi_n \pd_\mu \phi_m \pd^\mu \phi_\ell  \Delta_3(n,m,\ell) ] +c_2 M_m M_\ell [ \sqrt{2} \, \phi_0\phi_m\phi_\ell  \delta_{m\ell}  \nn\\
&  +  \phi_n\phi_m\phi_\ell  \widetilde{ \Delta}_3(n,m,\ell)  ] \bigg\}
,
\\[2mm]
\La_2[ \phi^4]  &=  \frac{\ka^2}{2\,}\! \sum_{n,m,\ell,k=1}^{\infty} \biggl\{  d_1\{  2\left(\phi_{0}\pd_{\mu} \phi_{0}\right)^{2} + 2[  \(\pd_\mu \phi_{0}\)^{2} \phi_{n} \phi_{m} \delta_{n m}+\phi_{0} \pd_{\mu} \phi_{0} \phi_{n} \pd^{\mu} \phi_{\ell} \delta_{n\ell}
	\nn\\
& \hspace{-10mm}
+ \phi_{0} \pd_{\mu} \phi_{0} \phi_{n} \pd^{\mu} \phi_{k} \delta_{nk}  +   \phi_{0} \pd_{\mu} \phi_{0} \phi_{m} \pd^{\mu} \phi_{k} \delta_{mk}  +  \phi_{0} \pd_{\mu} \phi_{0} \phi_{m} \pd^{\mu} \phi_\ell \delta_{m\ell} + \phi_{0}^{2} \pd_{\mu} \phi_\ell \pd^{\mu} \phi_{k}  \delta_{\ell k}  ]
\nn\\[2mm]
& \hspace{-10mm}
+ \sqrt{2}  [ \pd_{\mu} \phi_{0} \phi_{n} \phi_{m} \pd^{\mu} \phi_\ell  \Delta_{3}(n, m, \ell)  +  \pd_{\mu} \phi_{0} \phi_{n} \phi_{m} \pd^{\mu} \phi_{k} \Delta_{3}(n, m, k)+\pd_{\mu} \phi_{0} \phi_{n} \phi_\ell \pd^{\mu} \phi_{k} \Delta_{3}(n, \ell, k)
\nn\\[2mm]
& \hspace{-10mm}
+ \phi_{0} \phi_{m} \pd_{\mu} \phi_\ell \pd^{\mu} \phi_{k}  \Delta_{3}(m, \ell, k) ]  + \phi_{n} \phi_{m} \pd_{\mu} \phi_\ell \pd^{\mu} \phi_{k}  \Delta_{4}(n, m, \ell, k ) \}
+ d_2  M_\ell M_{k} [ 2 (\phi_{0} )^{2} \phi_\ell \phi_{k} \delta_{\ell k}
	\nn\\
& \hspace{-10mm}
+ \sqrt{2}  \phi_{0} \phi_{m} \phi_\ell \phi_{k}  \widetilde{\Delta}_{3}(m, \ell, k)    + \sqrt{2} \phi_{0} \phi_{n} \phi_\ell \phi_{k}  \widetilde{\Delta}_{3}(n, \ell, k)
+ \phi_{n} \phi_{m} \phi_\ell \phi_{k} \widetilde{\Delta}_4(n, m, \ell, k)   ] \biggr\},
\end{align}
\eeqs
where
\beqs
\begin{align}
\Delta_4^{}(n,m,\ell,k) = & \,\, \delta(n\!+\!m\!+\!\ell\!-\!k)+\delta(n\!+\!m\!-\!\ell\!-\!k)
+\delta(n\!-\!m\!+\!\ell\!-\!k)+ \delta(n\!-\!m\!-\!\ell\!-\!k)
\hspace*{13mm}
\nn\\
& +\delta(n\!-\!m\!-\!\ell\!+\!k) + \delta(n\!+\!m\!-\!\ell\!+\!k)
+\delta(n\!-\!m\!+\!\ell\!+\!k)\,,
\\[1mm]
\widetilde{\Delta}_4^{}(n,m,\ell,k) = & \,\, \delta(n\!+\!m\!+\!\ell\!-\!k)
-\delta(n\!+\!m\!-\!\ell\!-\!k)+\delta(n\!-\!m\!+\!\ell\!-\!k)
-\delta(n\!-\!m\!-\!\ell\!-\!k)
\nn\\
& +\delta(n\!-\!m\!-\!\ell\!+\!k)
- \delta(n\!+\!m\!-\!\ell\!+\!k) +\delta(n\!-\!m\!+\!\ell\!+\!k)
\,.
\end{align}
\eeqs
With these, we derive the cubic and quartic interaction vertices
of $\A\phi^2$, $\phi^3$, and $\phi^4$,
which are presented in Fig.\,\ref{fig:9new}.

\vspace*{2mm}

By explicit calculations,
we find that the scattering amplitudes of
$\,\pnnnn\,$ from the $\phi_{0}^{}$ ($\phi_{2n}^{}$) exchanges
and from the contact interaction are all of $\,\mO(\Mnn)\,$
under the high energy expansion, which do not contribute to
the LO Goldstone boson amplitude of $\mO(E^2)$\,.
This conclusion still holds for
the inelastic scattering process $\pnkml$\,.

\section{\hspace{-2mm}Massless Graviton Scattering and Double-Copy in 4d}
\label{app:E}

For the sake of comparison,
in this Appendix we compute the scattering amplitudes
of four gluons and of four gravitons at tree level in 4d,
by using the conventional Feynman techniques and
the reduced super-string amplitudes, respectively.
We also verify the double-copy construction
of massless graviton amplitudes from
the massless gluon amplitudes
by using the color-kinematics (CK) duality.
We find that the conversion constant
between the gauge boson coupling
and the graviton coupling in 4d differs from
what we have obtained
in the 5d KK theory analysis (section\,\ref{sec:5.2}).

\vspace*{-2mm}
\subsection{\hspace*{-2mm}Massless Graviton Scattering from Double-Copy Construction in 4d}
\label{app:E1}

For an SU(N) non-Abelian gauge theory, we can express
the four-gluon scattering amplitudes at tree level
as follows:

\begin{equation}
\T[gg \to gg] \,=\, -g^2 \, (\T_c+ \T_s + \T_t + \T_u)  \,,
\end{equation}
where the amplitude contains the contributions from a contact interaction diagram and the $(s,t,u)$-channel pole-diagrams whose amplitudes
are given by
\beqs
\begin{align}
\T_c	\ =&\ \ \CC_s \[(\ep_{1} \!\cdot \ep_{3})
(\ep_{2} \!\cdot \ep_{4})- (\ep_{1} \!\cdot \ep_{4})
(\ep_{2} \!\cdot \ep_{3}) \] + \CC_t
\[(\ep_{1} \!\cdot \ep_{2} ) (\ep_{3} \!\cdot \ep_{4}) -
(\ep_{1} \!\cdot \ep_{3}) (\ep_{2} \!\cdot \ep_{4}) \]
\hspace*{8mm}
\nn\\
& \ + \CC_u \[(\ep_{1} \!\cdot \ep_{4} ) (\ep_{2} \!\cdot \ep_{3})
- (\ep_{1} \!\cdot \ep_{2} ) (\ep_{3 } \!\cdot \ep_{4 }) \] ,
\\[1mm]
\T_s	\ =&\ \
\frac{\,\CC_s\,}{s} \[ \( p_{1}\!-p_{2} \)
(\ep_{1} \!\cdot \ep_{2}) + 2 (p_{2} \!\cdot \ep_{1})
\ep_{2}-2 (p_{1} \!\cdot \ep_{2} )\ep_{1} \]
\nn\\[1mm]
&\hspace{2em}\cdot
\[ \( p_{4}\!-p_{3} \) (\ep_{3} \!\cdot \ep_{4})
+ 2 (p_3 \!\cdot \ep_{4}) \ep_{3}
-2 (p_4 \!\cdot \ep_{3} )\ep_{4}  \] ,
\\[1mm]
\T_t	\ =&\ \
-\frac{~\CC_t\,}{t} \[ \( -p_{1} \!+ p_{4} \)
(\ep_{1} \!\cdot \ep_{4})
- 2 (p_{4} \!\cdot \ep_{1}) \ep_{4}
+ 2 (p_{1} \!\cdot \ep_{4} )\ep_{1} \]
\nn\\[1pt]
&\hspace{2em}\cdot
\[ \( -p_2 \!+ p_3  \) (\ep_{2} \!\cdot \ep_{3})
- 2 (p_3 \!\cdot \ep_{2}) \ep_{3}
+ 2 (p_2 \!\cdot \ep_{3} )\ep_{2}  \]   ,
\\[1mm]
\T_u	\ =&\ \
\frac{~\CC_u\,}{u} \[ \( -p_{1} \!+ p_{3}\)(\ep_{1} \!\cdot \ep_{3})
- 2 (p_{3} \!\cdot \ep_{1}) \ep_{3}
+ 2 (p_{1} \!\cdot \ep_{3} )\ep_{1} \]
\nn\\[1pt]
&\hspace{2em}\cdot
\[ \( -p_2 \!+ p_4  \) (\ep_{2} \!\cdot \ep_{4})
- 2 (p_4 \!\cdot \ep_{2}) \ep_{4}
+ 2 (p_2 \!\cdot \ep_{4} )\ep_{2}  \]  .
\end{align}
\eeqs
Here each external massless gauge boson (gluon) has two helicity states, as described by its two transverse polarization vectors
$\,\ep_{j \pm}^{\mu}\,$ ($j=1,2,3,4$):
\begin{alignat}{3}
\ep^{\mu}_{1+} = \ep^{\mu}_{2-} &= \frac{1}{\sqrt{2\,}\,} (0, 1, \ii , 0)  ,
&& \qquad \ep^{\mu}_{1-} = \ep^{\mu}_{2+} = \frac{1}{\sqrt{2\,}\,}(0, -1, \ii, 0),  \nn
\\[1mm]
\ep^{\mu}_{3+} = \ep^{\mu}_{4-} &= \frac{1}{\sqrt{2\,}\,}(0, \ii\ct ,1 , -\ii\st) ,
&& \qquad \ep^{\mu}_{3-} =\ep^{\mu}_{4+} = \frac{1}{\sqrt{2\,}\,} (0, \ii\ct , -1 , -\ii\st)   .
\label{eq:PolarizationYM}
\end{alignat}
Then, we compute the helicity amplitudes of the gauge boson scattering:
\beqs
\label{eq:T-Helicity}
\begin{align}
\label{eq:T-Helicity1}	
& \T[++++] =\T[----] \nn\\
&\hspace{2.05cm}  = g^2 \!\LB
\CC_s \(-2\ct\)+ \CC_t \[\! \frac{3-2\ct-\ctt}{1-\ct} \!\]+ \CC_u  \[\! \frac{-3-2\ct+\ctt}{1+\ct} \!\] \RB \!,
\\[1mm]
\label{eq:T-Helicity2}
& \T[+-+-] =\T[-+-+]\nn\\
&\hspace{2.05cm}  = g^2 \!\LB
\CC_t  \(-1-\ct  \)  +
\CC_u \[\! \frac{3+4 \ct +\ctt}{2 (1-\ct)}  \!\]\RB \!,  \\[1mm]
\label{eq:T-Helicity3}
& \T[+--+] =\T[-++-]\nn\\
&\hspace{2.05cm}  = g^2 \!\LB
\CC_t \[ \!\frac{-3+4 \ct - \ctt}{2 (1+\ct)}  \!\]+ \CC_u \( 1-\ct   \) \RB,
\end{align}
\eeqs
where $\,\ctt \!=\!\cos 2\theta$,\,
and all the helicity-flipped amplitudes vanish, which include
the amplitudes like $\T[+++-]$, $\T[++--]$, and so on.
For convenience, we rewrite the above amplitudes
\eqref{eq:T-Helicity1}-\eqref{eq:T-Helicity3} as follows:
\beqs
\begin{align}
\T[++++] \,=\, \T[----]
&\,=\, g^2 \!\[\!
\frac{\CC_s \NN_s}{s} + \frac{\CC_t \NN_t}{t} +	\frac{\CC_u \NN_u}{u}  \!\] \!,
\\[1mm]
\T[+-+-] \,=\,\T[-+-+]
&\,=\, g^2 \!\[\!
\frac{\CC_s \NN'_s}{s} + \frac{\CC_t \NN'_t}{t} +	
\frac{\CC_u \NN'_u}{u}  \!\] \!,
\\[1mm]
\T[+--+] \,=\,\T[-++-]
&\,=\, g^2 \!\[\!
\frac{\CC_s \NN''_s}{s} + \frac{\CC_t \NN''_t}{t} +
\frac{\CC_u \NN''_u}{u}  \!\] \!,
\end{align}
\eeqs
where the numerator parameters $(\NN_j^{},\,\NN_j',\,\NN_j'')$
are given by
\beqs
\label{eq:N-N'-N''}
\begin{alignat}{4}
\NN_s &=  -2 s \ct  \,, \hspace*{5mm}
\NN_t &=  \frac{\,s\,}{2} (-3+\!2\ct \!+ \ctt ) \,,
\hspace*{5mm} \hspace*{7mm}
\NN_u &=  \frac{\,s\,}{2} (3+\!2\ct\! - \ctt ) \,,
\\[1mm]
\NN'_s &= 0  \,,    \qquad \hspace*{5mm}
\NN'_t &=  \frac{\,s(2\!+\!\ct\!-2\ctt\!-\cttt)\,}{8(1\!-\!\ct)}
\,,  \hspace*{5mm} \hspace*{1.3mm}
\NN'_u &= \frac{\,s(-2\!-\ct\!+2\ctt\!+\!\cttt)\,}{8 (1\!-\!\ct)}\,,
\\[1mm]
\NN''_s &=  0  \,, \qquad \hspace*{5mm}
\NN''_t &=  \frac{\,s(2\!-\!\ct\!-\!2\ctt\!+\!\cttt)\,}
{8 (1\!+\!\ct)} \,, \hspace*{5mm} \hspace*{1.3mm}
\NN''_u &=  \frac{\,s(-2\!+\!\ct\!+\!2\ctt\!-\!\cttt)\,}
{8 (1\!+\!\ct)} \,.
\end{alignat}
\eeqs
Hence, we can readily verify that the numerators in
\eqrefe{eq:N-N'-N''} obey the kinematic Jacobi identity:
\\[-11mm]
\beqs
\begin{align}
\NN_s + \NN_t + \NN_u &\,=\, 0   \,,
\\[1mm]
\NN'_s + \NN'_t + \NN'_u &\,=\, 0   \,,
\\[1mm]
\NN''_s + \NN''_t + \NN''_u &\,=\, 0   \,.
\end{align}
\eeqs


Next, by using the double-copy approach with CK duality,
we reconstruct the massless graviton scattering amplitudes
as follows:
\beqs
\label{eq:AmpGra}
\begin{alignat}{3}
\label{eq:AmpGra++++}
\T_{\rm {DC}}^{} [++++] \,=\,\T_{\rm {DC}}^{} [----]
&\,=\, - 16 \tilde{c}_0 g^2 s \csc^2\!\theta
&&=\,
- \frac{\ka^2}{4} \frac{s^{3}}{\,t\hspace*{1.3pt}u\,} \,,
\\[1mm]
\label{eq:AmpGra+-+-}
\T_{\rm {DC}}^{}[+-+-] \,=\,\T_{\rm {DC}}^{}[-+-+]&\,=\,
-\, \tilde{c}_0 g^2 \frac{s\, s^6_{\theta}}{\,(1\!-\!\ct)^4\,}
&&=\, -\frac{\ka^2}{4}\frac{t^3}{\,s\hspace*{1.3pt}u\,}\,,
\\[1mm]
\label{eq:AmpGra+--+}
\T_{\rm {DC}}^{} [+--+] \,=\,\T_{\rm {DC}}^{}[-++-]
&\,=\, -\, \tilde{c}_0 g^2
\frac{s\(1-\ct\)^4}{s^2_{\theta} }
&&=\, -\frac{\ka^2}{4} \frac{u^3}{\,s\hspace*{1.3pt}t\,} \,,
\end{alignat}
\eeqs
where we have applied the conversion constant \eqref{eq:c0-4d}
for the amplitudes in the last equality of  Eqs.\eqref{eq:AmpGra++++}-\eqref{eq:AmpGra+--+} and $(+,-) \equiv (+2,-2)$.
The above reconstructed massless graviton scattering amplitudes
agree with the results of Refs.\,\cite{Berends:1975}\cite{Grisaru:1975}
which computed directly the graviton amplitudes by the conventional
Feynman techniques.

\vspace*{1mm}
\subsection{\hspace*{-2mm}Massless Graviton Scattering from Type-II Superstring Theory}
\label{app:E2}

The massless graviton scattering amplitudes can be computed
by the conventional Feynman technique in quantum field theory.
The 3-point and 4-point graviton interaction vertices were
derived by DeWitt\,\cite{DeWitt:1967}.
It is clear that by using
the conventional Feynman diagram approach,
the calculations of graviton scattering amplitudes are extremely
complicated and tedious. Hence, we will use the amplitudes
as computed within the type-II superstring theory
(SST-II)\,\cite{string}\cite{Schwarz:1982}\cite{Sannan:1986}.
The massless gravitons are described by closed strings and their
4-point scattering amplitude at tree level
in the SST-II is given by
\begin{equation}
\M [\lam_1^{} \lam_2 \lam_3 \lam_4] \,=\, \hka^2 C(\hhs,\hht,\hhu) \,
\vep_{\lam_1}^{\hat\mu\hat\al} \vep_{\lam_2}^{\hat\nu\hat\be} \vep_{\lam_3}^{\hat\rho\hat\ga} \vep_{\lam_4}^{\hat\sigma \hat\lam} \widehat{K}_{\hmn\hat\rho\hat\si} \widehat{K}_{\hab\hat\ga\hat\lam} \,,
\label{eq:AmpSST}
\end{equation}
where $\,\hka\,$ is the 10-dimensional gravitational coupling
constant, and the Mandelstam variables $(\hhs,\,\hht,\,\hhu)$ and
the Lorentz indices $(\hat\mu,\hat\nu,\cdots)$ are defined in 10d.
In \eqrefe{eq:AmpSST},
the $C$ function takes the following form:
\begin{equation}
C(\hhs,\hht,\hhu)\,=\, \frac{1}{\,128\,}
\frac{~\Ga(-\alp\hhs/2) \Gamma(-\alp\hht/2)\Ga(-\alp\hhu/2)~}
{\,\Ga(1\!+\!\alp\hhs/2)\Ga(1\!+\!\alp\hht/2) \Ga(1\!+\!\alp\hhu/2)\,}  \,,
\end{equation}
where we have set the Regge slope for the closed string to be $\,\alp\!=\!\fr{1}{4}\,$.

\vspace*{1mm}

Using the relation of gamma functions
$\,\Ga(1\!+\!z) \!=\! z\Ga(z)\,$,
we can rewrite the function $C(\hhs,\hht,\hhu)$\, as follows:
\begin{equation}
C(\hhs,\hht,\hhu) \,=\,-\frac{4}{\,\hhs\,\hht\,\hhu\,}
\frac{~\Ga(1\!-\!\alp\hhs/2)\,\Gamma(1\!-\!\alp\hht/2)\,
	 \Gamma(1\!-\!\alp\hhu/2)~}
{~\Ga(1\!+\!\alp\hhs/2)\,\Gamma(1\!+\!\alp\hht/2)\,
	\Ga(1\!+\!\alp\hhu/2)\,} \,.
\end{equation}
With Ref.\,\cite{Schwarz:1982}, we note that
by imposing compactification of \,$(10\!-\!d)$\,
spatial dimensions,
the amplitude defined in $d$-dimension has the same structure
as that of the original 10-dimension case at tree level.
In order to reduce the SST-II amplitude \eqref{eq:AmpSST}
to the amplitude in 4d\,,
we take the limit for Regge slope $\,\al' \ito 0$\,
and derive the reduced 4d graviton scattering amplitude:
\begin{equation}
\M[\lam_1^{}\lam_2^{}\lam_3^{}\lam_4^{}] \,=\,
-\frac{\,4\ka^2\,}
{\,s\hspace*{1.2pt}t\hspace*{1pt}u\,}
K_{\lam_1^{\pp}\lam_2^{\pp}\lam_3^{\pp}\lam_4^{\pp}}^2 ,
\end{equation}
where the $K_{\lam_1^{\pp}\lam_2^{\pp}\lam_3^{\pp}\lam_4^{\pp}}$ factor is given as
follows\,\cite{string}\cite{Schwarz:1982}\cite{Sannan:1986}:
\begin{align}
\label{eq:SSTKinematic}
\hspace*{-2mm}
K_{\lam_1^{\pp}\lam_2^{\pp}\lam_3^{\pp}\lam_4^{\pp}}
\!=&-\frac{1}{4}\big\{ [st (\ep_1 \!\cdot \ep_3)
(\ep_2 \!\cdot \ep_4)+su(\ep_1 \!\cdot \ep_4) (\ep_2 \!\cdot \ep_3)
+tu (\ep_1 \!\cdot \ep_2) (\ep_3 \!\cdot \ep_4)]
\nn\\[1mm]
&-2 s \, [(p_1\!\cdot \ep_4) (p_3\!\cdot \ep_2) (\ep_1\!\cdot \ep_3)
+ (p_1\!\cdot \ep_3) (p_4\!\cdot \ep_2) (\ep_1\!\cdot \ep_4)
+ (p_2\!\cdot \ep_3) (p_4\!\cdot \ep_1) (\ep_2\!\cdot \ep_4)
\nn\\
& + (p_2\!\cdot \ep_4) (p_3\!\cdot \ep_1) (\ep_2\!\cdot \ep_3)]
-2 t \, [(p_1\!\cdot \ep_2) (p_3\!\cdot \ep_4) (\ep_1\!\cdot \ep_3)
+ (p_1\!\cdot \ep_3) (p_2\!\cdot \ep_4) (\ep_1\!\cdot \ep_2)
\nn\\
&+(p_2\!\cdot \ep_1) (p_4\!\cdot \ep_3) (\ep_2\!\cdot \ep_4)
+ (p_3\!\cdot \ep_1) (p_4\!\cdot \ep_2) (\ep_3\!\cdot \ep_4)]
-2 u \, [(p_1\!\cdot \ep_2) (p_4\!\cdot \ep_3) (\ep_1\!\cdot \ep_4)
\nn\\
&+ (p_1\!\cdot \ep_4) (p_2\!\cdot \ep_3) (\ep_1\!\cdot \ep_2)
+(p_3\!\cdot \ep_2) (p_4\!\cdot \ep_1)
(\ep_3\!\cdot \ep_4)  + (p_3\!\cdot \ep_4) (p_2\!\cdot \ep_1)
(\ep_2\!\cdot \ep_3)] \big\}   \,,
\end{align}
where we denote the polarization vectors as
$\,\ep_j^\mu \equiv \ep_{\lam'_j}^\mu $\,.
We substitute the polarizations \eqref{eq:PolarizationYM} into
the above kinematic factor \eqref{eq:SSTKinematic}.
Thus, we can readily deduce
the 4d graviton scattering amplitudes at tree level:
\beqs
\label{eq:SSTAmp}
\begin{align}
\label{eq:SSTAmp++++}
\M^{}[++++] &
= \M^{}[----]
=   - \frac{\,\ka^2\,}{4} \,
\frac{s^{3}}{\,t\hspace*{1pt}u\,} \,,
\\[1mm]
\label{eq:SSTAmp+-+-}
\M^{}[+-+-] &=\M^{}[-+-+]
= -\frac{\,\ka^2\,}{4} \frac{t^3}{\,s\hspace*{1pt}u\,} \,,
\\[1mm]
\label{eq:SSTAmp+--+}
\M^{}[+--+] &=\M^{}[-++-]
= -\frac{\,\ka^2\,}{4} \frac{u^3}{\,s\hspace*{1.5pt}t\,} \,.
\end{align}
\eeqs
These fully agree with the amplitudes \eqref{eq:AmpGra} by the double-copy construction from the corresponding gauge boson amplitudes.

\vspace*{2mm}
\section{\hspace{-2mm}{Full Amplitudes of KK Gravitons and Goldstones in 5d GR}}
\label{app:F}


For completeness, we summarize the full elastic amplitudes of
the four longitudinal KK graviton scattering\,\cite{Chivukula:2020L}
and the four gravitational KK Goldstone boson scattering
(section\,\ref{sec:4.2}) as follows:
\beqs
\begin{align}
\label{eq:Amp4hL}
\M[4h_L^n] &= -\frac{\ka^2 M_n^2(X_0 + X_2 \ctt + X_4 \ctf + X_6 \cts ) \csc^2 \!\theta}{512 \bs (\bs-4) [ \bs^2 -(\bs-4)^2 \ctt + 24 \bs + 16]}  \,,
\\[1.mm]
\MT[4\phin] &=- \frac{\ka^2 M_n^2(\tX_0 + \tX_2 \ctt + \tX_4 \ctf + \tX_6 \cts ) \csc^2\! \theta}{512 \bs (\bs-4) [ \bs^2 -(\bs-4)^2 \ctt + 24 \bs + 16]}  \,,
\label{eq:Amp4Phi}
\end{align}
\eeqs
where parameters $X_j$ and $\tX_j$ are defined as
\beqs
\begin{align}
X_0 &= -2 (255\hs\bs^5 + 2824\hs\bs^4 - 19936\hs\bs^3
+ 39936\hs\bs^2 - 256\hs\bs + 14336),
\\[1mm]
X_2 &= 429\hs\bs^5-10152 \bs^4 +30816\hs\bs^3 -27136\hs\bs^2
-49920\hs\bs +34816\,,
\\[1mm]
X_4 &= 2 (39\hs\bs^5 - 312\hs\bs^4 - 2784\hs\bs^3
- 11264\hs\bs^2 + 26368\hs\bs - 2048),
\\[1mm]
X_6 &= 3\hs\bs^5 +40\hs\bs^4 +416\hs\bs^3 -1536\hs\bs^2
-3328\hs\bs -2048 \,,
\\[2mm]
\widetilde{X}_0 &=-2 (255\hs\bs^5 +8248\hs\bs^4 -4144\hs\bs^3
+79104\hs\bs^2 +642560\hs\bs+69632)\,,
\\[1mm]
\widetilde{X}_2 &= 429\hs\bs^5 +4152\hs\bs^4 +21216\hs\bs^3
-150016\hs\bs^2 +1142016\hs\bs +182272 \hs,
\\[1mm]
\widetilde{X}_4 &= 2(39\hs\bs^5 -1992\hs\bs^4 +17808\hs\bs^3
-58112\hs\bs^2 +70144\hs\bs-20480) \hs,
\\[1mm]
\tX_6 &= 3\hs\bs^5 -56\hs\bs^4 +416\hs\bs^3
-1536\hs\bs^2 +2816\hs\bs -2048\hs.
\end{align}
\eeqs
For our analyses in sections\,\ref{sec:4}-\ref{sec:5},
it is also useful to express the above amplitudes
in terms of the variable $\bs_0^{}$\,:
\beqs
\begin{align}
\M[4h_L^n] &= \frac{\,\ka^2 M_n^2(X^0_0 \!+\! X^0_2\ctt \!+\! X^0_4\ctf \!+\! X^0_6\cts )\csc^2\!\theta}{~512\bs_0^{}(\bs_0^{}\!+\!4) \[2 \bs_0^2 \st^2 \!+\! 32\bs_0^{} \!+\! 128 \]} \,,
\label{eq:FullAmphL}
\\[1.5mm]
\MT[4\phin] &=
\frac{\,\ka^2 M_n^2(\tX^0_0 \!+\! \tX^0_2\ctt \!+\! \tX^0_4\ctf \!+\! \tX^0_6\cts )\csc^2\!\theta}{~512\bs_0^{}(\bs_0^{}\!+\!4) \[2 \bs_0^2 \st^2 \!+\! 32\bs_0^{} \!+\! 128 \]} \,,
\label{eq:FullAmp4Phi}
\end{align}
\eeqs
where we have the following coefficients:
\beqs
\begin{align}
X^0_0 &= 2 (255\hs\bs_0^5 +7924\hs\bs_0^4 +66048\hs\bs_0^3
+235008\hs\bs_0^2 +411648\hs\bs_0 +360448)\hs,
\\[1mm]
X^0_2 &= -429\hs\bs_0^5 +1572\hs\bs_0^4 +62976\hs\bs_0^3
+357376\hs\bs_0^2 +837632\hs\bs_0 +786432\hs ,
\\[1mm]
X^0_4 &= 2 (-39\hs\bs_0^5-468\hs\bs_0^4+1536\hs\bs_0^3
+49664\hs\bs_0^2 + 227328\hs\bs_0^{} + 294912),
\\[1mm]
X^0_6 &= - (3\hs\bs_0^5+100\hs\bs_0^4 +1536\hs\bs_0^3
+9216\hs\bs_0^2 +18432\hs\bs_0)\hs,
\\[2mm]
\tX^0_0 &= 2 (255\hs\bs_0^5 +13348\hs\bs_0^4+168624\hs\bs_0^3
+984384\hs\bs_0^2 + 3514368\hs\bs_0+6012928) \hs,
\\[1mm]
\tX^0_2 &= -429\hs\bs_0^5 -12732\hs\bs_0^4-156288\hs\bs_0^3
-777728\hs\bs_0^2-2572288\hs\bs_0-5210112 \hs,
\\[1mm]
\tX^0_4 &= -2 (39 \bs_0^5 - 1212\hs\bs_0^4- 7824\hs\bs_0^3
- 10688\hs\bs_0^2) \hs,
\\[1mm]
\tX^0_6 &= - (3 \bs_0^5+4\bs_0^4) \hs.
\end{align}
\eeqs
Then, we make high energy expansion for the KK graviton
(Goldstone) scattering amplitudes
\eqref{eq:Amp4hL}-\eqref{eq:Amp4Phi} or
\eqref{eq:FullAmphL}-\eqref{eq:FullAmp4Phi} as follows:
\vspace*{-3mm}
\beqs
\begin{align}
\M[4\hLn] &\,=\,
\M_0[4h_L^n] + \dM[4h_L^n]  \,,
\\[1mm]
\MT[4\phin] &\,=\,
\MT_0[4\phin] + \dMT[4\phin] \,,
\end{align}
\eeqs
where the LO and NLO amplitudes are given by
$(\M_0^{},\,\MT_0^{})$ and $(\dM,\,\dMT)$, respectively.
For the high energy expansion in terms $\,1/\bs\,$,
we derive from expanding Eqs.\eqref{eq:Amp4hL}-\eqref{eq:Amp4Phi}
the following KK graviton (Goldstone) scattering amplitudes
at the LO and NLO:
\beqs
\begin{align}
\label{eq:dM-dMT-GR5xx}
\M_0^{}[4h_L^n] &\,=\, \MT_0^{}[4\phin] =
\frac{~3\ka^2\,}{~128~}\hs s\hs
( 7\hsm +\ctt )^2\hsm \csc^2\!\theta  \,,
\\[1.5mm]
\label{eq:Amp-E-2hLxx}
\da\M[4h_L^n] &\,=\, -\frac{~\ka^2 M_n^2~}{256}
(1810 \hsm +\hsm 93\hs\ctt \hsm +\hsm 126\hs\ctf \hsm +\hsm
19\hs\cts)\hsm \csc^4 \hsmx\theta  \,,
\\[2mm]
\label{eq:Amp-E-2phixx}
\delta\MT[4\phi_n] &\,=\, -
\frac{\,\ka^2M_n^2\,}{256}(-902+3669\hs\ctt - 714\hs\ctf - 5\hs\cts)\csc^4\!\theta \,.
\end{align}
\eeqs

For the high energy expansion in terms $\,1/\bsz\,$,
we can expand Eqs.\eqref{eq:FullAmphL}-\eqref{eq:FullAmp4Phi}
and derive the following KK graviton (Goldstone)
scattering amplitudes at the LO and NLO:
\beqs
\begin{align}
\label{eq:M0=MT0}
\M_0[4h_L^n] &\,=\,	\MT_0[4\phin] \,=\,
\frac{\,3\ka^2\,}{\,128\,}\sz\( 7+\ctt \)^2 \csc^2\!\theta
\hs ,
\\[1mm]
\delta \M[4h_L^n]  &\,=\,
-\frac{\,\ka^2 M_n^2\,}{128}(650+261\ctt\!+102\hs\ctf\!+11\cts )
\csc^4\!\theta   \hs ,
\label{eq:Amp-E-2hL}
\\[1mm]
\delta \MT[4\phin]  &\,=\,
-\frac{\,\ka^2M_n^2\,}{128}( -706+2049\hs\ctt - 318\hs\ctf - \cts)\csc^4\!\theta \hs ,
\label{eq:Amp-E-2phi}
\end{align}
\eeqs
where $\,\sz =s\hsm -\hsm 4\Mnn\hs$.
We note that the $\hs 1/\sz\hs$ expansion has shifted
a hidden $\mO(\Mnn)$ subleading term (contained in
$\hs s= \sz + 4\Mnn\hs)\hs$
from the LO amplitudes \eqref{eq:dM-dMT-GR5xx}
into the NLO amplitudes \eqref{eq:Amp-E-2hL}-\eqref{eq:Amp-E-2phi}.
But the difference between the above two NLO amplitudes remains
unchanged under this rearrangement in Eqs.\eqref{eq:M0=MT0}-\eqref{eq:Amp-E-2phi}.
Hence, we can derive the contribution of the residual terms
by computing the following amplitude-difference either from
Eqs.\eqref{eq:Amp-E-2hLxx}-\eqref{eq:Amp-E-2phixx} or from
Eqs.\eqref{eq:Amp-E-2hL}-\eqref{eq:Amp-E-2phi}:
\begin{align}
\label{eq:RTerm-NLO-GR}
\delta \M[4h_L^n] -  \delta  \MT[4\phin]
~=~
-\frac{\,3\ka^2M_n^2\,}{2}\!\(\!\frac{\,39\,}{2}+\ctt\hsmx\)\!.
\end{align}

\section{\hspace{-2mm}Extending KLT Construction to KK Amplitudes}
\label{app:G}

In this Appendix, we extend the KLT\,\cite{KLT} relation
to studying the double-copy construction of the KK amplitudes,
in comparison with the extended BCJ approach used
in sections\,\ref{sec:5.2}-\ref{sec:5.3}.
The KLT relation was derived to connect the product of the scattering amplitudes of two open strings to that of the closed string
at tree level. The KLT kernal may be further reinterpreted as
the inverse amplitude of a bi-adjoint scalar theory
in QFT (\`{a} la CHY)\,\cite{CHY}.

\vspace*{1mm}

We summarize the LO and NLO amplitudes for $A^n_L$ and $A^n_5$
as well as their difference:
\beqs
\label{eq:TL-dTL-T5-dT5-Tv}
\begin{align}
\T_{0L}^{} &=\,g^2 \sum_{j}\!\frac{\,\CC_j \NN_j^0\,}{s_{0j}} \,,
\qquad
\da\T_L^{} =\, g^2 \sum_{j}\! \frac{\,\CC_j \da \NN_j\,}{s_{0j}} \,,
\\[1mm]
\tT_{05}^{} &=\, g^2 \sum_{j} \!\frac{\,\CC_j \NNt_j^0\,}{s_{0j}} \,,
\qquad
\da\tT_5^{} =\, g^2 \sum_{j}\!\frac{\,\CC_j \da\NNt_j\,}{s_{0j}} \,,
\\[1mm]
\Delta\T &\equiv \da\T_L^{} \!- \da\tT_5^{}
	\,=\, g^2 \sum_{j}\!\frac{\,\CC_j (\da \NN_j \!-\! \da \NNt_j)\,}{s_{0j}} \,,
\end{align}
\eeqs
where $j\in (s,t,u)$\, and their numerators satisfy
the kinematic Jacobi identities:
\begin{equation}
\sum_j \NN_j^{0} = \sum_j \NNt_j^{0} = 0\,,~~~~~~~
\sum_j (\da\NN_j^{} \!-\!\da\NNt_j^{}) = 0 \,.
\end{equation}

Then, we expand the color factors ($\CC_j$) in terms of traces of
group generators:
\beqs
\label{eq:Color-decompose}
\begin{align}
\CC_s &\,=\, \fr{1}{2} \! \( -\Tr[1234]  + \Tr[1243] + \Tr[2134] - \Tr[2143] \)  \,,
\\[1mm]
\CC_t &\,=\,  \fr{1}{2} \! \(  -\Tr[1423]  + \Tr[1432] + \Tr[4123] - \Tr[4132] \) \,,
\\[1mm]
\CC_u &\,=\,  \fr{1}{2} \! \(  -\Tr[1342]  + \Tr[1324] + \Tr[3142] - \Tr[3124] \) \,,
\end{align}
\eeqs
where the abbreviation
$\,\{1,2,3,4\} \!=\! \{T^a, T^b ,T^c ,T^d \}$
is used.
Thus, each full four-particle scattering amplitude $\,\T_4^{}\,$
can be decomposed into the sum of color-ordered partial amplitudes in terms of the trace of group factors:
\begin{equation}
\T_4^{} ~=\, g^2\!\! \sum_{\mathcal{P}(234)}\! \A_4^{}[1234]\,
\Tr\!\[\!T^a T^bT^cT^d\!\]\!.
\end{equation}
We may further write the $n$-point color-ordered partial amplitudes
in the following general form\,\cite{Elvang:2013}\cite{Vaman:2010ez},
\begin{equation}
\A_i^{} ~=~ g^2 \sum_{j=1}^{(n-2)!}
\Theta_{ij}^{}\, \hat{n}_j^{} \,,
\end{equation}
where the quantities $\,\{\Theta_{ij}^{}\}\,$ form a
$\,(n\!-\!2)! \!\times\! (n\!-\!2)!$\, matrix containing massless
scalar propagators, and the numerators
$\,\hat{n}_j^{}\,$ include the kinematic information.

\vspace*{1mm}

For the case of four-particle scattering ($n=4$),\,
we choose the partial amplitudes with ordering
$\A[1234]$ and $\A[1243]$ in the Kleiss-Kuijf basis\,\cite{Kleiss-Kuijf}\cite{DelDuca:1999rs}.\footnote{%
\baselineskip 15pt
Alternatively, one may choose $\A[1324]$ instead of $\A[1243]$
in the basis, because the U(1) decoupling identity gives
$\A[1234] + \A[1243]  + \A[1324]=0$\,.}\,
Thus, the amplitudes in  Eq.\eqref{eq:TL-dTL-T5-dT5-Tv}
can be reexpressed as follows:
\begin{equation}
\(\! \begin{aligned}
\A [1234]
\\[-1mm]
\A [1243]
\end{aligned}\)
=~ g^{2}\,\Theta \!\times\!\(\! \begin{aligned}
\hat{n}_s
\\[-1mm]
\hat{n}_t
\end{aligned} \) \!,
\end{equation}
where $\,\A = \T_{0L}^{}\,, \, \tT_{05}^{} \,, \, \Delta\T$ and
$\,\hat{n}_j^{}= \NN^0_j,\, \NNt^0_j, \,
\da\NN_j^{}\!-\!\da\NNt_j^{}\,$.
The above propagator matrix $\,\Theta\,$ takes the form:
\begin{equation}
\Theta ~= \(\! \begin{aligned}
- \frac{1}{\,\sz} \hspace*{.45cm}  &\hspace*{.5cm}
\frac{1}{\,\tz}
\\[1mm]
\frac{1}{\,\sz} \!+\! \frac{1}{\,\uz}  &\hspace*{.5cm}
\frac{1}{\,\uz}	
\end{aligned} \,\) \!,
\end{equation}
where we can readily check $\,\det \Theta = 0\,$.
Then, we can derive the color-ordered LO amplitudes:
\beqs
\begin{align}
\T_{0L}^{}[1234] &\,=\,
g^2\!\(\!-\frac{\NN_s^0}{\sz}+\frac{\NN_t^0}{\tz}\)\!,
\hspace*{-20mm}
& \T_{0L}^{}[1243] &\,=\,
g^2\!\(\!\frac{\NN_s^0}{\sz}-\frac{\NN_u^0}{\uz}\)\!,
\\[1.5mm]
\tT_{05}^{}[1234] &\,=\,
g^2\!\(\!-\frac{\NNt_s^0}{\sz}+\frac{\NNt_t^0}{\tz}\)\!,
\hspace*{-20mm}
& \tT_{05}^{}[1243] &\,=\,
g^2\!\(\!\frac{\NNt_s^0}{\sz}-\frac{\NNt_u^0}{\uz}\)\!,
\end{align}
\eeqs
and the color-ordered NLO amplitudes:
\beqs
\begin{align}
\Delta\T [1234] &\,=\,
g^2\!\(\!-\frac{\,\dNN_s^{}\!-\!\dNNt_s^{}\,}{\sz}
+\frac{\,\dNN_t^{}\!-\!\dNNt_t^{}\,}{\tz}\)\!,
\\[1.5mm]
\Delta\T [1243] &\,=\,
g^2\!\(\!\frac{\,\dNN_s^{}\!-\!\dNNt_s^{}\,}{\sz}
-\frac{\,\dNN_u^{}\!-\!\dNNt_u^{}\,}{\uz}\)\!.
\end{align}
\eeqs

\vspace*{1mm}

With the above, we extend the KLT double-copy construction and
compute the KK graviton scattering amplitudes at the LO:
\beqs
\begin{align}
\M_0^{} [1234] &\,=\, \frac{\ka^2}{\,24\,} \,\sz\,
\T_{0L}^{} [1234] \,\T_{0L}^{} [1243]
\,=\,  \frac{\,3\ka^2}{\,128\,} \,s_0^{}\,
(7\!+ \ctt )^2 \csc^2 \!\theta \,,
\label{eq:App-M}
\\[1mm]
\MT_0^{} [1234] &\,=\,
\frac{\ka^2}{\,24\,}\,\sz\,\tT_{05}^{}[1234]\,\tT_{05}^{}[1243]
\,=\,  \frac{\,3\ka^2}{\,128\,} \,s_0^{}\,
(7\!+ \ctt )^2 \csc^2 \!\theta  \,.
\label{eq:App-Mt}
\end{align}
\eeqs
Then, with the definitions of
$(\Delta\M_1^{},\,\Delta\M_2^{})$
in Eqs.\eqref{eq:DM1}-\eqref{eq:DM2},
we construct the KK graviton amplitudes at the NLO:
\beqs
\begin{align}	
\Delta\M_1^{} [1234] &\,=\,
\frac{\ka^2}{\,12\,} \sz\,\tT_{05}^{}[1234]\,
\Delta\T [1243] = \frac{\,\ka^2}{\,12\,}\,\sz\,
\Delta\T [1234]\,\tT_{05}^{} [1243]
\nn\\[1.5mm]
&\,=\, -\ka^2 M_n^2\,(7\!+  \ctt ) \,, \label{eq:App-DM1}
\\[1.5mm]
\Delta\M_2^{} [1234] &= \frac{\,\ka^2}{\,12\,}\,\sz\,
\T_{0L}^{} [1234]\, \Delta\T [1243]
\,=\, \frac{\,\ka^2}{\,12\,} \,\sz\,
\Delta\T [1234]\, \T_{0L}^{} [1243]
\nn\\[1.5mm]
&\,=\, - \ka^2 M_n^2\, (7\!+\ctt )  \,.
\label{eq:App-DM2}
\end{align}
\eeqs
From the above, we see that the amplitudes \eqref{eq:App-M}-\eqref{eq:App-Mt} and \eqref{eq:App-DM1}-\eqref{eq:App-DM2} agree with the amplitudes
\eqref{AmpDC-phinnnn-2} and \eqref{eq:DM1}-\eqref{eq:DM2}
which we derived by using the improved BCJ construction
in the case of four-point amplitudes.
We will consider generalizing the present analysis to
the KK graviton (Goldstone) amplitudes with five or more external
lines in the future work.


\end{document}